\documentclass[aps,prd,preprintnumbers,twocolumn]{revtex4-1}
\usepackage{graphicx}
\usepackage{amsmath,amssymb}
\usepackage{float}
\usepackage{xcolor}
\usepackage{bm}
\usepackage{cleveref}

\newcommand\beq{\begin{equation}}
\newcommand\eeq[1]{\label{#1}\end{equation}}
\newcommand{\bea}{\begin{eqnarray}}
\newcommand{\eea}{\end{eqnarray}}

\def\slashchar#1{\setbox0=\hbox{$#1$}           % set a box for #1 
   \dimen0=\wd0                                 % and get its size
   \setbox1=\hbox{/} \dimen1=\wd1               % get size of /
  \ifdim\dimen0>\dimen1                        % #1 is bigger
 \rlap{\hbox to \dimen0{\hfil/\hfil}}      % so center / in box
  #1                                        % and print #1
 \else                                        % / is bigger
    \rlap{\hbox to \dimen1{\hfil$#1$\hfil}}   % so center #1
    /                                         % and print /
 \fi}                                         %

\definecolor{hu-berlin-blue}{RGB}{0,65,137} % HEX 004189

\definecolor{hu-berlin-blue}{RGB}{0,65,137} % HEX 004189

\begin{document}

\title{Static quark anti-quark interactions at non-zero temperature from lattice QCD}

\author{Dibyendu Bala$^a$, Olaf Kaczmarek$^a$, Rasmus Larsen$^b$, Swagato Mukherjee$^c$, Gaurang Parkar$^b$, Peter Petreczky$^c$, Alexander Rothkopf$^b$,
Johannes Heinrich Weber$^d$
}

\affiliation{
$^a$ Fakult\"at f\"ur Physik, Universit\"at Bielefeld, D-33615 Bielefeld, Germany\\
$^b$ Faculty of Science and Technology, University of Stavanger, NO-4036 Stavanger, Norway,\\
$^c$ Physics Department, Brookhaven National Laboratory, Upton, New York 11973, USA\\
$^d$Institut f\"ur Physik \& IRIS Adlershof, Humboldt-Universit\"at zu Berlin, D-12489 Berlin, Germany
}

\collaboration{HotQCD Collaboration}

\preprint{HU-EP-21/32-RTG}

\date{\today}

\begin{abstract}
We study the interactions of a static quark antiquark pair at non-zero temperature using realistic 2+1 flavor lattice QCD 
calculations. The study consists of two parts: the first investigates the properties of Wilson line correlators in Coulomb gauge and compares to predictions of hard-thermal loop perturbation theory. As a second step we extract the spectral functions underlying the correlators using four conceptually different methods: spectral function fits, a HTL inspired fit for the correlation function, Pad\'e rational approximation and the Bayesian BR spectral reconstruction. We find that our high statistics Euclidean lattice data are amenable to different hypotheses for the shapes of the spectral function and we compare the implications of each analysis method for the existence and properties of a well defined ground state spectral peak. 
\end{abstract}
\maketitle

\section{Introduction}
\label{sec:intro}

Heavy quarkonia, the bound states of a heavy quark and anti-quark are a unique laboratory of the strong interactions. At the same time they constitute a central tool in the investigation of the primordial state of matter created in relativistic heavy-ion collisions, the quark-gluon plasma. In turn elucidating the properties of these strongly interacting bound states in extreme condition remains a central focus of experimental and theoretical research (see  Refs.~\cite{Rothkopf:2019ipj,Aarts:2016hap,Mocsy:2013syh,Bazavov:2009us} for reviews).

Interest in heavy quarkonium in relativistic heavy-ion collisions erupted with the seminal paper by Matsui and Satz \cite{Matsui:1986dk}. Their paper put forward two key ideas: on the one hand it argues that the formation of the deconfined medium in heavy ion collisions will interfere with the binding of the heavy quarks through color screening and thus prevent formation of a bound state. The second idea states that such an absence of bound states in the medium will lead to a suppression of quarkonium yields.

In case that in a heavy-ion collision only a few heavy quark pairs are produced, the suppression Matsui and Satz envisioned has been clearly established by experiment for both charmonium at RHIC and bottomonium at LHC. At increasing energies, where a wealth of heavy quarks may be produced in the initial state, it has been observed that quarkonium yields, in particular charmonium at LHC, can be replenished. This phenomenon is attributed to recombination.

The question of whether or how color screening (and in general the interactions with a (non-)thermal medium) affect the survival of heavy quarkonium states, remains an open research question. Using lattice QCD simulations it has so far been established that the free energy of static quark anti-quark pair is indeed screened at large separations  (see e.g. ref.~\cite{Bazavov:2020teh} for a recent review). The most recent lattice analysis of the static $Q \bar Q$ free energy shows that the interactions are screened beyond distances larger than $0.4/T$ \cite{Bazavov:2018wmo}.

Matsui and Satz took the idea of static color screening and applied it to quarkonium with finite mass constituents. In fact, their idea of melting from color screening relies on a non-relativistic potential picture of quarkonium binding. At zero temperature such a potential picture has been highly successful in describing the phenomenology of the ground and excited states below the open heavy flavor threshold (see e.g. Ref. \cite{Brambilla:2004wf}). The lattice calculations
of quarkonium Bethe-Salpeter amplitudes at zero temperature 
are also consistent with the potential model \cite{Kawanai:2011xb,Kawanai:2011jt,Kawanai:2013aca,Nochi:2016wqg,Larsen:2020rjk}.

The past two decades have seen significant progress in our understanding of heavy quarkonium systems based on the concept of effective field theory. Such a systematic approximation of QCD allows us to clarify the concept of a potential in the context of heavy (but not static) quarks. At zero temperature there exist three distinct energy scales: $M \gg M v \gg M v^2$ with $M$ being
the heavy quark mass and $v$ the relative velocity of the heavy quarks inside the bound state. By focusing on physical processes involving energies smaller than $M$, we may cast the description of the quark anti-quark pair in terms of non-relativistic Pauli spinors (pair creation at the scale $M$ is not explicitly treated but remains present as four-fermi interaction). This process of \textit{integrating out} the so called hard scale $M$ leads to the theory of non-relativistic QCD (NRQCD) \cite{Caswell:1985ui}, which is valid at scales up to $M v$. We may further restrict our focus on e.g.~the binding properties of the heavy quark antiquark pair at the ultrasoft scale $Mv^2$, which leads (as long as the same degrees of freedom as in NRQCD can be identified) to a theory of color singlet $S$ and octet $O$ wavefunctions, called potential NRQCD or pNRQCD for short \cite{Brambilla:1999xf}.
The Lagrangian of pNRQCD has the form
\begin{widetext}
\begin{align*}
\nonumber {\cal L}_{\rm pNRQCD}=\int d^3\mathbf{r} {\rm Tr}\Big[& S^\dagger \big[ i\partial_0 - \big(   \frac{\mathbf{D}^2}{2 M} + V_S^{(0)} + \frac{V_S^{(1)}}{m_Q} + \ldots \big) \big]S + O^\dagger \big[ iD_0 + \frac{\mathbf{D}^2}{2 M} + V_O^{(0)} + \frac{V_O^{(1)}}{m_Q} +\ldots\big) \big]O\Big]\\
+&V_A(r){\rm Tr}\Big[  O^\dagger \mathbf{r} g \mathbf{E} S + S^\dagger \mathbf{r} g \mathbf{E} O \Big]+V_B(r){\rm Tr}\Big[  O^\dagger \mathbf{r} g \mathbf{E} O + O^\dagger O\mathbf{r} g \mathbf{E}  \Big] +{\cal O}\big(r^2,\frac{1}{m_Q^2}\big)+{\cal L}_{light~quarks,gluons}\label{eq:pNRQCDcont}
\end{align*}
\end{widetext}
The singlet and octet fields depend on the label $\mathbf{r}$ that corresponds to the distance between the heavy quark and anti-quark, and the ultrasolft gluon fields depend on the center of mass coordinate, $\mathbf{R}$
of the heavy $Q\bar Q$ pair.
This theory of pNRQCD puts the ideas of previous non-relativistic potential models on a more solid footing. Its Lagrangian tells us that the propagation of color singlet and octet d.o.f. depends on two mechanisms. 
The first is encoded in the  Schr\"odinger-like part of the Lagrangian (top line), in which the physics of the integrated-out gluons appears as Wilson coefficients in the form of time-independent potential terms $V$. Note that the static potentials $V^{(0)}_{S/O}$ are but the first terms in a systematic expansion in powers of the inverse rest mass. In contrast to the naive potential model we see that the presence of ultrasoft gluons (on the scale $Mv^2$) on the other hand introduces transitions between the singlet and octet wave functions, due to the Wilson coefficients $V_A$ and $V_B$. These transitions in general cannot be summarized in terms of a simple potential and are referred to as non-potential effects (see also the discussion in ref.~\cite{Rothkopf:2019ipj}). The potential picture thus describes only the lowest order (tree level) of pNRQCD. Even if we are interested in the static potential $V^{(0)}_S$, we therefore have to take care to distinguish between the concept of a potential (Wilson coefficient of pNRQCD) and the static energy of the quark-antiquark pair (which refers to the energy of the lowest lying excitation in its spectrum).

In some instances, depending on the specific separation of scales and a level of coarse graining in time, the terms referred to above as non-potential terms may be absorbed into additional time-independent potential terms, allowing for a simple potential description based on a Schr\"odinger equation. E.g. it has been shown that in vacuum if 
$M v^2 \ll \Lambda_{QCD}$, the static energy and the potential agree, i.e. the 
static energy accessible from lattice QCD correlation functions can be used as a potential \cite{Brambilla:1999xf}.

The situation at finite temperature is much more involved, as additional energy scales come into play. These are related to the thermal medium 
%and which at weak coupling exhibit the hierarchy $T \gg m_D \sim gT  \gg g^2 T$.
and exhibit the hierarchy $T \gg m_D \sim gT  \gg g^2 T$ at weak coupling. The thermal physics may both influence the potential and non-potential contributions to pNRQCD. 
%In the context of the potential, one may expect the medium to modify the potential, 
In the context of pNRQCD, one may expect the medium to modify the potential, but again, this only holds true for particular scale hierarchies. E.g. when considering deeply bound quarkonium states with very small spatial extent, the real-part of the potential relevant for their physics remains effectively Coulombic. In some scale hierarchies the physics of the singlet and octet transitions may be summarized in additional contribution to the potential, leading e.g. to the emergence of an imaginary part \cite{Laine:2006ns,Brambilla:2008cx,Beraudo:2007ky} related to dissipative effects in the medium.

The static in-medium potential has been studied non-perturbatively on the lattice via spectral function reconstruction and model spectral function fits. Based on the Bayesian BR method \cite{Burnier:2013nla} for spectral reconstruction the static potential has so far been investigated in quenched QCD \cite{Rothkopf:2011db,Burnier:2016mxc} and in full QCD simulations based on the legacy asqtad action \cite{Burnier:2014ssa,Burnier:2015tda}. Recently HTL motivated decomposition of APE smeared Wilson loop in symmetric and anti-symmetric parts has also been used to extract the thermal potential in the quenched approximation \cite{Bala:2019cqu}. These studies concluded that the real-part of the potential eventually becomes screened in the deconfined phase and have identified hints for the existence of an imaginary part once one simulates above the crossover temperature. Concurrently the potential has been extracted by fitting modified HTL spectral functions to Euclidean correlators in \cite{Bazavov:2014kva} and deploying a skewed or non-skewed Lorentzian fit in \cite{Petreczky:2017aiz}. In both cases values for the real-part were obtained that are significantly larger than those extracted via the direct spectral function reconstruction lying closer to the $T=0$ results.

Concurrent to the development of the EFT approach, the past five years have seen rapid progress in understanding the dynamical evolution of heavy quarkonium in the context of open quantum systems (see \cite{Brambilla:2016wgg,Brambilla:2017zei,Brambilla:2019tpt,Brambilla:2020qwo,Rothkopf:2019ipj,Akamatsu:2020ypb} for recent reviews). In particular the role of the imaginary part of the potential has been elucidated and its relation to wavefunction decoherence \cite{Kajimoto:2017rel,Miura:2019ssi} highlighted. It has been shown how a separation of scales in terms of energy scales is connected to a separation of time scales. Using different scale separation scenarios (and different time coarse graining prescriptions), various so called master equations for the real-time evolution of the reduced density matrix of heavy quarkonium in a medium have been derived, revealing e.g. the subtle interplay between screening and decoherence in a hot QCD medium.

One central goal, both in the EFT and open-quantum systems community, is to go beyond the weak coupling considerations, on which many of the arguments related to scale separations are anchored on. In order to make progress e.g.~in the phenomenologically relevant temperature regime just above the QCD crossover transition, it is therefore necessary to explore whether a potential picture can be established non-perturbatively and if so, what the functional form of such a potential is.

As a starting point we therefore set out in this study to investigate the interactions of static quark-antiquark pairs at $T>0$ using realistic state-of-the-art lattice QCD calculations.
To this end, in \cref{sec:gencon} we will present general considerations on the real-time dynamics of static color sources and their study from Euclidean lattice simulations. The first part of our study is presented in \cref{sec:latcorr}, 
where 
after discussing the lattice setup in \cref{sec:setup}, we investigate the lowest three cumulants of the correlation function in \cref{sec:corrmom},
%in \cref{sec:corrmom} we investigate the low lying cumulants of the correlation function 
and compare them in \cref{sec:corrHTLcmp} to predictions from hard thermal loop perturbation theory (HTL). 
In \cref{sec:spectra} 
%to
we present
the investigation of the underlying spectral structure of the correlators using four different methods: spectral model fits \cref{sec:potfit}, the HTL-motivated approach \cref{sec:BalaDatta}, Pade rational approximations \ref{sec:Pade} and the Bayesian BR method \cref{sec:Bayes}. We conclude with a discussion in \cref{sec:conclusion}.

\section{General considerations}
\label{sec:gencon}
In order to connect the EFT description of quarkonium to QCD we have to carry out a matching procedure. I.e. correlation functions with the same physics content in both languages need to be identified. Once we demand that their values agree at a certain matching scale it allows us to fix the Wilson coefficients of the effective theory. In the static limit $m_Q\to\infty$, it has been shown that the Wilson loop is the appropriate QCD quantity which we can identify with the unequal time correlation function of two color singlet fields in pNRQCD \cite{Brambilla:1999xf}. The matching condition 
%in 
at 
the leading order in multipole expansion reads \cite{Brambilla:1999xf}:
\begin{align}
    W_\square(r,t,T) &= \langle {\rm exp}[ig \int_\square dz^\mu A_\mu]\rangle_{\rm QCD}\\
   \nonumber  &{\equiv} \langle S(r,0)S^\dagger(r,t) \rangle_{\rm pNRQCD}.
\end{align}
The Wilson loop in QCD itself emerges self consistently from the static limit of the retarded $Q\bar{Q}$ meson correlator. 
By matching with different quantities related to the singlet and octet sector, the ultimate goal here lies in identifying individually the potential ($V_S$,$V_O$) and non-potential contributions ($V_A$,$V_B$,$\ldots$) that govern the Wilson loop evolution in Minokwski-time.

Let us focus on the singlet sector. Instead of studying the evolution of $W_\square(r,t)$ in the real-time domain, it is advantageous to go over to its Fourier transform
\begin{align}
    \rho_r(\omega,T)=\int dt W_\square(r,t,T) e^{-i\omega t}.\label{eq:RTwilsonspecdec}
\end{align}
This Fourier transform, as shown in \cite{Rothkopf:2009pk}, also coincides with the positive definite spectral function of the Wilson loop. This fact is relevant, as in Euclidean lattice simulations we do not have direct access to the real-time Wilson loop but can exploit its spectral function as bridge between the imaginary and real-time domain. The Euclidean Wilson loop, which we can simulate on the lattice has a spectral decomposition, housing the same spectral functions as in \cref{eq:RTwilsonspecdec}, which here is related to the lattice observable by a Laplace transform
\begin{align}
    W_\square(r,\tau,T)=\int d\omega e^{-\omega \tau} \rho_r(\omega,T).
    \label{eq:ETwilsonspecdec}
\end{align}
We may thus gain insight in the real-time evolution of the Wilson loop by studying the spectral function encoded in its Euclidean counterpart. The inversion of \cref{eq:ETwilsonspecdec} however constitutes an ill-posed inverse problem, which we will attack with four different and complementary 
numerical strategies in \cref{sec:spectra}.

At zero temperature in a finite volume the spectral function 
consists of a ground state (lowest lying) delta peak
separated by an energy gap from many excited state delta peaks (hybrid potential, static-light mesons etc.).
In the infinite volume limit some of these excited state contributions form a continuum.
The excited state contributions will be seen as deviation from a single exponential behavior
of the correlator at small $\tau$.
Performing a spectral decomposition of the non-zero temperature Euclidean time correlator in Eq. \ref{eq:ETwilsonspecdec} in a finite volume by inserting
a complete set of energy eigenstates  one can see
that in addition to the ground state delta peak additional peaks in its proximity
will appear. 
This is shown in Appendix \ref{app:spec_decomp}. The coefficients
of these additional delta functions are proportional to
Boltzmann factors and therefore, their relative weight will increase with 
increasing temperature. 
Thus, we will see a broadening of the zero temperature ground state peak. 
At finite temperature there will be some additional peaks in the $\omega$ region corresponding
to excited states, but these will not change the overall shape of
the spectral function significantly, because of the already large
density of states. Therefore, any possible modifications in that region  should not have a significant effect on the Euclidean correlator.
Thus the most interesting part of the finite temperature spectral
function is the position, $\Omega(r,T)$, and the effective width, $\Gamma(r,T)$, of 
this dominant broadened peak.
Furthermore, as also shown in Appendix \ref{app:spec_decomp},~
%that 
the finite temperature
spectral function could be non-zero even for $\omega \ll \Omega(r,T)$. We call this part
of the spectral function the low energy tail. Thus we expect that the spectral
function of static $Q \bar Q$ pair should consist of ground state peak, a high
energy part, which to a good approximation is temperature independent and the low
energy tail. 

The goal of this study is modest. 
Using for the first time finite temperature lattices with realistic pion masses, we set out to elucidate the lowest lying peak in the spectral function. 
We will refrain from making a quantitative connection of peak structures in the spectral functions to Wilson coefficients ($V_S$,$V_O$,$V_A$,$V_B$,$\ldots$) and solely attempt to 
%constraint 
constrain the values of $\Omega(r,T)$ and $\Gamma(r,T)$ as reliably as possible, given the currently available lattice data.

We also note that the overall form of the spectral function depends on the choice of our static meson operator, i.e. on the choice of the spatial part of the Wilson
loop. 
As mentioned above,
on the lattice at finite volume, the spectral function consists of a sum of delta peaks. Choosing between e.g. the Wilson loop with straight spatial lines, with deformed spatial lines, smearing the links from which to build the Wilson loop or taking instead Wilson line correlators in a particular gauge, such as Coulomb gauge, will change the amplitudes of the peaks in the spectral function but not their position. At $T=0$ where one encounters well separated peaks, and only their position is of interest, the tuning of operators is a common procedure to optimize the signal to noise ratio in the determination of these peak positions (see also the discussion in \cite{Jahn:2004qr,Bazavov:2008rw}). 
At finite temperature, where multiple peaks may congregate around a dominant central value, the changes introduced in the envelope of amplitudes by modifying the operator are less straightforward to predict. 
However, the position of the dominant peak and its width should be largely independent
of the choice of the static meson operator. Here we may gain some intuition e.g. from HTL perturbation theory. It was shown that in the leading order of HTL perturbation theory the central position of the lowest lying spectral peak remains unaffected by the choice of either considering the Wilson loop or the Wilson line correlator in Coulomb gauge \cite{Burnier:2013fca}. 
At the same time a clear difference was found in the structures surrounding the lowest lying peak.  In quenched QCD an example has been given in Ref.~\cite{Rothkopf:2019ipj} that while the overall values of the Wilson line correlator are gauge dependent, its slope at intermediate imaginary time (thus corresponding to the position of its dominant lowest lying spectral peak) is virtually unaffected by the gauge transformation. I.e. there are indications that the properties of the lowest lying spectral structure may be extracted in an operator-independent fashion from Euclidean correlators. On the other hand the high energy part of
the spectral function seems to be strongly dependent on the choice of operator. The same is true
for the low energy tail of the spectral function at non-zero temperature, see Appendix \ref{app:spec_decomp}.

\section{Study of the lattice correlation function}
\label{sec:latcorr}
\begin{figure}
    \centering
    \includegraphics[width=8cm]{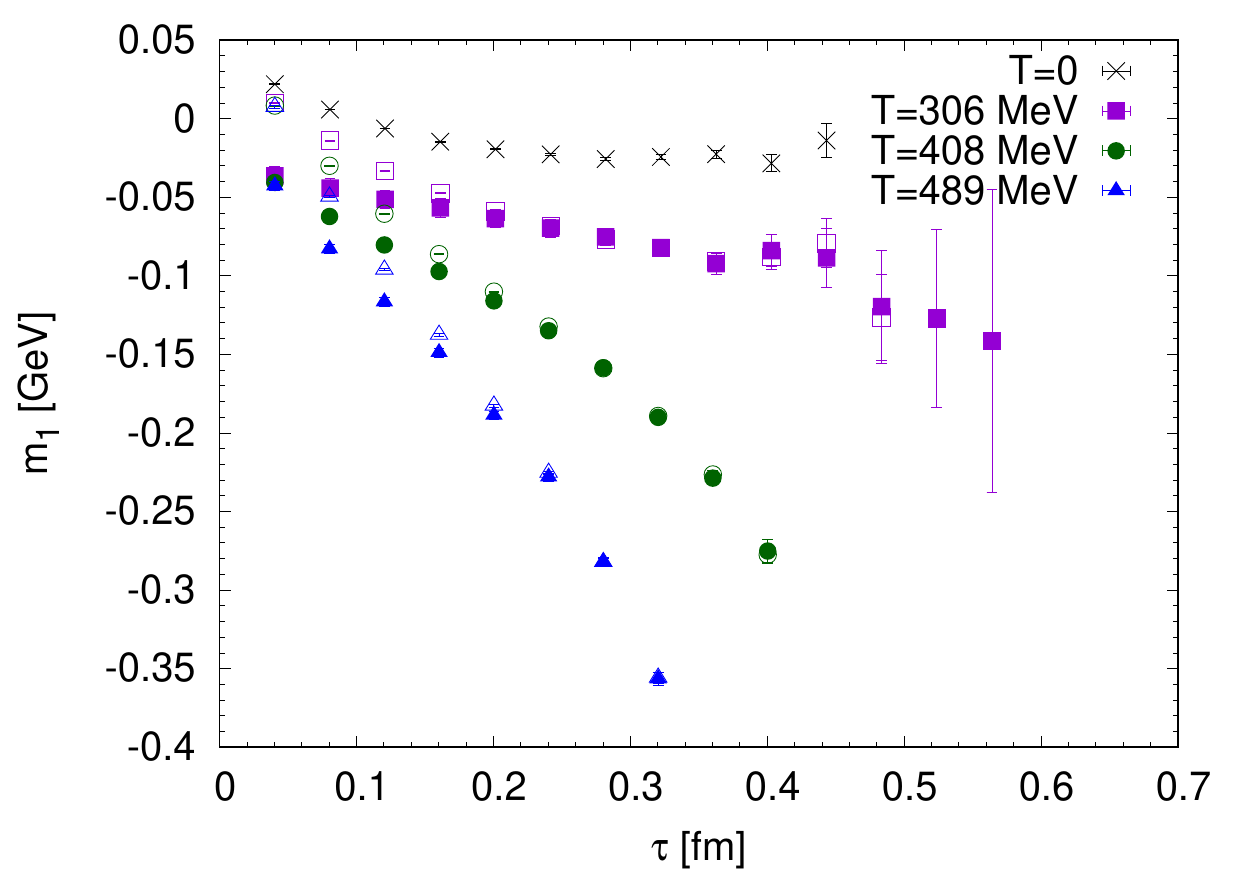}
    \caption{The first cumulant calculated at $r=0.24$fm for  $\beta=7.825$ and $N_{\tau}=64,~16,~12$ and $10$,
    corresponding to $T\simeq 0,~306,~408$ and $489$ MeV, respectively. The filled symbols
    correspond to the subtracted correlator, while the open symbols to unsubtracted correlator, see text.}
    \label{fig:demo_m1}
\end{figure}

\begin{figure*}
\includegraphics[width=8.4cm]{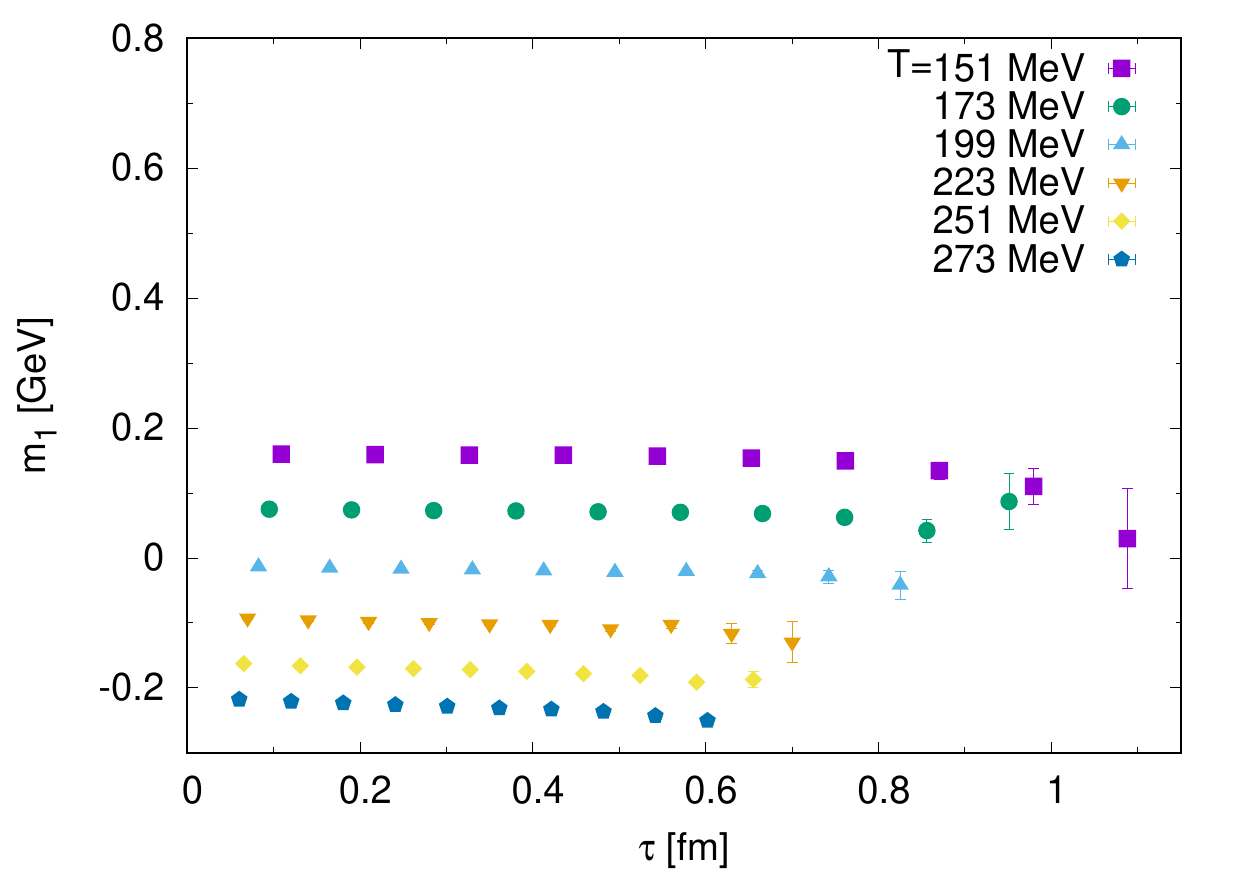}
\includegraphics[width=8.4cm]{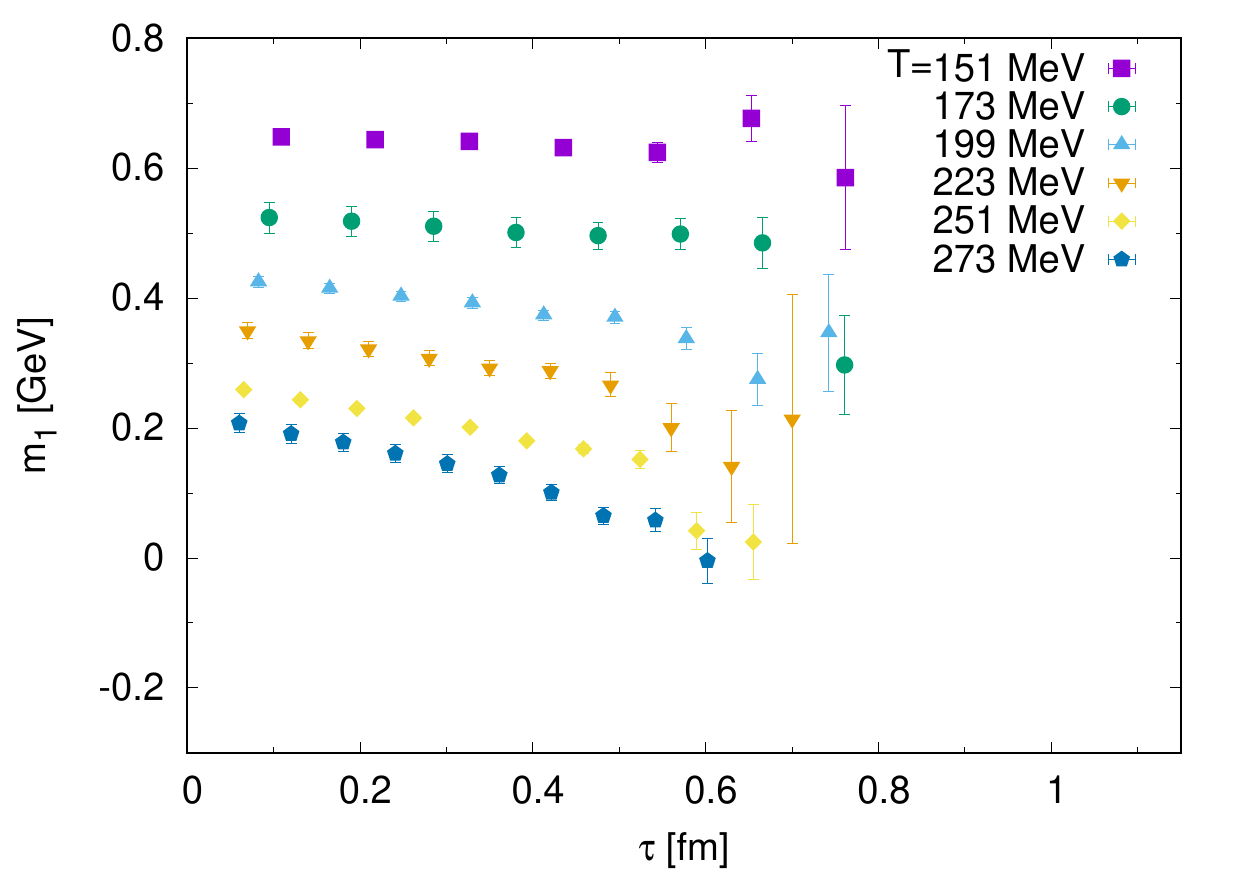}
\caption{The first cumulant as function of $\tau$ obtained on $N_{\tau}=12$ lattices
for $rT=1/4$ (left) and $rT=1/2$ (right) at different temperatures.}
\label{fig:demo_m1_ext}
\end{figure*}
\subsection{Lattice setup}
\label{sec:setup}

We performed calculations of Wilson loops and correlators of Wilson lines 
in Coulomb gauge at non-zero temperature in (2+1)-flavor
QCD with physical strange quark mass using gauge configurations generated by HotQCD
and TUMQCD collaborations with L\"uscher-Weisz gauge action and highly improved staggered quark action 
\cite{Bazavov:2011nk,Bazavov:2013uja,Bazavov:2014pvz,Ding:2015fca,Bazavov:2017dsy,Bazavov:2018wmo,Bazavov:2019qoo}. The Wilson
line correlator is defined by,
\begin{equation}
W(r, \tau, T)=\frac{1}{3} \langle Tr(L(0,\tau) L^\dagger(r, \tau))\rangle_{T}
\end{equation}
where $L(r,\tau)=\exp(i\int_0^{\tau} A_{4}(r,\tau^{\prime}) d\tau^{\prime})$.
We used 
$N_{\sigma}^3 \times N_{\tau}$ lattices with
$N_{\tau}=10,~12$ and $16$, to control lattice spacing effects, and $N_{\sigma}/N_{\tau}=4$ ~\footnote{While a few ensembles actually have $N_z=2N_\sigma$, this is irrelevant for the considerations in the following.}. 
Previous experience shows that the aspect ratio $N_{\sigma}/N_{\tau}=4$ is large enough
to control finite volume effects. The light ($u$ and $d$) quark mass was set to $m_s/20$, which
in the continuum limit corresponds to pion mass of $161$ MeV. At high temperatures, $T>300$ MeV
we also performed calculations with light quark mass equal to $m_s/5$, as quark mass
effects are expected to be small in this region. The calculations performed here were part of 
a larger campaign by TUMQCD collaboration to study the interaction of static quarks
at non-zero temperature and to extract the strong coupling 
constant \cite{Bazavov:2018wmo,Bazavov:2019qoo}. 
As in the previous studies the lattice spacing has been fixed using the $r_1$ scale defined in
terms of the static $Q\bar Q$ energy at zero temperature $V(r)$
\footnote{It is usually referred to as the potential in the lattice literature.}
\begin{equation}
\left . r^2\frac{d V}{d r}\right|_{r=r_1}=1.
\end{equation}
The values of $r_1/a$ 
as well as the zero temperature Wilson loops and Wilson line correlators 
for (2+1)-flavor HISQ
configurations have been determined in
Refs. \cite{Bazavov:2011nk,Bazavov:2014pvz,Bazavov:2017dsy}. 
We use the parametrization given in Ref. \cite{Bazavov:2017dsy} to obtain $a/r_1$ and 
the value $r_1=0.3106$ fm \cite{Bazavov:2010hj}.
Our calculations cover a large temperature range from temperature as low as $120$ MeV
to about $2$ GeV. This allows us to perform comparisons to the weak coupling calculations.
The parameters of the calculations, including the temperature values, the bare gauge
coupling $\beta=10/g^2$ and the corresponding statistics are summarized in the Appendix \ref{app:lat}; an account of the zero temperature ensembles is given there as well.
For Wilson loops we used 3D-HYP smeared links in the spatial direction to improve the signal.
We used zero, one, two, or
%and 
five steps of HYP smearings. In what follows we will use the notation
$W(r,\tau,T)$ for both Wilson line correlators and Wilson loops.

The Wilson line correlators require multiplicative renormalization. 
This renormalization corresponds to additive
renormalization of the static $Q\bar Q$ energy
at zero temperature. 
As in our previous studies with HISQ action
we choose
the renormalization scheme which corresponds
to the choice $V(r=r_0)=0.954/r_0$, with $r_0$
being the Sommer scale \cite{Sommer:1993ce}.
The renormalization constants corresponding
to this choice have been first calculated
in Refs. \cite{Bazavov:2011nk,Bazavov:2014pvz}
for $\beta \le 7.825$ and later extended to
larger $\beta$ values and also refined using  result on the free energy of
a static quark \cite{Bazavov:2016uvm,Bazavov:2018wmo}. Here
we use the value of the renormalization constants given in Tab. X of Ref. \cite{Bazavov:2018wmo} for $\beta\ge 7.15$ and in Tab. V of Ref. \cite{Bazavov:2016uvm} for smaller $\beta$ values. 
\subsection{Cumulant analysis of the correlation functions}
\label{sec:corrmom}
To understand the main features of our lattice results 
and to what extent these can constrain the spectral
function of a static meson it is useful to consider
the $n$-th cumulants of the correlation functions defined as
\begin{eqnarray}
&
m_1(r,\tau,T)=-\partial_{\tau} \ln W(r,\tau,T),\\
&
m_n=\partial_{\tau} m_{n-1}(r,\tau,T), n>1.
\label{eq:m_n}
\end{eqnarray}
The first cumulant $m_1$ is nothing but the effective mass, which at non-zero lattice
spacing is defined  as
\begin{equation}
m_1(r,\tau,T)=\frac{1}{a} \ln\frac{W(r,\tau,T)}{W(r,\tau+a,T)}.
\end{equation}
\begin{figure*}
    \centering
    \includegraphics[width=8.5cm]{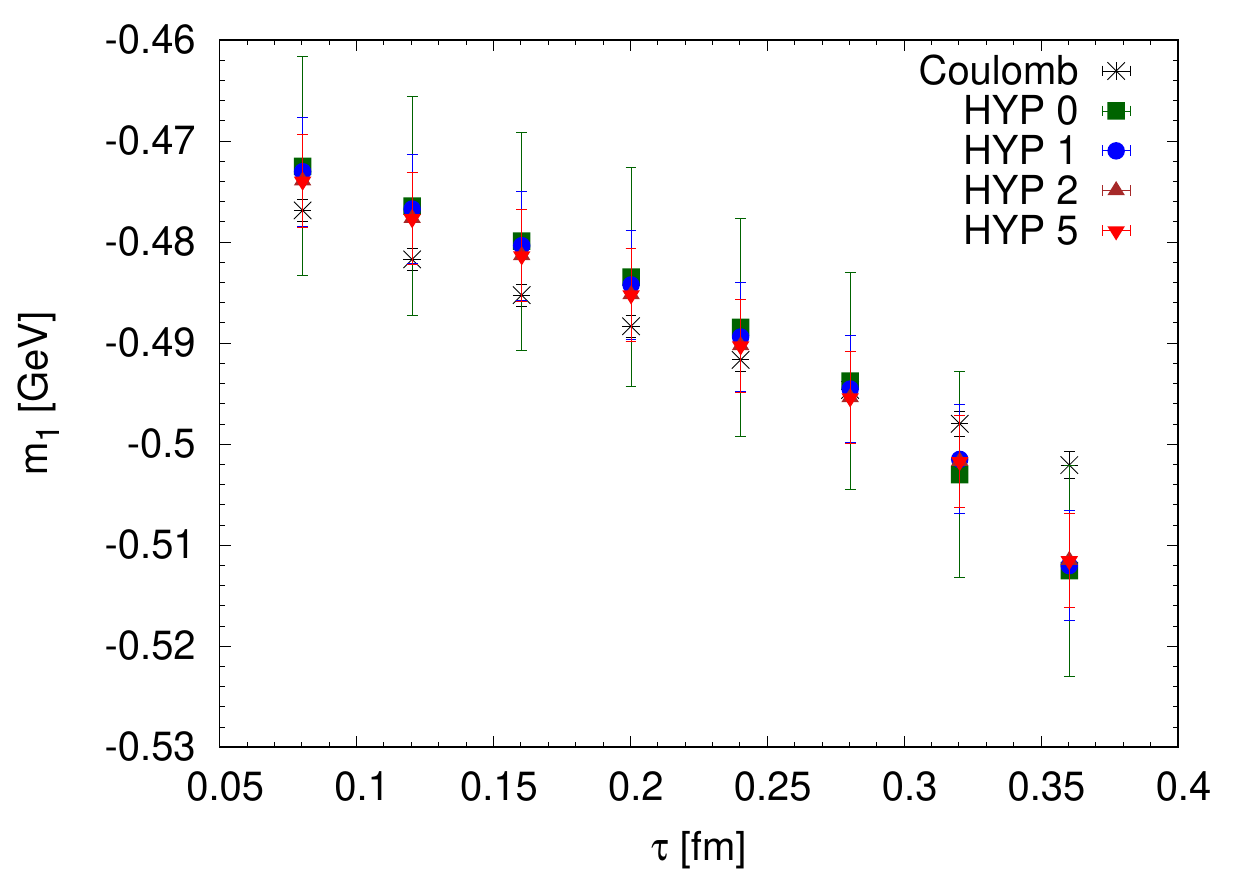}
    \includegraphics[width=8.5cm]{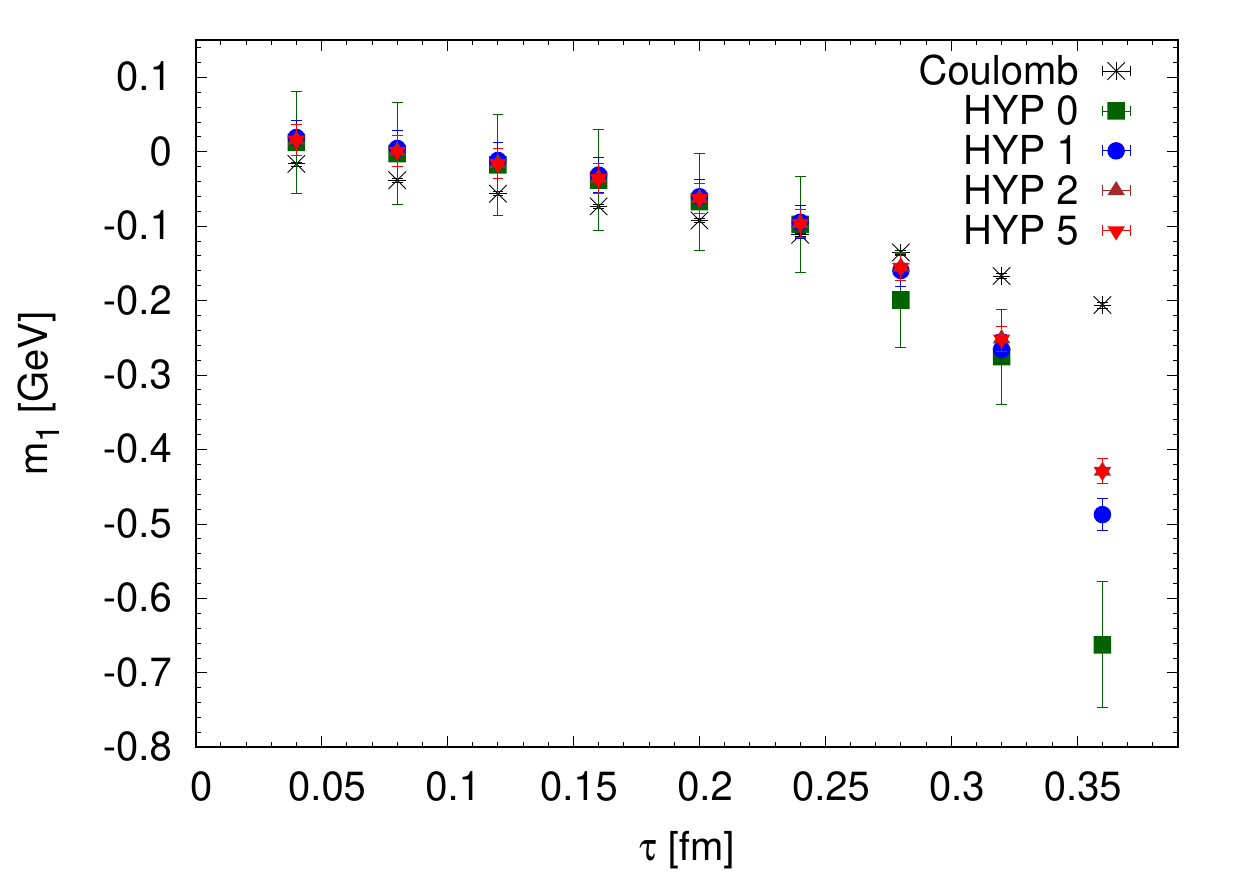}
    \caption{Comparison of the subtracted first cumulant
    from the Wilson line correlators in Coulomb gauge and smeared Wilson loops with
    different levels of HYP smearing at $T=408$ MeV with $N_{\tau}=12$ and $\beta=7.825$.
    The left panel shows the results for $rT=1/4$, while the right panel shows the result for $rT=1/2$.}
    \label{fig:comp2hyp}
\end{figure*}

The first cumulant needs an additive renormalization which is the same as the additive
renormalization of static $Q\bar Q$ energy or the free energy. In what
follows we will present the renormalized first cumulant using the known
renormalization constants as discussed above.

Since some of the calculations in the high temperature region are performed with light
quark masses significantly larger than the physical value we have to make sure that
this does not affect our results. In Appendix \ref{app:lat} 
we compared the calculations performed
at $m_l=m_s/20$ and $m_l=m_s/5$ and see no light quark mass dependence within statistical
errors. Since we have
different $N_{\tau}$ values we can check the size of the cutoff effects. This is also
discussed in the Appendix \ref{app:lat}. The size of the cutoff dependence turn out 
smaller than our statistical errors. Because of this we mostly focus our discussion
on $N_{\tau}=12$ data. For this data set we have relatively small statistical errors
and sufficient number of data points in the Euclidean time direction. When appropriate
we also show the $N_{\tau}=10$ and $16$ data.

In Fig. \ref{fig:demo_m1} we show the first cumulant from Wilson 
line correlators at $r=0.24$ fm for $\beta=7.825$ and $N_{\tau}=16,12$ and $10$ corresponding to temperatures
for $T=306$, $408$ and $T=489$ MeV, respectively and compared to the zero temperature first cumulant.
At $T=0$ the first cumulant approaches a plateau for $\tau>0.2$ fm. On
the other hand the non-zero temperature cumulant decreases monotonically.
At small $\tau$ the difference between the zero temperature and the finite temperature
first cumulant is very small and increases monotonically as $\tau$ increases.
The slope of the first cumulant increases  with increasing the temperature.
This means that the in-medium modifications of the spectral function are larger at larger
temperature, as expected. For the lowest temperature, $T=306$ MeV the decrease in
the first cumulants is approximately linear in $\tau$~around $\tau \sim 1/(2T)$, while for the higher 
temperatures this linear trend is only seen for 
%small $\tau$
smaller $\tau$, corresponding to the reduction in $1/(2T)$.

The small $\tau$ behavior of the Wilson line correlators is dominated by the high omega
part of the spectral function.
The high $\omega$ part of the spectral function is largely temperature independent, as discussed in the previous section,
and therefore, it is not very interesting from the point
of view of studying the in-medium effects on the static $Q \bar Q$ pair. On the other hand
it complicates the analysis and so it would be nice to get rid of it.
Let us assume ~-- following the arguments in Appendix \ref{app:spec_decomp} --~that we can decompose the spectral function as $\rho_r(\omega,T)=\rho_r^{tail}(\omega,T)+\rho_r^{med}(\omega,T)+
\rho_r^{high}(\omega)$, with $\rho_r^{med}(\omega,T)$ containing only the spectral structures of interest, in particular the dominant peak, and $\rho_r^{high}(\omega)$ describing the well separated UV behavior
of the spectral function. At zero temperature $\rho_r^{med}(\omega,T)$ is a single delta function
describing the ground state of static $Q \bar Q$ pair for our choice of static meson operator.
Therefore, assuming that the higher lying peaks are well separated from the ground state, we may isolate it at $T=0$ and subtract it from $W(r,\tau,T=0)$. This gives us 
%the 
an estimate for the contribution
of the high $\omega$ part of the spectral function. Once evaluated at zero temperature it can
be used to subtract off an estimate of the high omega contribution at $T>0$ at the same value of $\beta$. We calculated
the first cumulant from the subtracted correlator and the results are also shown in 
Fig. \ref{fig:demo_m1}. At large $\tau$ the subtraction has no effect, however, at small $\tau$ 
the subtracted first cumulant at $T>0$ shows a weaker $\tau$-dependence.
At the same time it shows visible temperature dependence already for
small $\tau$'s. At these small $\tau$ values we see an approximately linear
$\tau$-dependence of $m_1$ at non-zero temperature~with a slope similar to the $\tau \sim 1/(2T)$ region. 

In Fig. \ref{fig:demo_m1_ext} we show the subtracted first cumulants at lower temperatures
for two distances, $rT=1/4$ and $rT=1/2$. We consider the distances scaled by the temperature since with increasing temperatures the medium modification of
the correlator will manifest at shorter and shorter distances. The form of $\tau$ dependence of $m_1$ will 
scale with $rT$. We see that for fixed $rT$ the decrease of the first cumulant with $\tau$ is stronger at higher temperatures. Furthermore, this decrease is larger for larger $rT$.
We again see an approximate linear dependence of the first cumulants in $\tau$, except for the few largest
$\tau$ values. This feature of the first cumulants, which is a necessary consequence of the existence of the low energy tail (see Appendix ~\ref{app:spec_decomp}), will play an important role when modeling the
spectral function of the static meson. We also point out that the behavior of the 
first cumulant shown in Figs. \ref{fig:demo_m1} and \ref{fig:demo_m1_ext} is similar to the behavior
of the bottomonium first cumulants in NRQCD at non-zero temperature when extended meson operators are
used \cite{Larsen:2019bwy,Larsen:2019zqv}.

It is interesting to compare the results of the Wilson line correlators in Coulomb gauge 
with the ones obtained from Wilson loops. Both types of static meson correlators 
have been used to obtain the static energy at zero temperature \cite{Bernard:2000gd,Cheng:2007jq,Bazavov:2011nk,Bazavov:2014pvz}.
The first cumulants for Wilson line correlators and from  smeared or unsmeared Wilson loops
have been compared in Ref. \cite{Bazavov:2019qoo} for zero temperature. It turned out that both approach
to the same plateau value for sufficiently large $\tau$.
At small $\tau$ the first cumulants for Wilson loops are systematically larger 
than those for the Wilson line correlators. The first cumulants for the Wilson line
correlators 
approach the plateau at smaller Euclidean time separation and have smaller errors
\cite{Bazavov:2019qoo}. 
In this sense the Wilson line correlators in Coulomb gauge are very good in 
projecting to the ground state, while there are significant excited state contribution
in the Wilson loops.
We performed a similar comparison of Wilson line correlators in Coulomb gauge and Wilson loops
with different levels of HYP smearings for $T=411$ MeV ($\beta=7.825$) and $N_{\tau}=12$.
As in the zero temperature case there is a significant difference in the first cumulants
for Wilson loops and Wilson line correlators at small Euclidean time as in the zero 
temperature case due to the excited state contamination, or equivalently due to 
$\rho_r^{high}(\omega)$.  Therefore, in Fig. \ref{fig:comp2hyp} we show the comparison
of the Wilson loops and Wilson line correlators in Coulomb gauge 
at two distances, $rT=1/4$ and $rT=1/2$, in terms of the subtracted first cumulants. 
At the smaller distance the first cumulant from Wilson line correlator and for Wilson
loops with different levels of HYP smearing agree within errors. For the larger distance,
$rT=1/2$ the two correlators agree at small $\tau$, where we see a nearly linear decrease 
of the first cumulants, while at large $\tau$, the non-linear behavior in $\tau$
of the first cumulants depends on the number of HYP smearings, and is also different
for the Wilson line correlator in Coulomb gauge. Thus the large Euclidean time
behavior depends on the choice of the static meson operator.
This is to be expected as explained above.
The situation is similar to the case of bottomonium correlator in NRQCD at non-zero temperature when different
extended meson operators are used \cite{Larsen:2019bwy,Larsen:2019zqv}.
Since the behavior of the first cumulant
at $\tau$ close to $1/T$ depends on the choice
of the static meson operators it is non-trivial to obtain physical
information from $W(r,\tau,T)$ in this $\tau$ region.

\begin{figure*}
    \centering
    \includegraphics[width=5.5cm]{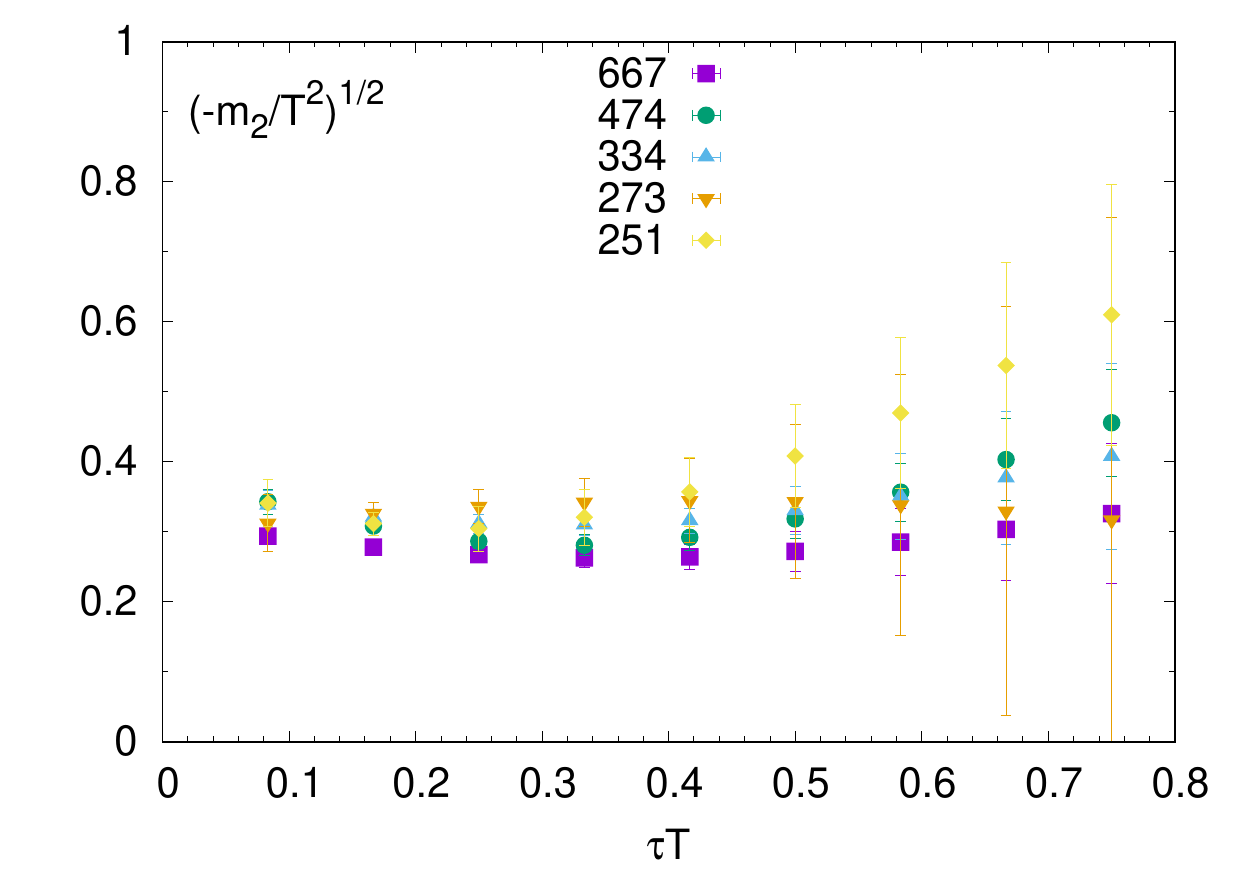}
    \includegraphics[width=5.5cm]{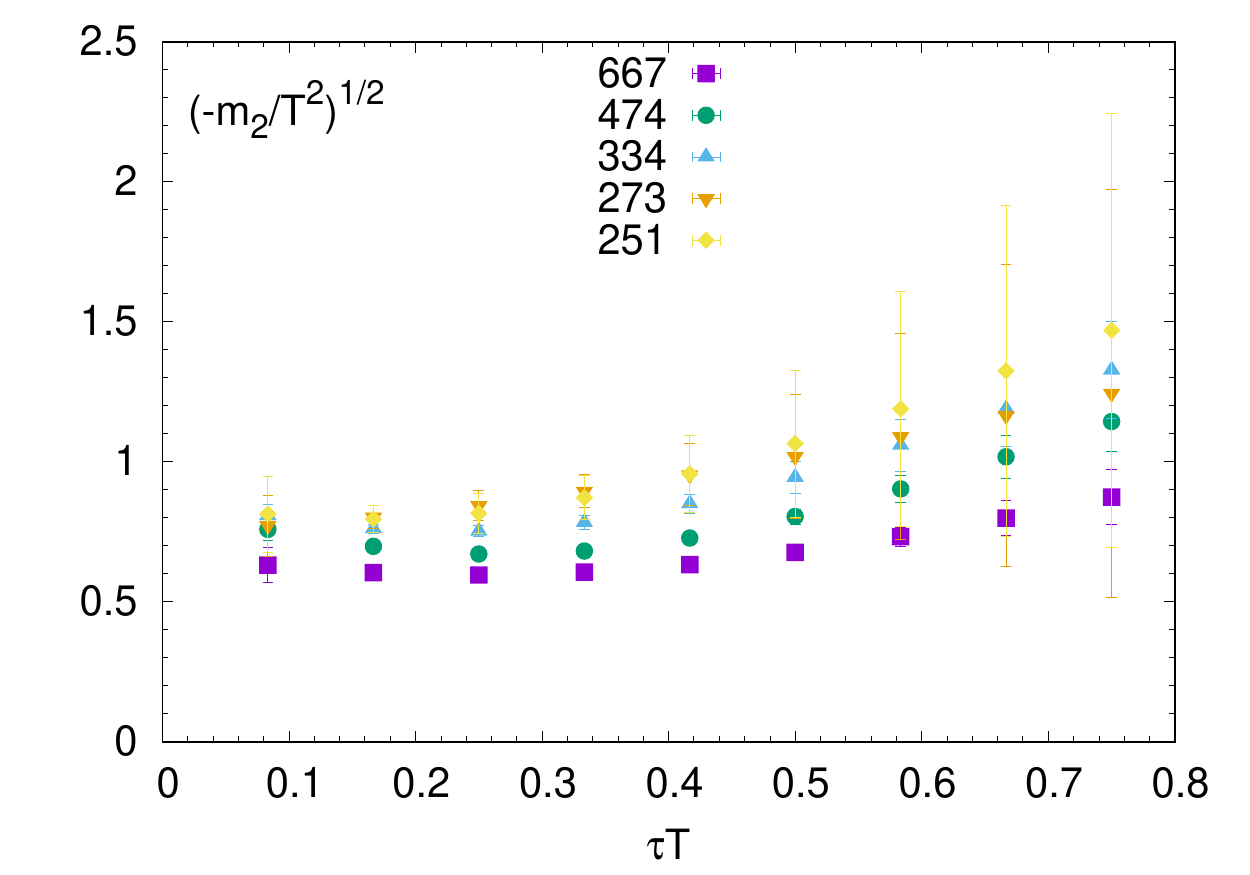}
    \includegraphics[width=5.5cm]{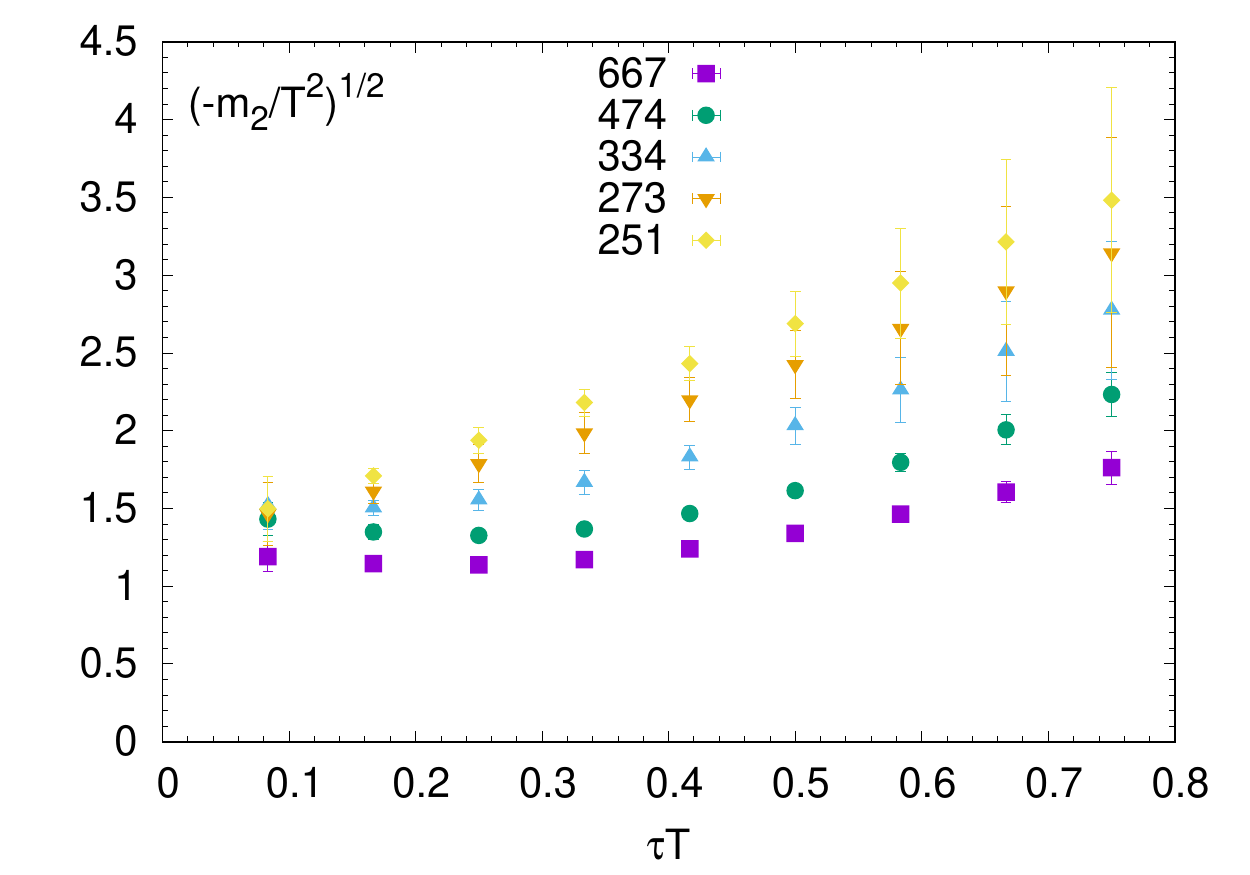}
    \caption{The second cumulants $m_2$, obtained from a fourth order polynomial fit to the first cumulant $m_1$, of the subtracted static meson correlator on $N_{\tau}=12$ lattices for $rT=1/4$ (left), $rT=1/2$ (middle) and $rT=1$ (right) for several temperatures. The different symbols correspond to different temperatures given in MeV.}
    \label{fig:m2}
\end{figure*}

In order to better understand our numerical results on the first cumulants and see to what
extent these can constrain the spectral function of a static meson it is helpful to calculate
higher cumulants of the correlator. In the following we consider the cumulants of
the subtracted correlator as we are interested in exploring the $\tau$-dependence caused
by thermal broadening of the dominant peak.
To evaluate higher cumulants 
we performed fits of the first cumulants of the subtracted correlator
using fourth order polynomials,
and estimated the higher cumulants by taking the
derivatives of the resulting polynomial.
The results for the second cumulants for three distances,
$rT=1/4,~1/2$ and $1$ at several temperatures are shown in Fig. \ref{fig:m2} for $N_{\tau}=12$. 
The errors on the cumulants have been estimated using
the jackknife procedure.
Since the second cumulant is negative, and
the square root of the negative second cumulant may
be related to the width, as discussed later, in the figure we show
$\sqrt{-m_2}$ in temperature units.
We see that the errors on the second
cumulants increase with decreasing temperatures. At short
distances, the second cumulant is approximately constant
for small $\tau$ and then starts to increase rapidly with
increasing $\tau$. For $rT=1$ the almost constant behavior 
of $m_2$ is only seen for the highest two temperatures.
The results for $T<251$ MeV are not shown as these have
much larger errors. However, within these large errors
the second cumulant is compatible with a constant at
these temperatures. 
\begin{figure*}
    \centering
    \includegraphics[width=5.5cm]{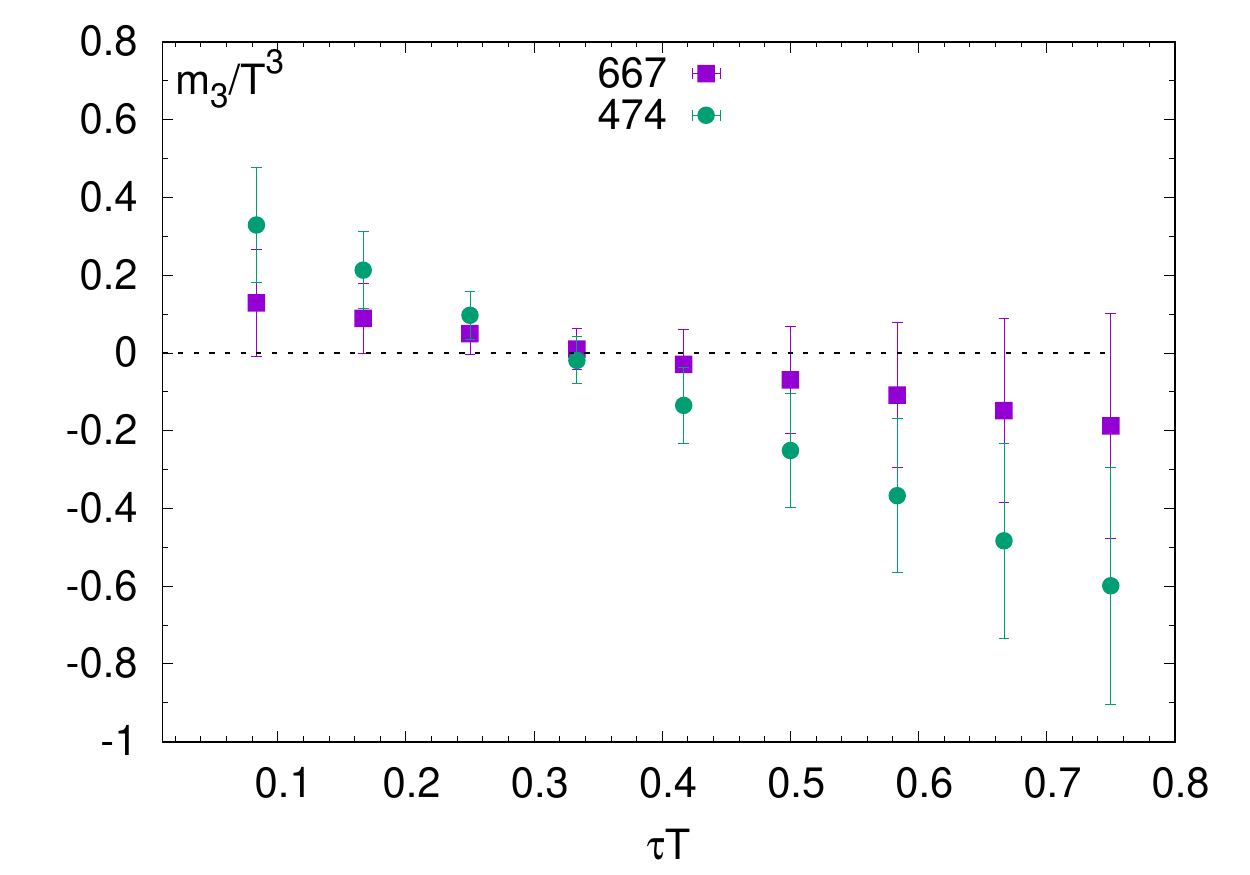}
    \includegraphics[width=5.5cm]{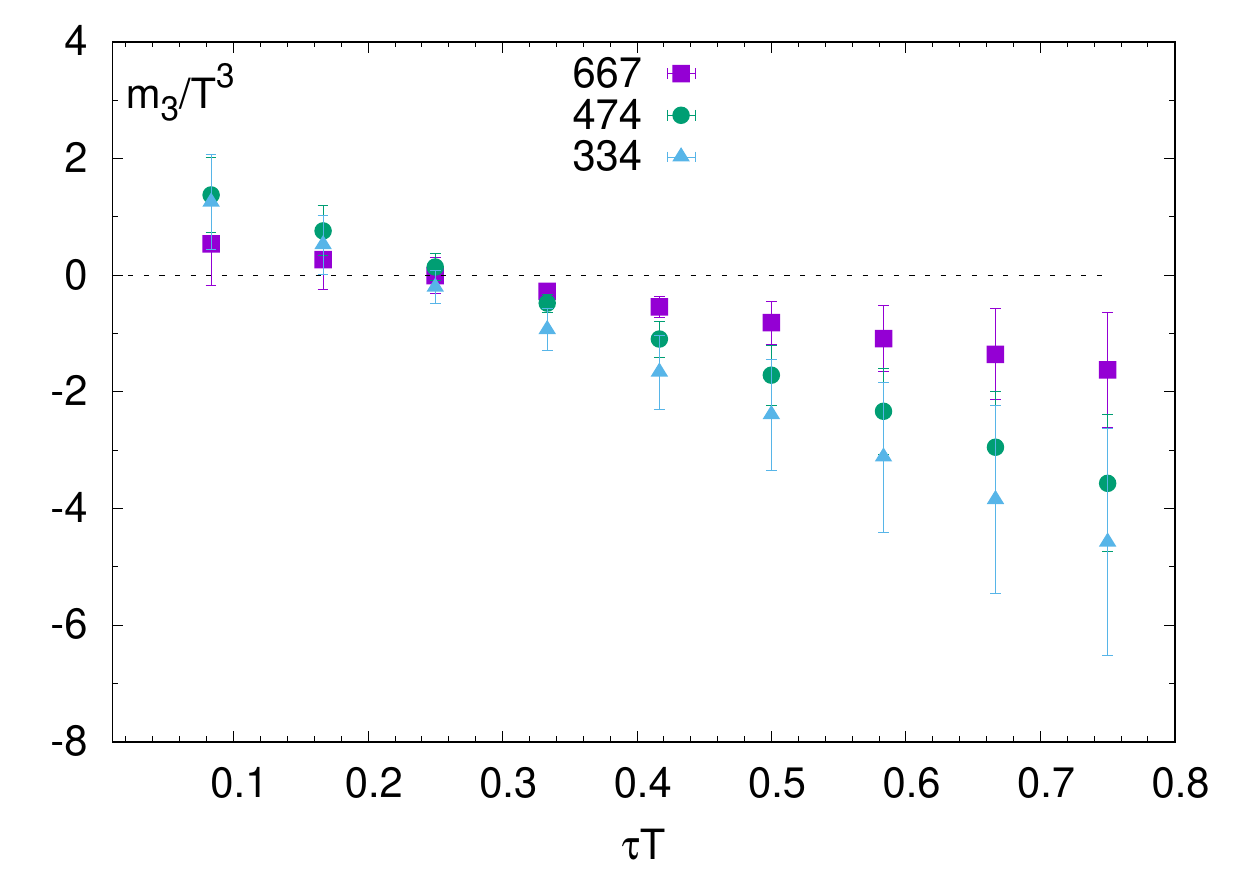}
    \includegraphics[width=5.5cm]{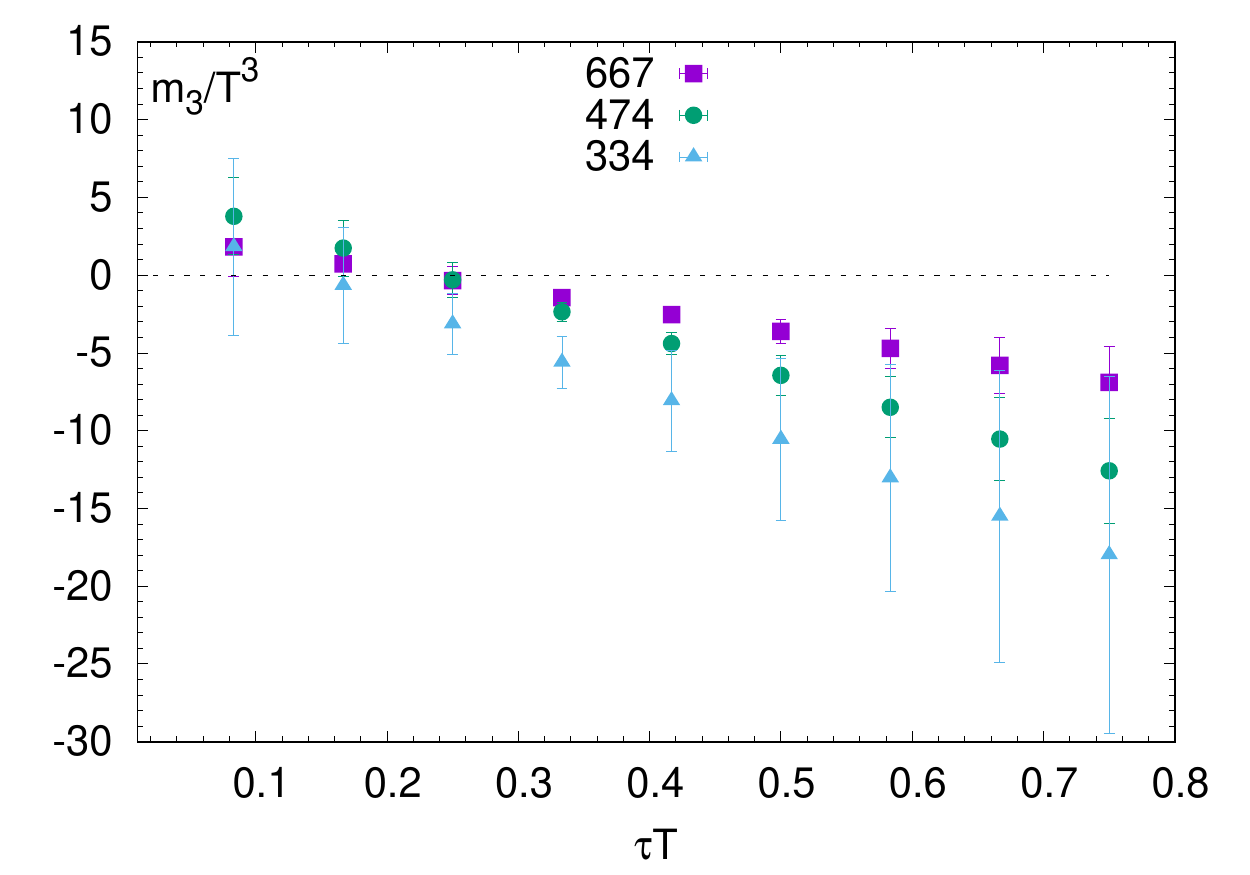}
    \caption{The third cumulants $m_3$, obtained from a fourth order polynomial fit to the first cumulant $m_1$, of the subtracted static meson correlator on $N_{\tau}=12$ lattices for $rT=1/4$ (left), $rT=1/2$ (middle) and $rT=1$ (right) for several temperatures. The different symbols correspond to different temperatures given in MeV.}
    \label{fig:m3}
\end{figure*}

In Fig. \ref{fig:m3} we show the third cumulant of the Wilson
line correlator, obtained from a fourth order polynomial fit to the first cumulant $m_1$, as function of $\tau T$ in temperature 
units. The results are shown for three representative
distances, $rT=1/4$, $rT=1/2$ and $rT=1$. We only show
our findings for the third cumulant for $T \ge 334$ MeV as
at lower temperatures the errors are too large to extract meaningful information from them. Furthermore,
for $rT=1/4$ the errors are already very large for $T=334$ MeV. The absolute value of the third cumulant increases rapidly with increasing
$rT$ and decreases with increasing temperature. These features can be already deduced by looking at the result for
the second cumulant. For $\tau T>0.35$ the third cumulant is
negative, while for  
$\tau<0.3$ it is positive but small given the errors. 
The small positive third cumulant at small $\tau$ is equivalent to 
having a nearly constant second cumulant.

From
Fig. \ref{fig:m3} it is clear that estimating the fourth and higher order cumulants from the present lattice results 
is very challenging. This will be important when considering
parametrization of the spectral function of a static meson, as the data can only constrain such a limited amount of parameters.
Hence any such parametrization should not contain more than three or four parameters.

\subsection{Comparison to HTL predictions}
\label{sec:corrHTLcmp}
At high temperatures, it is expected that the Wilson loops and  the Wilson line correlators 
in Coulomb gauge can be described in the weak coupling approach. 
The Wilson loops and Wilson line correlators have been calculated at leading order in Hard 
Thermal Loop (HTL) perturbation theory 
\cite{Laine:2006ns,Burnier:2013fca}. The HTL approximation is valid when $r \sim 1/m_D$ \cite{Brambilla:2008cx}, with
$m_D$ being the leading order Debye mass in QCD.
At distances $r \ll 1/m_D$ this approximation
is not expected to work.
In HTL approximation the logarithm of the Wilson loop or the Wilson 
line correlator can be written as
\begin{align}
\label{lnW_htl}
    \log W(r,\tau,T)=-{\rm Re} V(r,T)\times \tau+\nonumber\\
    \int_{-\infty}^{\infty}\frac{d \omega}{2 \pi} 
    (e^{-\omega \tau}+e^{-\omega (\beta-\tau)}) (1+n_B(\omega)) \sigma_r(\omega,T)+\mathrm{const},
\end{align}

where $n_B(\omega)=(\exp(\omega/T)-1)^{-1}$. The spectral function $\sigma_r(\omega,T)$
is related to the HTL spectral functions of the transverse and longitudinal
gluons and is distinct from the spectral function $\rho_r(\omega,T)$.
For the Wilson line correlator, it only depends on the spectral function
of the longitudinal gluons.
 
The important feature of this correlator is that the static energy exists,
\begin{equation}
\label{htl_energy}
\begin{split}
E^{HTL}_{s}(r, T)=\lim_{t\rightarrow\infty} i \frac{\partial \log W(r,\tau=it,T)}{\partial t}\\
	=\mathrm{Re}\,V(r, T)-i\,\mathrm{Im}\,V(r, T).
\end{split}
\end{equation}
At leading order, the real or imaginary parts are given as
\begin{equation}
\label{htl_energy LO}
\begin{split}
\mathrm{Re}\,V(r, T)  =  - \frac{g^2 C_F}{4 \pi} \left(\frac{e^{- m_{D} r}}{r} + m_{D}\right) \\
\mathrm{Im}\,V(r, T) = \frac{g^2 C_F}{4\pi} T \int\limits_0^\infty dz \,
\frac{2 z}{\left(z^2+1\right)^2} \left[ 1 - \frac{\sin z m_D r}{z m_D r} \right].
\end{split}
\end{equation}
%\JHW{JHW: The first relation in the Eq.~\eqref{htl_energy} is generally true in HTL approximation, the latter is only true at leading order. Hence, I've split the equations into two. Please check that the splitting didn't break anything.}

The real part of the potential $\mathrm{Re}\,V(r, T)$, which in this approximation, is at leading order identical to the singlet free energy in Coulomb gauge \cite{Laine:2006ns,Brambilla:2008cx}. We observe the $\tau$ dependence in the above correlator consists of linear and periodic part in $\tau$. This particular $\tau$ dependence of the HTL correlator along with the fact that $\sigma_{r}(\omega, T)$ has a $1/\omega$ singularity allows us to have a well-defined limit in Eq.(\ref{htl_energy}). In \cref{sec:BalaDatta}, while calculating the static energy non-perturbatively we will parametrize the correlator as a combination of linear and periodic parts in $\tau$. The obvious consequence of a parametrization as in Eq.(\ref{lnW_htl}) is that the first cumulant of $W(r,\tau,T)$ is anti-symmetric around the mid-point $\tau=1/(2 T)$.

\begin{figure}
\includegraphics[width=7cm]{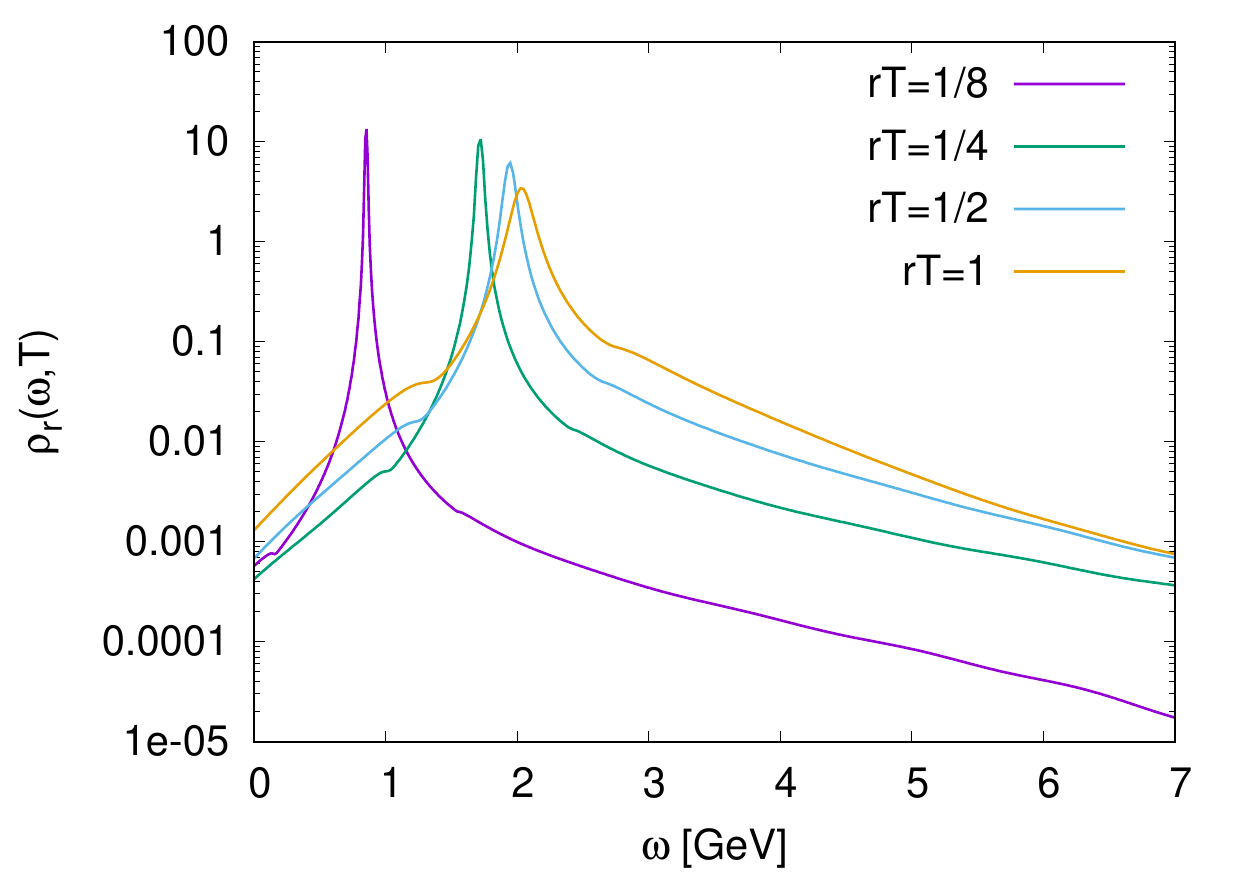}
\caption{The HTL spectral function for $T=667$ MeV for different $r$.}
\label{fig:spf_htl_T667}
\end{figure}

Since we study the Wilson line correlators in a large temperature range, including
high-temperature values it makes sense to compare the lattice results with the
weak-coupling ones. Comparison with the HTL perturbative result is also important since it
gives some insight into the general features of the spectral function and how these
features manifest in the cumulants of the Euclidean time correlator.
Therefore, in Fig.\ref{fig:spf_htl_T667} we show the spectral functions corresponding
to Wilson line correlators for different $r$ at $T=667$ MeV in the HTL approximation. Note 
that we use a different renormalization prescription compared to Ref. \cite{Burnier:2013fca}
as well as the two-loop running of the coupling constant with
$\Lambda_{\overline{MS}}^{n_f=3}=332$ MeV \cite{Petreczky:2020tky}.
We will use this choice for the gauge coupling throughout this paper.
We see a peak in the spectral function at $\omega={\rm Re} V(r, T)=F_S(r, T)$, which can be well 
described by skewed Lorentzian for frequencies around the location of the peak \cite{Burnier:2013fca}. 
Far away from the peak position, the spectral function is described by different structures, distinct 
from the Lorentzian. 

In general, one has to
%would
expect that the non-perturbative spectral function such as the one calculated on a lattice 
%has an additional structure. 
contains further structures. 
In particular, the lattice spectral function has a large UV continuum part 
along with a tail at very low $\omega$. 
The HTL feature of the lattice spectral functions could 
%can 
only
come from the medium dominated part of the spectral function ($\rho_{r}^{med}(\omega, T)$). Therefore, we could 
%we can 
possibly see the HTL-like features in the non-perturbative correlator near the $\tau\sim 1/(2T)$ region~if these are sufficiently separated from any further structures.

We consider the comparison between the lattice and HTL result for the Wilson line correlator in terms of its first cumulant, $m_1$. The first cumulant 
should be sensitive to the peak position of the spectral functions. In absence of peak width, i.e. 
when $\rho_{r}^{med}(\omega, T)\sim \delta(\omega-E_{0}(r))$, $m_1$ should approach
the energy of static $Q \bar Q$ pair $E_0$ at intermediate $\tau$. It is known 
that the leading order perturbative result does not provide an accurate description of the static $Q\bar Q$ energy at zero temperature.

%Therefore 
For this reason, we naturally expect that at finite temperature non-perturbative real and imaginary parts of 
%this 
the complex static energy defined through the parametrization of Eq.~\eqref{lnW_htl} will be different from the expressions given in Eq.~(\ref{htl_energy LO}).
Therefore, non-perturbative investigation of this complex static energy is very important, about which
we will discuss in detail in \cref{sec:BalaDatta}.

Furthermore, the static energy needs to
be renormalized, and the renormalization condition used on the lattice is different
from the one in the $\overline{MS}$ scheme. Connecting these two renormalization schemes is a non-trivial
task. We also know that the 
real part of the static energy is given by the so-called singlet $Q \bar Q$ free energy,
$F_S(r,T)$ in the HTL approximation \cite{Brambilla:2008cx}, as discussed above.
Therefore, when comparing the lattice results on $m_1$ to the HTL results we will assume 
that the peak position is similar to $F_S(r, T)$ and subtract the latter from the first cumulant.

As mentioned above the HTL calculation is not expected to describe the spectral function at large $\omega$.
Therefore, we should use the subtracted first cumulant when
comparing the lattice and HTL results, or simply ignore the data points at small $\tau$ in the comparison. But the HTL feature in the correlator could only appear in the data points around $\tau\sim 1/2T$, where the effect of this subtraction is small.

\begin{figure*}
\includegraphics[width=8cm]{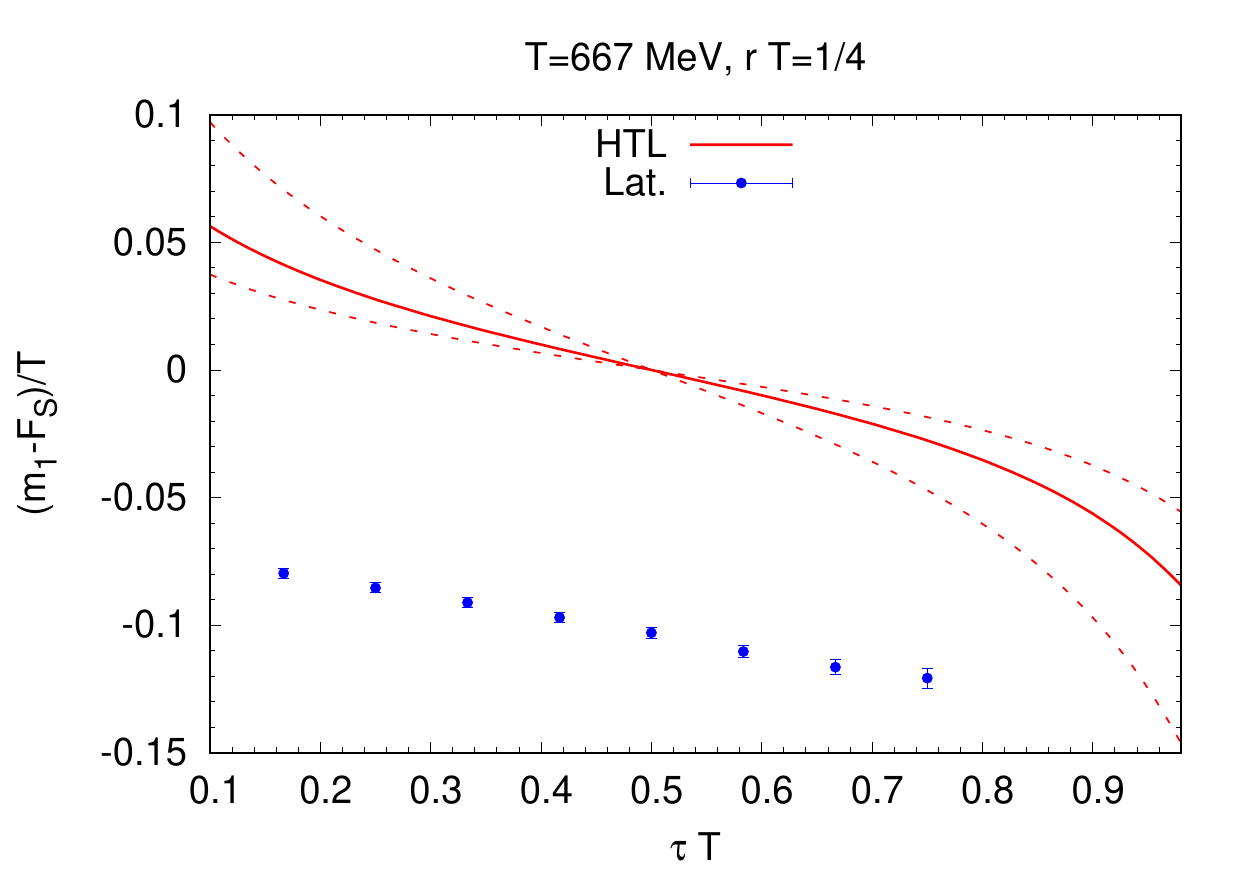}
\includegraphics[width=8cm]{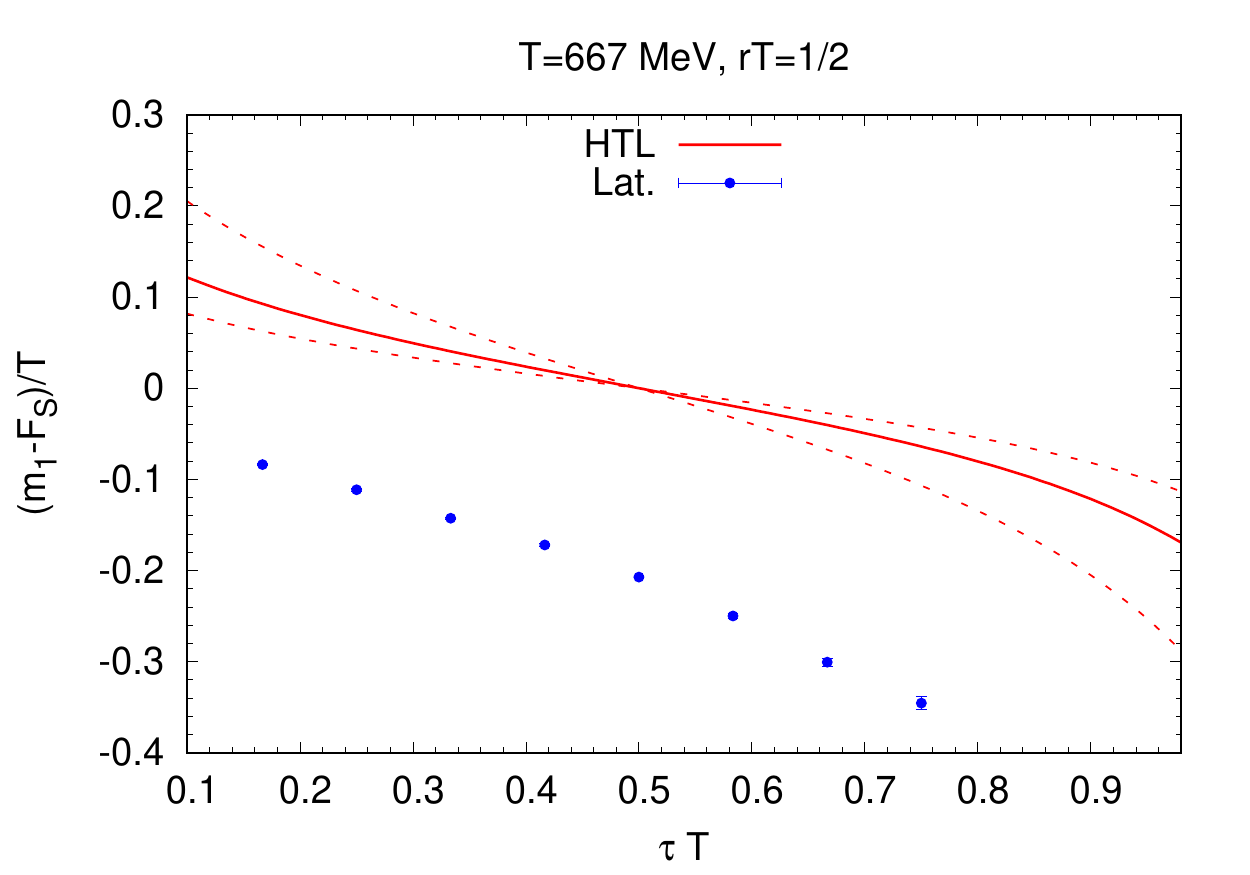}
\includegraphics[width=8cm]{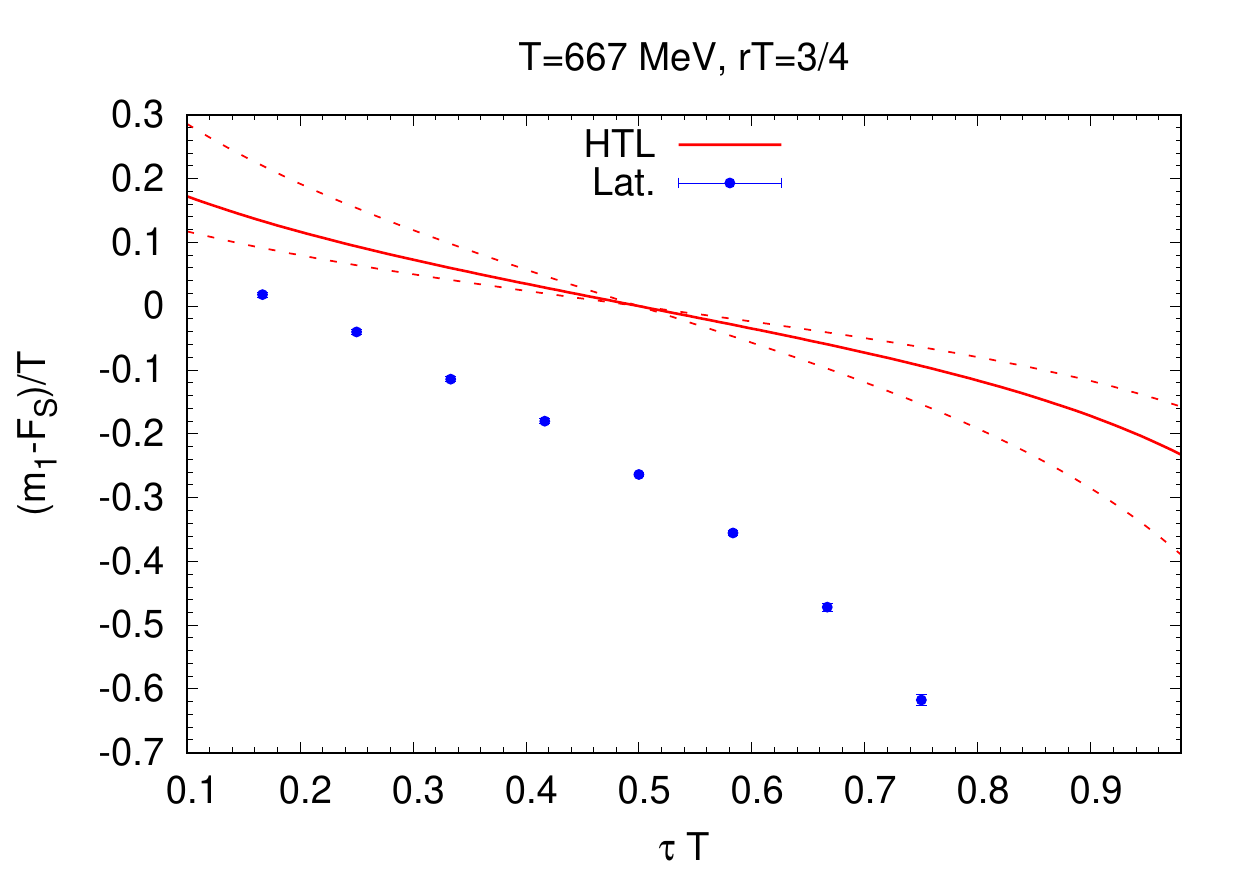}
\includegraphics[width=8cm]{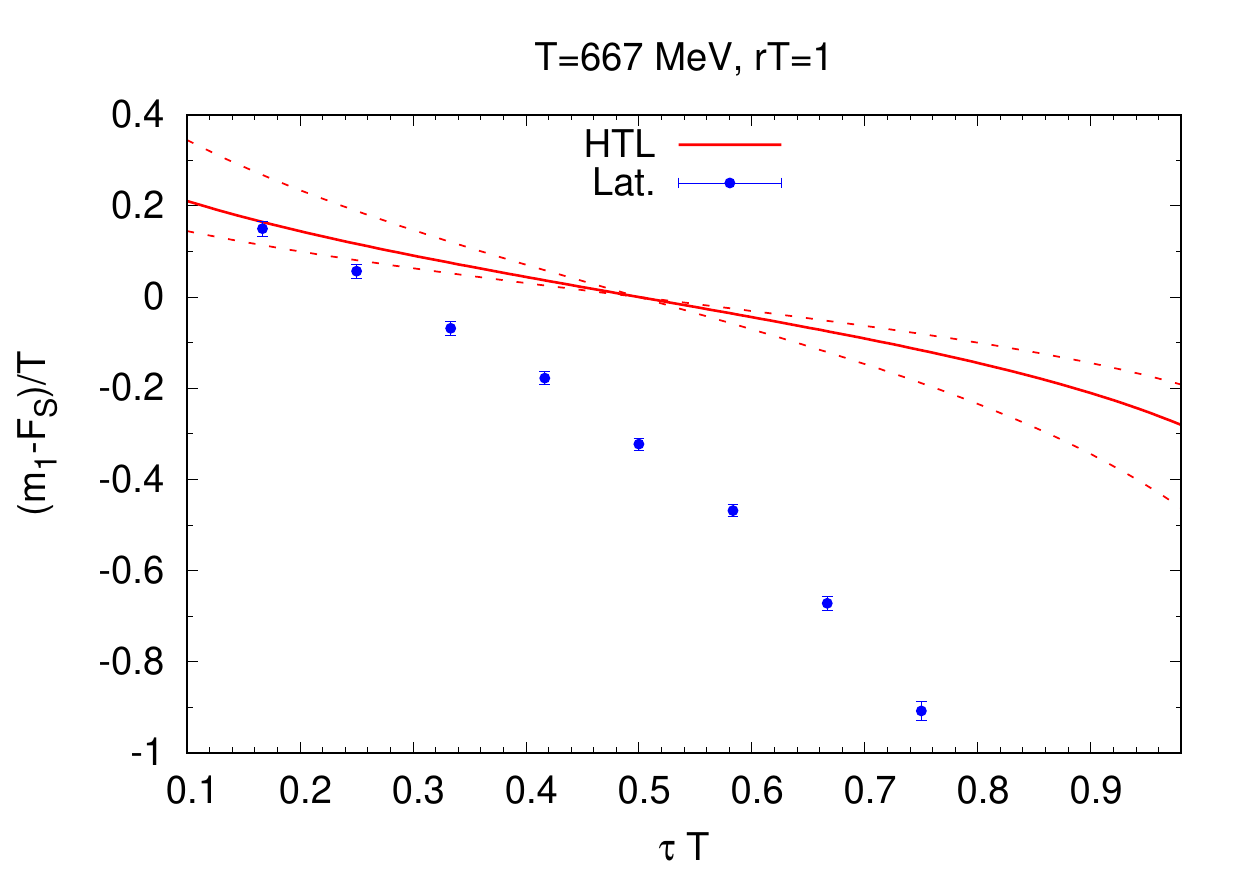}
\caption{
The comparison of $m_1-F_S$ on the lattice (subtracted)  with HTL results at $T=667$ MeV for $rT=1/4,~1/2,~3/4$ and $1$.
The HTL result for $\mu=2 \pi T$ are shown as solid lines. The dashed lines
correspond to variation of the scale $\mu$  by a factor of two.}
\label{fig:comp_htl_667}
\end{figure*}

We performed a comparison of the lattice results on the subtracted first cumulant
with leading-order HTL calculations for $T=474$ MeV and $T=667$ MeV. In the HTL calculations we used three values 
of the renormalization scale $\mu=\pi T,~2 \pi T$ and $4 \pi T$. The comparison is shown in Fig.\ref{fig:comp_htl_667}
for $T=667$ MeV and four representative distances $rT=1/4,~1/2,~3/4$ and $1$. 
The lattice and the HTL results for $m_1$ share some qualitative features, namely
they decrease monotonically with increasing $\tau$. 
This decrease of $m_1$ with $\tau$ around $\beta/2$ comes from the fact that the spectral
function is not a delta function but rather a broad peak (see Fig. \ref{fig:spf_htl_T667})
and the slope of $m_1$ is loosely related to the width of the peak.

The HTL curve
%corrector 
is antisymmetric with respect to $1/(2T)$ 
%, however, the lattice result is not antisymmetric for the whole range of $\beta$. 
over the whole $\tau$ range, while the lattice data do not show the same antisymmetry.
This is expected as lattice correlator gets contributions
both from $\rho_{r}^{high}(\omega, T)$ and $\rho_{r}^{tail}(\omega, T)$. 
%There is no reason that these two contributions would appear in an antisymmetric fashion.  Given the discussion of their respective physical origins in Appendix~\ref{app:spec_decomp} it is implausible that these two contributions could give rise to antisymmetric contributions.
However, as we will see in \cref{sec:BalaDatta}, non-perturbative data are compatible with a small
%is consistent with a small 
antisymmetric region around $\tau\sim 1/(2T)$.

%We find that the lattice result for $m_1$ is different from the \JHW{leading-order} HTL result.
We mentioned earlier that the leading-order HTL results for the real part of static energy and the singlet free energy agree exactly; hence, 
%observed that at $\tau =1/(2T)$ 
%for the perturbative correlator
the corresponding $m_1-F_{S}$ vanishes at $\tau =1/(2T)$. 
%This correspond to the fact that for leading order perturbative correlator the real part of static energy and singlet free energy are exactly equal. 
However, the lattice result for 
%non-perturbative correlator 
$m_1-F_s$ is non-zero at $\tau =1/(2T)$. 
This implies that
%we will find that 
%in \cref{sec:BalaDatta} this implies that non-perturbatively 
the real part of the static energy that will be determined in \cref{sec:BalaDatta} must be different from the singlet free energy. 
The slope of $m_1$ around $\tau \sim  1/(2T)$ is much larger for the lattice correlator than for the leading-order HTL curve.
This corresponds to the fact that non-perturbative imaginary part 
%calculated on the lattice 
determined from the lattice data in \cref{sec:BalaDatta} 
is significantly different from the expression given in Eq.(\ref{htl_energy LO}).

The comparison turned out to be similar for $T=474$ MeV.
We also performed a comparison between the lattice and HTL calculations at the highest temperature
available, $T=1938$ MeV. Since we do not have the corresponding zero temperature result, here the comparison
is performed in terms of the un-subtracted cumulants. 
We see that also at the highest temperature  
the lattice data differs from the perturbative HTL data. 
The first cumulant calculated on the lattice has stronger dependence on $\tau T$ correspond to the fact that
non-perturbative imaginary part is much higher than the perturbative imaginary part even at that very high temperature.
\begin{figure*}
\includegraphics[width=8cm]{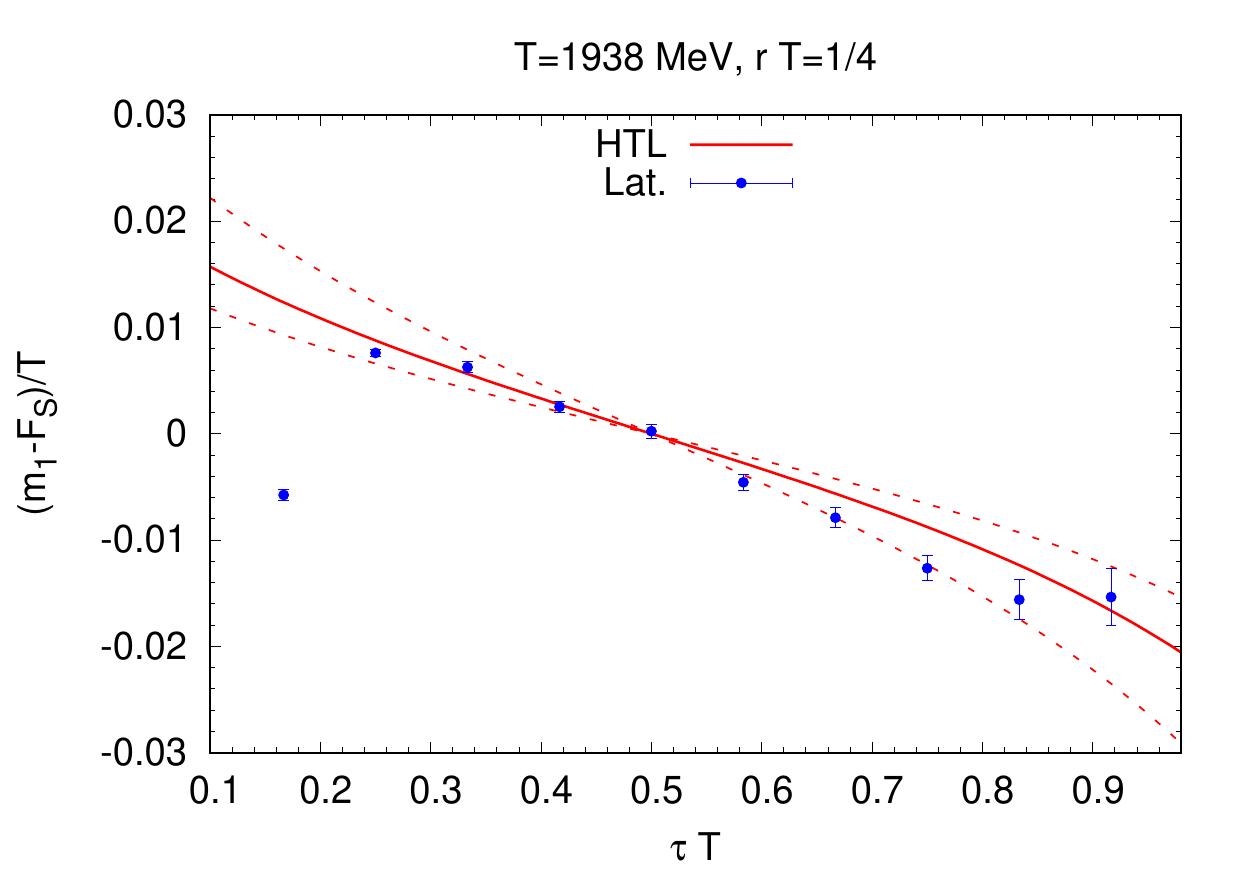}
\includegraphics[width=8cm]{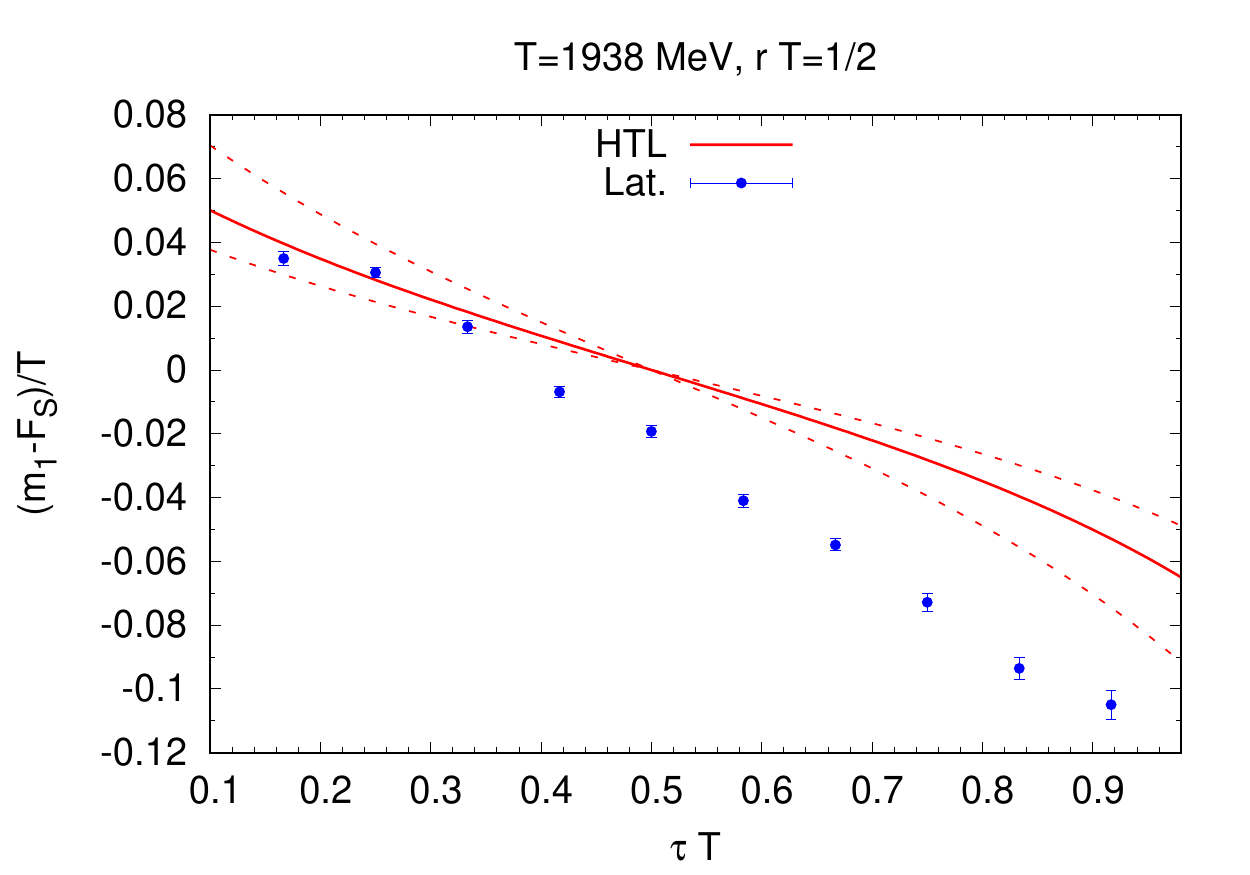}
\includegraphics[width=8cm]{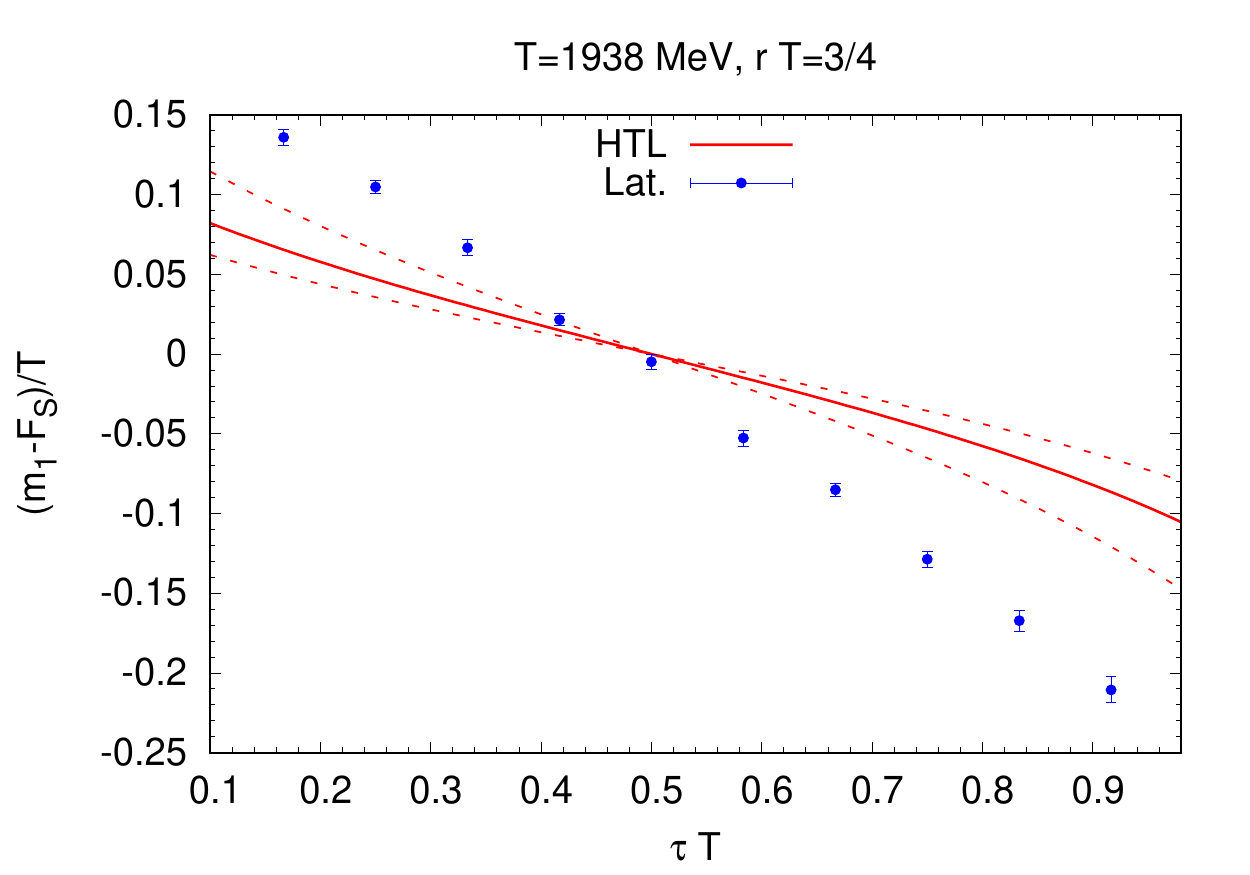}
\includegraphics[width=8cm]{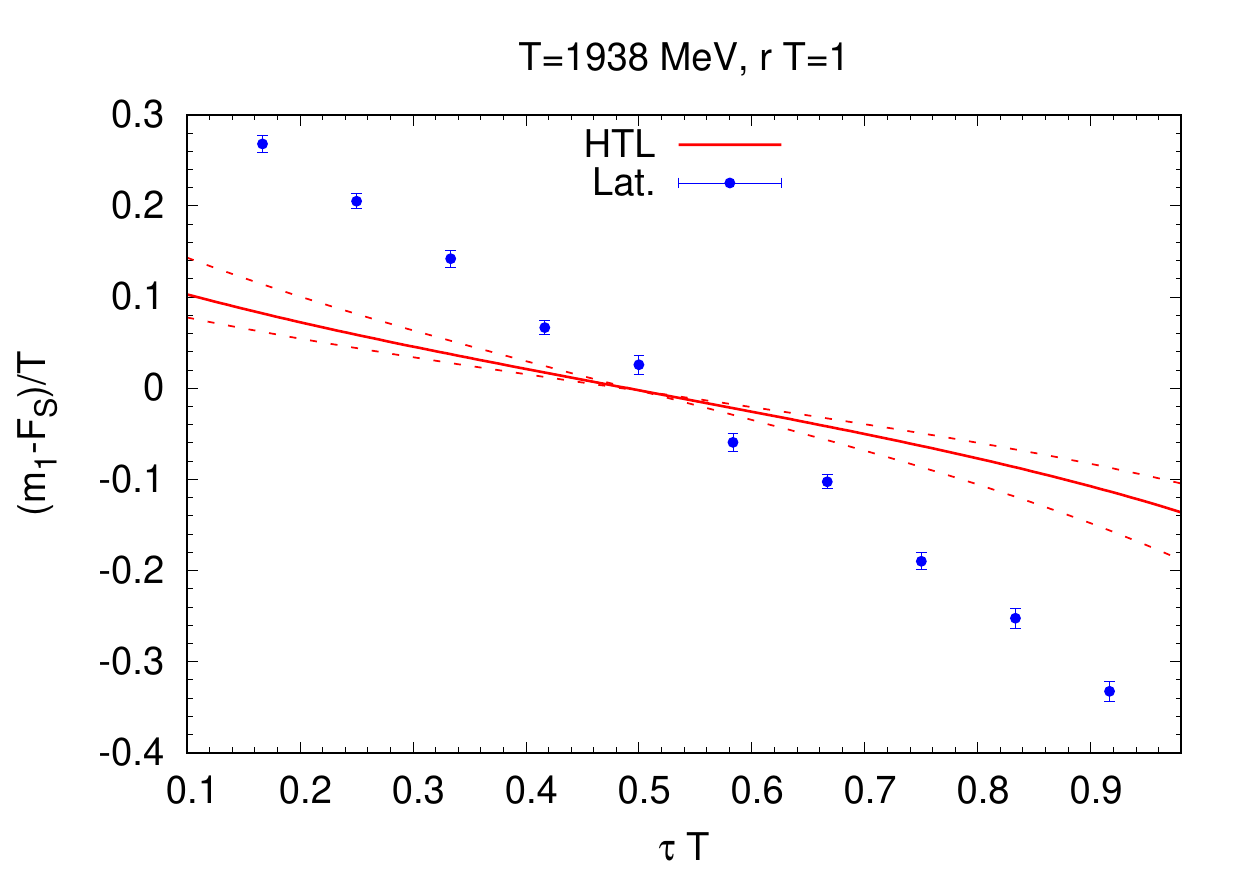}
\caption{
The comparison of $m_1-F_S$ calculated on the lattice with HTL results at $T=1938$ MeV for $rT=1/4,~1/2,~3/4$ and $1$.
The HTL result for $\mu=2 \pi T$ are shown as solid lines. The dashed lines
correspond to variation of the scale $\mu$  by a factor of two.}
\label{fig:comp_htl_1938}
\end{figure*}

\begin{figure*}
\includegraphics[width=5.8cm]{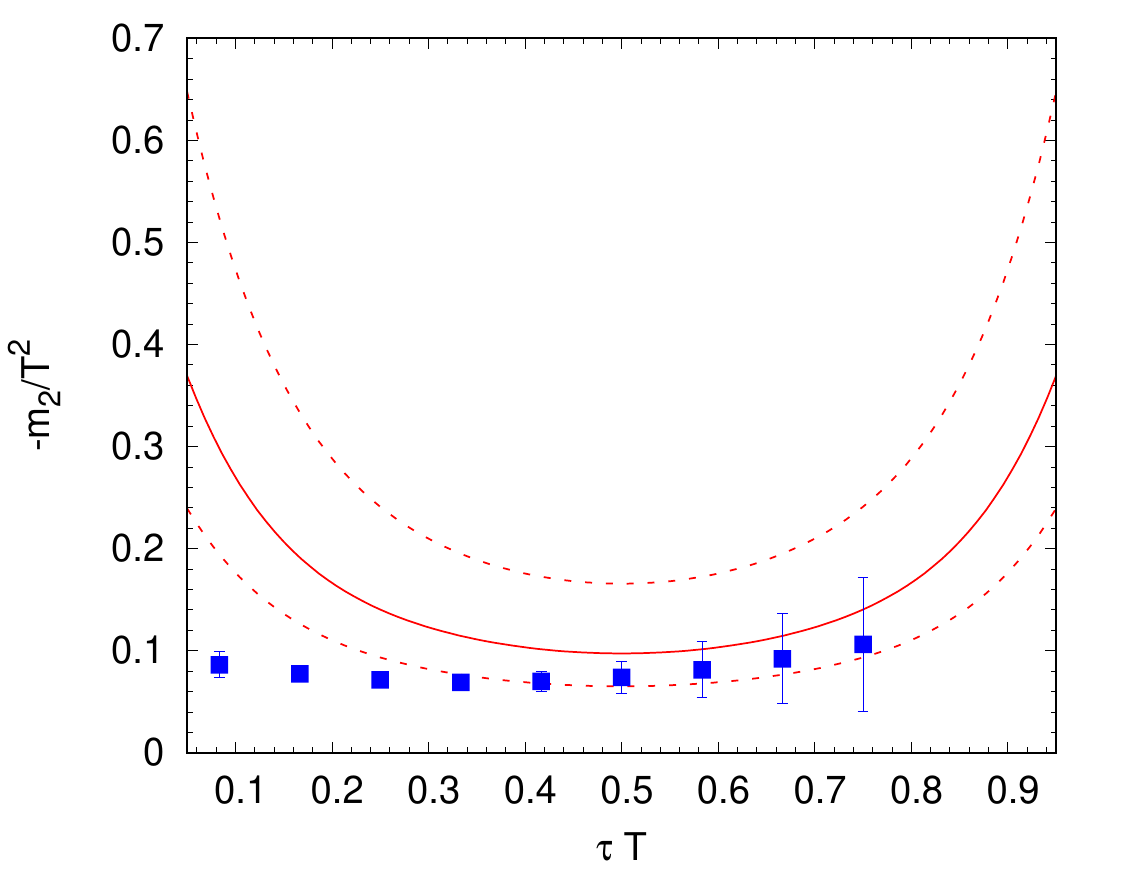}
\includegraphics[width=5.8cm]{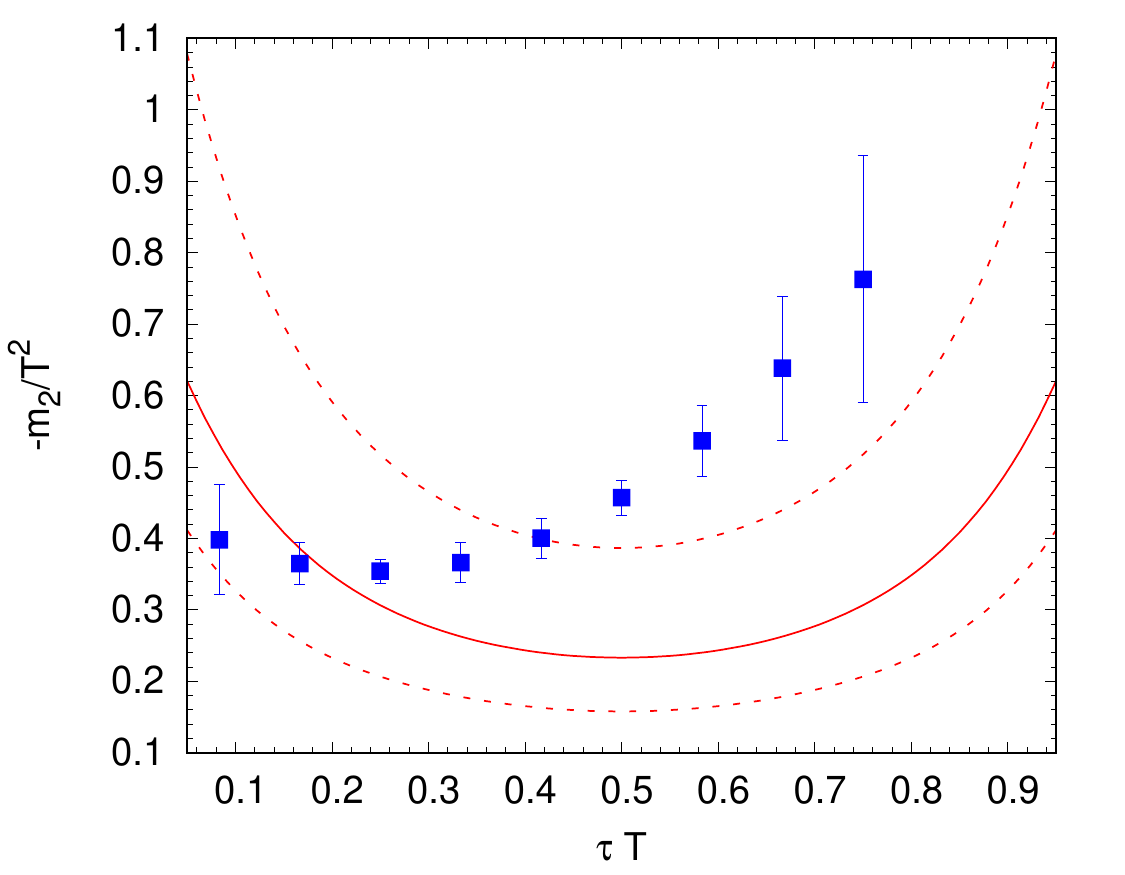}
\includegraphics[width=5.8cm]{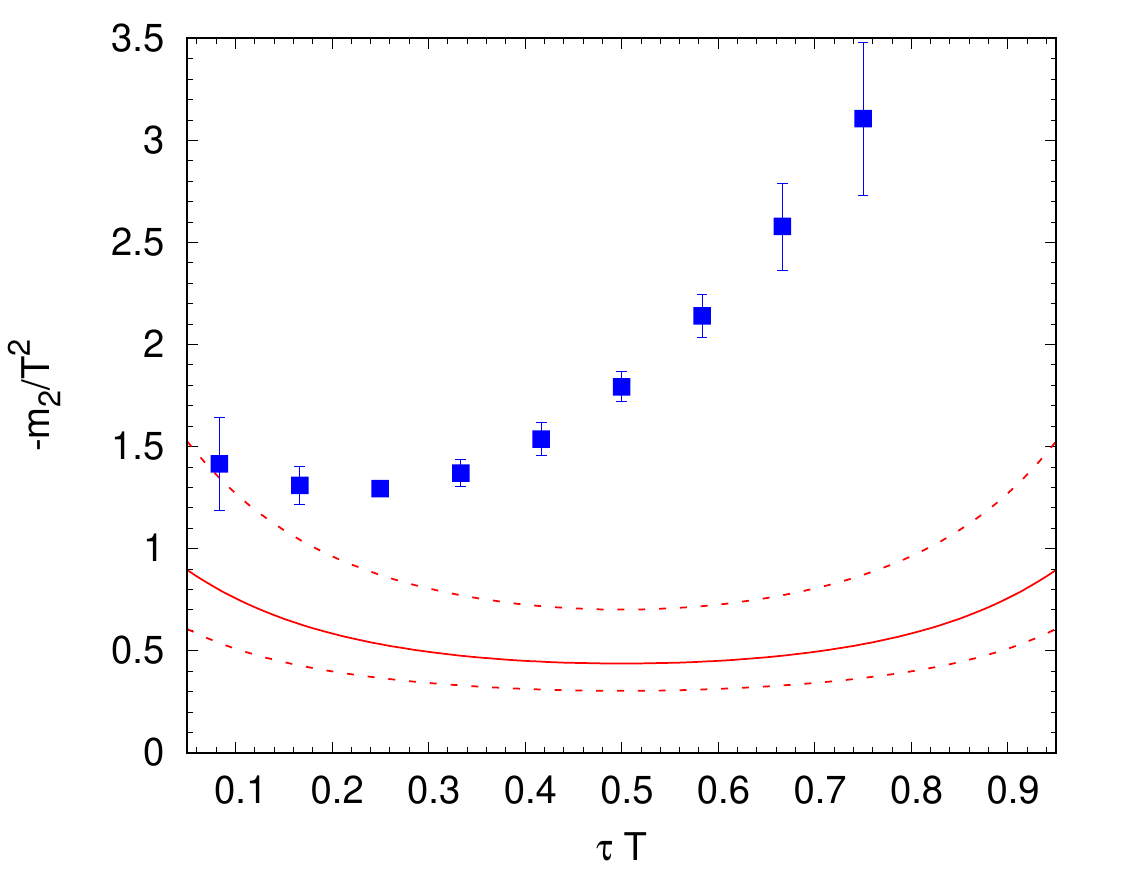}
\caption{The second cumulant, obtained from a fourth order polynomial fit to the first cumulant $m_1$, of the subtracted Wilson line correlators as function of $\tau$ at $T=667$ MeV calculated
on $N_{\tau}=12$ lattices and in HTL perturbation theory (lines) for
$rT=1/4$ (left), $rT=1/2$ (middle) and $rT=1$ (right). The solid line corresponds to the choice
of the renormalization scale $\mu=2 \pi T$, while the dashed lines correspond to the scale choice
$\mu=\pi T$ and $4 \pi T$.}
\label{fig:m2htl}
\end{figure*}

In Fig.~\ref{fig:m2htl} we show the comparison of the lattice and HTL results for the second cumulant, $m_2$
for $T=667$ MeV and three representative distances, $rT=1/4,~1/2$ and $1$. 
We choose the renormalization scale in the HTL calculation to be $\mu=2 \pi T$ and
vary it by a factor of two around this value. 
Unsurprisingly, there is no quantitative agreement between the lattice results and leading-order HTL as already visible from Fig.~\ref{fig:comp_htl_667}.
The fact that $m_2$ on the lattice is constant at small $\tau$ is due to subtraction of UV contribution, while the fact that $m_2$ is not a constant at large $\tau$ is due to the low-energy tail. 
Since neither of these are present in the leading-order HTL result, the mismatch between lattice and leading-order HTL is particularly pronounced in these regions.

\section{Study of the spectral function and its ground state peak}
\label{sec:spectra}
We will in the subsections of this section make four different attempts to analyze the lattice results. The four methods are: 
fit with a finite width spectral function (\ref{sec:potfit}), 
fit with HTL Ansatz (\ref{sec:BalaDatta}), Pade fit (\ref{sec:Pade}), 
and fit using Bayesian methods (\ref{sec:Bayes}). We will outline here the basic idea behind each method and the pros and cons of each choice, and leave a more technical description for each subsection. 
We clearly state that each among the first three approaches (Bayesian methods are an exception) only aims at identifying and parametrizing the lowest, dominant spectral feature. 
Thus, the inability of their results to  reproduce the input data over the complete $\tau$ range has to be expected. 
We stress that this is nothing unsual -- the same applies to almost any analysis of zero-temperature lattice correlators that may have to leave out the first few time steps due to not having enough independent information in this range to fully constrain the complex UV structure affecting those data.
The aim
%hope 
is to leave it to the reader to judge each method based on the results put forth in this paper and let them decide 
%themself on what to use. 
which one they prefer.

The first method used in section \ref{sec:potfit} is a simple fit, 
using a model spectral function
%with a spectral function 
with a well defined position and width. This choice comes from the observation that when one uses zero temperature results to remove contributions coming from higher energy excitations, the first cumulant
%effective mass 
takes a form that is well approximated by a low order polynomial, sometimes even that of a straight line. The benefit of this method is that it gives a precise answer to the question ``What is the position and width of the the dominant feature in the spectral function?'' with a fit that works very well, excluding the first and last point. The downside is that for it to work, the high energy contributions to the correlator has to be removed using zero temperature results, a procedure which might not be well defined. Also the actual 
%spectral functions form is not decided, only the width of it is
shape of the dominant spectral feature is not determined, but only its position and effective width.

%The second method is in section \ref{sec:BalaDatta} and uses the ansatz from HTL to make a fit form that extract physical relevant information. As shown earlier in this paper, HTL does not explain the full spectral function, but the fit does not rely on a full fit on all data. Instead a fit is done around the middle, where the behavior can be approximated by an analytic form. The positive of this method is that one can interpret the parameters of the fit in terms of a static energy, while the negative about this method is that it does not fit the full data, and therefore have to rely on a subsection of the data.

The second method is in~\cref{sec:BalaDatta} and uses 
an Ansatz motivated by HTL to fit in a narrow range around $\tau \sim 1/(2T)$ to extract physically relevant information. Namely, 
%the ansatz from HTL to make a fit form near $\tau \sim 1/(2T)$ that extracts physical relevant information. In this method, 
one parametrizes the  $\tau$ dependence around $1/(2T)$ 
%similar to 
exactly like the $\tau$ dependence in Eq.~(\ref{lnW_htl}). %The importance of this method is that 
In this method
the peak position and peak width can be interpreted as a real and imaginary part of thermal static energy. As shown earlier in this paper, HTL 
%does not 
cannot
explain the full spectral function. %however this fit form
The HTL-motivated fit only attempts to describe the dominant feature of the spectral function which is responsible for the thermal static energy.

The third method used in section \ref{sec:Pade} is the Pad\'e interpolation. This approach operates on the Fourier transformed lattice data, instead of directly on the correlator. The rational interpolation of the Matsubara frequency correlator is subsequently rotated to real time. In our mock data tests, this method has shown to give reasonable results for the position of the peaks, but failed to reliably estimate the width of the spectral function. By construction the Pad\'e approach does not need to reproduce the input data and we find that the reconstructed spectral function indeed does not fulfill the original spectral decomposition. The method requires a high quality of the input data, as it does not contain a regulator of the statistical noise. 

The fourth and last method we use is the Bayesian Reconstruction (BR) method described in section \ref{sec:Bayes}, which is based on Bayesian inference. The basic idea here is to regularize a $\chi ^2$ fit with an additional functional, which encodes how compatible the fitted spectral function is with prior knowledge one possesses on the spectrum, such as its positivity. 
This method is constructed such that it will always reproduce the Euclidean input data points within their statistical uncertainty. It has shown to outperform the Maximum Entropy Method in the reproduction of sharp peaked features but may suffer from ringing artifacts in the reconstruction of extended spectral features, and requires high precision data. 
However, it was realized that correlators on the finer lattices with improved gauge action contain non-negligible contributions with negative weights that render the BR method inapplicable.

\subsection{Determination of the ground state peak from spectral function model fits}
\label{sec:potfit}
%\begin{figure}
%    \centering
%    \includegraphics[width=7.5cm]{GaussianFit_b7825_r9.pdf}
%    \caption{The first cumulant and the subtracted
%     first cumulant at non-zero temperature and non-zero temperature for
%    $N_{\tau}=12, beta=7.825$. The corresponding zero temperature results are %also shown. 
%    The lines show the fit results.}
%    \label{fig:GaussianFit_b7825}
%\end{figure}
\begin{figure*}
    \centering
    \includegraphics[width=7.5cm]{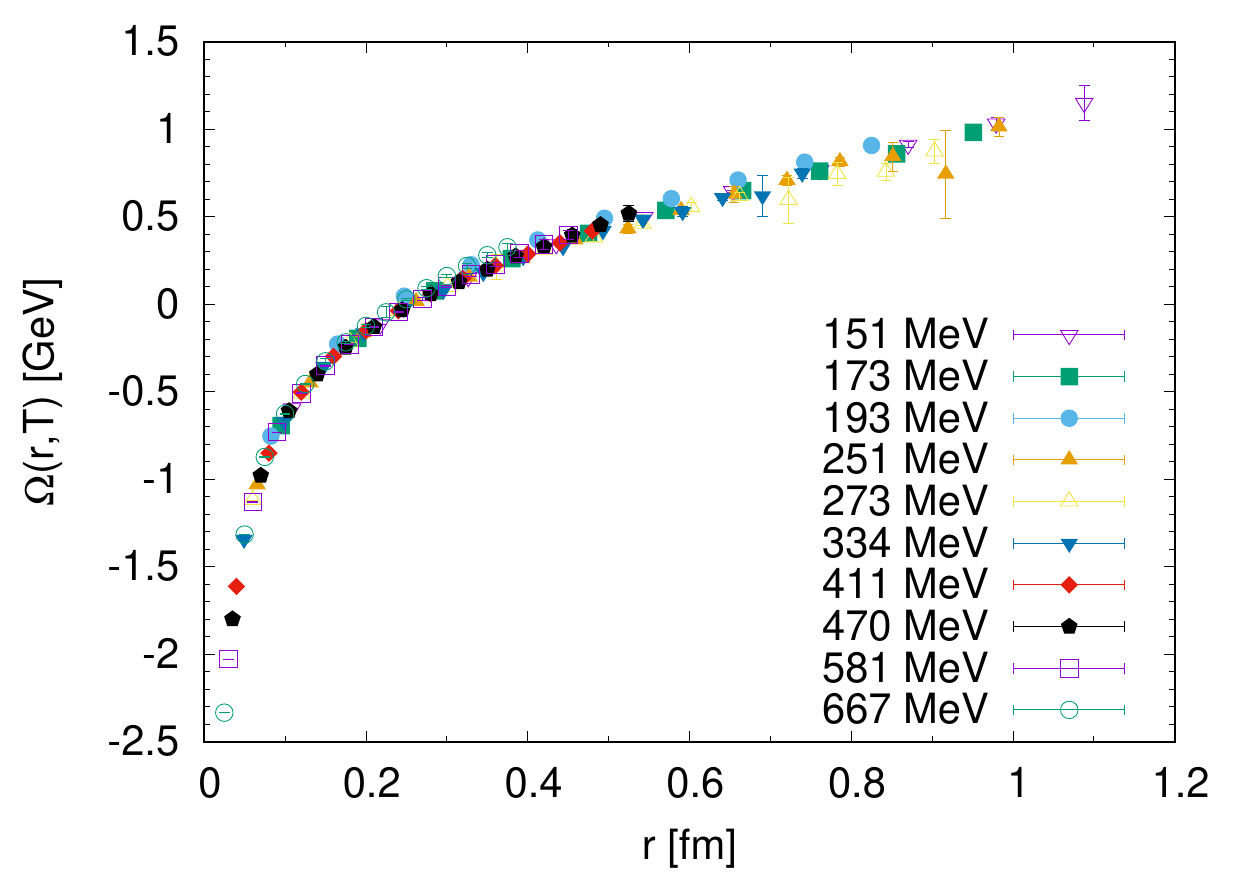}
    \includegraphics[width=7.5cm]{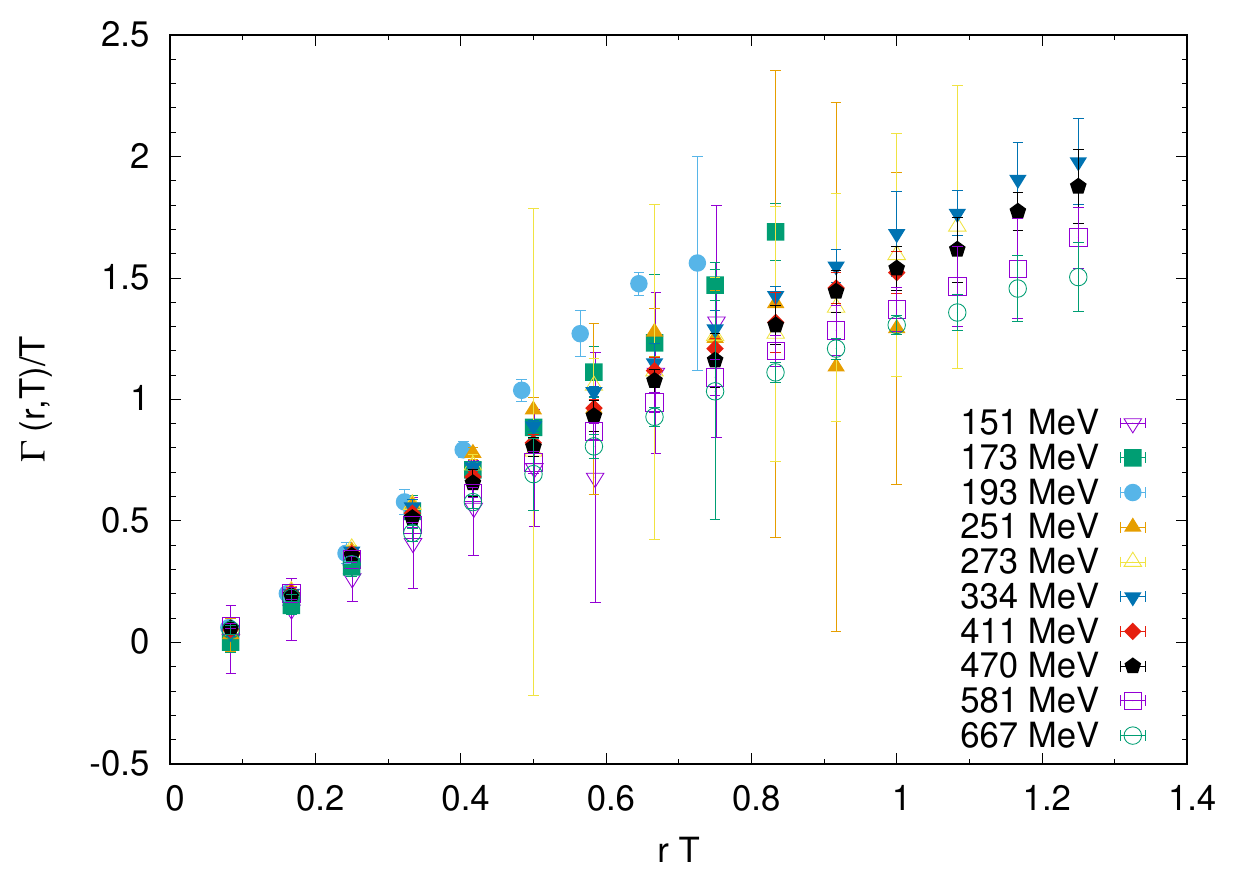}
    \caption{The peak position of the spectral function (left figure) and
    the width  (right figure) as function
    of the separation $r$ obtained from Gaussian fits of the $N_{\tau}=12$ data.}
    \label{fig:pot}
\end{figure*}

In order to constrain the spectral function $\rho_r(\omega,T)$ from limited data on Euclidean time correlation functions  we need to assume
some functional form for it. As for the analysis of the cumulants we assume that the spectral function can be written
$\rho_r(\omega,T)=\rho_r^{tail}(\omega,T)+\rho_r^{med}(\omega,T)+\rho_r^{high}(\omega)$, 
with $\rho_r^{high}(\omega)$ assumed to be temperature independent
high frequency part of the spectral function and $\rho_r^{med}(\omega,T)$ containing the dominant peak structure. Based on
general grounds and EFT arguments it is natural to assume that $\rho_r^{med}(\omega,T)$ has a Lorentzian form. However, for
a Lorentzian form the integral in 
%Eq. (\ref{eq:RTwilsonspecdec}) 
Eq. \eqref{eq:ETwilsonspecdec}
will not converge at the lower integration limit. We also do not expect that the Lorentzian form can describe the spectral function well below
$\omega=\Omega(r,T)$. This follows from the general properties
of the spectral function discussed in Appendix \ref{app:spec_decomp}.
In the case of HTL spectral function we have seen that 
while around the peak the spectral function appears to be Lorentzian, different structures dominate
the spectral function far away from the peak, in particular at very low frequency, see 
Fig. \ref{fig:spf_htl_T667}. Thus in addition to the parametrization of the peak of the spectral function,
we also need to parametrize the behavior of the spectral function at very low frequency, i.e. the 
%low $\omega$ tail. 
low energy tail. 
This part of the spectral function will affect the correlation function at large
values of $\tau$. Unfortunately, we do not have a well motivated form for this part of the spectral
function. Furthermore, for calculations in finite volume the spectral function is not a continuous function
but a discrete sum of delta functions with an envelope function of certain shape.  For small volumes
as used in the present calculations there could be significant distortion of the envelope function, since
the number of low lying energy levels and the corresponding number
of $\delta$ peaks is quite limited. 
This is especially the case for the  low $\omega$ tail as it extends over a large $\omega$-range below the dominant peak
position, including negative $\omega$ values.

The information we have on the different structures in the spectral function is also quite limited.
At small $\tau$ values only the first two cumulants  can be determined with the third cumulant being 
zero within the estimated errors. Therefore, at small $\tau$ the lattice data are only sensitive
to the position and the effective width of the dominant peak, and a Gaussian
form provides a simple  parametrization for this that avoids convergence problem in Eq. (\ref{eq:ETwilsonspecdec}). At larger $\tau$ the correlation function
is sensitive to the 
%low $\omega$ tail, 
low energy tail, 
i.e. the region $\omega \ll \Omega(r,T)$. In the previous
section we have seen that in this region also the third cumulant is non-zero, but cumulants
beyond the third one cannot be constrained by our lattice data.
While it would be tempting to parametrize the low $\omega$ tail of the spectral function by 
a series of delta functions avoiding any bias, in practice it is impossible 
to constrain all the corresponding parameters. We need to approximate this part 
of the spectral function by a single delta function
Thus a simple parametrization of the Wilson
line correlator
function consistent with the above observations is the following:
\begin{align}
    W(r,\tau,T) =& A_P(r,T) \exp(-\Omega(r,T)\tau+\Gamma_G(r,T) ^2 \tau ^2/2)+\nonumber\\&\label{GAnsatz}  A_{cut}(r,T) \exp(-\omega_{cut}(r,T)\tau),
\end{align}
with $A_{cut} \ll A_P$ and $\omega_{cut}\ll \Omega$
The first cumulant corresponding to this form will decrease linearly at small $\tau$, while
exhibiting a non-linear behavior for large $\tau$ as observed in our lattice results.   We performed correlated fits of our lattice data using 
Eq.~(\ref{GAnsatz}) and determined the parameters $A_P,~\Omega,~\Gamma_G,~A_{cut}$ and $\omega_{cut}$.
%Sample fits are shown in Fig. \ref{fig:GaussianFit_b7825} in terms of the %effective masses. 
The fits describe the lattice data very well, with possible exception of
the data at the smallest $\tau$ value. The details of these fits are
discussed in Appendix \ref{app:fit}.

The peak position, $\Omega(r,T)$ is shown in Fig. \ref{fig:pot} as function of the distance $r$ for
different temperatures. It shows no temperature dependence and agrees with the zero temperature static energy.
The fact that $\Omega$ is close to the zero temperature static energy can be easily understood from 
Fig. \ref{fig:demo_m1}. The subtracted first cumulant at smallest $\tau$ is already close
to the zero temperature plateau and shows and linear behavior at small $\tau$. A linear extrapolation
naturally gives the zero temperature static energy. The width of the dominant peak depends on
the specific parametrization of the spectral function and the Gaussian form has no physical
motivation. A parametrization independent definition
of the effective width could be the width at the half maximum. 
For a Gaussian form this means $\Gamma=\Gamma_G \sqrt{2 \ln 2}$. In Fig. \ref{fig:pot} we also show
the effective width $\Gamma$ as function of the distance, $r$ at different temperatures. We see
that $\Gamma$ increases with increasing $r$. We also see that when plotted as function of $rT$
the effective width in temperature units shows very little temperature dependence. This is expected
at very high temperature, but not in the temperature range studied by us. 
For the other two fit parameters we find that $\omega_{cut} \ll \Omega$ and $A_{cut} \ll A_P$
in accordance with our expectations. The same parametrization of the spectral function has been
used in the analysis of NRQCD bottomonium correlators at non-zero temperature \cite{Larsen:2019bwy,Larsen:2019zqv}. It has been observed that different bottomonium states have thermal width, but no significant mass shift has been
observed. Furthermore, the thermal width turned out to be larger for higher
lying bottomonium states that have larger size \cite{Larsen:2019bwy,Larsen:2019zqv}. Thus the thermal modification of static
$Q\bar Q$ states and bottomonium is quite similar. 
Furthermore, the bottomonium
Bethe-Salpeter amplitudes also do not show large
temperature modifications \cite{Larsen:2020rjk}. 
Using this result a potential model analysis resulted
in a potential that has a real part which is identical to 
the zero temperature static energy \cite{Shi:2021qri}.

\subsection{Determination of the ground state peak via the HTL-motivated method}
\label{sec:BalaDatta}
In this section, we will use the method of \cite{Bala:2019cqu} to obtain the position and the width of the dominant peak
of the static $Q\bar Q$ spectral function. In this method, the peak position $\Omega(r, T)$ and peak width $\Gamma(r, T)$ of the dominant peak are interpreted as the real and imaginary part of thermal static energy $E_{s}(r, T)$.
Quantitatively, one assumes the following limit exists 
\begin{equation}
E_{s}(r,T)=\lim_{t\rightarrow\infty} i\frac{\partial \log W(r,t,T)}{\partial t}=\Omega(r,T)-i\Gamma(r,T).
\label{p-def}
\end{equation}
Here, $W(r,t,T)$ is the real-time correlator obtained as the
%by doing 
Fourier transform of the spectral function $\rho_{r}(r,\omega)$.

Below the crossover temperature 
%from transfer matrix argument 
$W(r,\tau,T=0)\sim \exp (-\Omega \tau)$ for  $0\ll\tau\ll1/T$
follows from a transfer matrix argument, 
and therefore the above limit exists trivially. However, above the crossover temperature, 
the existence of the limit in Eq.~\ref{p-def} 
is a non-trivial statement
%. The existence of static energy at finite temperatuKajimoto:2017relre is
whose consequences are 
important for various applications like in open quantum systems \cite{Kajimoto:2017rel} or the construction of vector current spectral function \cite{Burnier:2007qm}. 

The definition of the static energy involves a $Q\bar Q$ correlator, whose large-time behavior %at large Minkowski time. The large Minkowaski time behavior of the $Q\bar Q$ correlator 
is governed by the lowest, dominant feature of $\rho_{r}^{med}(\omega, T)$. As a consequence,
%, as a result for the determination of static thermal energy, 
it is sufficient to determine the structure of the dominant peak of the spectral function to obtain the thermal static energy. 
%In this section therefore 
In this HTL-motivated method, we model the dominant peak of the spectral function such that the above limit exists. 

In \cref{sec:corrHTLcmp} we mentioned that for the leading-order HTL 
%perturbative 
correlator the limit in Eq.~(\ref{p-def}) exists. We
%and we 
observed that this is possible because the correlator can be written as a combination of a part linear in $\tau$ and a part periodic in $\tau$,  cf.~Eq.~\eqref{lnW_htl}. 
Now let us see whether the non-perturbative data near $\tau\sim 1/(2T)$ could
%can 
be parameterized by a combination of periodic and linear parts in $\tau$. 
Motivated by this we write the following %general 
parameterization of the $Q\bar Q$ correlator near $\tau\sim {1}/{(2T)}$,

\begin{align}
\log\,W(r,\tau,T)=-\Omega(r,T)\,\tau+\nonumber\\
\int_{-\infty}^{\infty} d\omega \Sigma_r(\omega,T) (e^{-\omega \tau}+e^{-\omega (\beta-\tau)})\nonumber\\
	+\mathrm{const}
\label{htl_param}
\end{align}

The condition for the limit in Eq.~(\ref{p-def}) is then 
\begin{align}
	\lim_{t\rightarrow \infty}\int_{-\infty}^{\infty} d\omega \Sigma_{r}(\omega) \omega 
	(e^{-i\omega t}-e^{-\omega (\beta-it)})
	=\mathrm{const}.
\end{align}
Using the fact  %$\lim_{t\rightarrow \infty}[\exp(i\omega t)-\exp(\omega(\beta-i t))]=-2\pi i \omega \delta(\omega)$, 
$\lim\limits_{t\rightarrow \infty}(e^{-i\omega t}-e^{-\omega (\beta-it)})=-2\pi i \omega \delta(\omega)$,
we observe that the above limit will exist only 
%when 
if 
$\Sigma_r(\omega)\sim \frac{1}{\omega^2}$ as $\omega \rightarrow 0$.

Without loss of generality we can introduce a factor $(1+n_b(\omega))$ and write
\begin{align}
\Sigma_r(\omega,T)=(1+n_b(\omega))\,\eta_{r}(\omega,T).
\label{htl_spf_exp}
\end{align}

Since the function $(1+n_{b}(\omega))$ already contains a factor of $1/\omega$, the $\eta_{r}(\omega, T)$ should also contain $1/\omega$ at small $\omega$. 
Then Eq.~(\ref{htl_param}) becomes,

\begin{align}
\log\,W(r,\tau,T)=-\Omega(r,T)\,\tau+\nonumber\\
	\int_{-\infty}^{\infty} d\omega \eta_r(\omega,T) \frac{\exp(\omega \tau)+\exp(\omega (\beta-\tau))}{\exp(\omega \beta)-1}\nonumber\\
        +\mathrm{const}.
\label{htl_param1}
\end{align}

$\eta_{r}(\omega, T)$ can only be an odd function $\omega$. 
The most general expansion of the function $\eta_{r}(\omega, T)$ consistent with the existence of thermal static energy can then be written as 
\begin{align}
\eta_r(\omega,T)=\frac{c_0(r,T)}{\omega}+c_1(r,T) \omega + c_2(r,T) \omega^3+ \dots .
\label{htl_spf_exp}
\end{align}

We emphasize that Eq.~(\ref{htl_param1}) is a completely non-perturbative parametrization. The only information we take from HTL perturbation theory is the possible $\tau$ dependence, which can give rise to the limit in Eq.~(\ref{p-def}).

Using this form of $\eta_r(\omega)$, the $\tau$ dependence of the integration in Eq.~(\ref{htl_param}) can be computed and we obtain the following expression for the Wilson line correlator,
\begin{align}
\log W(r,\tau,T)=-\Omega(r,T)\tau+\frac{c_0(r,T)}{\pi} \log[\sin(\pi \tau T)]+ \nonumber\\
	\sum_{l=1}^{\infty} \frac{c_{l+1}(r,T)}{\pi} (2l-1)! T^{2l} \left( \zeta\left(2l,\tau T\right)+\zeta\left(2l,1-\tau T\right) \right)
	\nonumber\\+\mathrm{const}(r,T).
\label{htl_m1_exp}
\end{align}

Using this HTL parametrization it is easy to check that the limit in Eq.~(\ref{p-def}) exists
and the static energy is given by
\begin{equation}
E_{s}(r,T)=\lim_{t\rightarrow\infty} i\frac{\log W(r,t,T)}{\partial t}=\Omega +i c_{0} T.
\end{equation}

From this equation we can identify the imaginary part of static energy $\Gamma(r,T)=-c_{0} T$. Furthermore, at large Minkowski time higher-order terms, $l\ge 1$, do not contribute, i.e. the large Minkowski time behavior of $W(r,t, T)$ is determined by $\Omega$ and $c_0$, which in this case can be identified with the real and imaginary part of the static energy. 

It has been found in \cite{Bala:2019cqu} for quenched QCD with unimproved gauge action and large $N_{\tau}$ 
that with the expression in Eq.~(\ref{htl_m1_exp}), which is motivated from leading-order HTL perturbation theory, 
a reasonable number of data points around $\tau \sim 1/(2T)$ could
indeed be described by the above expression. 
However, with $N_{\tau}=12$ the number of data points available for fitting  near 
%$\tau/a\sim\frac{1}{2Ta}$ 
$\tau \sim 1/(2T)$ 
becomes small.

If we focus on the narrow region around $\tau=1/(2 T)$, and if the higher order terms in Eqs. (\ref{htl_spf_exp}) and (\ref{htl_m1_exp}) can be neglected, 
%and 
we
can fit the lattice results on the first moment with the form
\begin{align}
m_{1}(r,n_\tau=\tau/a)a=\log\left(\frac{W(r,n_\tau,N_{\tau})}{W(r,n_\tau+1,N_\tau)}\right)\nonumber\\
		=\Omega(r,T)\,a-\frac{\Gamma(r,T)a N_\tau }{\pi }\,\log\left[\frac{\sin(\pi n_\tau/N_\tau)}{\sin(\pi  (n_\tau+1)/N_\tau)}\right]
\label{BDfit}
\end{align}
We performed fits of our $N_{\tau}=12$ lattice data for $m_1$ 
for $\tau/a=5,6,7$ using Eq.~(\ref{BDfit}) to determine $\Omega(r,T)$ and $\Gamma(r,T)$. The details of these fits can be found in Appendix \ref{app:fit}. 

\begin{figure}
    \centering
    \includegraphics[width=8cm]{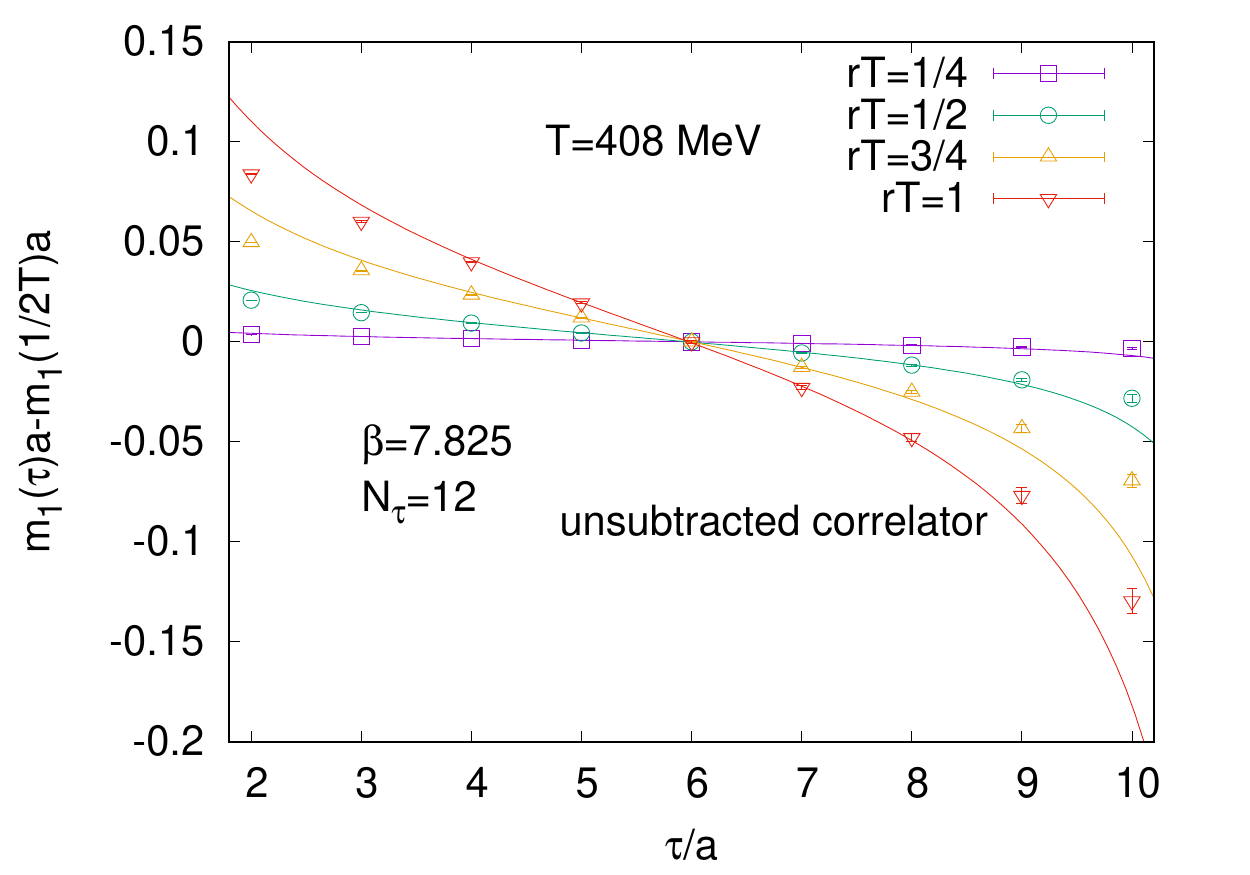}
    \includegraphics[width=8cm]{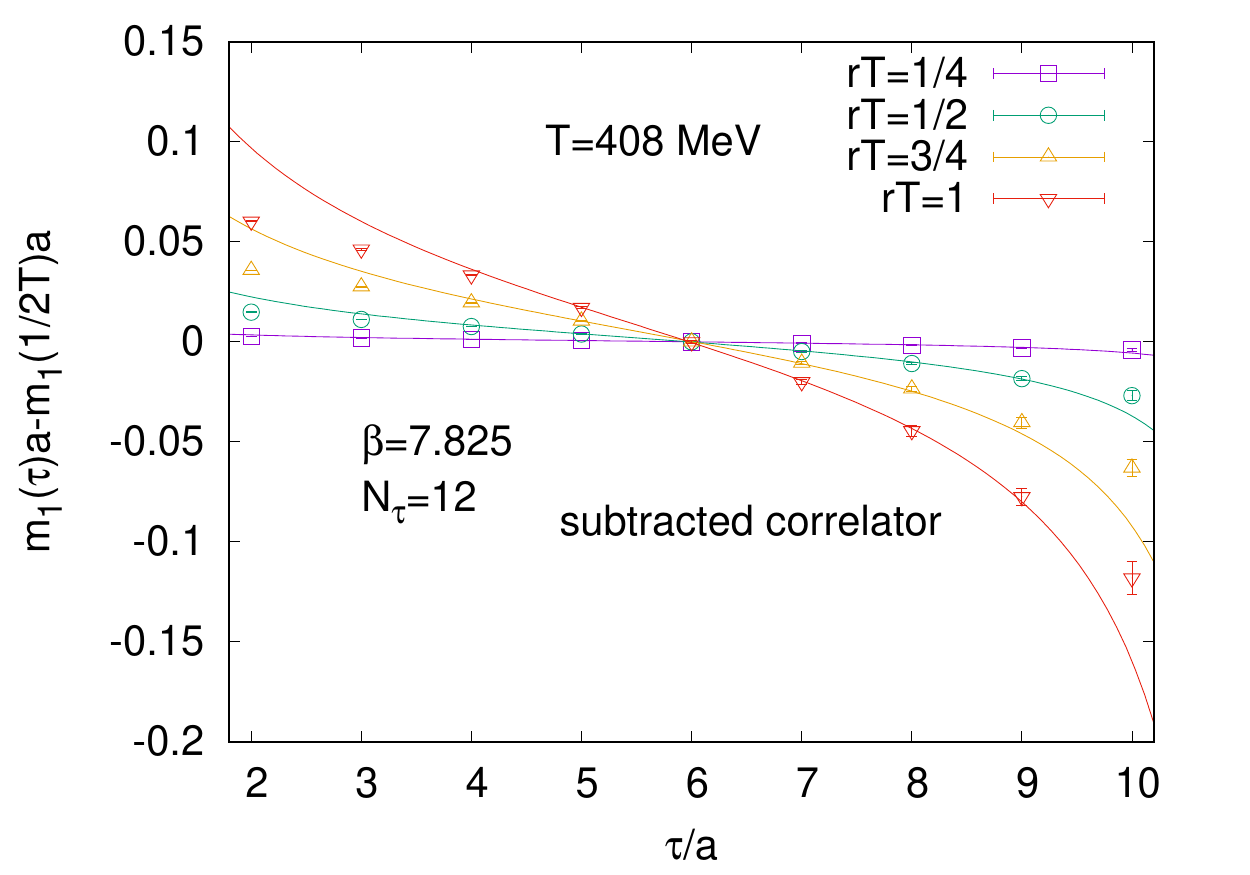}
	\caption{Sample fit of the lattice result (Top) unsubtracted correlator and (Bottom) subtracted correlator
    to Eq. (\ref{BDfit}), see text.}
    \label{fig:BDfit}
\end{figure}

\begin{figure*}
    \centering
    \includegraphics[width=8cm]{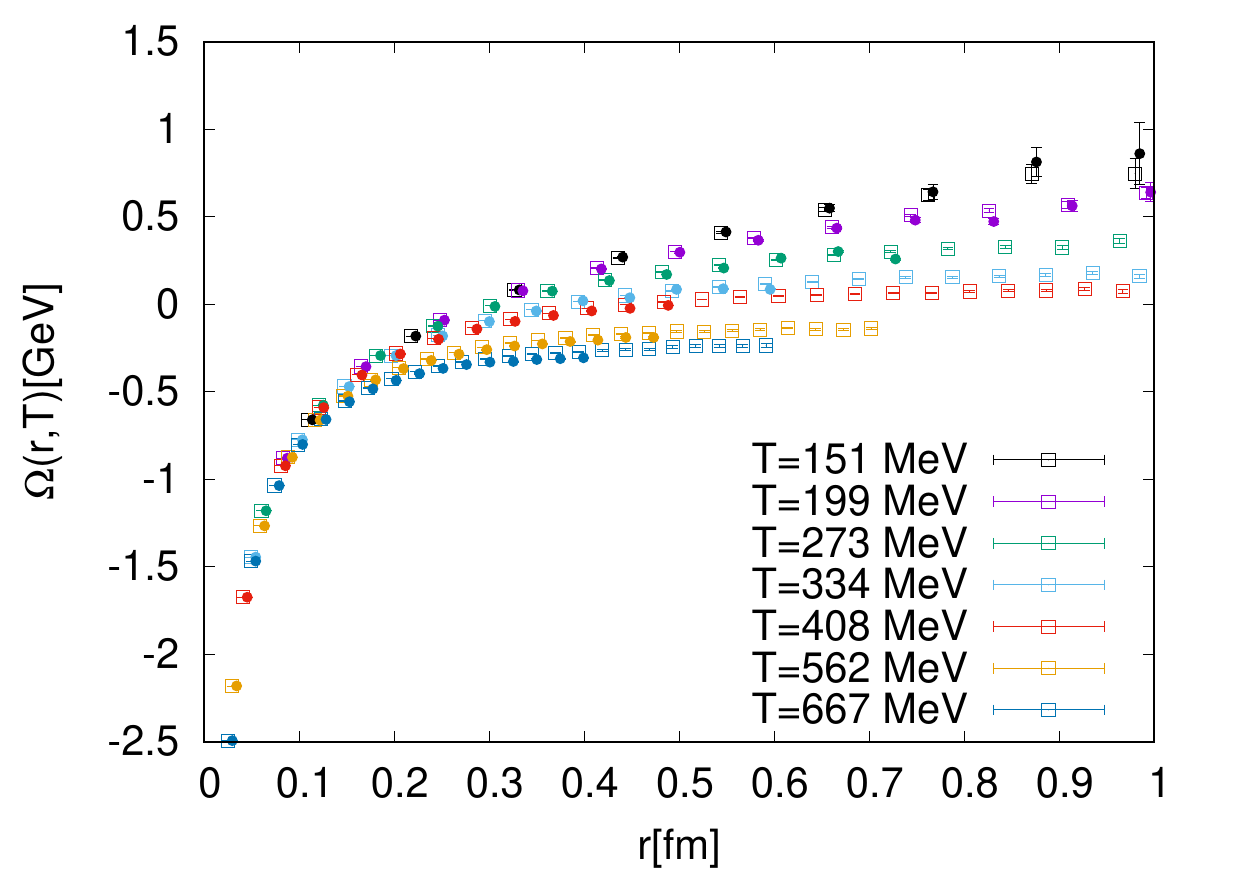}
	\includegraphics[width=8cm]{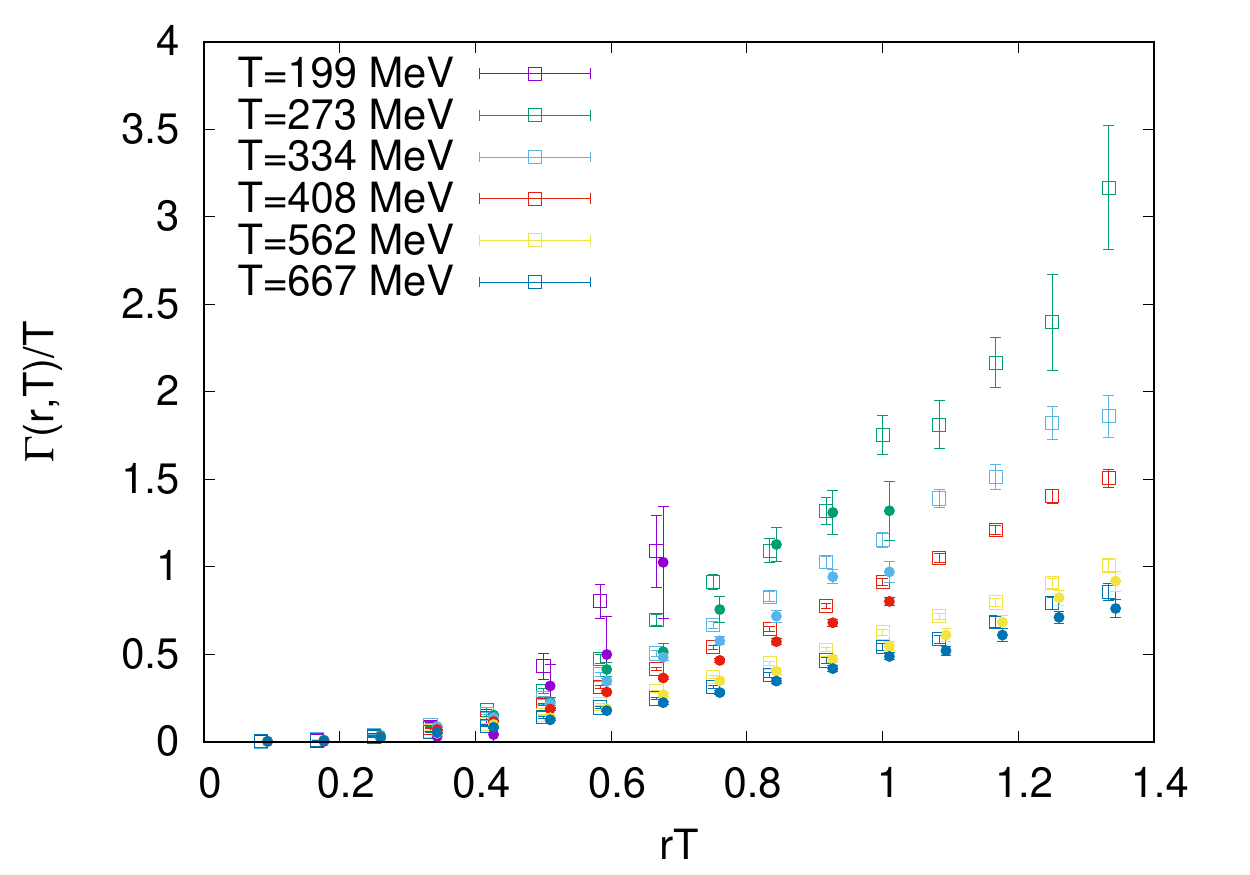}
    \caption{The peak position (left) and the width (right) from the HTL motivated method as function of $r$ at different temperatures.
    The open(closed) symbols corresponds to real and imaginary part from unsubtracted(subtracted) correlator.}
    \label{fig:BDpot}
\end{figure*}

A sample fit for both unsubtracted and subtracted data has been shown in Fig.~\ref{fig:BDfit}. The ansatz also describes some data points outside the fitting range.
The smaller $\tau$ and larger $\tau$ behavior are not expected to be described by the above Ansatz, as it only describes the dominant peak of the spectral function.
In Fig.~\ref{fig:BDpot} we show $\Omega(r,T)$ and $\Gamma(r,T)$ from these fits as function
of $r$ at different temperatures. The peak position $\Omega(r, T)$  and width $\Gamma(r, T)$ for 
subtracted and unsubtracted correlators are very close to each other. 
This is expected because we only consider $\tau$ values around
$1/(2T)$, where the contribution of
the high $\omega$ part of the spectral function
is small.  The peak position, $\Omega(r, T)$ shows significant temperature
dependence and differs from the zero temperature potential. The width of the peak, $\Gamma(r, T)$
increases with increasing $r$. Furthermore, $\Gamma(r, T)$ does not scale with the temperature unlike
in the case of Gaussian fits in the temperature range explored by us. We also find
that in the temperature region studied by us  $\Gamma(r, T)$ is larger than the HTL result.

\begin{figure}
    \centering
    \includegraphics[width=8cm]{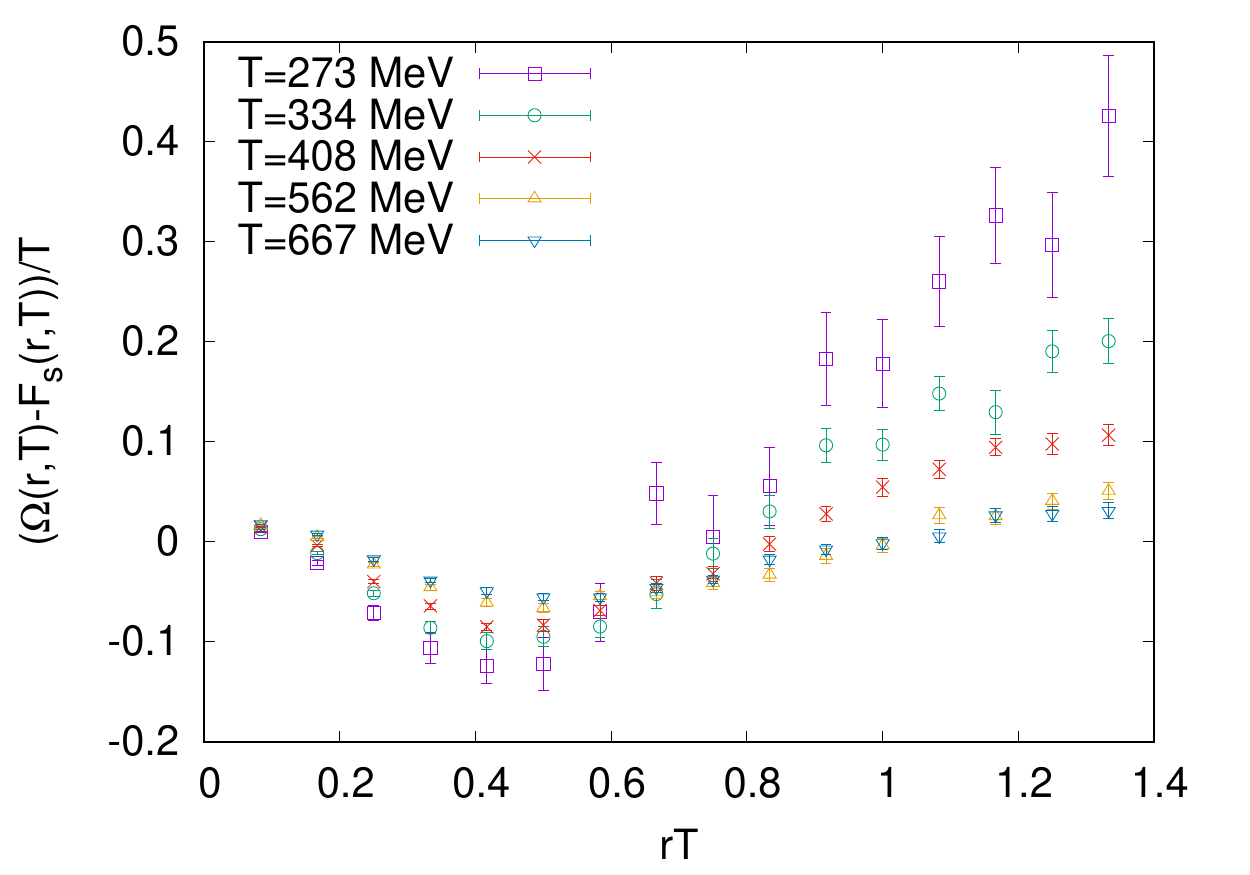}
	\caption{The difference between the peak position $\Omega(r,T)$ and singlet free energy $F_s(r,T)$ at different temperatures.}
    \label{fig:f_real}
\end{figure}

Another widely studied quantity at finite temperature is the singlet free energy $F_S(r, T)$, see e.g. \cite{Bazavov:2018wmo}.
As mentioned above in leading-order HTL perturbation theory, the singlet free energy and the real part of the static energy are 
the same. From Fig. \ref{fig:f_real} we see that even non-perturbatively the difference between $\Omega(r,T)$ and $F_S(r,T)$ is 
very small, while the difference between the zero temperature static energy and $F_S(r,T)$ is even smaller for $rT < 0.4$~\cite{Bazavov:2018wmo}. 
%.
This is very similar to the findings of the calculations in quenched QCD, where 
smeared Wilson loops have been used \cite{Bala:2019cqu}.

\begin{figure}
    \centering
    \includegraphics[width=8cm]{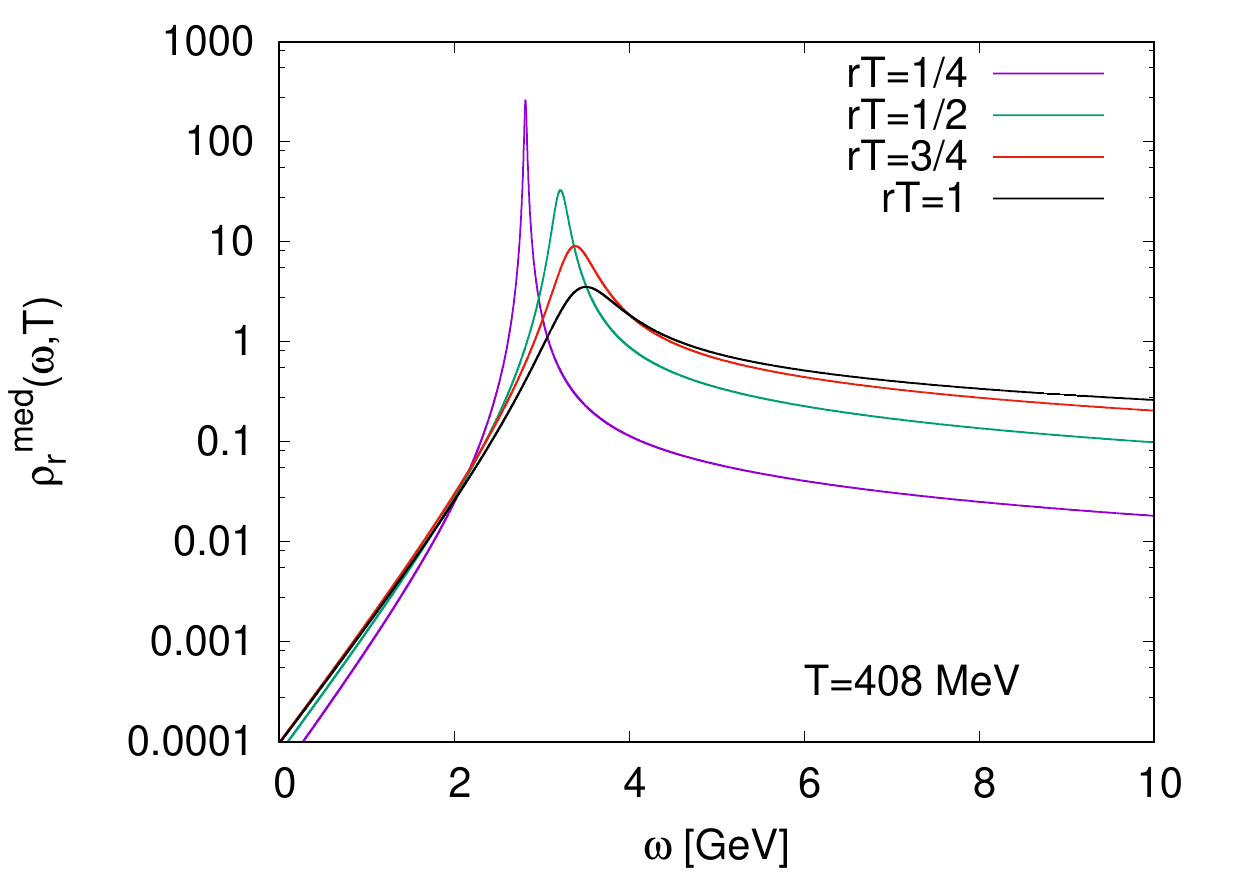}
        \caption{Dominant peak of spectral function for T=408 MeV at various distances from HTL-motivated method.}
    %\label{fig:spf_htl}
    \label{fig:spf_baladatta}
\end{figure}

It is straightforward to continue the parametrization
of the Wilson line correlator given by Eq. (\ref{htl_m1_exp}) to Minkowski time and 
then calculate the dominant peak of the spectral function $\rho_r^{med}(\omega,T)$,
%:
\begin{equation}
\rho_r^{med}(\omega,T)=\int_{-\infty}^{\infty} W(r,t,T)\exp(i\omega t) \,dt 
,
\label{htl_mot_sp}
\end{equation} 
%The dominant peak of this spectral function 
which has been plotted in Fig. \ref{fig:spf_baladatta}. The dominant peak shows a qualitatively similar feature with the leading-order HTL spectral function, see Fig. \ref{fig:spf_htl_T667}.  

We would like to again mention that the spectral feature $\rho_r^{med}(\omega,T)$ plotted in the figure is not the full spectral function $\rho_{r}(\omega, T)$, but rather the 
dominant peak of $\rho_r^{med}(\omega,T)$ due to the thermal static energy. $\rho_r^{med}(\omega,T)$ is quite different from the full spectral function $\rho_{r}(\omega, T)$ for $\omega$ far away from its peak at $\Omega(r,T)$.
%part of the spectral function that is responsible for the thermal static energy to exist. 
%Therefore the spectral function in Fig.~\ref{fig:spf_baladatta} is not valid at very high $\omega$ and as well as at very small $\omega$. 
A similar situation arises also while calculating the $Q\bar Q$ potential in hadronic phase. In this case it is well known that the dominant peak of the spectral function is the Dirac delta function, and this describe only the plateau region of $m_1$. 

The integration in Eq.(\ref{htl_mot_sp}) can be performed exactly \cite{Bala:2020tdt} and near the peak, where $\rho_r^{med}(\omega,T)$ describes the spectral function reliably,
%the spectral function is reliable, 
it can be approximated by
\begin{align}
\rho_r^{med}(\omega,T) &\approx \sqrt{\frac{2}{\pi}} \ \frac{\Gamma(
r,T)}{(\Omega(r,T) - \omega)^2 \; +
    \; \Gamma(r,T)^2} \nonumber \qquad \\
	&{} |\Omega(r,T) - \omega|, \;
\Gamma(r,T) \ll T \nonumber .\\
\end{align}
This is expected as we already assumed the limit in Eq.~(\ref{p-def}) exists.

\subsection{Determining the ground state peak via the Pade rational approximation}
\label{sec:Pade}
So far we have used two methods to extract the properties of the ground state spectral peak, which both required some form of modeling input. In the spectral fit approach it amounts to the choice of fitting with a spectral function for which the second moment is much larger than the higher moments. A Gaussian shape for the dominant peak and a number of supporting delta peaks is one possible such choice. In the Bala-Datta method of Ref.~\cite{Bala:2019cqu} the applicability of a non-standard spectral representation (see Eq. (\ref{htl_param})) is assumed. Here we attempt to extract the potential with a model independent approach, based on the Pad\'e rational approximation. One reason to deploy the Pad\'e is that for the Symanzik gauge action, the spectral function of the Wilson
line correlator (or Wilson loop) is not positive for large $\omega$ when
the separation, $r$ is small as discussed in Ref. \cite{Bazavov:2019qoo}.
Thus Bayesian approaches designed to operate on positive spectral functions, deployed in the past, may not be reliable on this dataset. We see that at high temperatures an small separations distances the Bayesian approaches indeed fail when applied to the raw data. 

In the Pad\'e approach we first transform the Euclidean correlator data into Matsubara frequency space, after which we carry out a projection of the data onto a set of rational basis functions. It is these rational functions, which are then analytically continued. From the ensuing correlation functions in real-space frequencies, the spectral functions are obtained by taking the negative of the imaginary part, drawing on the analytic properties of the Lehmann representation
\begin{align}
\displaystyle 
    W(r,\tilde \omega_n,T)=\int d\omega \frac{1}{ \omega-i \tilde \omega_n} \rho_r(\omega,T).\label{eq:lehmannrep}
\end{align}
The Wilson line correlator is particularly well suited in this context. Since it does not contain the cusp divergences that plague the Wilson loop, approximations that exploit analyticity, such as the Pad\'e, are expected to work well. 

The Pad\'e approximation so far is not commonly deployed for spectral function reconstruction (for recent work see e.g. \cite{Tripolt:2018xeo}), since it is known to require extremely precise data to yield robust results, often beyond what a lattice simulation can provide in practice. In addition it is known that the Pad\'e does not respect the spectral representation of the input data (see e.g. \cite{Cyrol:2018xeq}), i.e. the reconstructed spectral function inserted into \cref{eq:lehmannrep} does not necessarily reproduce the input data.

All direct projection methods, such as e.g.\ those by Cuniberti \cite{Burnier:2011jq}, suffer from the fact that, in contrast to the Bayesian approach, the influence of the data uncertainty on the projection is not regularized. On the other hand the Pad\'e method has an advantage over the Bayesian approach in that it can exploit much more efficiently the smallness of statistical errorbars. In the Bayesian approach, reducing the statistical uncertainty, while leaving the number of datapoints fixed may result in increased ringing artifacts in practice, an issue that the Pad\'e does not suffer from in the same manner. And it is the exceptionally high statistics of the ensembles present in this study, which promise that a meaningful Pad\'e approximation can be carried out, as we will show in the following.

\begin{figure}[t]
\centering
\includegraphics[scale=0.58]{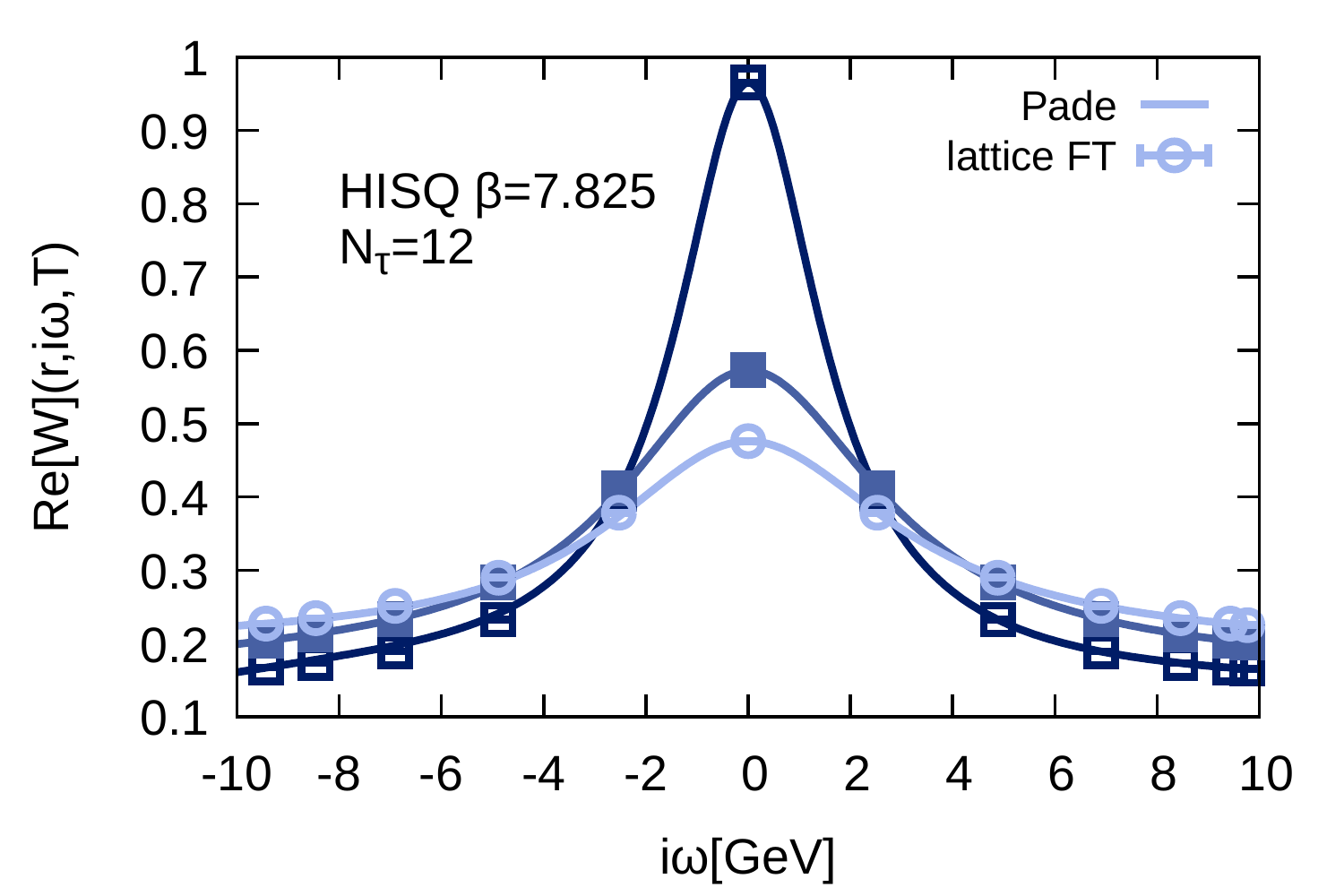}\vspace{-0.28cm}
\includegraphics[scale=0.58]{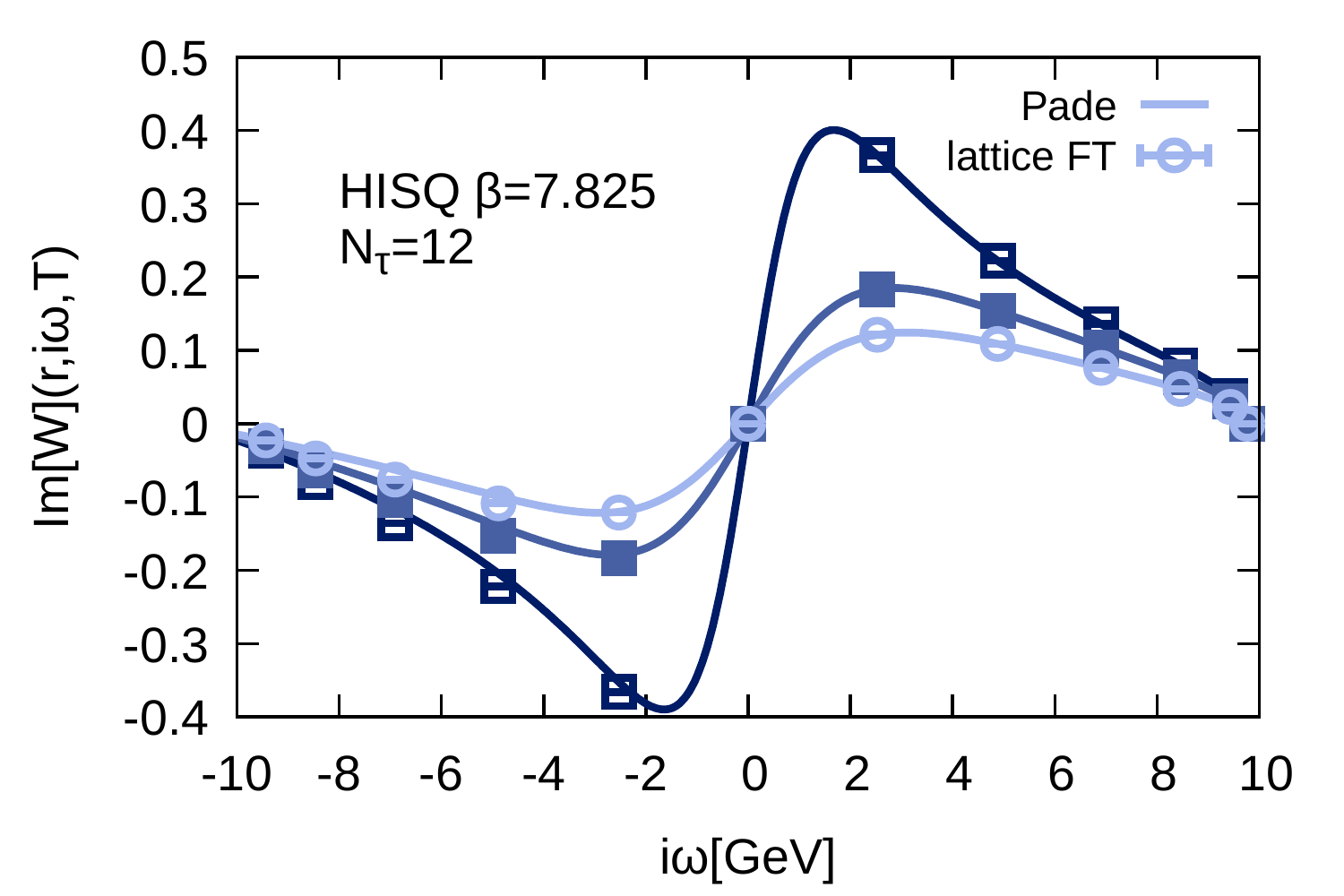}
\caption{Discrete Fourier transform of the $T>0$ Wilson line correlators at $T=407$MeV ($\beta=7.825$, $N_\tau=12$) at three spatial separation distances $r=0.03872$ fm, $r=0.1758$ fm and $r=0.2964$ fm . The top panel shows its real part, while the lower panel its imaginary part as colored symbols. The solid lines denote the Pad\'e approximation based on eight data points, which is subsequently used in the analytic continuation.}
\label{Fig:LatCorrFiniteTPade}
\end{figure}

In this study we implement the Pad\'e approximation in the form of a continued fraction according to the Schlessinger prescription \cite{Schlessinger:1968}. This particular approach amounts to a Pad\'e approximation in which the polynomial in the denominator carries at least the same order as that in the numerator or one higher order, leading to an expression that is able to robustly reproduce functions that decay at large frequencies, which is just the case for the Wilson line correlator \cite{Burnier:2013fca}. Note that it actually amounts to an interpolation of the data, which in contrast to a fitted rational approximation does not require us to carry out a costly minimization. We deploy the approximation on the Wilson line correlators in imaginary frequency space
\begin{align}
\displaystyle 
W(r,\tilde \omega_n,T)=\sum_{j=0}^{N_\tau-1}e^{ia \tilde \omega_n j } W(r,j a,T),~\tilde \omega_n=2 \pi n/aN_{\tau}. \label{eq:specdec}
\end{align}
A representative example is shown in Fig. \ref{Fig:LatCorrFiniteTPade}, where we plot as discrete data points in the top panel the real and in the bottom panel the imaginary part of the correlator at $T = 408$ MeV ($\beta = 7.825$ $N_\tau=12$) at three spatial distances $r=0.0387$ fm,$r=0.176$ fm and $r=0.296$ fm (dark blue to light blue).

Note that the discrete Fourier transform (DFT) applied to Eq. (\ref{eq:specdec}) does not reproduce the continuum Lehmann kernel but introduces corrections related to both the finite lattice spacing and available grid size. Since our subsequent strategy to extract the spectral function will rely on the continuum form of the Lehmann representation, we need to compensate for these artifacts, which we do in the spirit of the tree-level corrections of the lattice artifacts in
the static $Q\bar Q$ energy \cite{Necco:2001xg}, I.e. instead of using the naive Fourier frequencies $\tilde \omega_n$, we instead assign the Matstubara correlator datapoints to the eigenvalues of the discrete frequency operator
\begin{align}
\tilde \omega_n \rightarrow \omega_n=2 {\rm sin}\big(\frac{\pi n}{ N_\tau}\big)/a
\end{align}
The $\omega_n$ absorb the distortion of the frequency Brillouin zone in the UV and we may interpret the correlator as being expressed otherwise in its continuum form (we have checked that taking into account the DFT artifacts improves the stability of the Pade extraction using mock data). The deployment of the corrected frequencies also means that our correlators are plotted on non-equidistant frequency values in figure \ref{Fig:LatCorrFiniteTPade}.

With the Wilson line correlators not being symmetric in Euclidean time, their discrete Fourier transform is in general complex valued. The complex data along the corrected imaginary frequencies $\omega_n$ is interpolated by a continued fraction $C_{N_\tau}$ of the form
% \begin{align}
% \nonumber &C_{N_\tau}(r,i\omega,T)=\frac{W(r,\omega_0,T)}{1+} \frac{a_0(r,\omega-\omega_0,T)}{1+}\\ 
% &\frac{a_1(r,\omega-\omega_1,T)}{1+}\ldots\frac{a_{N_\tau-1}(r,\omega-\omega_{N_\tau-1,T})}{1+}.\label{Eq:ContFrac}
% \end{align}
\begin{align}
\nonumber&C_{N_\tau}(r,i\omega,T)=\frac{W(r,\omega_0,T)}{1+} \frac{a_0(r,T)[\omega-\omega_0]}{1+}\frac{a_1(r,T)[\omega-\omega_1]}{1+}\\
 &\ldots\frac{a_{N_\tau-2}(r,T)[\omega-\omega_{N_\tau-2}]}{1+}a_{N_\tau-1}(r,T)[\omega-\omega_{N_\tau-1}].\label{Eq:ContFrac}
\end{align}
For better readability the above continued fraction is expressed in the following way: each subsequent level of the continued fraction, instead of being written in the denominator of the preceding term, is listed as separate fraction to the left. The expression $1+$ in the denominator therefore indicates that the following fraction should be considered the next level of the continued fraction, concretely $\frac{A}{1+}\frac{B}{1+}C\equiv (A/(1+(B/(1+C))))$.

The complex coefficients are determined recursively by demanding that the rational approximation exactly reproduces the input data at each available frequency, leading to the following prescription 
% \begin{align}
% a_l(& r,\omega_{l+1}-\omega_l,T)=-\Big\{ 1+\\
% \nonumber &\frac{a_{l-1}(r,\omega_{l+1}-\omega_{l-1},T)}{1+}\frac{a_{l-2}(r,\omega_{l+1}-\omega_{l-2},T)}{1+}\cdots \\ \nonumber &\cdots\frac{a_{0}(r,\omega_{l+1}-\omega_{0},T)}{1-[W(r,\omega_0,T)-W(r,\omega_{l+1},T)]}\Big\}.
% \end{align}
\begin{align}
a_l(& r,T)(\omega_{l+1}-\omega_l)=-\Big\{ 1+\\
\nonumber &\frac{a_{l-1}(r,T)[\omega_{l+1}-\omega_{l-1}]}{1+}\frac{a_{l-2}(r,T)[\omega_{l+1}-\omega_{l-2}]}{1+}\cdots \\ \nonumber &\cdots\frac{a_{0}(r,T)[\omega_{l+1}-\omega_0]}{1-[W(r,\omega_0,T)/W(r,\omega_{l+1},T)]}\Big\}.
\end{align}
Applying this formula directly on complex valued imaginary frequency data amounts to a generalization of the resonances via Pad\'e method used e.g.\ in Ref. \cite{Tripolt:2016cya} to non-symmetric correlators in Euclidean time. The evaluation of a continued fraction is prone to accumulation of rounding errors, which is why we compute the $a_i$'s with at least 30 digits accuracy. This need for accuracy is independent of the amount of noise present in the underlying data and is a well known drawback of direct projection methods \cite{Burnier:2011jq}. The outcome of the interpolation, based on a subset of eight input datapoints (the seven positive Matsubara frequency data points and the one at the smallest negative available frequency), is shown as solid colored lines in Fig. \ref{Fig:LatCorrFiniteTPade}.
Substituting in the continued fraction the Euclidean frequencies by their Minkowski counterparts $C_{N_\tau}(r,i\omega,T)\to C_{N_\tau}(r,\omega,T)$ we explicitly implement the analytic continuation.

There are two equivalent ways to proceed. We may either compute the spectral function from $C_{N_\tau}(r,\omega,T)$ via the real-time relation
\begin{align}
\rho_r(\omega,T)\approx-\frac{1}{\pi}{\rm Im}[ C_{N_\tau}(r,\omega,T) ].
\end{align}
and carry out a similar analysis as that in previous studies based on Bayesian  spectral reconstructions. In that approach we locate the lowest lying peak structure in $\rho_r(\omega,T)$ and fit it with a skewed Lorentzian, embedded in a polynomial background of the form derived in Ref. \cite{Burnier:2012az}
\begin{align}
&\rho_r(\omega,T)\propto \\ \notag &\frac{|\Gamma(r,T)|{\rm cos}[{\rm Re}{\sigma_\infty}(r,T)]-(\Omega(r)-\omega){\rm sin}[{\rm Re} {\sigma_\infty}(r,T)]}{ \Gamma(r,T)^2+ (\Omega(r)-\omega)^2}\label{skewedrho}\\ \notag&+{c_0}(r,T)+{c_1}(r,T)(\Omega(r,T)-\omega) \\ \notag &+{c_2}(r,T)(\Omega(r,T)-\omega)^2\ldots. 
\end{align}

On the other hand we may ask whether the information encoded in the dominant spectral peak may be read-off from the real-time correlator directly in a more simple fashion. Indeed the peaks of the spectral function are but a projection of the pole structure of the underlying correlation function. Since we are in possession of the rational function approximation of the correlator, we can compute the pole structure explicitly from the roots of the polynomial in the denominator. The number of poles present, obviously depends on the degree of the Pade interpolation, but we find that varying the number of input points does not change the fact that one of the poles lies significantly closer to the real frequency axis than all other poles. This pole in turn leads to the dominant peak structure seen in the spectral function. 
\begin{figure}[t]
\centering
\includegraphics[scale=0.58]{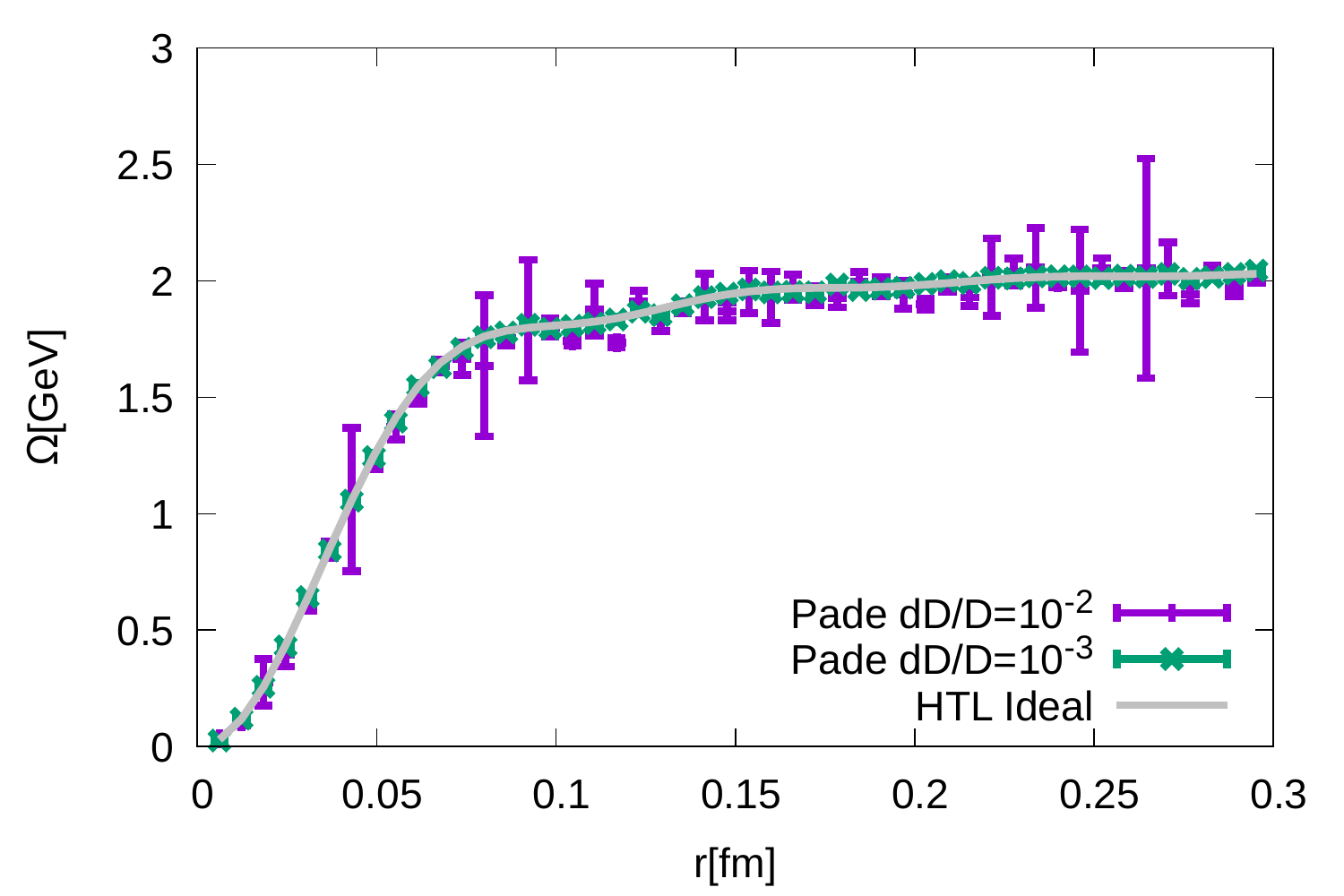}\vspace{-0.28cm}
\includegraphics[scale=0.58]{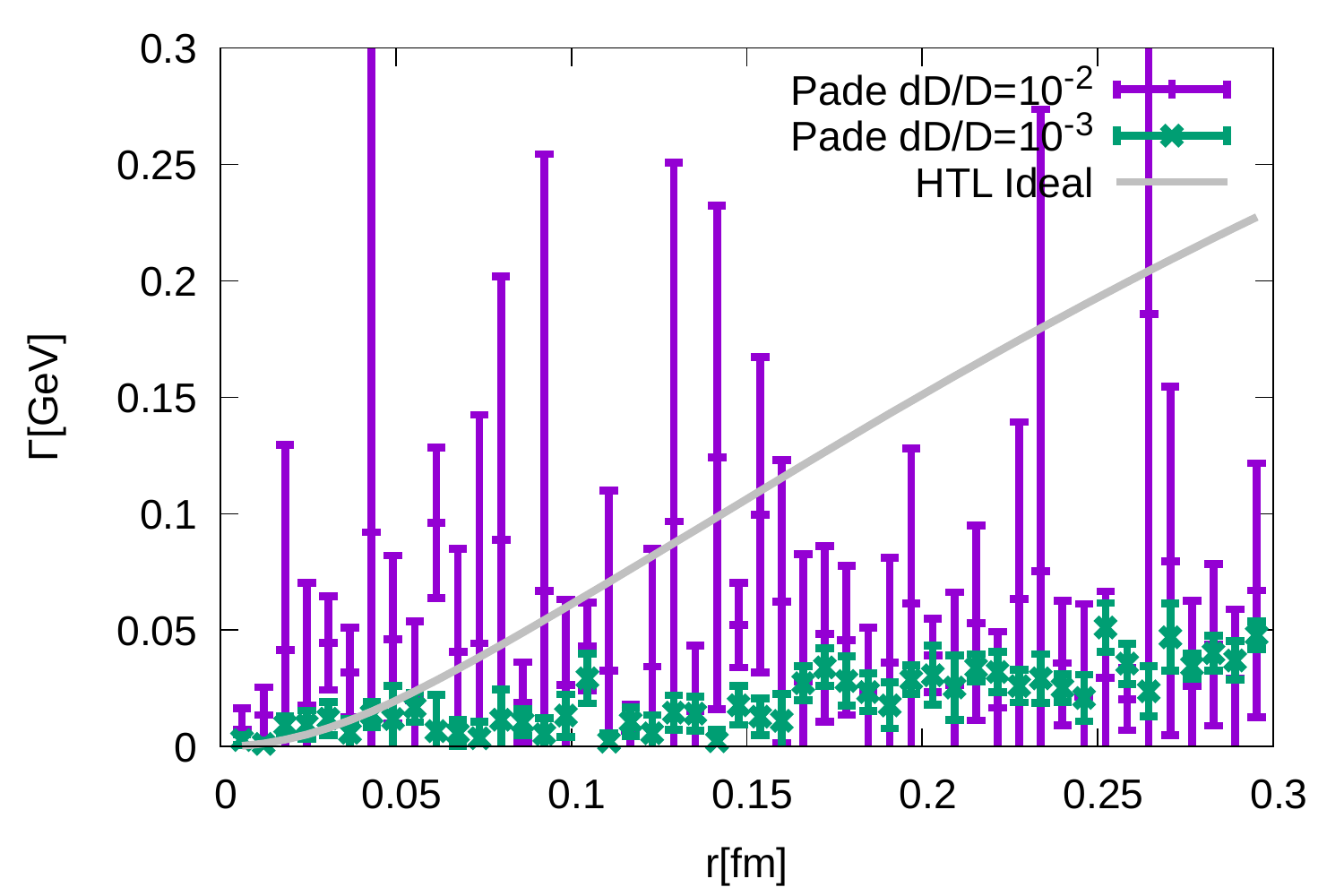}\vspace{-0.28cm}
\caption{Extraction of spectral position $\Omega$ and width $\Gamma$ of the dominant peak, based on Hard Thermal Loop mock data for $dD/D = 10^{-2}$ and $dD/D = 10^{-3}$ for $T=667$MeV. The error bars are obtained from Jackknive resampling.}
\label{Fig:HTL_PotPade}
\end{figure}
We have checked that both approaches give numerically consistent results for the position and width of the dominant spectral peak structure and therefore in the remainder of the study will analyze the spectrum directly via the poles.

In order to ascertain, whether the Pad\'e is a viable method for the exploration of spectral structures in practice we must assess its reliability in a realistic test scenario.  To this end  we carry out mock data tests based on HTL correlation functions. We deploy as starting point the ideal correlators computed for $T=667$MeV, discretized on $N_\tau=12$ points. This ideal data is distorted by Gaussian noise. Here 1000 samples of the correlator are generated, such that their mean exhibits constant relative uncertainties of either $\Delta D/D=10^{-2}$ or $\Delta D/D=10^{-3}$. Since the data in our lattice study is precise down to sub-percent level, the choice of one-percent relative error corresponds to a worse case scenario for the Pade analysis, while the one-permille error represents the best-possible scenario. 

We carry out the Pade interpolation and pole analysis based on a selection of eight noisy input data points, starting with the correlator at positive Matsubara frequencies. We have checked that adding or removing two datapoints does not significantly change the results, as well as have checked that a reordering of the datapoints in the construction of the continued fraction does not have any relevant effects. Note that when adding more and more datapoints in the construction of the Pad\'e, it will eventually become unreliable. The reason is that the redundancy of the Matsubara input data (symmetry of ${\rm Re}[W]$, anti-symmetry of ${\rm Im}[W]$) requires subtle cancellations to take place in the continued fraction. The optimal choice for stability we found lies at using $N_\tau/2+2$ datapoints.

The real and imaginary part of the dominant pole are plotted in the top and bottom panel of Fig. \ref{Fig:HTL_PotPade} respectively as colored data points. The analytically known values for the peak position $\Omega$ and its width $\Gamma$ are shown as gray solid lines. The errorbars here arise from a combination of Jackknife uncertainty, the differences in changing the number of datapoints by one or two, as well as the reordering in the construction of the continued fraction.

\begin{figure}[t]
\centering
\includegraphics[scale=.38]{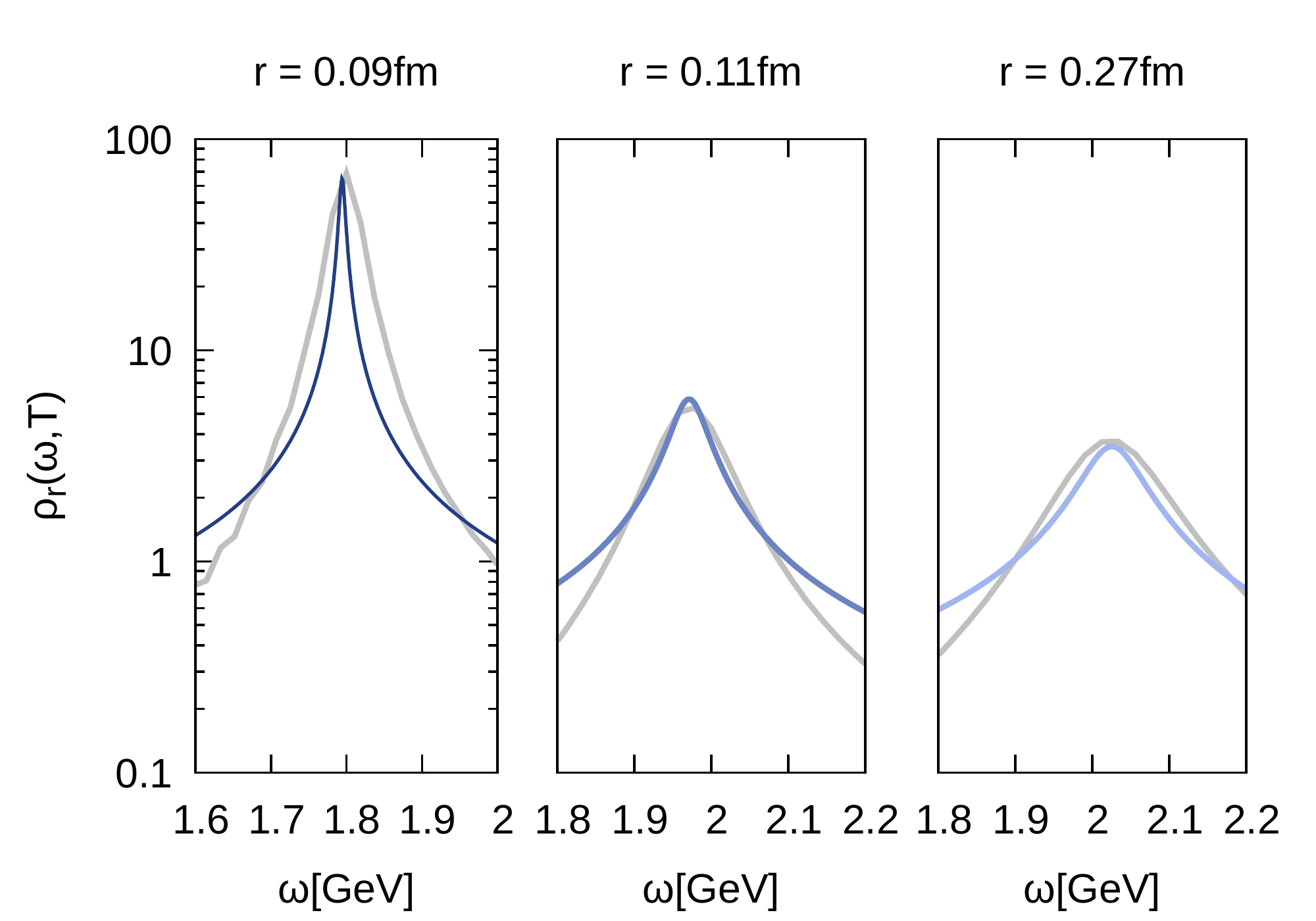}\vspace{-0.28cm}
\caption{Representative selection of spectral functions extracted from Hard thermal loop ideal data for $dD/D = 10^{-3}$ at $r =0.09,0.11$ and $0.27$ fm respectively using Pade (colored lines) vs. the analytic HTL result (gray lines).}
\label{Fig:HTL_SpectraPade}
\end{figure}
The HTL Pade pole analysis is very encouraging in that even under the adverse circumstances of relatively large statistical uncertainty of $\Delta D/D=10^{-2}$ it allows us to recover the position of the dominant peak well within uncertainties. For $\Delta D/D=10^{-3}$ the results are spot-on.

In Fig. \ref{Fig:HTL_SpectraPade} we have also computed several spectral functions for $\Delta D/D=10^{-3}$. We can see that the peak position is very well estimated. As expected the determination of the spectral width $\Gamma$ on the other hand is much more difficult and for the small number of datapoints present here ($N_\tau=12$), the results are not yet robust at $\Delta D/D=10^{-2}$ and tend to significantly underestimate the true value even for $\Delta D/D=10^{-3}$. Thus for the application to actual lattice data we will focus on 
%the 
extracting $\Omega$ in the following.

Having checked the limitations of the Pad\'e method in a non-trivial realistic test case, we proceed to apply it to our HISQ lattice data. We have carried out the pole analysis for Pad\'e interpolations based on different number of input datapoints. On $N_\tau=12$ lattices the results are unaffected by changing between seven to eleven input points and we arbitrarily decide to show the results based on eight. The uncertainty budget represented by the error bars includes the Jackknife errors, as well as variation due to change in the ordering when composing the continued fraction.

\begin{figure}
\centering
\includegraphics[scale=0.58]{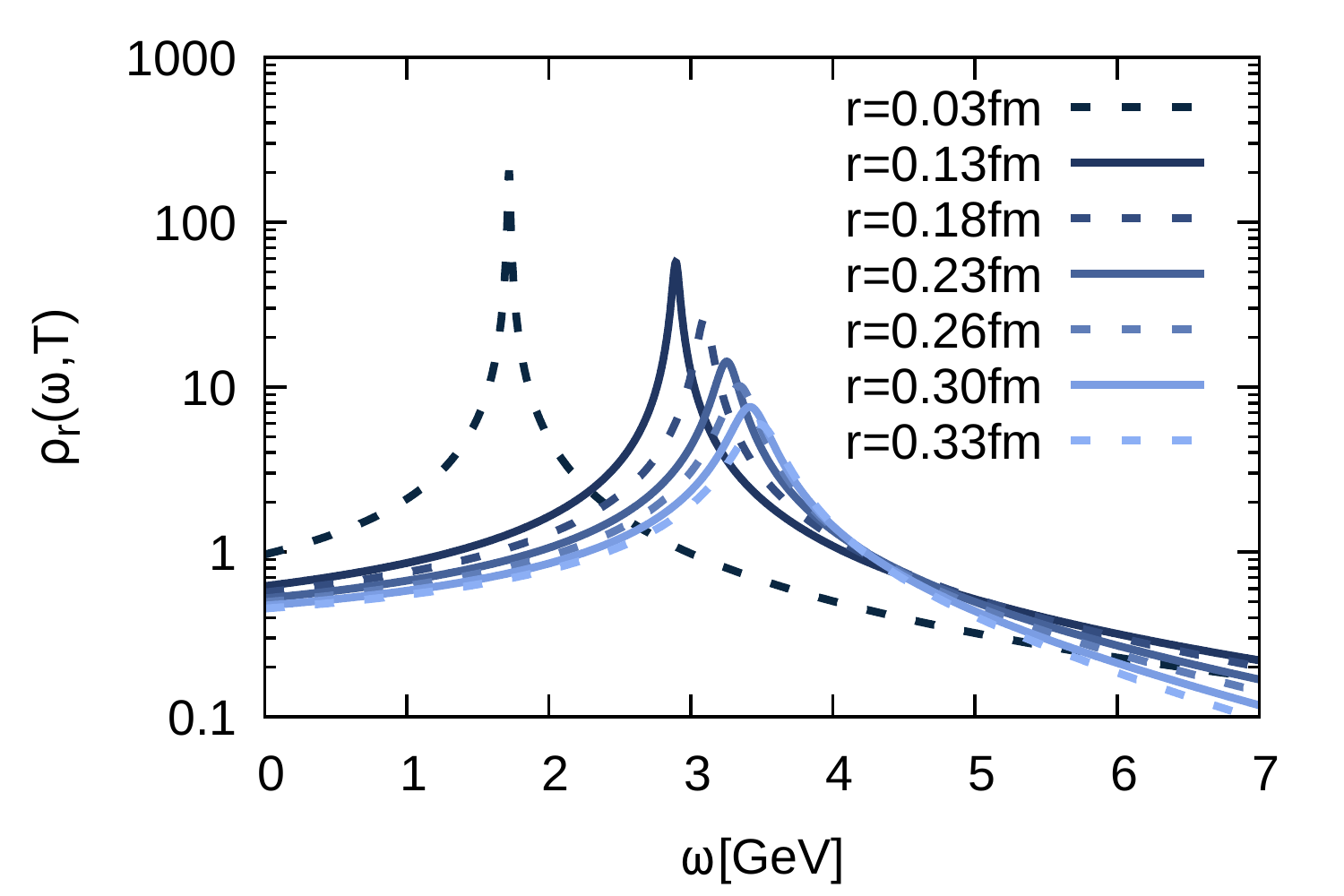}\vspace{-0.28cm}
\caption{Representative spectral functions obtained from the Pad\' e interpolation at $T=407$~MeV ($\beta = 7.825$ $N_\tau = 12$) for different separation distances. A single well defined peak structure of skewed Lorentzian form emerges from the analysis.}
\label{Fig:PadeSpectra}
\end{figure}

\begin{figure}[t]
\centering
\includegraphics[scale=0.58]{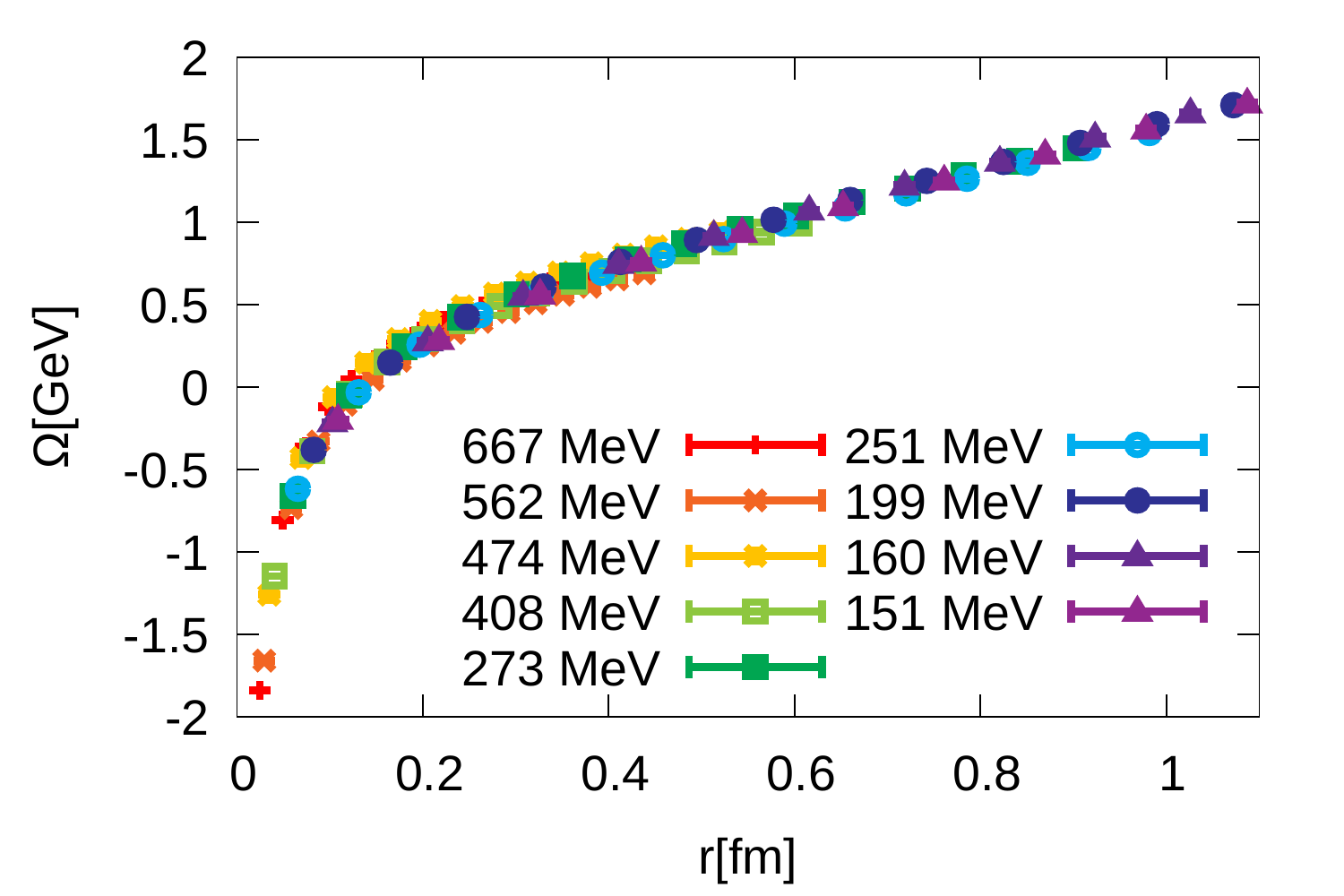}
\includegraphics[scale=0.58]{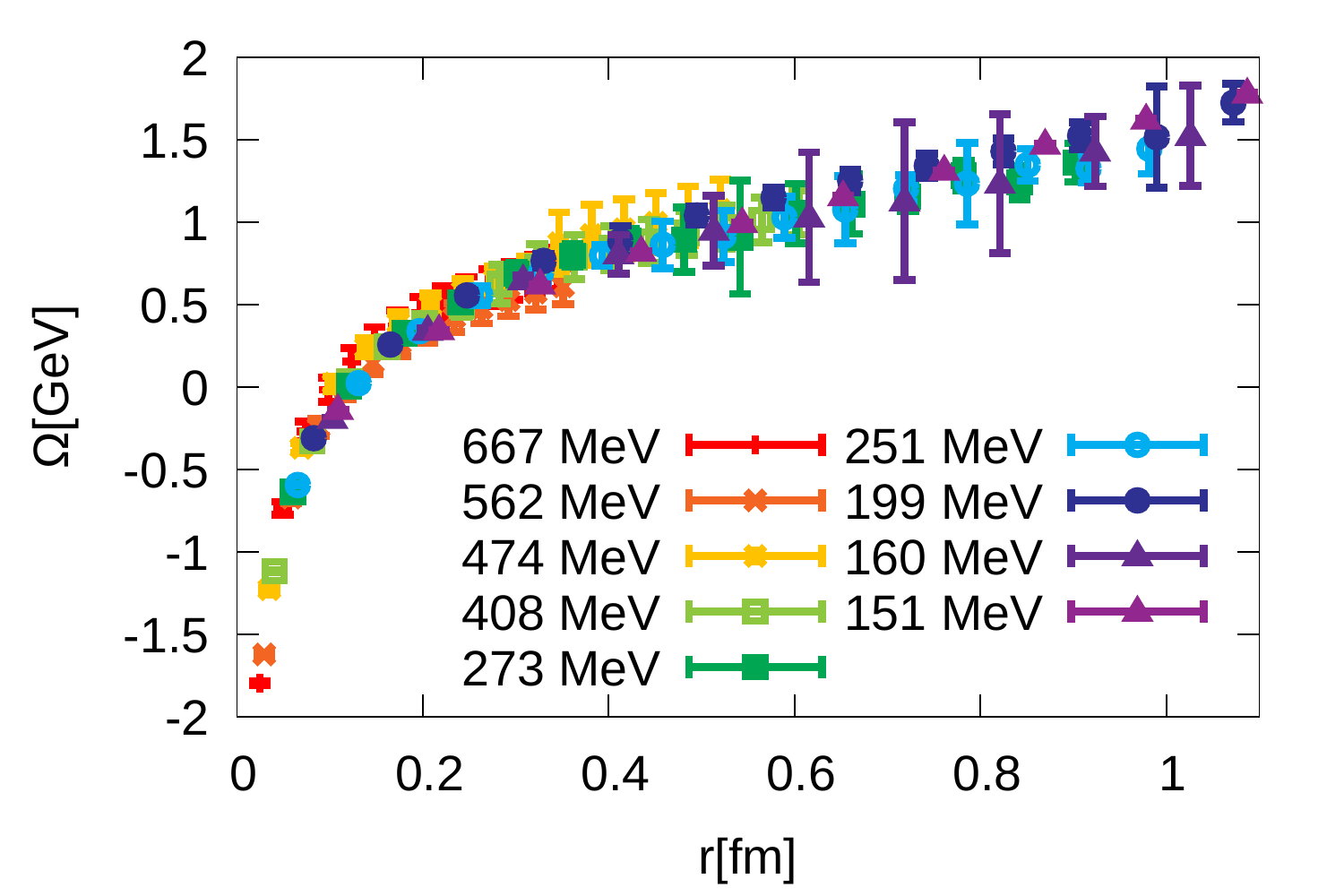}
\caption{$\Omega$ as a function of separation distance for different temperatures obtained from a Pad\'e pole analysis on $N_\tau = 12$. The figure on the top is obtained by using the unsubtracted correlator and the figure on the bottom is obtained using the subtracted correlator.}
\label{Fig:RealVPadeHISQ}
\end{figure}

\begin{figure}[t]
\centering
\includegraphics[scale=0.58]{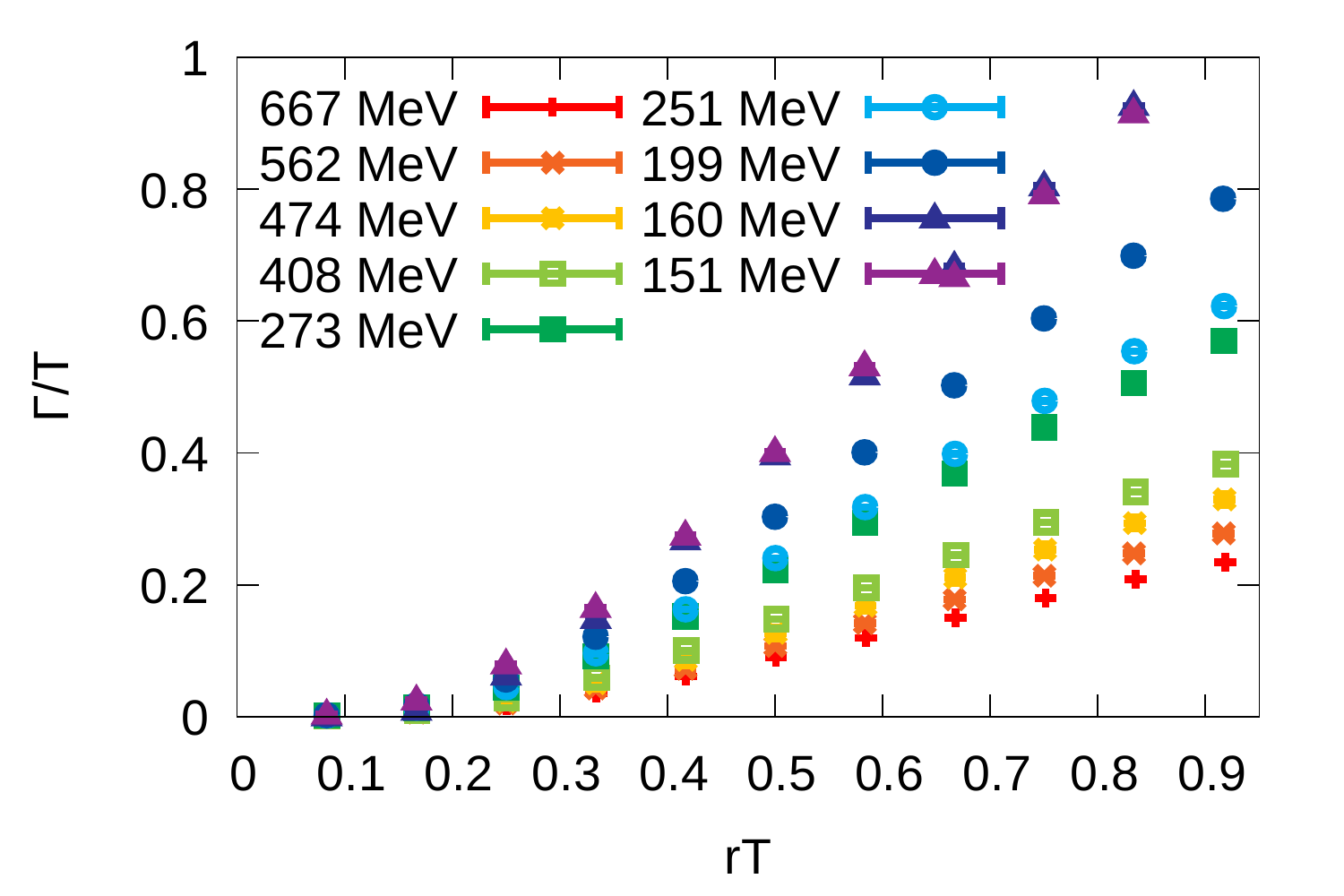}
\caption{ Width $\Gamma$  as a function of separation distance for different temperatures obtained from Pad\'e pole analysis for $N_\tau = 12$. The analysis is done using unsubtracted correlators.}
\label{Fig:ImVPadeHISQ}
\end{figure}

For the $N_\tau=12$ lattices we investigated, the Pad\'e interpolation yields one dominant pole close to the real axis manifesting itself as a well-defined skewed Lorentzian peak in the spectral function, as shown in Fig. \ref{Fig:PadeSpectra}. Reading off the values of the real-part of the pole as estimate for $\Omega$ we obtain the values plotted in Fig. \ref{Fig:RealVPadeHISQ}. The corresponding values for the imaginary part as estimate of $\Gamma$ are shown in Fig.  \ref{Fig:ImVPadeHISQ}. As we are cautioned about the quantitative reliability of the extraction of $\Gamma$ from the mock data analysis, we here present its values simply for completeness. We have carried out the analysis on both the subtracted and unsubtracted correlators (see \cref{sec:corrmom}) and found that the subtracted correlators are computed to a statistical precision which unfortunately is not high enough for the Pad\'e  to extract the value of $\Gamma$ with even statistical reliability.  

\begin{figure}[t]
\centering
\includegraphics[scale=0.58]{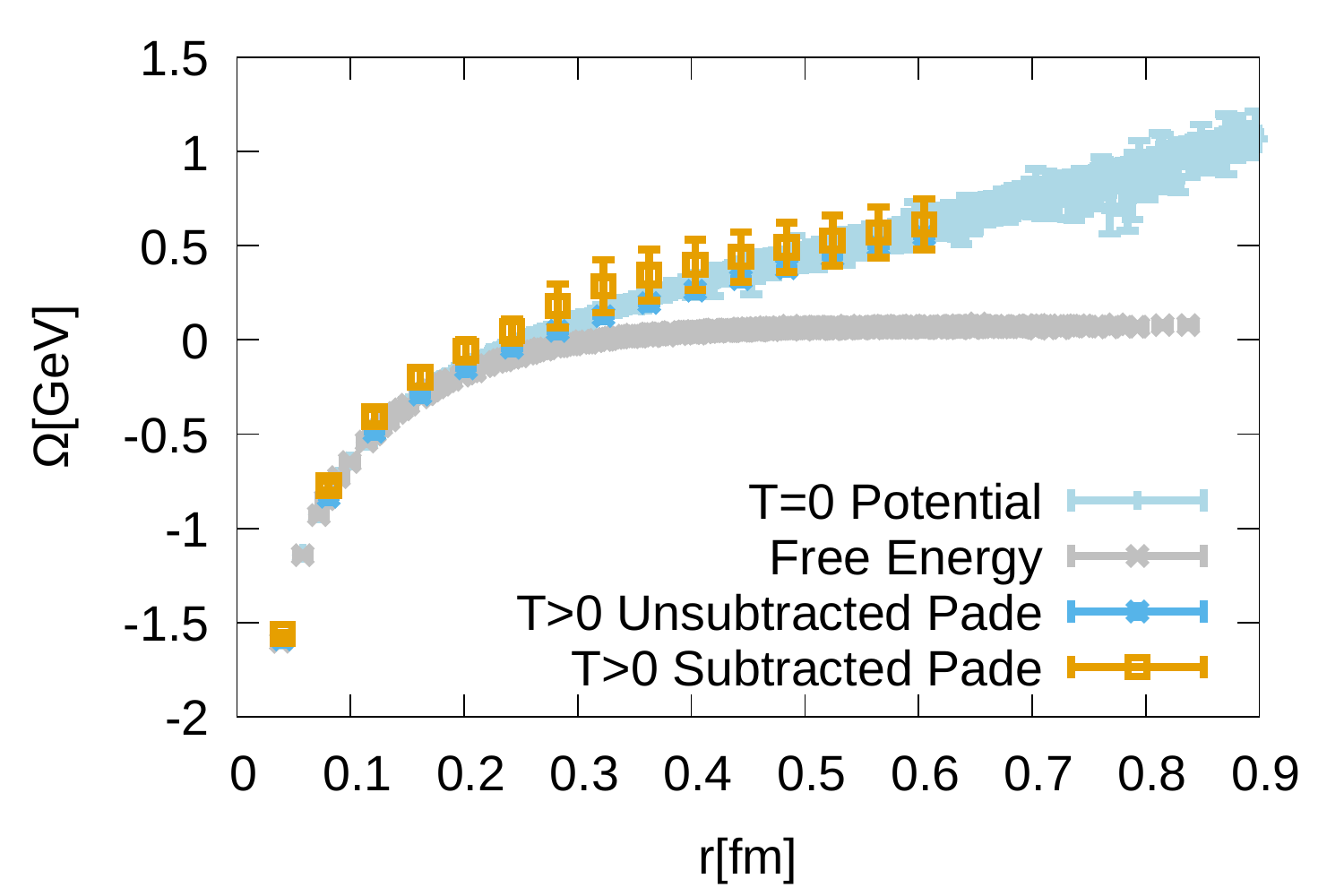}
\caption{Comparison of extracted $\Omega$ using subtracted and unsubtracted correlators using Pad\'e pole analysis with the T=0 effective mass and colour singlet free energy at $T=408$~MeV ($\beta = 7.825$, $N_\tau = 12$).}
\label{Fig:CompVPadeHISQ7825}
\end{figure}
The values the Pade analysis yields for $\Omega$ on the HISQ Wilson line correlators are similar to the results obtained from the model spectral function fits deployed in section \ref{sec:potfit}. We find that the values do not show any significant changes over a large temperature range. 

In Fig. \ref{Fig:CompVPadeHISQ7825} we pick out the results at $T=407$~MeV for a closer inspection. We plot $\Omega$, based on the subtracted and unsubtracted Euclidean correlator Pad\'e analysis at $T>0$ (orange and dark blue data points), along with the $T=0$ static energy (light blue datapoints) and the colour singlet free energy. The results obtained are in stark contrast to those of the method by Bala and Datta, in which at temperatures inside the QGP phase one does observe a deviation from the linear rise present in the hadronic phase. Our Pade results appear also in stark contrast to previous analyses of the spectral functions of Wilson lines from both quenched \cite{Burnier:2013nla,Burnier:2016mxc} and dynamical QCD \cite{Burnier:2014ssa,Burnier:2015tda} based on Bayesian methods. There a discernable change of $\Omega$ with temperature was found, more similar to the results of HTL-motivated method in this study. A previous Pade analysis of a subset of the HISQ data was discussed in Ref. \cite{Petreczky:2018xuh}. That analysis showed relatively large uncertainties, arising from the fact that less statistics was available and that the improved frequencies were not deployed. Within its sizable uncertainties, these results were consistent with the Bayesian studies but within $2\sigma$ would also encompass the result obtained here.

\subsection{Determining the ground state peak via Bayesian reconstruction}
\label{sec:Bayes}

\begin{figure}[t]
\centering
\includegraphics[scale=0.58]{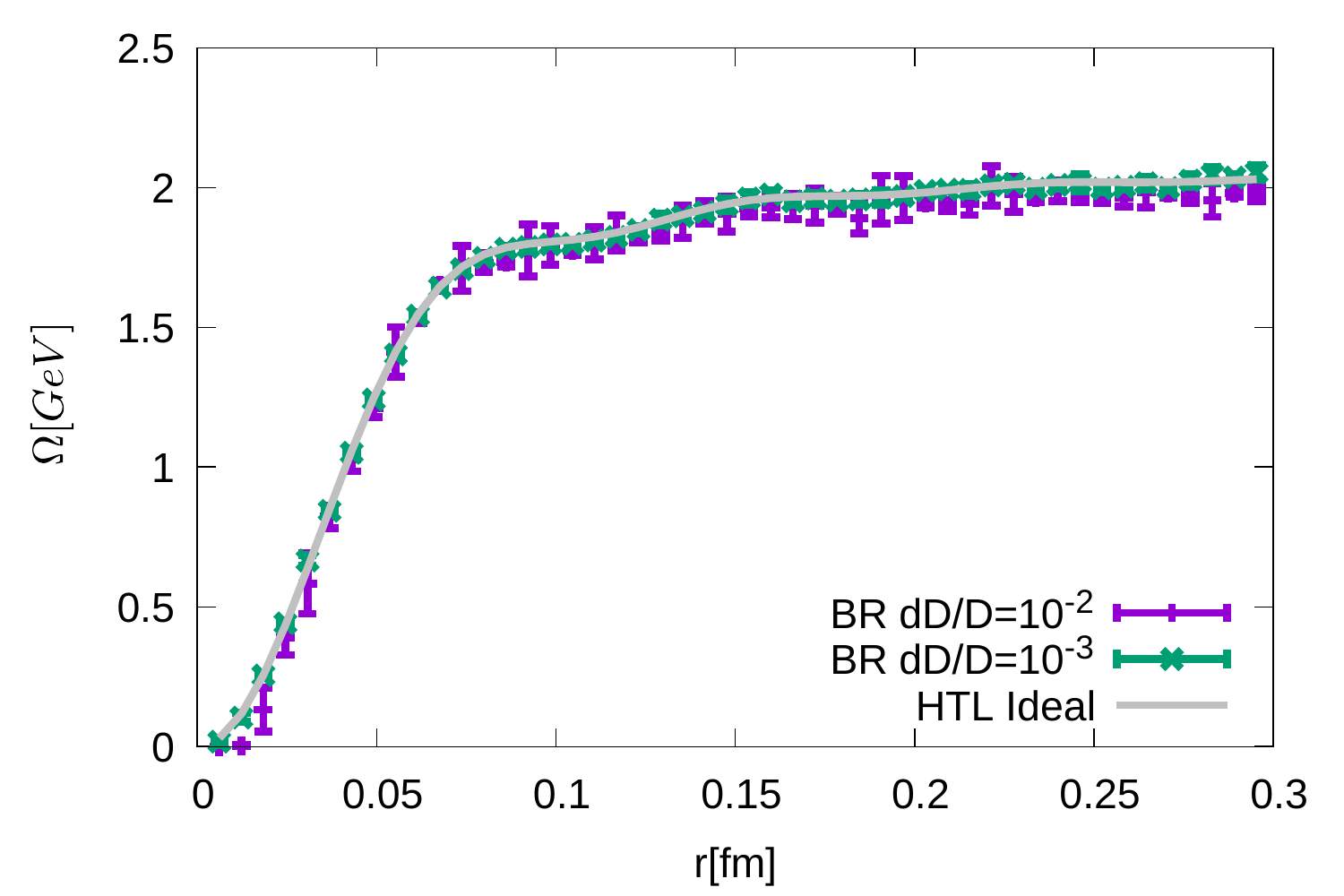}
%\vspace{-0.28cm}
\includegraphics[scale=0.58]{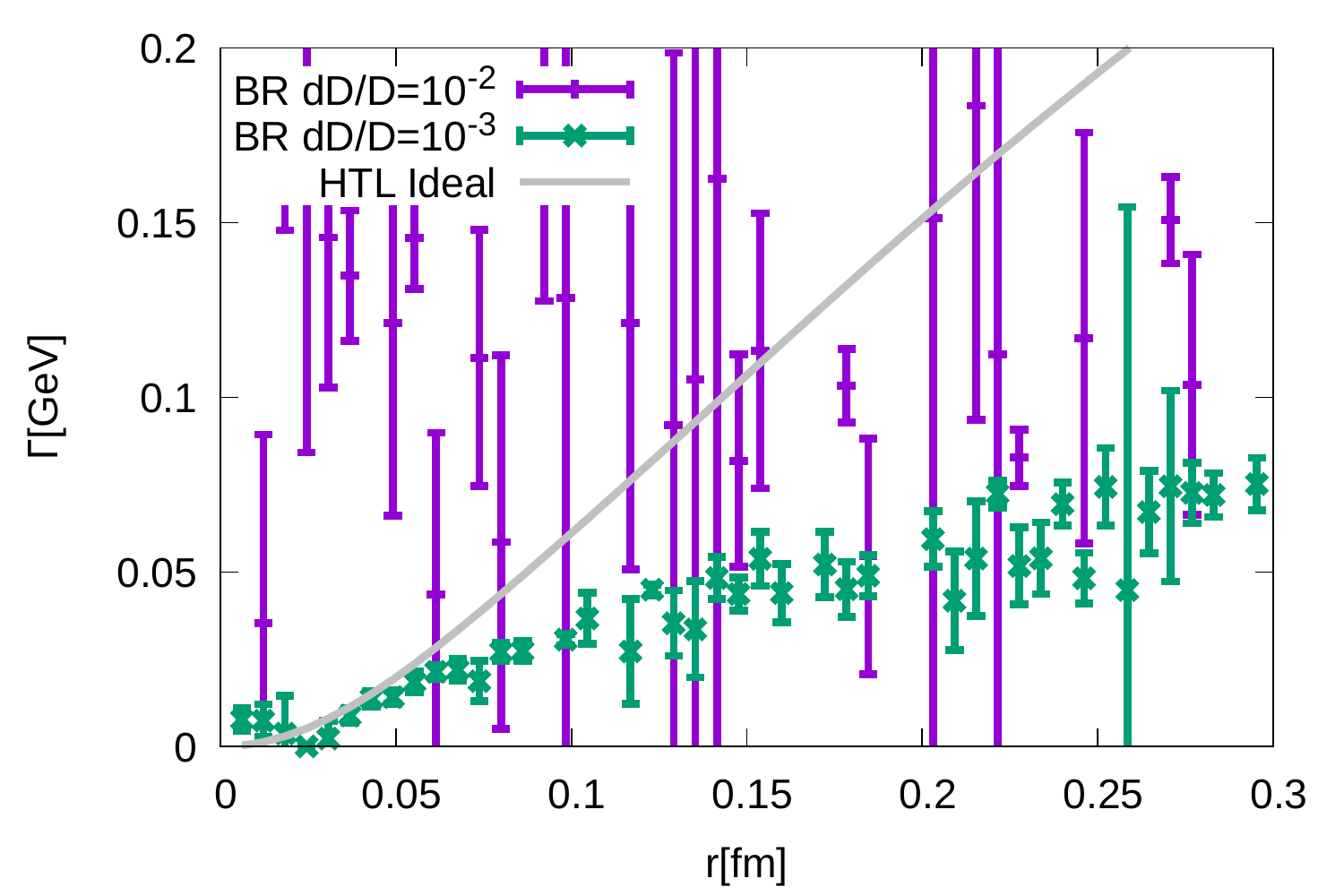}\vspace{-0.28cm}
\caption{Extraction of $\Omega$ and width $\Gamma$ for Hard Thermal Loop ideal data for $dD/D = 10^{-2}$ and $dD/D = 10^{-3}$ for $T=667$MeV using the BR method. The error bars are obtained from Jackknife resampling.}
\label{Fig:HTL_PotBR}
\end{figure}

\begin{figure}[t]
\centering
\includegraphics[scale=0.58]{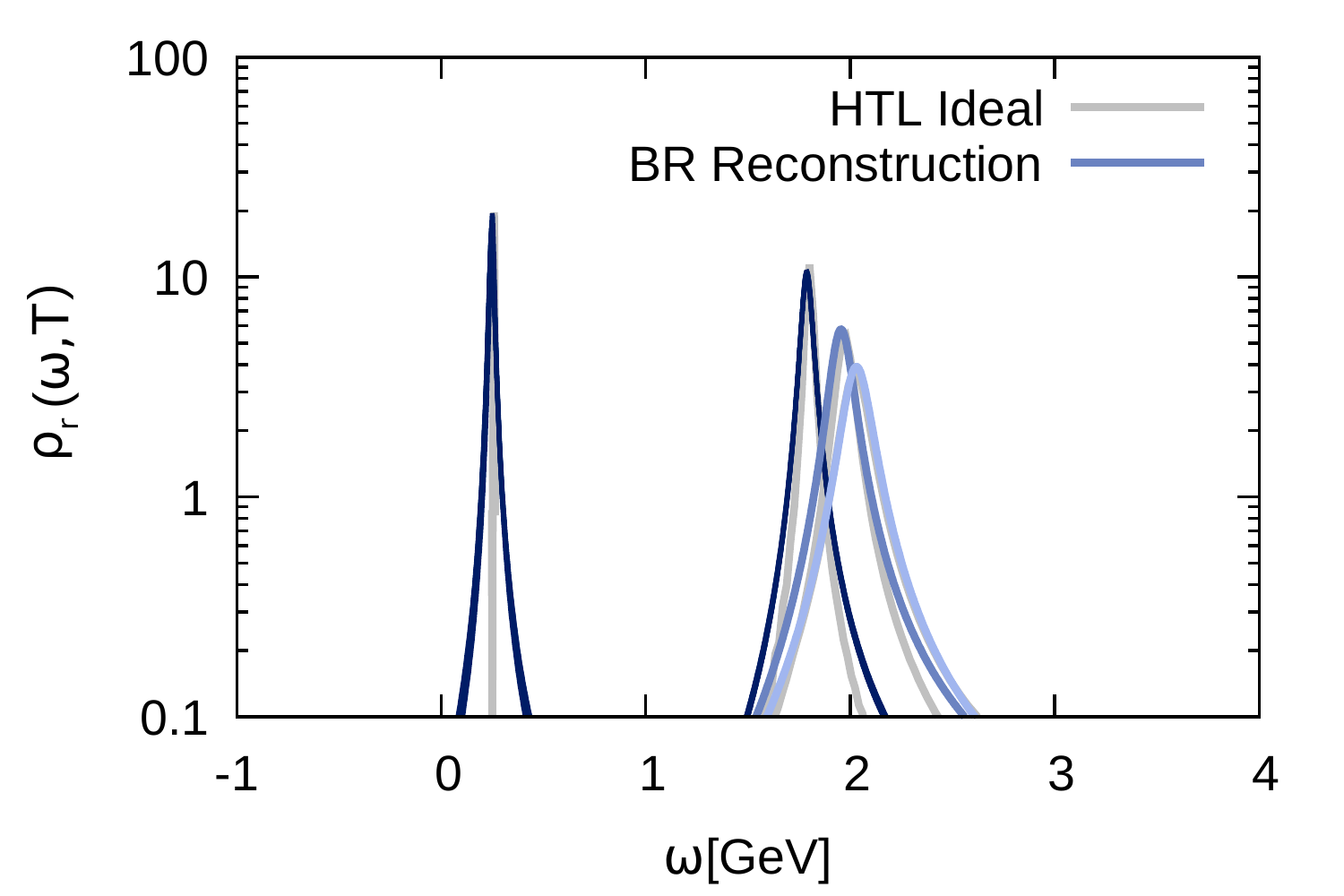}\vspace{-0.28cm}
\caption{A selection of spectral functions extracted from Hard thermal loop data for $dD/D = 10^{-3}$ $r = 0.01,0.09,0.11$ and $0.27$ fm respectively (dark blue to light blue) using Bayesian Reconstruction vs. the analytic result (solid gray).}
\label{Fig:HTL_SpectraBR}
\end{figure}

The last type of methods to be deployed in the study of the Wilson line spectral function are Bayesian spectral reconstructions. While the most well known variant, the Maximum Entropy Method \cite{Asakawa:2000tr} faces challenges, as it does not easily reproduce Lorentzian structures encoded in peaks, the more recently developed Bayesian reconstruction (BR) method \cite{Burnier:2013nla} has been deployed successfully in the extraction of such structures, both from mock data as well as from lattice QCD data.

All Bayesian methods exploit Bayes' theorem 
\begin{align}
P[\rho|D,I]\propto P[D|\rho,I]P[\rho|I] = {\rm exp}[-L+\alpha S_{\rm BR}],
\end{align}
to systematically regularize the inversion problem. They amend the simulation data $D$ by additional so called prior information $I$. The posterior probability $P[\rho|D,I]$ for a test function $\rho$ denotes the probability for $\rho$ to be the correct spectrum, given simulation data and prior information, which in turn is written as the product of the likelihood $ P[D|\rho,I]$ and prior probability $P[\rho|I]$. The former states how compatible $\rho$ is with the simulation data, and is nothing but the usual quadratic distance functional used in $\chi^2$ fitting
\begin{align}
L=\frac{1}{2}\sum_{i,j=1}^{N_d}(D_i-D^\rho_i)C_{ij}^{-1}(D_j-D^\rho_j).
\end{align}
Here $C_{ij}$ denotes the standard unbiased covariance matrix. It is amended by the prior probability $P[\rho|I]={\rm exp}[\alpha S_{\rm BR}]$, which acts as a regulator to the many flat directions of the likelihood functional
\begin{align}
S_{\rm BR}=\int d\omega \big( 1- \frac{\rho(\omega)}{m(\omega)} + {\rm log}\big[ \frac{\rho(\omega)}{m(\omega)} \big]\big).
\end{align}
The function $m(\omega)$ denotes the default model and by definition corresponds to the correct spectrum in the absence of data. In this work, we choose to use a default model, which implements a $1/\omega$ falloff at large frequencies as $m(\omega)\propto 1/(a\omega-a\omega_{\rm min}+1)$. When estimating the error budget for the spectral features we include the variation between results based on different default models, including the constant one, as well as $m\propto \omega, \omega^2$ and $m\propto 1/(a\omega-a\omega_{\rm min}+1)^2$.

In the original formulation of the BR method, the hyperparameter $\alpha$, which weighs the influence of prior information and data is marginalized from the posterior by assuming no knowledge of its values $P[\alpha]=1$ so that
\begin{align}
P[\rho|D,I,m]\propto P[D|\rho,I]\int_0^{\infty} d\alpha P[\rho | m,\alpha] P[\alpha]. \label{Eq:IntOutAlpha}
\end{align}
In this study we deploy a different handling of $\alpha$, which is akin to the Morozow criterion in classic regularization, i.e. we simply tune the hyper parameter such that the likelihood takes on the value $L=N_\tau/2$. The motivation for this choice is to avoid very large or very small values of $\alpha$ to contribute to the end results, as would be the case when integrating over $P[\alpha]=1$. In turn the occurrence of ringing artifacts is expected to be diminished. In the present study no significant differences between different choices of $\alpha$ handling were found.

After specifying the likelihood and prior, we numerically search for the most probable spectrum in the Bayesian sense by locating the unique extremum of the posterior $P[\rho|D,I,m]$ via a quasi-Newton optimization algorithm, the LBFGS method.  In practice we resolve the spectrum along $N_\omega=1000$ points in a frequency interval of $\omega a = [0:15]$.  We take into account all Euclidean data points except at $\tau=0$ and $\tau=1/T$. This in particular excludes the point from which the color singlet free energies are defined. 

In previous studies it has been observed that the BR method is well suited to extract strongly peaked spectral features with high accuracy. On the other hand, if only a small number of datapoints are available $O(10)$, the regulator $S_{\rm BR}$ is unable to avoid ringing artifacts in the reconstruction if the encoded spectrum contains broad structures. Improving the regulator functional to retain its resolving capability, while preventing ringing is work in progress (see e.g. \cite{Fischer:2017kbq}).

We can benchmark the reliability of the spectral reconstruction at high temperatures by using the non-trivial mock data computed in hard-thermal loop perturbation theory (HTL) similarly as for the Pad\'e reconstruction. In this case we add Gaussian noise with constant relative error $\Delta D/D=\kappa={\rm const.}$ directly onto the Euclidean data and supply it to the reconstruction algorithm. As shown in Fig. \ref{Fig:HTL_PotBR} we find that for the worst-case test with $\kappa=10^{-2}$, we are able to reproduce the position of the lowest lying peak with higher precision than what the Pade allows us to do. For the width $\Gamma$, the BR results at $\kappa=10^{-2}$ are equally disappointing as with the Pade method. However if we go to the best case scenario of $\kappa=10^{-3}$, the BR method shows its strength in being able to much more closely recover the correct imaginary part compared to the Pade.

It is important to note that in none of the HTL mock data tests have we encountered any signs of ringing artifacts, which, combined with the good performance regarding the imaginary part of the HTL potential, bodes well for the application of the BR method to the task of extracting the spectral features from data, which features positive spectral functions. A representative selection of reconstructed spectral functions obtained with the BR method from HTL mock data is shown in Fig. \ref{Fig:HTL_SpectraBR}.

\begin{figure}[t]
\centering
\includegraphics[scale=0.58]{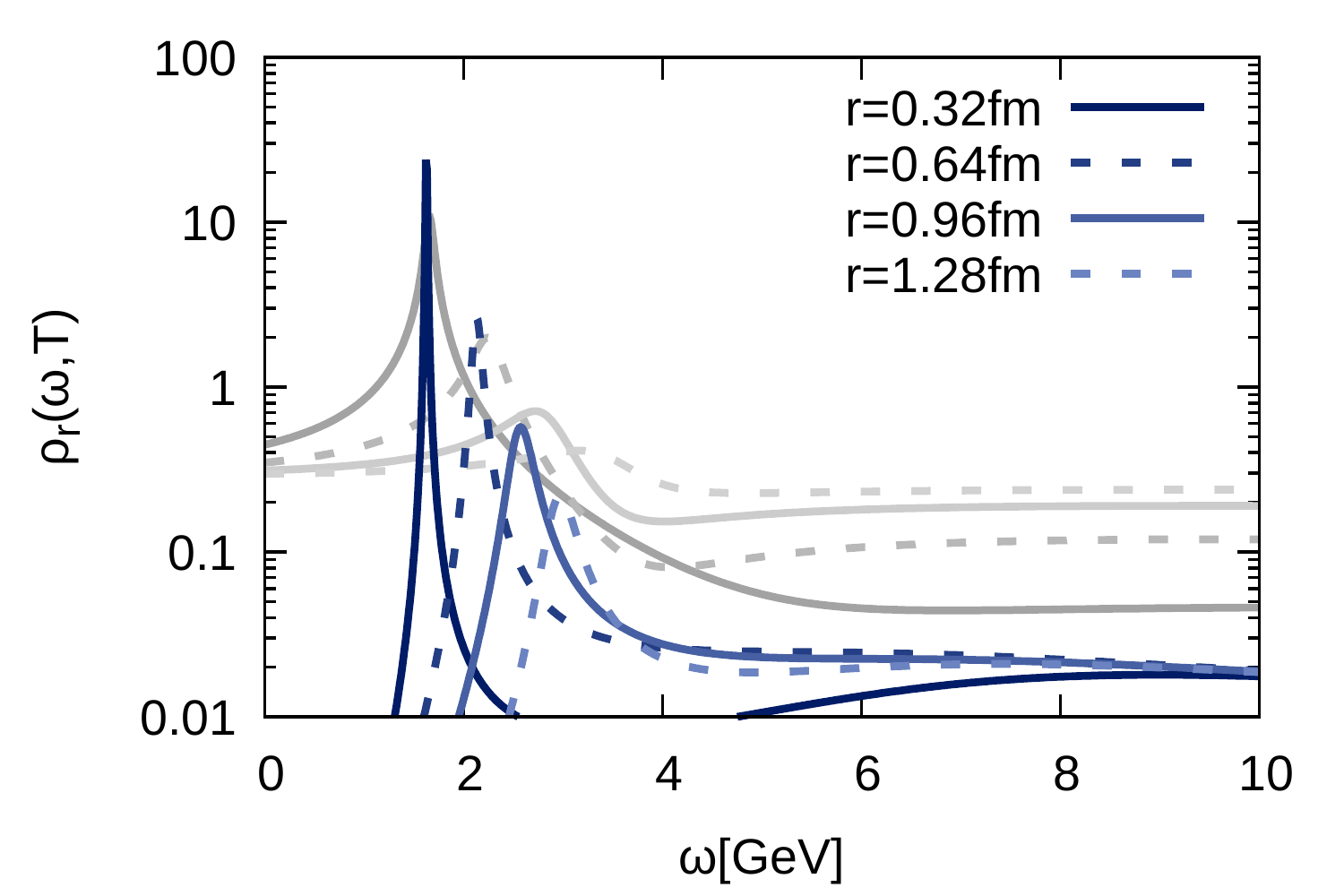}\vspace{-0.28cm}
\caption{Comparison of reconstructed spectra using Pade (grey) and BR (blues) method at  $T=151$~MeV ($\beta = 6.740$, $N_\tau =12$) at different separation distances $r = 0.32, 0.64,0.96$ and $1.28$ fm.}
\label{Fig:Pade_BR_spectra}
\end{figure}

At low temperatures, e.g. at $T=151$~MeV, the Euclidean correlators do not yet show signs of positivity violation (i.e. we obtain effective masses that are monotonic in Euclidean time) and the BR method succeeds in reconstructing their spectral function. By construction, the result reproduces the input Euclidean data points within their statistical errors. A selection of these spectra for $r = 0.32, 0.64,0.96$ and $1.28$ fm is shown in Fig. \ref{Fig:Pade_BR_spectra} (solid dark blue to lighter blue) compared to the outcome of the Pad\'e reconstruction (gray solid). 

We find important differences between the two approaches. The BR method reconstructions, as expected from the effective mass analysis, shows a single well defined lowest lying peak. Towards the origin that peak rapidly decays in an exponential fashion, qualitatively similar to the behavior observed in HTL spectral functions. In contrast the Pad\'e reconstruction assigns significant weight to the low frequency region. This difference is among the reasons, why the spectral function of the Pad\'e reconstruction does not fulfill the spectral decomposition of the original Euclidean data, a known drawback of the Pad\'e reconstruction method.

At larger frequencies than its maximum, the BR spectral function shows a tail, which eventually behaves as $\propto 1/\omega$ per choice of the default model. We have checked that changing the default model to different powers $\alpha$ as $m\propto \omega^\alpha$ does not change the peak structure significantly. The central peak obtained by the BR method agrees in position with the Pad\'e result at small separation distances but the Pad\'e eventually seems to smear out significantly with the center of the bump lying at a higher frequency than the BR spectra peak. 

\begin{figure}[t]
\centering
\includegraphics[scale=0.58]{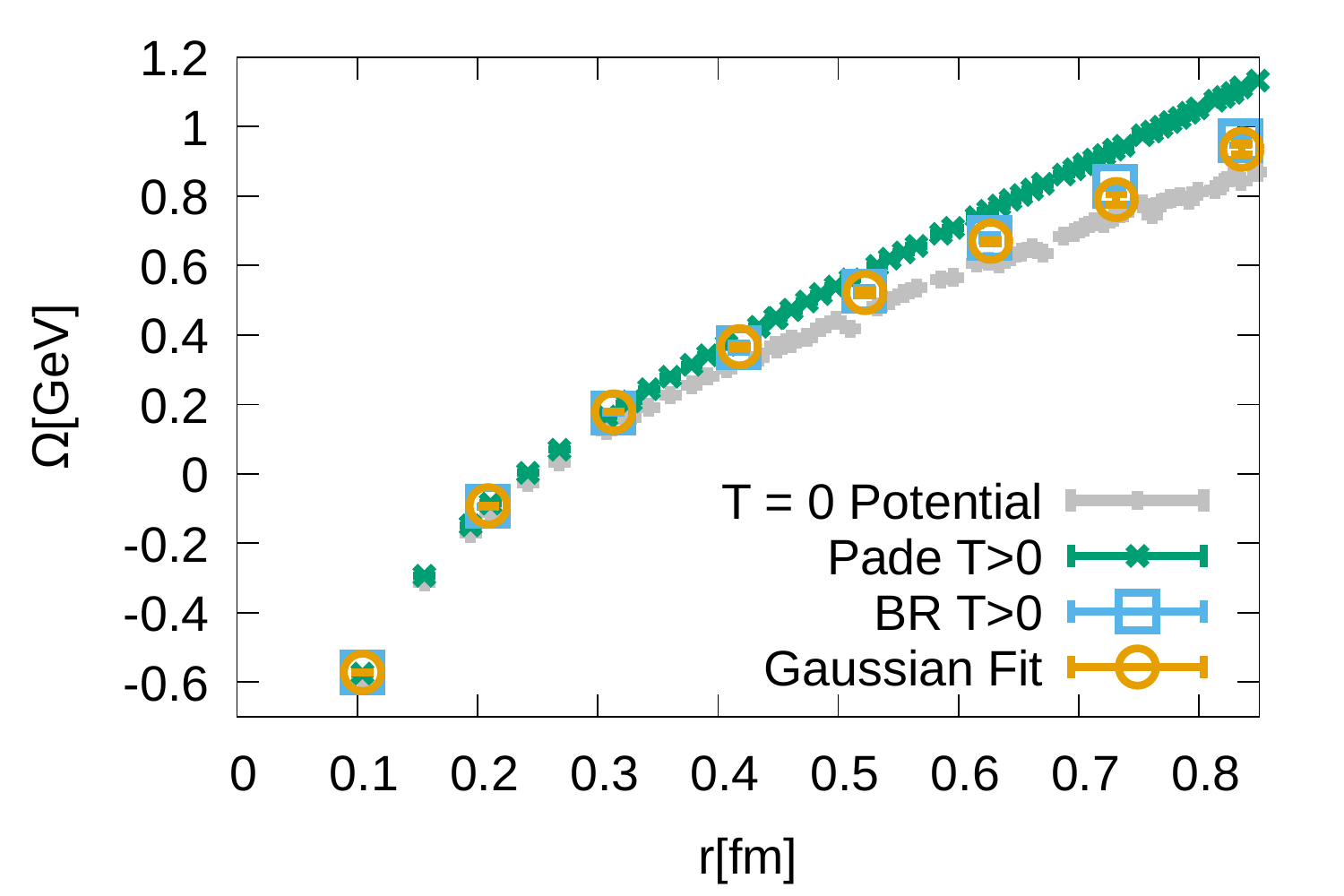}\vspace{-0.28cm}
\caption{Comparison of $\Omega$ using the Pad\'e, BR method and Gaussian Fit at $\beta = 6.740$ with $N_\tau =12$ ($T=151$ MeV). The $T=0$ potential for the same $\beta$ are given as grey data points.}
\label{Fig:ReV_Comp}
\end{figure}

Let us compare $\Omega$ obtained from the Pade pole analysis (magenta) and the skewed Breit-Wigner fit of the BR spectral reconstruction (light blue) in Fig. \ref{Fig:ReV_Comp}. In addition we provide the values of the peak position $\Omega$ at $T=0$ in green. Since $T=151$~MeV still lies close to the crossover transition, the effect of the medium on the static potential is expected to be weak. We find this intuition reflected in the agreement between the zero temperature static energy and the BR result for $\Omega$. Interestingly all three results agree up to around $r=0.3$fm. However as we consider larger distances we find that the Pad\'e reconstruction shows a systematic tendency to lie above the zero temperature static energy. The effect becomes significant around $r=0.5$fm and remains visible up to $r=0.85$fm, beyond which the signal of the $T=0$ static energy is lost. The BR result lies much closer to the $T=0$ effective masses over the whole range of distances.

The success of the BR reconstruction at $T=151$~MeV tells us that the data is compatible with a dominant skewed Breit-Wigner peak structure in the spectral function. 
At higher temperatures the BR method cannot be reliably applied to the extraction of spectral functions from the raw correlators, due to the presence of non-positivity in the underlying spectral functions. At $T=407$~MeV, for example, the effective masses at small distances show explicitly non-monotonic behavior \cite{Bazavov:2018wmo}. 
However, the spectral density may not
be positive definite even if the effective masses decrease monotonically.
We see that also at intermediate distances, the BR method fails to converge successfully.

While in principle we could proceed by investigating the UV-subtracted finite temperature correlators, we have found that the statistical uncertainties introduced by the $T=0$ subtraction dominate over those inherent in the $T>0$ data, thus preventing us from a precision analysis of the spectral function at higher temperatures.

\section{Conclusions}
\label{sec:conclusion}
In the first part of the study we have investigated the Wilson line correlation functions obtained from the numerical simulations directly in imaginary time. 
Computing the first three cumulants, as defined in Eq.~(\ref{eq:m_n}), of imaginary-time correlation functions. The n$^\mathrm{th}$ cumulant of the imaginary-time correlation function at $\tau=0$ is equivalent to the n$^\mathrm{th}$ moment of the corresponding spectral functions (see Eq.~(\ref{eq:ETwilsonspecdec})), assuming that the moment of the underlying spectral functions is finite.
We found that they differ beyond statistical errors from the predictions of resummed HTL perturbation theory at all temperatures investigated, including the highest temperature,
$T=1938$ MeV. Furthermore, even at qualitative level there is a difference between
the lattice results and HTL result. The first cumulant calculated in leading order
HTL perturbation theory is antisymmetric around $\tau=1/(2T)$, but the lattice
results do not show such feature except near $\tau\sim1/(2T)$.
In addition we checked that our datasets allow for a meaningful determination of up to the third cumulant of the correlation function in Euclidean time. For higher moments the signal to noise does not suffice. In turn we understand that our input data will be able to constrain spectral information only within the limitations placed by these three moments.

The second part of the study was concerned with extracting the position $\Omega$ and width $\Gamma$ of the dominant spectral peak structure encoded in the Wilson line correlators. We deployed four different approaches: spectral function model fits where the dominant peak is described by a Gaussian, the HTL-inspired fit of Bala and Datta, the Pade approximation and, where positivity allowed, the Bayesian BR method.

In essence each of the four methods introduces certain prior information in order to regularize the ill-posed inversion problem to gain access to the spectral function. It turns out that the Euclidean data scrutinized in the first part of our study is amenable to different possible hypotheses, which in turn lead to different outcomes for $\Omega$ and $\Gamma$.

%The spectral function fits assumes on the one hand that the ground state peak of %the zero temperature spectral function is well enough separated from any other %structures, so that after the subtraction procedure, laid out in %sec.\ref{sec:potfromfits} the peak at finite temperature is not affected. On the %other hand the spectral function fit assumes that the effect of tail structures %of the lowest lying peak can be accurately approximated by the addition of one %delta peak to a dominant Gaussian peak.
The spectral function fits assume that the high energy part of the spectral
function has negligible temperature dependence, and that the observed temperature
dependence of the Wilson line correlators is determined by the dominant peak
structure. 
%The dominant peak is chosen to have a second moment much larger than higher moments, for which a Gaussian for the dominant peak, and the low energy tail that is approximated by a single delta function is the simplest choice.
Since the correlator is found to have a second cumulant much larger than its higher cumulants, a Gaussian for the dominant peak, and a single delta function for the low energy tail are the simplest, permissible choices for parametrizing the data.
The Gaussian spectral function model shows a value of $\Omega$, which is virtually independent of temperature and a width, which scales trivially with the temperature.

In order to extract the values of $\Omega$ via the HTL-inspired fit, one assumes that the correlation functions are amenable to a certain non-standard spectral decomposition, similar to the one encountered in leading-order HTL perturbation theory. This spectral decomposition leads to a first cumulant that is anti-symmetric around $\tau=1/(2T)$. Because of 
small $N_\tau$, the fits can be performed only in a small region around $\tau=1/(2T)$.
This fit yields an $\Omega$, which shows clear temperature dependence and signs of asymptotic flattening in the QGP phase. The width that the method computes shows a non-trivial scaling with the temperature, which is weaker than linear in the temperature.

The third method we deployed is the Pad\'e rational approximation. The only assumption it makes is that the correlation function represents an analytic function. However it suffers from the drawback that its outcome is known to violate the spectral decomposition of the input data. I.e. the Pad\'e spectrum, when reinserted into the Lehmann representation, does not reproduce the original correlator. We have however tested the Pade method under non-trivial settings in HTL perturbation theory and found that for the temperature and spatial separation distances probed, the position of the lowest lying peak structure was well reproduced. Applied to genuine lattice data we obtained results that were robust under changes in the number of input points and a reordering when constructing the Pade approximation of the Matsubara domain input data. The outcome of the extraction of $\Omega$ based on the Pade method yields values, which similarly to the Gaussian model fit, shows virtually no temperature dependence. While the mock data tests tell us to take the outcome of the width with a significant grain of salt, we find small statistical errorbars and a behavior that qualitatively agrees with that of the Bala-Datta method, i.e. $\Gamma$ scales weaker than linear with the temperature.

Last but not least we also deployed the Bayesian BR method, where positivity allowed. The BR method has been extensively tested on HTL mock data and has been shown to outperform other Bayesian methods, such as the MEM in the accurate reconstruction of the lowest lying peak from Wilson line correlators, a finding reproduced in this study.  As the BR method is designed to reproduce the Euclidean input data within their uncertainty, its reconstructed spectra denote a valid hypothesis for the actual underlying spectrum. The BR method possesses an explicit default model dependence, which however can and is assessed by repeating reconstructions for different functional forms of 
%$m$. 
the default model.
And while the BR method is known to be susceptible to ringing artifacts, as its regulator is weakest among the reconstruction methods on the market, no signs of ringing have been observed in this study, neither in the HTL mock test nor in the reconstruction of genuine lattice data. 

As a crucial limitation in the context of the current study, the BR method is only applicable to positive definite spectral functions. If effective masses show non-monotonicity it indicates that the BR method cannot be deployed. However even if the effective masses are monotonous, positivity violation may persist, which explains that the BR method fails to converge successfully for higher temperatures on the raw Euclidean correlators. The outcome of the extraction of $\Omega$, based on the BR method at low temperatures such as $T=151$~MeV yields a real-part which agrees well with the 
%effective mass analysis, 
static energy from (multi-state) exponential fits,
also applicable on those lattices. We find that the spectral functions show well defined Breit-Wigner like peaks, which get exponentially cut off close to the origin, similar to what is seen in HTL perturbation theory at much higher temperatures.
%, as expected from HTL perturbation theory. 
Comparing the BR result to the Pad\'e we find that the Pad\'e incorrectly assigns too much weight to the low frequency regime and at the same time produces a less and less well-defined peak, which is consistently located at a higher position than the BR peak. Agreement between the BR and the effective masses and the tension with the Pade method starting around $r=0.5$fm seem to indicate that the Pade tends to overestimate the values of $\Omega$ when applied to our lattice data.

\begin{figure}
    \centering
    \includegraphics[scale=0.58]{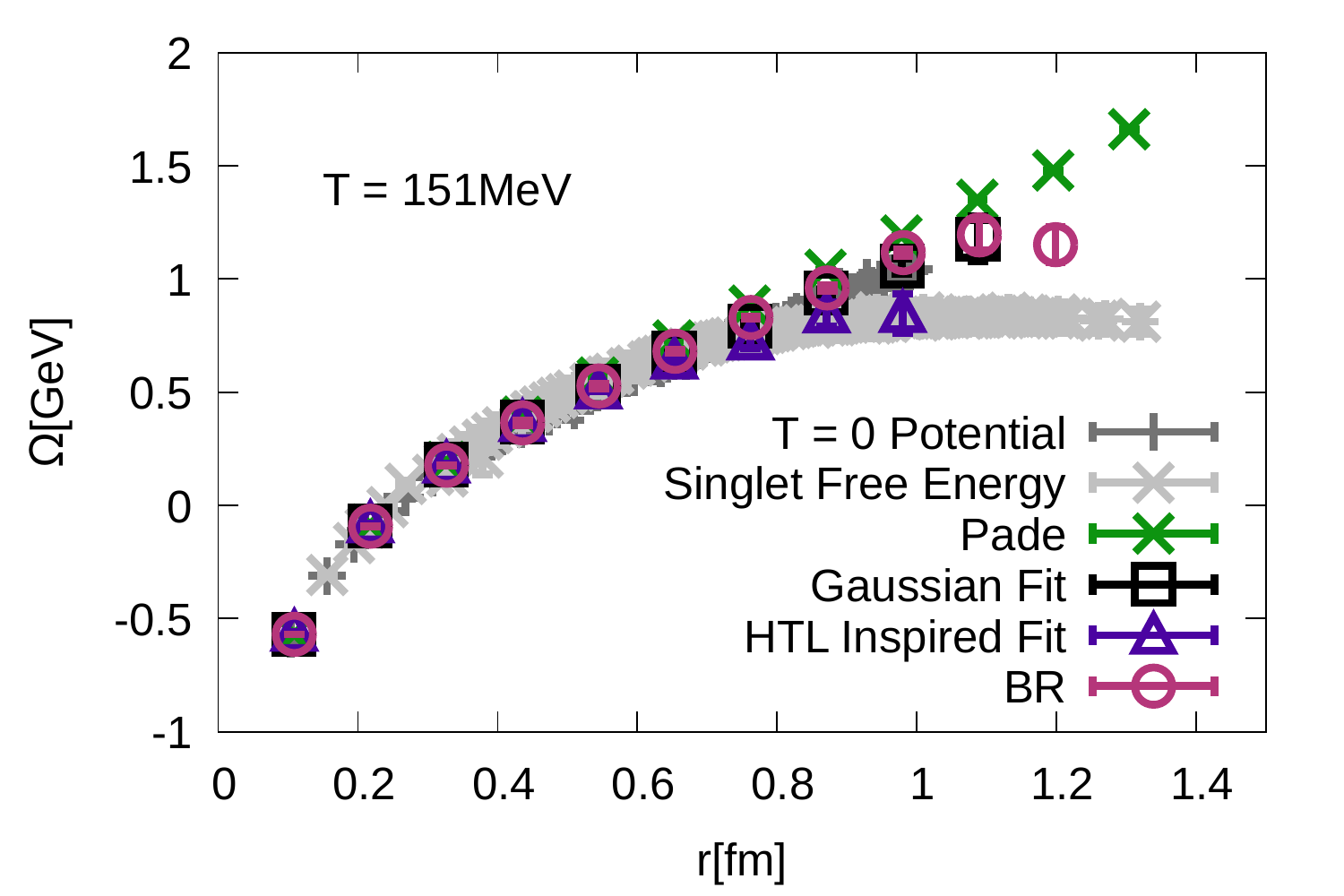}\vspace{-0.28cm}
    \includegraphics[scale=0.58]{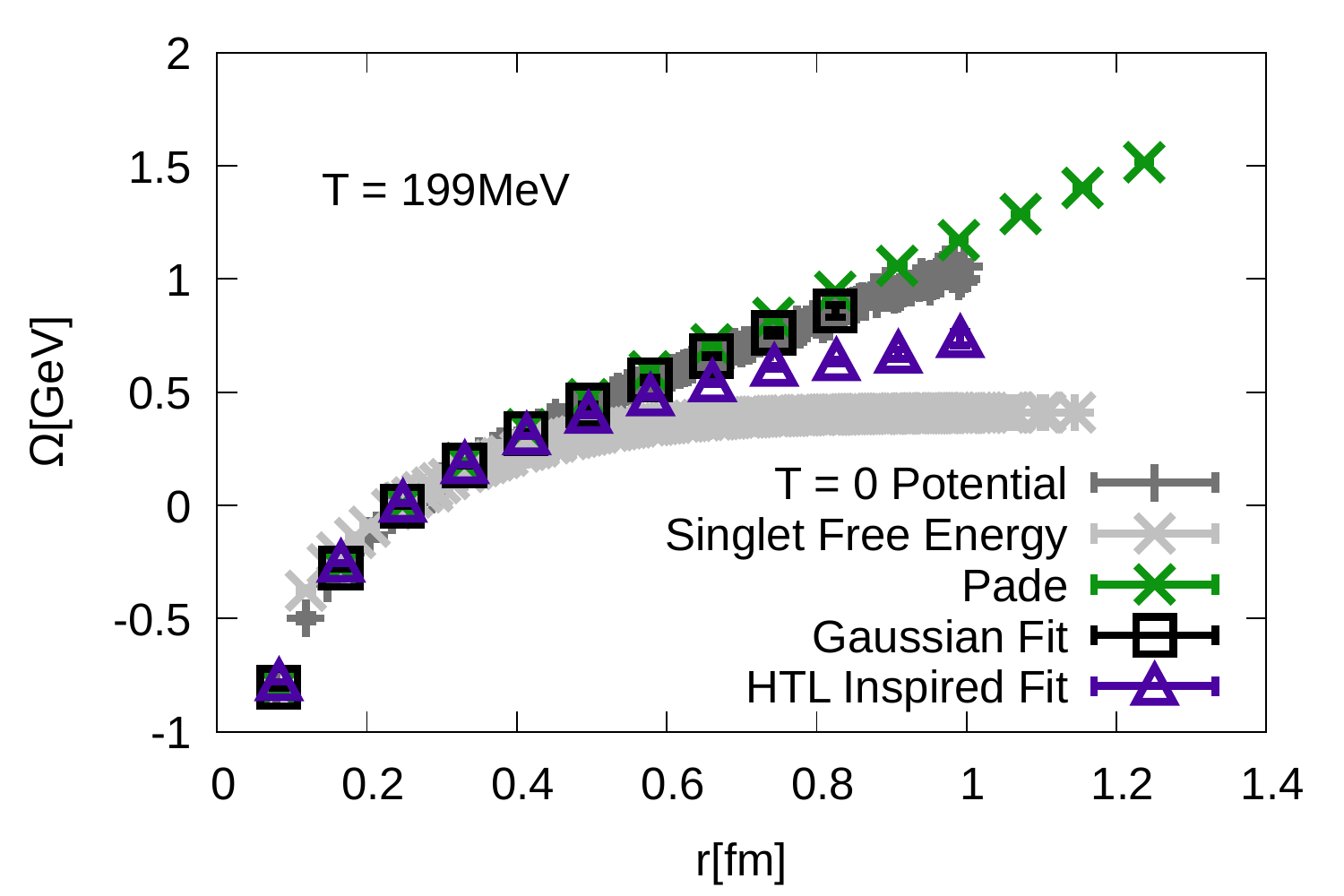}\vspace{-0.28cm}
    \includegraphics[scale=0.58]{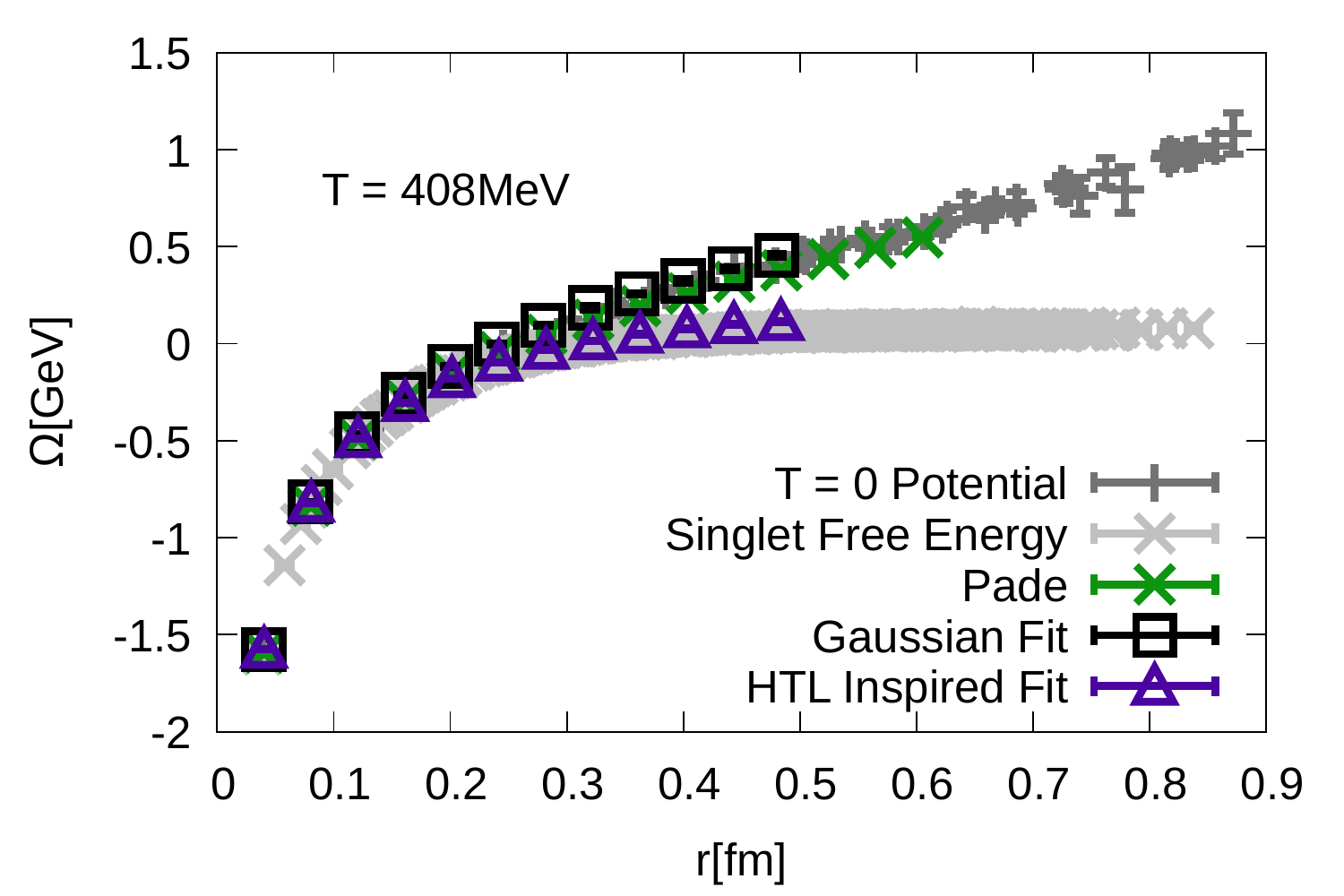}\vspace{-0.28cm}
    \caption{Comparison of $\Omega$ as a function of separation distances for three different temperatures 151,199 and 408 MeV obtained from different methods discussed in the text. We have also shown the T=0 potential (dark grey) for all temperatures and the free energy (light grey) for high temperature (408 MeV).}
    \label{fig:ReV_conc}
\end{figure}

\begin{figure}
    \centering
    \includegraphics[scale=0.58]{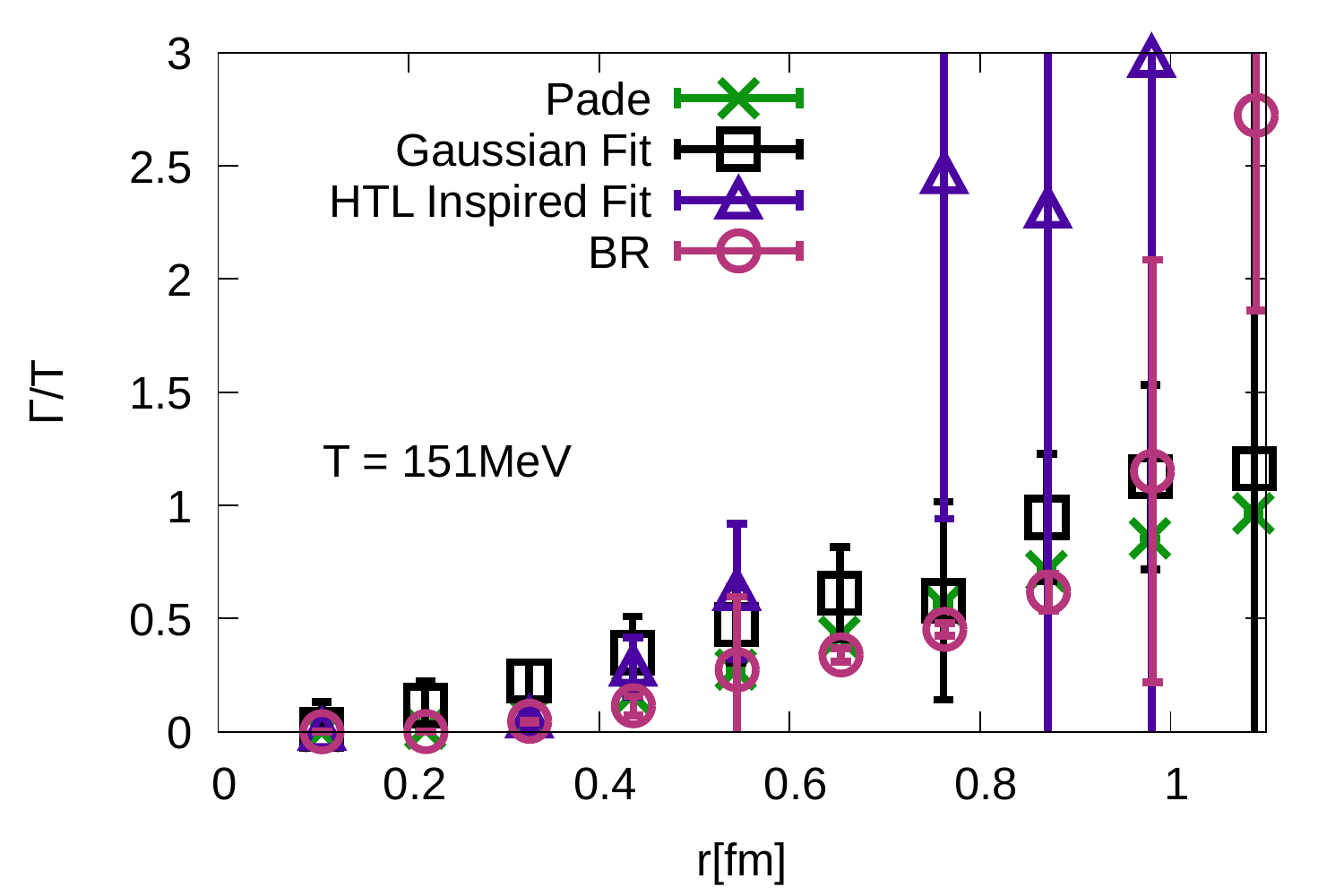}\vspace{-0.28cm}
    \includegraphics[scale=0.58]{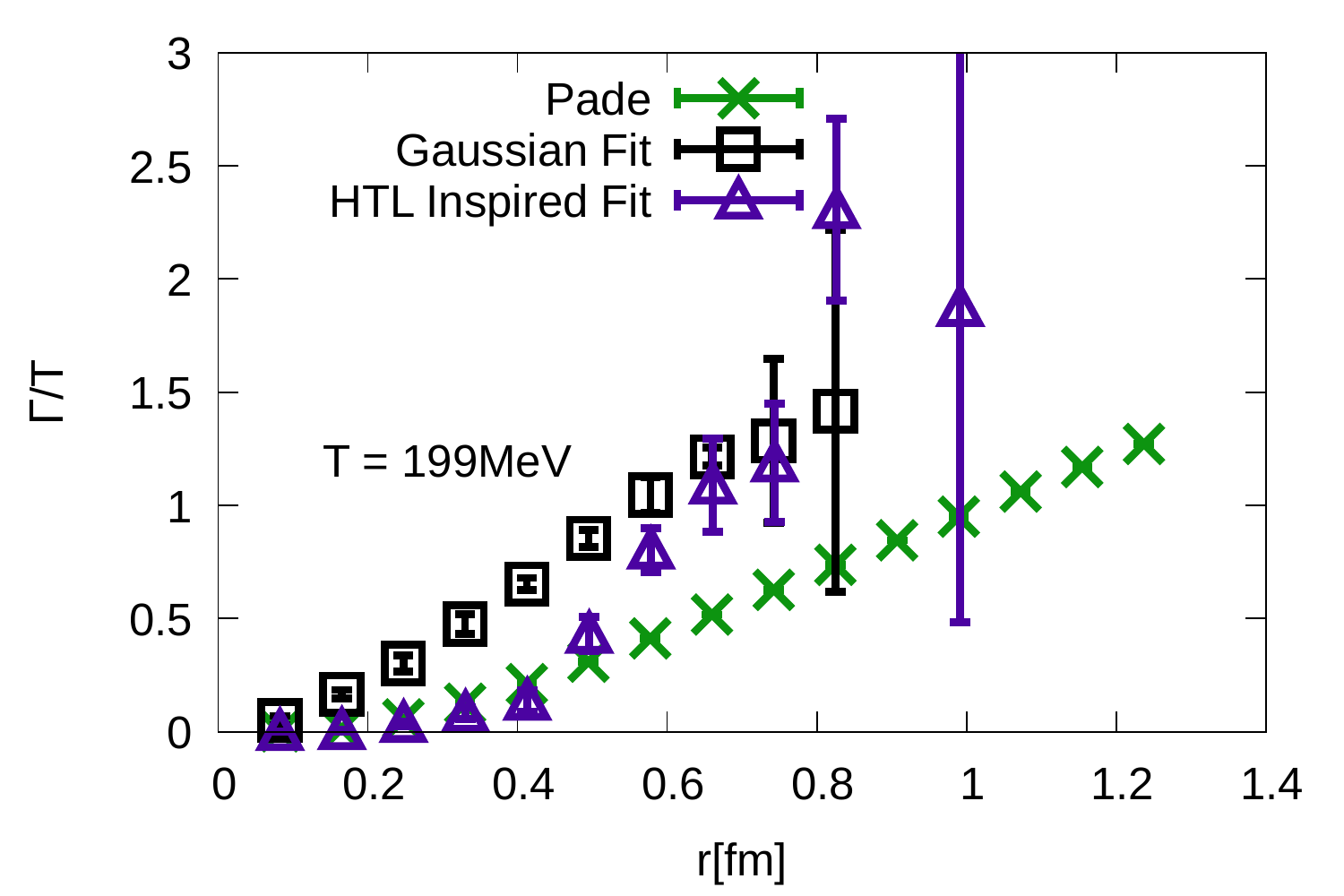}\vspace{-0.28cm}
    \includegraphics[scale=0.58]{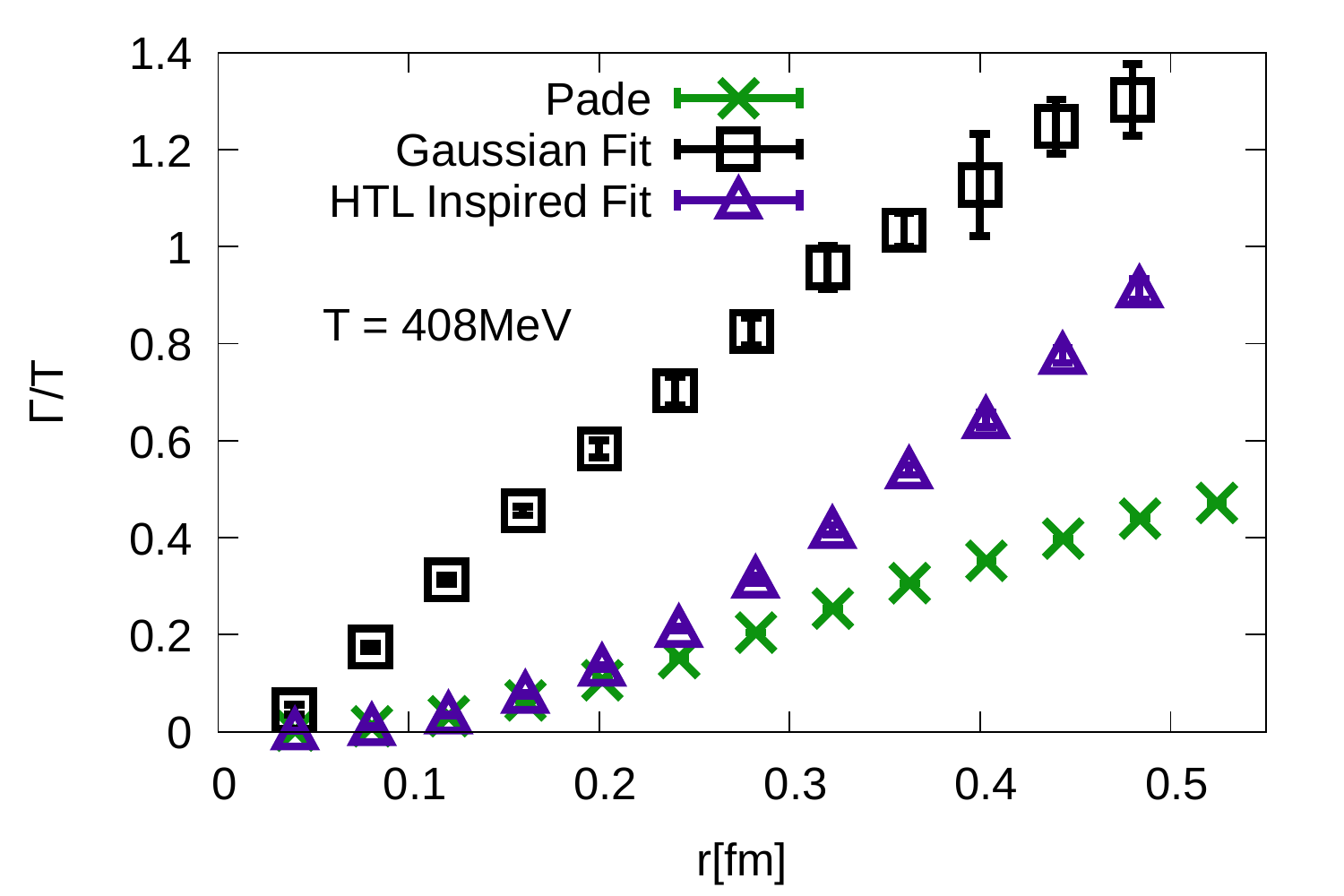}\vspace{-0.28cm}
    \caption{Comparison of $\Gamma/T$ as a function of separation distances for three different temperatures 151, 199 and 408 MeV obtained from different methods discussed in the text. }
    \label{fig:ImV_conc}
\end{figure}

The comparison of different methods of the spectral reconstruction in terms
of the peak position, $\Omega$ of the dominant peak and its width, $\Gamma$
is summarized in Fig. \ref{fig:ReV_conc} and Fig. \ref{fig:ImV_conc}, respectively for three temperatures, $T=151$ MeV (just below the chiral crossover), $T=199$ MeV (the typical temperature most relevant for RHIC),
and $T=408$ MeV deep in QGP.
The present study sheds new light onto the extraction of $\Omega$ and $\Gamma$.
While different methods often lead to quantitatively different
results some general features are the same. The width $\Gamma$
is significant compared to the temperature scale and increases with distance $r$ for all temperatures. In fact, for the lowest
temperature all method give consistent results for $\Gamma$.
For temperatures $150~{\rm MeV} < T < 200~{\rm MeV}$, the Gaussian fits and HTL fits lead to
similar width for large $r$, while at small $r$ the HTL fit gives
a smaller width.
The Pade method always gives a smaller $\Gamma$ than Gaussian and HTL fits
at large $r$, but agrees with the HTL result at small $r$,
c.f. Fig. \ref{fig:ImV_conc}.  
The $r$ dependence of the peak position
turns out to be similar for the Gaussian fits and the Pade
method, indicating an apparent absence of the screening effects.
Furthermore in the temperature range $150~{\rm MeV} < T < 200~{\rm MeV}$ and at intermediate distances, all the explored methods give a peak position that is slightly larger than the singlet free energy, see Fig. \ref{fig:ReV_conc}.
For these temperatures, which are the most relevant ones
for RHIC, the spread
of the results is not too large in order to have an impact on
the phenomenological studies. At higher temperatures, which are
of interest for quarkonium phenomenology in heavy ion collisions
at LHC our results are inconclusive at present, and lattice
calculations with larger $N_{\tau}$ and smaller statistical
errors are needed. Increasing the temporal extent of the lattice
will be possible in the coming years. At the same time accumulation of statistics at $T=0$ will also enable a high precision subtraction, which in turn will enable us to use the BR method above the crossover temperature.

All data from our calculations, presented in the figures of this paper, can be found in \cite{data}.

\section*{Acknowledgement}

This material is based upon work supported by the U.S. Department of Energy, Office of Science, Office of Nuclear Physics through the (i) Contract No. DE-SC0012704, and  (ii) Scientific Discovery through Advance Computing (SciDAC) award  Computing the Properties of Matter with Leadership Computing Resources. (iii) R.L., G.P. and A.R. acknowledge funding by the Research Council of Norway under the FRIPRO Young Research Talent grant 286883. (iv) J.H.W.’s research was funded by the Deutsche Forschungsgemeinschaft (DFG, German Research Foundation) - Projektnummer 417533893/GRK2575 ``Rethinking Quantum Field Theory''. (v) D.B. and O.K. acknowledge support by the Deutsche Forschungsgemeinschaft (DFG, German Research Foundation) through the CRC-TR 211 'Strong-interaction matter under extreme conditions'– project number 315477589 – TRR 211.

This research used awards of computer time provided by: (i) The INCITE and ALCC programs at Oak Ridge Leadership Computing Facility, a DOE Office of Science User Facility operated under Contract No. DE-AC05- 00OR22725. (ii) The National Energy Research Scientific Computing Center (NERSC), a U.S. Department of Energy Office of Science User Facility located at Lawrence Berkeley National Laboratory, operated under Contract No. DE-AC02- 05CH11231. (iii) The PRACE award on JUWELS at GCS@FZJ, Germany. (iv) The facilities of the USQCD Collaboration, which are funded by the Office of Science of the U.S. Department of Energy. (v) The UNINETT Sigma2 - the National Infrastructure for High Performance Computing and Data Storage in Norway under project NN9578K-QCDrtX "Real-time dynamics of nuclear matter under extreme conditions".

The computations in this work were performed using the SIMULATeQCD suite \cite{Mazur:2021zgi,Altenkort:2021fqk}.

\appendix
\section{The spectral decomposition of static $Q\bar Q$ correlators}
\label{app:spec_decomp}
In this appendix we discuss the general features of the spectral
decomposition of static meson correlation function. Let 
$O_r=\bar Q(x) U(\mathbf{x}-\mathbf{y}) Q(y)$ be a static meson 
operator with $r=|\mathbf{x}-\mathbf{y}|$. The gauge connection
is $U(\mathbf{x}-\mathbf{y})$ runs along the straight line connecting
the spatial points $\mathbf{x}$ and $\mathbf{y}$ and may be constructed 
from smeared links on the lattice. More complicated paths are also 
possible. In Coulomb gauge we set $U(\mathbf{x}-\mathbf{y})=1$.
The static meson correlation function at non-zero temperature
is defined as:
\begin{align}
W(r,\tau,T) 
&=\langle O_r(\tau) O_r(0) \rangle 
\nonumber\\[2mm]
&= \frac{1}{Z(T)} {\rm Tr} [O_r(0)e^{-\tau H}O_r(0) e^{-(\beta-\tau) H}],\nonumber\\[2mm]
& \beta=1/T,~Z(T)=\sum_n e^{-\beta E_n}.
\end{align}
Using the energy eigenstates to evaluate the trace in the above
expression and inserting a complete set of energy eigenstates between
the operators $O_r$
the spectral decomposition of the static meson correlation function can be written as
\begin{equation}
W(r,\tau,T)=\frac{1}{Z(T)} \sum_{n,n'} e^{-(\beta -\tau) E_n} e^{-E_n' \tau} |\langle n|O_r(0)|n'\rangle|^2,
\end{equation}
where $|n\rangle$ are states without the static $Q\bar Q$ pair and $|n'\rangle$ are the states that also contain
the static $Q\bar Q$  pair at distance $r$. 
Since static quarks are not part of the thermal system in the evaluation of
the trace we only take into account the states $|n\rangle$. Hereafter, we introduce $C_{nm}^r=|\langle n |O_r| m'\rangle|^2$.
Let us take into account
the few lowest terms in the spectral decomposition, $n=0,~1,~2$ and  $n=0',~1',~2'$:
\begin{align}
\label{w2state}
%&
%\displaystyle
 Z(T) W(r,\tau,T)
 &=e^{-(\beta-\tau) E_0} e^{-E_{0'}(r) \tau} 
 C_{00} 
 %|\langle 0 |O_r| 0'\rangle|^2
 \nonumber\\[2mm]
%&
%\displaystyle
&+ e^{-(\beta-\tau) E_0} e^{-E_{1'}(r) \tau} 
 C_{01}^r 
 %|\langle 0 |O_r| 1'\rangle|^2
 \nonumber\\[2mm]
%&
%\displaystyle
&+ e^{-(\beta-\tau) E_1} e^{-E_{0'}(r) \tau} 
 C_{10}^r 
 %|\langle 1 |O_r| 0'\rangle|^2
 \nonumber\\[2mm]
%&
%\displaystyle
&+ e^{-(\beta-\tau) E_1} e^{-E_{1'}(r) \tau} 
 C_{11}^r
 %|\langle 1 |O_r| 1'\rangle|^2
 \nonumber\\[2mm]
%&
%\displaystyle
&+ e^{-(\beta-\tau) E_0} e^{-E_{2'}(r) \tau} 
 C_{02}^r 
 %|\langle 0 |O_r| 2'\rangle|^2 
 \nonumber\\[2mm]
%&
%\displaystyle
&+ e^{-(\beta-\tau) E_2} e^{-E_{0'}(r) \tau} 
 C_{20}^r
 %|\langle 2 |O_r| 0'\rangle|^2
 \nonumber\\[2mm]
%&
%\displaystyle
&+ e^{-(\beta-\tau) E_2} e^{-E_{2'}(r) \tau} 
 C_{22}^r 
 %|\langle 2 |O_r| 2'\rangle|^2
 .
\end{align}
Now we assume that the energy of the vacuum is zero,  while the energy of the vacuum state plus a static $Q \bar Q$ is
the static $Q\bar Q$ energy: $E_0=0$ and $E_{0'}(r)=V(r)$. 
We ignore the contribution of hybrid potentials for now. For the energy of the states with $Q\bar Q$ pair plus a hadronic state we 
write 
\begin{equation}
E_{n'}(r)=V(r)+E_n+\Delta E_n(r),
\end{equation}
where $\Delta E_n(r)$ is the interaction energy of $Q\bar Q$ pair with hadron state $n$, and it could
be either positive or negative.
For small $r$ we expect $\Delta E_n(r)\ll E_n$ because hadrons do not interact significantly with a tiny color singlet dipole.
Now let us assume that $|1\rangle$ has non-vacuum quantum number, 
i.e. $C_{01}^r=C_{10}^r=0$ and
state $|2\rangle$ has vacuum quantum number, i.e. $C_{20}^r\ne 0$ and
$C_{20}^r \ne 0$. Then we also have $C_{21}^r=C_{12}^r=0$ and
the corresponding terms have been already omitted in Eq. (\ref{w2state}).
Note that
the matrix elements $C_{02}^r$ and 
$C_{20}^r$ could be small since
the operator $O_r$ is optimized to create a single static meson state,
while $| 2'\rangle$ is an extended state of a static meson and a multi-hadron
state with vacuum quantum numbers.
Now we can write
\begin{eqnarray}
&
\displaystyle
Z(T) W(r,\tau,T)=\nonumber\\[2mm]
&
\displaystyle
C_{00}  e^{-V(r) \tau}+C_{11} e^{-\beta E_1} e^{-(V(r)+\Delta E_1(r))\tau}+\nonumber\\
&
\displaystyle
C_{22} e^{-\beta E_2} e^{-(V(r)+\Delta E_2(r))\tau}\nonumber\\[2mm]
&
\displaystyle
C_{02} e^{-(V(r)+E_2+\Delta E_2(r)) \tau}+C_{20} e^{-(\beta-\tau) E_2} e^{-V(r) \tau}. \nonumber\\
&
\end{eqnarray}
The first term is the contribution of the vacuum $Q \bar Q$ energy. The second and third terms are due to thermal effects, they correspond
to broadening of the spectral function and vanish in the zero temperature limit.
The fourth term is just an excited state contribution to the correlator 
at $T=0$ corresponding 
%of 
to a static $\bar Q Q$ pair of size $r$ and a hadron
state with vacuum quantum numbers, e.g. isospin zero $\pi \pi$ states.
The last term is Boltzmann suppressed for all $\tau$, except
for $\tau$ very close to $\beta$. This is the term that causes a rapid decrease of effective masses when $\tau \sim \beta$.
The spectral function corresponding to the above correlator can be written as
\begin{eqnarray}
&
\rho_r(\omega,T)=\rho^{med}_r(\omega,T)+\rho_r^{high}(\omega,T)+\rho_r^{tail}(\omega,T),\nonumber\\[2mm]
& \\[2mm]
& 
\displaystyle
\rho_r^{med}(\omega,T)=\frac{1}{Z(T)}\Bigg[C_{00}^r \delta(\omega-V(r))+\nonumber\\[2mm]
&
\displaystyle
C_{11}^r e^{-\beta E_1} \delta(\omega-V(r)-\Delta E_1(r))+\nonumber \\[2mm]
&
\displaystyle
C_{22}^r e^{-\beta E_2} \delta(\omega-V(r)-\Delta E_2(r))\Bigg],\\[2mm]
&
\displaystyle
\rho_r^{high}(\omega,T)=\frac{1}{Z(T)} C_{02}^r \delta(\omega-V(r)-\Delta E_2(r)-E_2)\nonumber
\\[2mm]
&
\\[2mm]
&
\displaystyle
\rho_r^{tail}(\omega,T)=\frac{1}{Z(T)} C_{20}^r e^{-\beta E_2} \delta(\omega-V(r)+E_2).
\end{eqnarray}
We see that $\rho^{med}$ is a discrete version of a narrow broadened peak, while $\rho_r^{tail}$  can have support for $\omega$ well below the ground state peak.
Note that the above considerations are valid not only for 
%a 
static meson
correlators but also for non-relativistic meson correlators, e.g.
extended bottomonium correlators in NRQCD considered in Refs. \cite{Larsen:2019bwy,Larsen:2019zqv,Larsen:2020rjk}.

It is straightforward to generalize the above spectral function to include all states. First, we note that $E_{0'}$ does not only
correspond to the ground state $Q\bar Q$ energy but also to hybrid potentials as well as static-light mesons. We denote these additional
states by $V_{\alpha}(r),~\alpha\ge 1$. If we use index $k$ for all hadron states with vacuum quantum number (e.g. isospin zero two pion states)
for the spectral function we can write:
\begin{eqnarray}
&
\rho_r(\omega,T)=\rho^{med}_r(\omega,T)+\rho_r^{high}(\omega,T)+\rho_r^{tail}(\omega,T),\nonumber\\[2mm]
& \\[2mm]
&
\displaystyle
\rho_r^{med}(\omega,T)=\frac{1}{Z(T)}\Bigg[C_{00}^r \delta(\omega-V(r))+\nonumber\\[2mm]
&
\displaystyle
\sum_n
C_{nn}^r e^{-\beta E_n} \delta(\omega-V(r)-\Delta E_n(r))\Bigg],\\[2mm]
&
\displaystyle
\rho_r^{high}(\omega,T)=\frac{1}{Z(T)} \Bigg[\sum_k C_{0k}^r \delta(\omega-V(r)-\Delta E_k(r)-E_k) 
\nonumber\\[2mm]
&
\displaystyle
+\sum_{\alpha} C_{00}^{r \alpha} \delta(\omega-V_{\alpha}(r))
\nonumber\\
&
\displaystyle
+\sum_{n,\alpha} C_{nn}^{r \alpha} e^{-\beta E_n} \delta(\omega-V_{\alpha}(r)-\Delta E_n(r))\Bigg]\\[2mm]
&
\displaystyle
\rho_r^{tail}(\omega,T)=\frac{1}{Z(T)} \Bigg[ 
\sum_k C_{k0}^{ r} e^{-\beta E_k} \delta(\omega-V(r)+E_k)+ \nonumber\\
&
\displaystyle
\sum_{\alpha} \sum_k C_{k0}^{r\alpha} e^{-\beta E_k} \delta(\omega-V_{\alpha}(r)+E_k)+ \nonumber\\
&
\displaystyle
\sum_{n,m \neq 0} C_{nm}^r e^{-\beta E_n} \delta(\omega-V(r)-\Delta E_{m}(r)-E_{m}+E_n)+
\nonumber\\
&
\displaystyle
\sum_{\alpha}\sum_{n, m \neq 0} C_{nm}^{r\alpha} e^{-\beta E_n} \delta(\omega-V_{\alpha}(r)-\Delta E_{m}(r)-E_{m}+E_n) \Bigg]
\nonumber\\[2mm]
&
\end{eqnarray}
Since $E_k$ can be very large the $\rho_r^{tail}(\omega,T)$ can be non-zero at very small $\omega$.
We see that the spectral function of static $Q\bar Q$ pair has three contribution: a dominant peak that is broadened
at non-zero temperature, a high $\omega$ part reflecting the excited state contributions and a low $\omega$ tail that mostly contributes at $\tau=\beta$.
This is exactly what we see in our lattice results.

\section{Parameters of the calculations, quark mass and cutoff effects}
\label{app:lat}

In this Appendix we summarize the parameters used in our
lattice calculations, including the bare gauge coupling $\beta=10/g_0^2$,
the quark masses, the corresponding temperatures values, as well as the corresponding statistics in terms of molecular dynamic time units (TU). Measurements of the Wilson loops and Wilson line correlators have been performed every 10 TUs.  We also discuss some systematic errors in our calculations such as quark mass effects and cutoff effects.

The parameters and statistics for the zero temperature lattices are summarized in Table~\ref{tab:tzero}. 
The parameters for
$N_{\tau}=16,~12$ and $10$ lattice are shown in Tables \ref{tab:nt16},
\ref{tab:nt12} and \ref{tab:nt10}, respectively.

\begin{table}
\parbox{.98\linewidth}{
  \begin{tabular}{|c|c|c|c|c|c|c|}
    \hline
    \multicolumn{7}{|c|}{\(m_l=m_s/20\):} \\
    \hline
    $ \beta $ & a\,(\rm{fm}) & \(N_\sigma,N_\tau\) & $am_s$ & $m_\pi L$ & \#TUs & Ref. \\
    \hline
   6.740 & 0.109 & \(48^4\) & 0.0476 & 4.2 & 1350 & \cite{Bazavov:2011nk} \\
   6.800 & 0.103 & \(32^4\) & 0.0448 & 2.7 & 5650 & \cite{Bazavov:2011nk} \\
   6.880 & 0.095 & \(48^4\) & 0.0412 & 3.7 & 1400 & \cite{Bazavov:2011nk} \\
   6.950 & 0.089 & \(32^4\) & 0.0386 & 2.3 & 10830 & \cite{Bazavov:2011nk} \\
   7.030 & 0.083 & \(48^4\) & 0.0356 & 3.2 & 1355 & \cite{Bazavov:2011nk} \\
   7.150 & 0.074 & \(64^3\times 48\) & 0.0320 & 2.9 & 1458 & \cite{Bazavov:2011nk} \\
   7.280 & 0.066 & \(64^3\times 48\) & 0.0284 & 2.5 & 1734 & \cite{Bazavov:2011nk} \\
   7.373 & 0.060 & \(64^3\times 48\) & 0.0250 & 2.3 & 4623 & \cite{Bazavov:2014pvz} \\
   7.596 & 0.049 & \(64^4\) & 0.0202  & 2.6 & 4757 & \cite{Bazavov:2014pvz} \\
   7.825 & 0.040 & \(64^4\) & 0.0164  & 2.0 & 4768 & \cite{Bazavov:2014pvz} \\
    \hline
    \hline
    \multicolumn{7}{|c|}{\(m_l=m_s/5\):} \\
    \hline
    $ \beta $ & a\,[\rm{fm}] & \(N_\sigma,N_\tau\) & $am_s$ & \(m_\pi L\) & \#TUs & Ref. \\
    \hline
    8.000 & 0.035 & \(64^4\) & 0.01299 & 3.6 & 4616 & \cite{Bazavov:2017dsy} \\
    8.200 & 0.029 & \(64^4\) & 0.01071 & 3.1 & 4616 & \cite{Bazavov:2017dsy} \\
    8.400 & 0.025 & \(64^4\) & 0.00887 & 2.6 & 4616 & \cite{Bazavov:2017dsy} \\
    \hline
  \end{tabular}
  \caption{\label{tab:tzero}
  Parameters of the calculations on zero temperature lattices for $m_l=m_s/20$  (upper part)
and $m_l=m_s/5$ (lower part).
  }
}
\end{table}

\begin{table}[t]
\parbox{1.0\linewidth}{
  \begin{tabular}{|c|c|c|c|}
    \hline
    \multicolumn{4}{|c|}{\(m_l=m_s/20\):} \\
    \hline
    $ \beta $  & $am_s$ & T (MeV) & \#TUs \\
    \hline
    7.825 & 0.00164  & 306  & 67960 \\
    8.570 & 0.008376 & 577  & 10400 \\
    8.710 & 0.007394 & 650  & 10190 \\
    8.850 & 0.006528 & 731  &  4480 \\
    9.060 & 0.004834 & 872  & 41870 \\
    9.230 & 0.004148 & 1005 &  3610 \\
    9.360 & 0.003691 & 1121 &  3530 \\
    9.490 & 0.003285 & 1250 &  6790 \\
    9.670 & 0.002798 & 1454 & 42060 \\
    \hline
    \hline
    \multicolumn{4}{|c|}{\(m_l=m_s/5\):} \\
    \hline
    $ \beta $  & $am_s$ & T (MeV) & \#TUs  \\
    \hline
    8.000 & 0.001299 & 356  & 11460 \\
    8.200 & 0.001071 & 422  & 10660 \\
    8.400 & 0.000887 & 500  & 64370 \\
    \hline
  \end{tabular}
  \caption{
\label{tab:nt16}
  Parameters of the $64^3 \times 16$ calculations for the two different light quark masses.
  The lower part of the table corresponds to $m_l=m_s/5$, while the upper part corresponds
  to $m_l=m_s/20$.
  }
}
\end{table}

\begin{table}
\parbox{.98\linewidth}{
  \begin{tabular}{|c|c|c|c|}
    \hline
    \multicolumn{4}{|c|}{\(m_l=m_s/20\)} \\
    \hline
    $ \beta $ & $am_s$ & T (MeV) & \#TUs \\
    \hline
    6.515 & 0.0604  &  122 & 32500 \\
    6.608 & 0.0542  &  133 & 19990 \\
    6.664 & 0.0514  &  141 & 45120 \\
    6.700 & 0.0496  &  146 & 15900 \\
    6.740 & 0.0476  &  151 & 29410 \\
    6.770 & 0.0460  &  156 & 15530 \\
    6.800 & 0.0448  &  160 & 36060 \\
    6.840 & 0.0430  &  166 & 17370 \\
    6.880 & 0.0412  &  173 & 46350 \\
    6.950 & 0.0386  &  185 & 50550 \\
    7.030 & 0.0378  &  199 & 65940 \\
    7.100 & 0.0332  &  213 & 9640 \\
    7.150 & 0.0320  &  223 & 9600 \\
    7.200 & 0.0296  &  233 & 4010 \\
    7.280 & 0.0284  &  251 & 58210 \\
    7.373 & 0.0250  &  273 & 85120 \\
    7.596 & 0.0202  &  334 & 98010 \\
    7.650 & 0.0202  &  350 & 3230  \\
    7.825 & 0.0164  &  408 & 134600 \\
    8.000 & 0.0140  &  474 & 3110  \\
    8.200 & 0.01167 &  562 & 30090  \\
    8.400 & 0.00975 &  667 & 29190  \\
    8.570 & 0.008376 &  770 & 6320 \\
    8.710 & 0.007394 &  866 & 6490 \\
    8.850 & 0.006528 &  974 & 6340 \\
    9.060 & 0.004834 & 1162 & 7430 \\
    9.230 & 0.004148 & 1340 & 7280 \\
    9.360 & 0.003691 & 1495 & 7910 \\
    9.490 & 0.003285 & 1667 & 9780 \\
    9.670 & 0.002798 & 1938 & 7650 \\
    \hline
    \multicolumn{4}{|c|}{\(m_l=m_s/5\)} \\
    \hline
    $ \beta $  & $am_s$ & T (MeV) & \#TUs  \\
    \hline
    8.000 & 0.01299  &  474 & 71670  \\
    8.200 & 0.01071  &  563 & 71390  \\
    8.400 & 0.00887  &  667 & 71170  \\
    \hline
  \end{tabular}
  \caption{\label{tab:nt12}
Parameters of the calculations on $48^3 \times 12$ lattices for $m_l=m_s/20$  (upper part)
and $m_l=m_s/5$ (lower part).}
}
\end{table}

\begin{table}
\parbox{.98\linewidth}{
  \begin{tabular}{|c|c|c|c|c|c|}
    \hline
    \multicolumn{4}{|c|}{\(m_l=m_s/20\)} \\
    \hline
    $ \beta $ & $am_s$ & T (MeV) & \#TUs \\
    \hline
    6.285 & 0.0790  &  116 & 9260 \\
    6.341 & 0.0740  &  123 & 39190 \\
    6.423 & 0.0670  &  133 & 10360 \\
    6.488 & 0.0620  &  142 & 102690 \\
    6.515 & 0.0604  &  146 & 107530 \\
    6.575 & 0.0564  &  155 & 106020 \\
    6.608 & 0.0542  &  160 & 112890 \\
    6.664 & 0.0514  &  169 & 155440 \\
    6.740 & 0.0476  &  181 & 200250 \\
    6.800 & 0.0448  &  192 & 279830 \\
    6.880 & 0.0412  &  208 & 341490 \\
    6.950 & 0.0386  &  222 & 243480 \\
    7.030 & 0.0378  &  239 & 137730 \\
    7.150 & 0.0320  &  267 & 145440 \\
    7.280 & 0.0284  &  301 & 105990 \\
    7.373 & 0.0250  &  328 & 50840 \\
    7.596 & 0.0202  &  400 & 51710 \\
    7.825 & 0.0164  &  489 & 54000 \\
    8.000 & 0.0140  &  569 & 6780  \\
    8.200 & 0.01167 &  675 & 27500 \\
    8.400 & 0.00975 &  800 & 7540 \\
    8.570 & 0.008376 & 924 & 3000 \\
    8.710 & 0.007394 & 1039 & 15320 \\
    8.850 & 0.006528 & 1169 & 7690  \\
    9.060 & 0.004834 & 1395 & 15490 \\
    9.230 & 0.004148 & 1608 & 7630 \\
    9.360 & 0.003691 & 1794 & 15800 \\
    9.490 & 0.003285 & 2000 & 7990 \\
    9.670 & 0.002798 & 2326 & 15760 \\
    \hline
    \multicolumn{4}{|c|}{\(m_l=m_s/5\)} \\
    \hline
    $ \beta $ & $am_s$ & T (MeV) & \#TUs \\
    \hline
    8.000 & 0.01299 & 569 & 82770 \\
    8.200 & 0.01071 & 675 & 72180 \\
    8.400 & 0.00887 & 800 & 72770 \\
    \hline
  \end{tabular}
  \caption{\label{tab:nt10}
  Parameters of the calculations on $40^3 \times 10$ lattices for $m_l=m_s/20$  (upper part)
and $m_l=m_s/5$ (lower part).
  }
}
\end{table}

\begin{figure}
    \centering{
    \includegraphics[width=7cm]{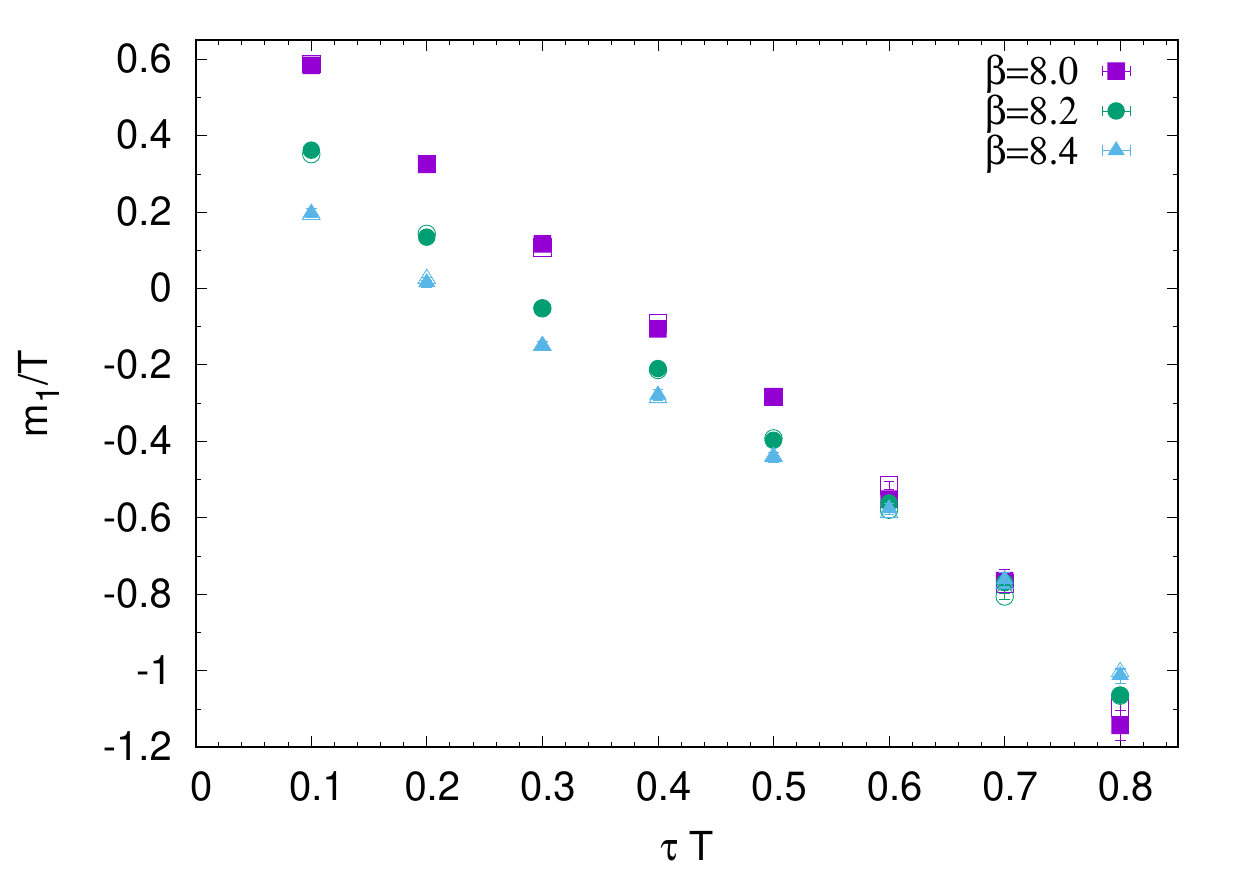}
    \includegraphics[width=7cm]{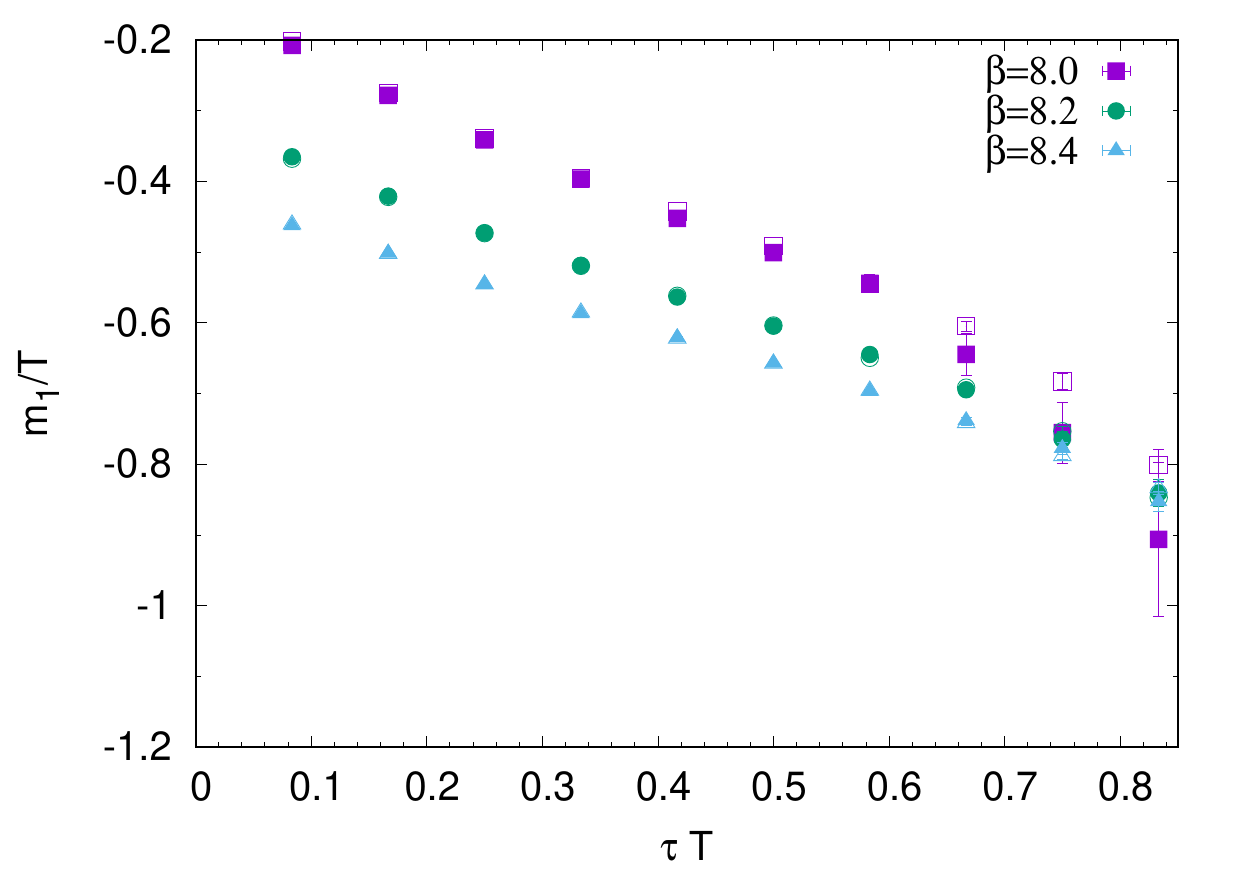}
    }
    \caption{The first cumulants calculated for $N_{\tau}=10$, $rT=1$ (top) and $N_{\tau}=12$, $rT=1/2$ (bottom) for
    $\beta=8.0,~8.2$ and $8.4$ for different quark masses.
    The open symbols correspond to $m_s/5$ calculations, while the filled symbols correspond to $m_s/20$ calculations.}
    \label{fig:meff_ml}
\end{figure}

As one can see from the tables for $N_{\tau}=10$
and $N_{\tau}=12$ lattices we have calculations
at the same values of $\beta=8.0,~8.2$ and $8.4$ but with different quark masses.
In Fig. \ref{fig:meff_ml} we show the comparison of the effective masses
calculated at different sea quark quark masses for $N_{\tau}=10$ and $12$. 
We see from the figures that there is no statistically
significant difference between the calculations performed
for $m_l=m_s/5$ and $m_s/20$. Thus we conclude that
the quark mass effects are small for temperatures
$T>474$ MeV.

\begin{figure}
    \centering{
    \includegraphics[width=7cm]{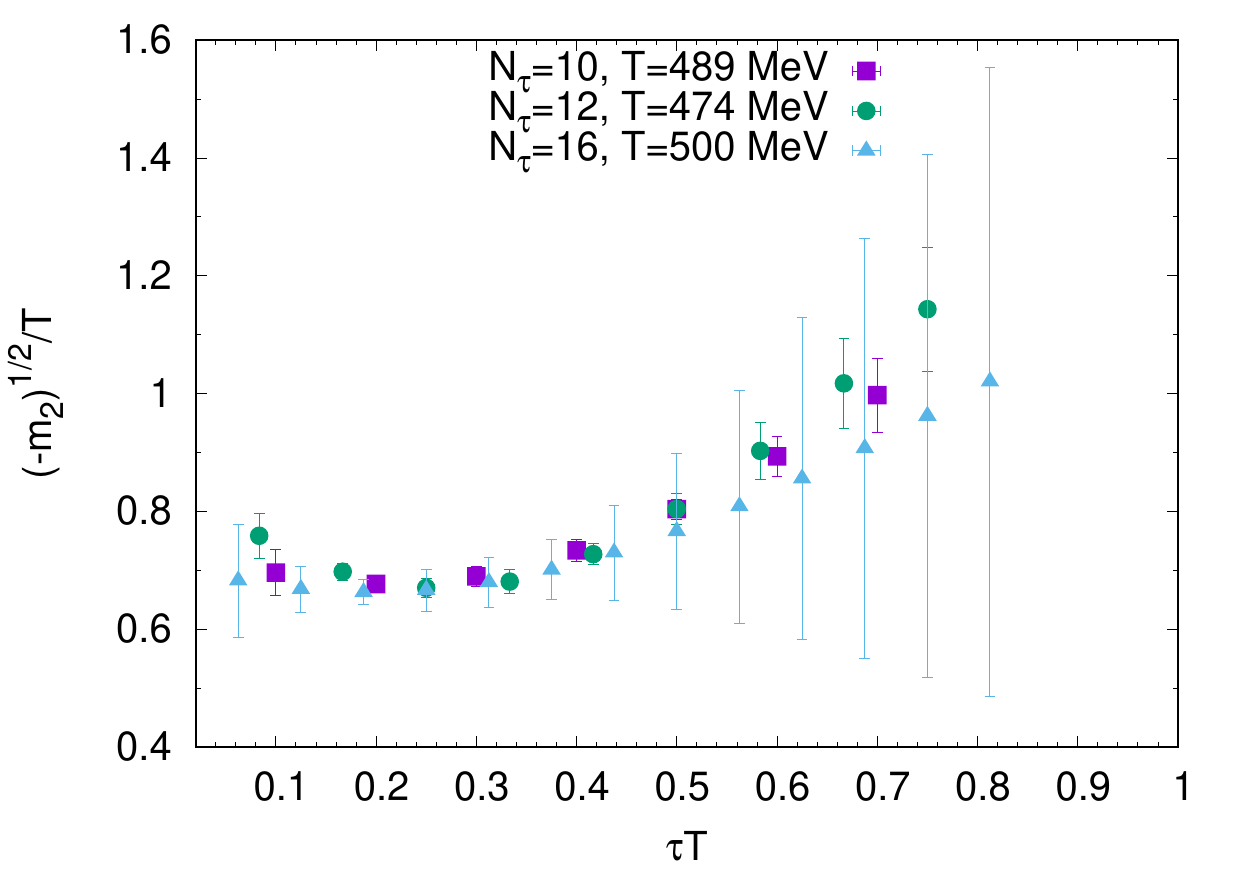}
    \includegraphics[width=7cm]{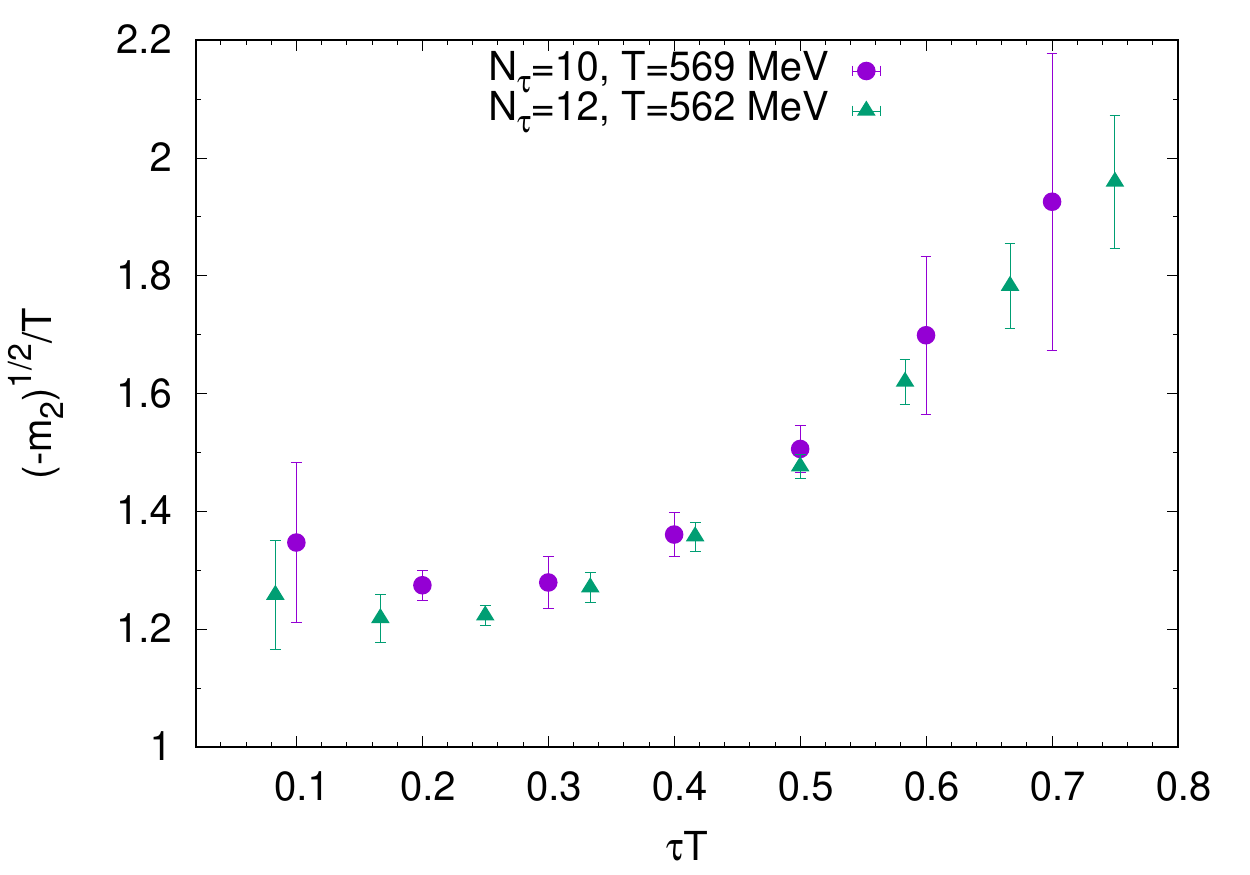}
    }
    \caption{The cutoff dependence of the second  cumulant
    for temperatures around $500$ MeV (top) and around $570 $ MeV (bottom) for distances $rT=1/2$ and $rT=1$, respectively.}
    \label{fig:m2_Ntdep}
\end{figure}

We study cutoff effects in terms of the second cumulant,
$m_2$. In Fig. \ref{fig:m2_Ntdep} we show the $N_{\tau}$
dependence of the square root of negative $m_2$ 
as function of $\tau T$ for temperatures which are very close. We do not see statistically significant $N_{\tau}$ dependence.

\section{Quality of the fits of the $Q\bar Q$ correlation function}
\label{app:fit}

\begin{figure*}
\includegraphics[width=5cm]{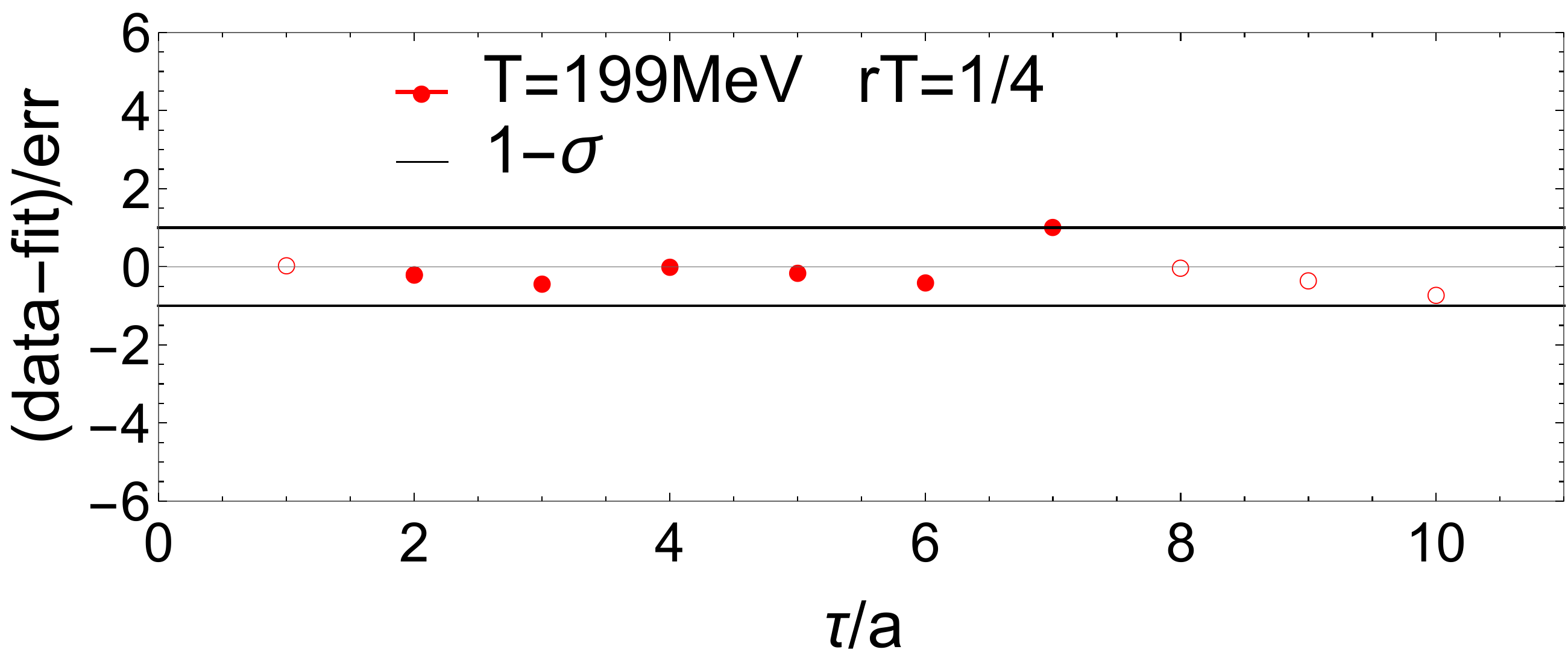}
\includegraphics[width=5cm]{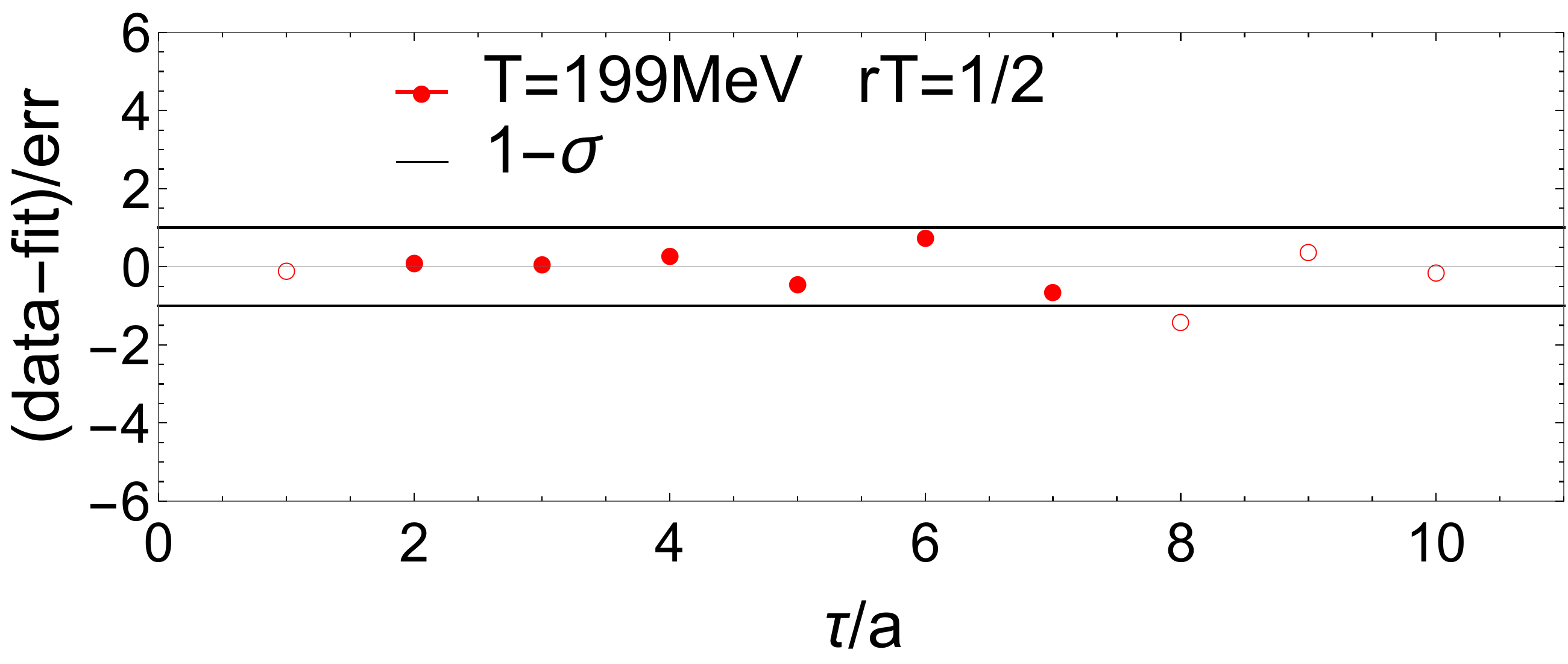}
\includegraphics[width=5cm]{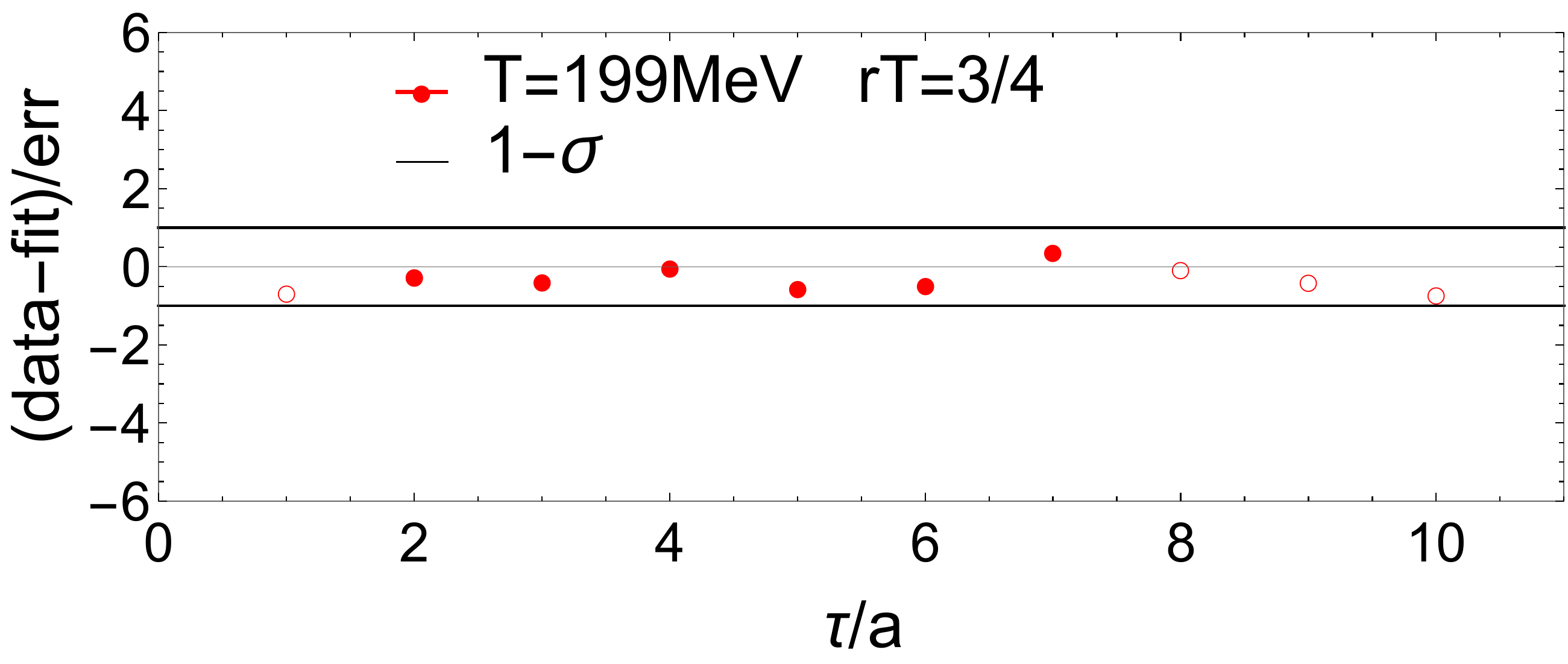}
\includegraphics[width=5cm]{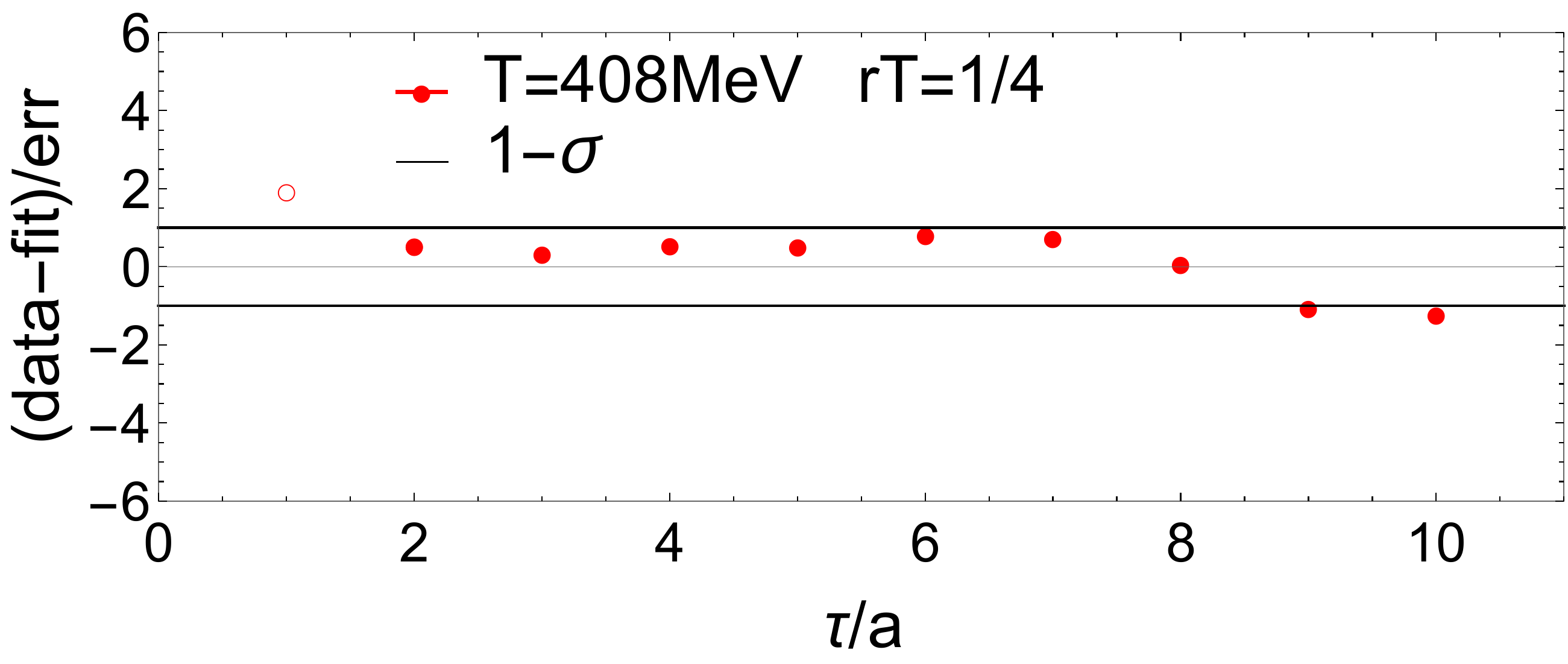}
\includegraphics[width=5cm]{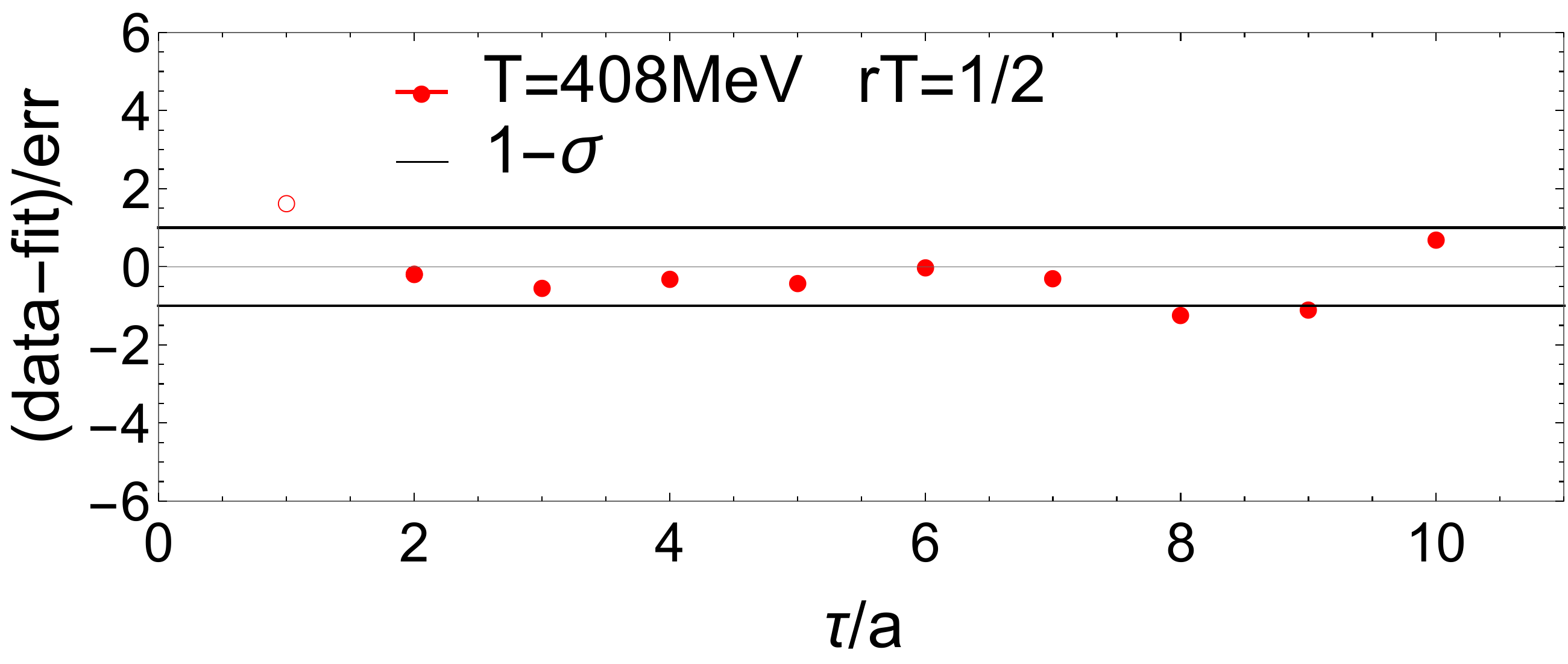}
\includegraphics[width=5cm]{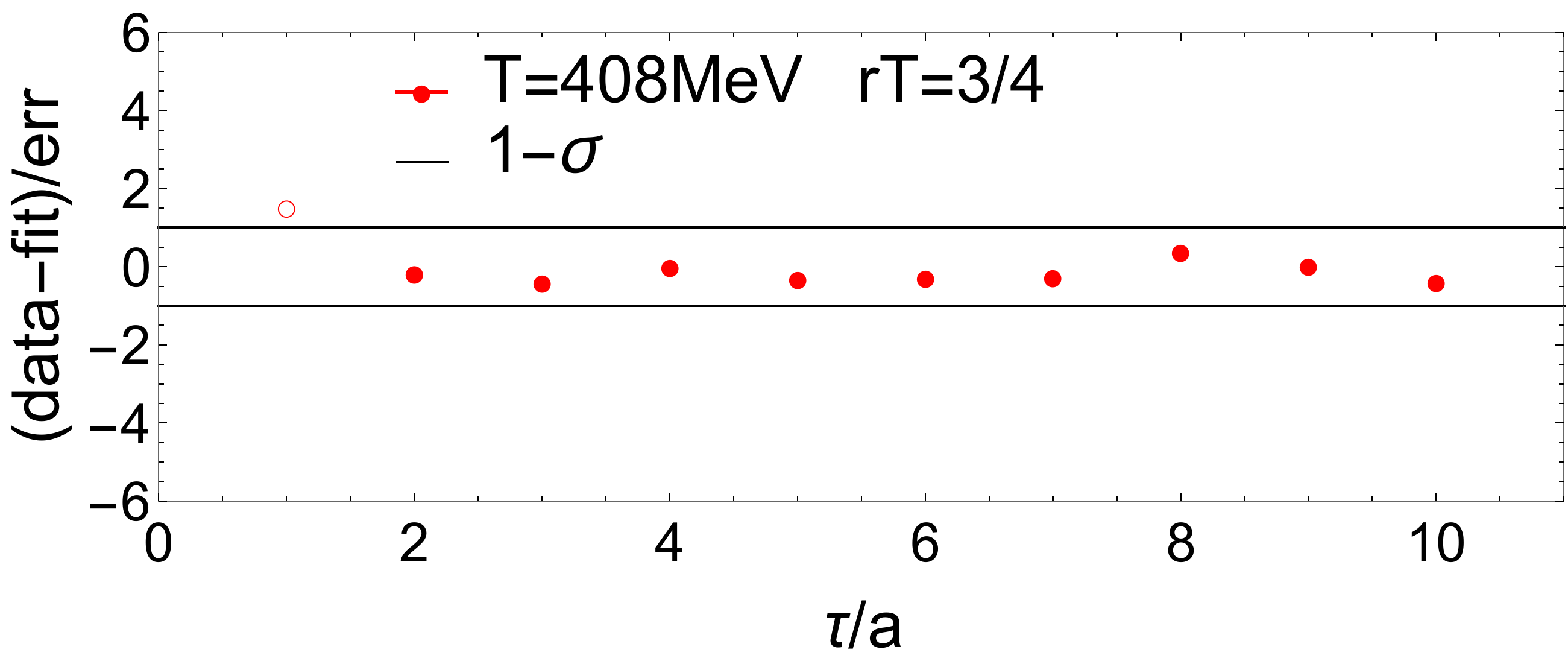}
\includegraphics[width=5cm]{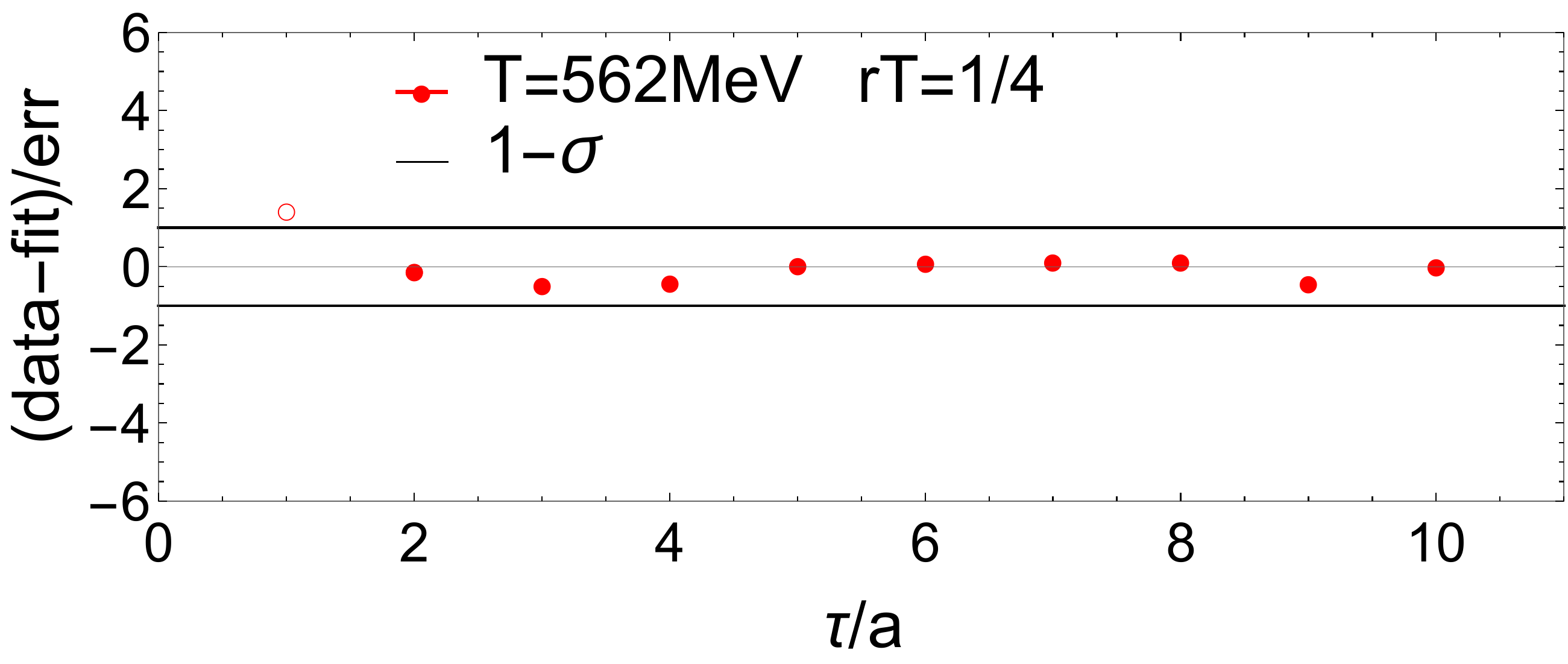}
\includegraphics[width=5cm]{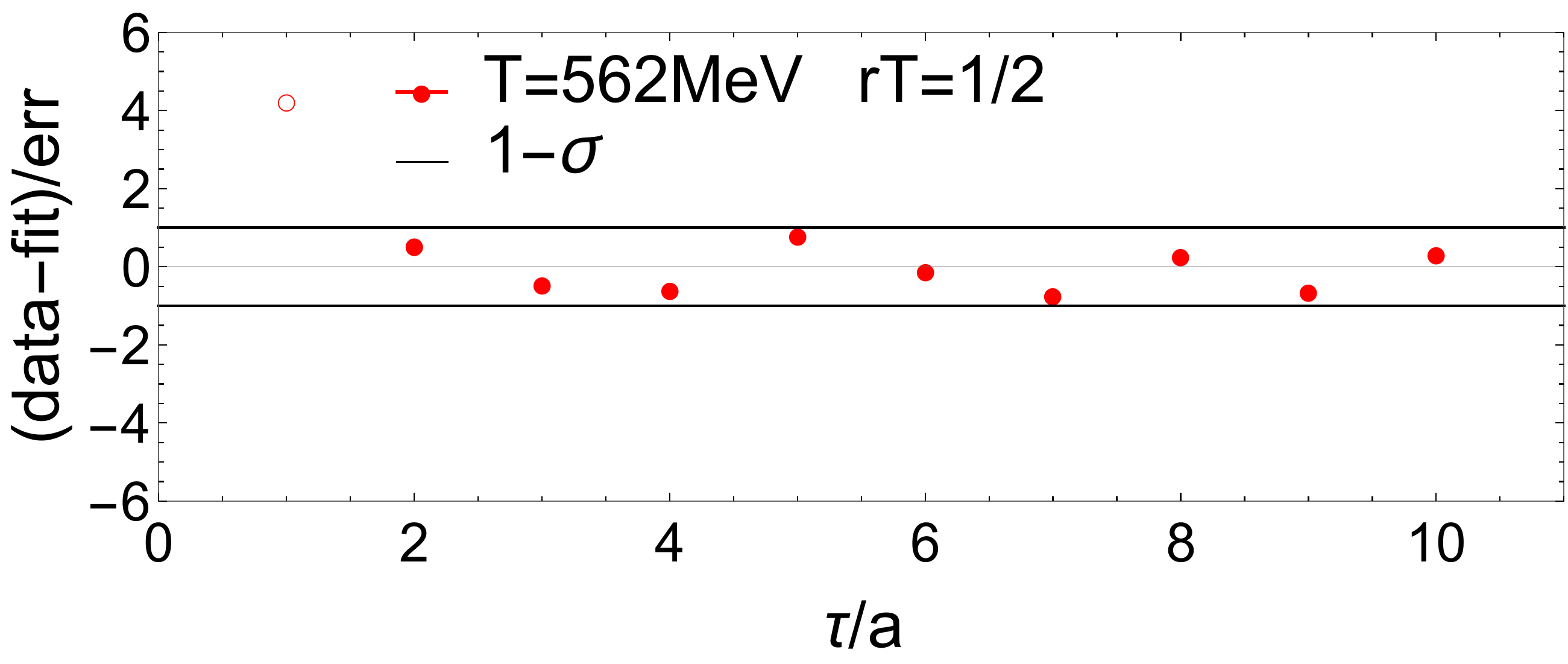}
\includegraphics[width=5cm]{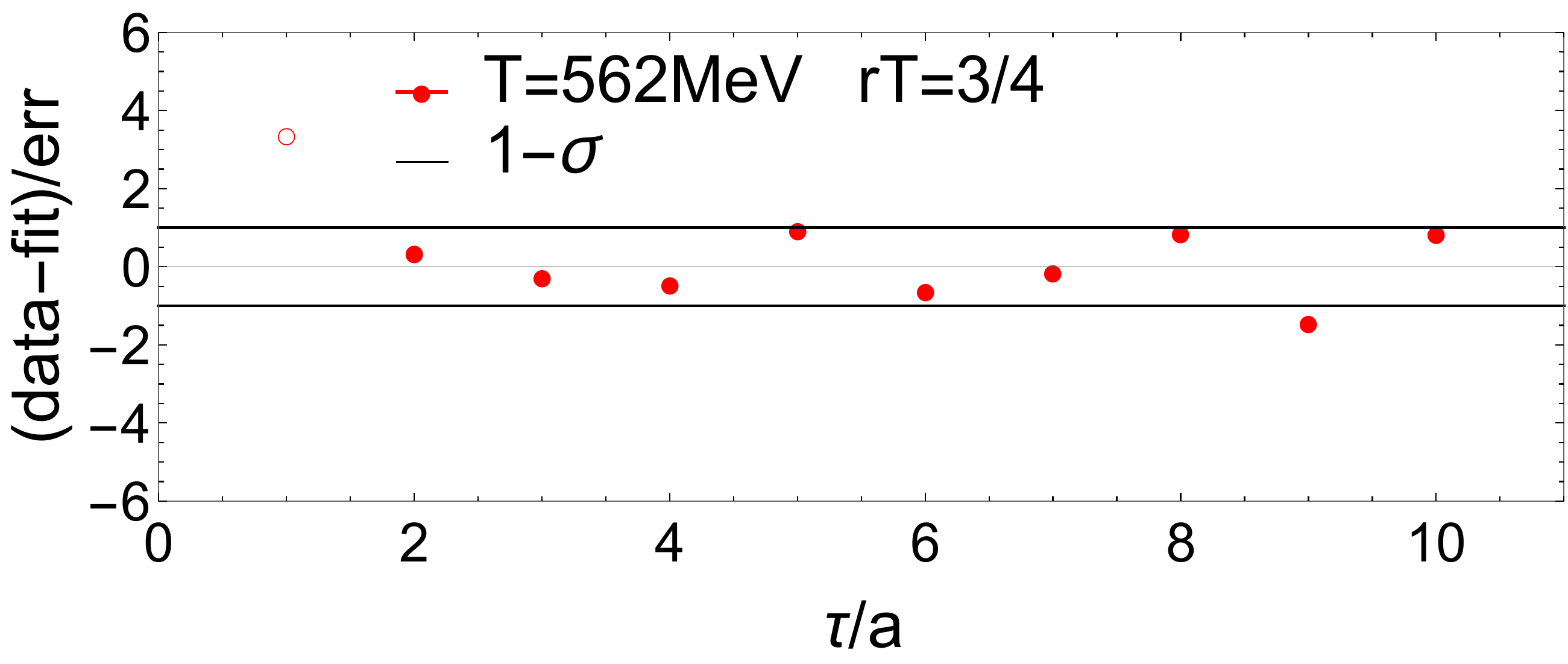}
\caption{The difference between the Gaussian fit and the lattice data on the first cumulant for $N_{\tau}=12$ as function of $\tau$ normalized
by the statistical errors for $rT=1/4$ (left), $rT=1/2$ (middle) and $rT=3/4$ (right). The top
row shows the result for $T=199$ MeV, the middle row shows the result for $T=408$ MeV, while the bottom
row corresponds to $T=562$ MeV. The open symbols correspond to the data points not included in
the fit.}
\label{fig:Gaussfit_res}
\end{figure*}

In this appendix we discuss the quality of the fit of the static $Q\bar Q$ correlation
function in Coulomb gauge with the Gaussian form given by Eq. (\ref{GAnsatz}) and with the HTL inspired (Bala-Datta) form given by Eq. (\ref{BDfit}).
In Fig. \ref{fig:Gaussfit_res} we show the residue of the Gaussian 
fits in terms of the first cumulant for three representative temperatures and 
$rT=1/4,~1/2$ and $3/4$.
As one can see from the figure
the Gaussian form describes the Wilson line correlator very well in the entire $\tau$-range, with possibly an exception of $\tau=a$ data point at high temperatures.

\begin{figure*}
\includegraphics[width=5cm]{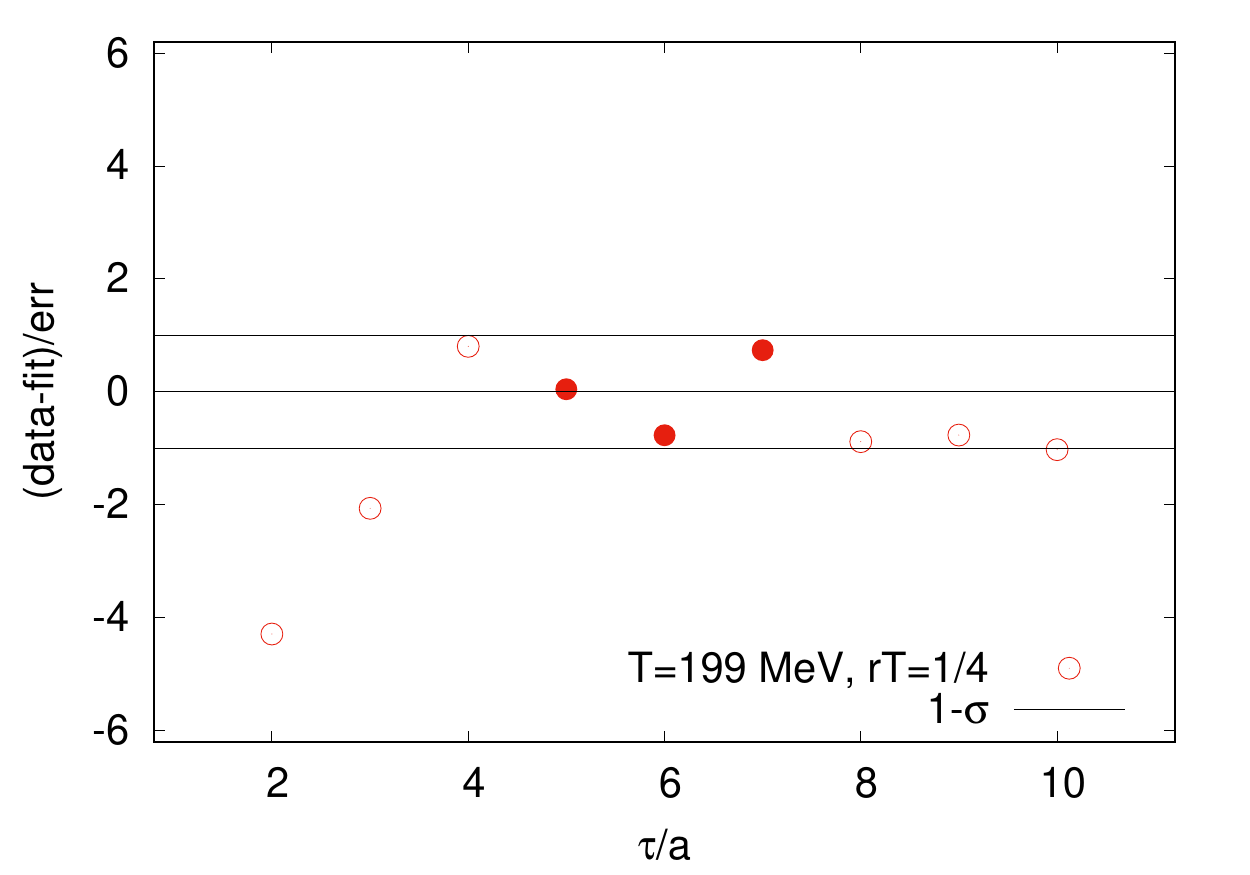}
\includegraphics[width=5cm]{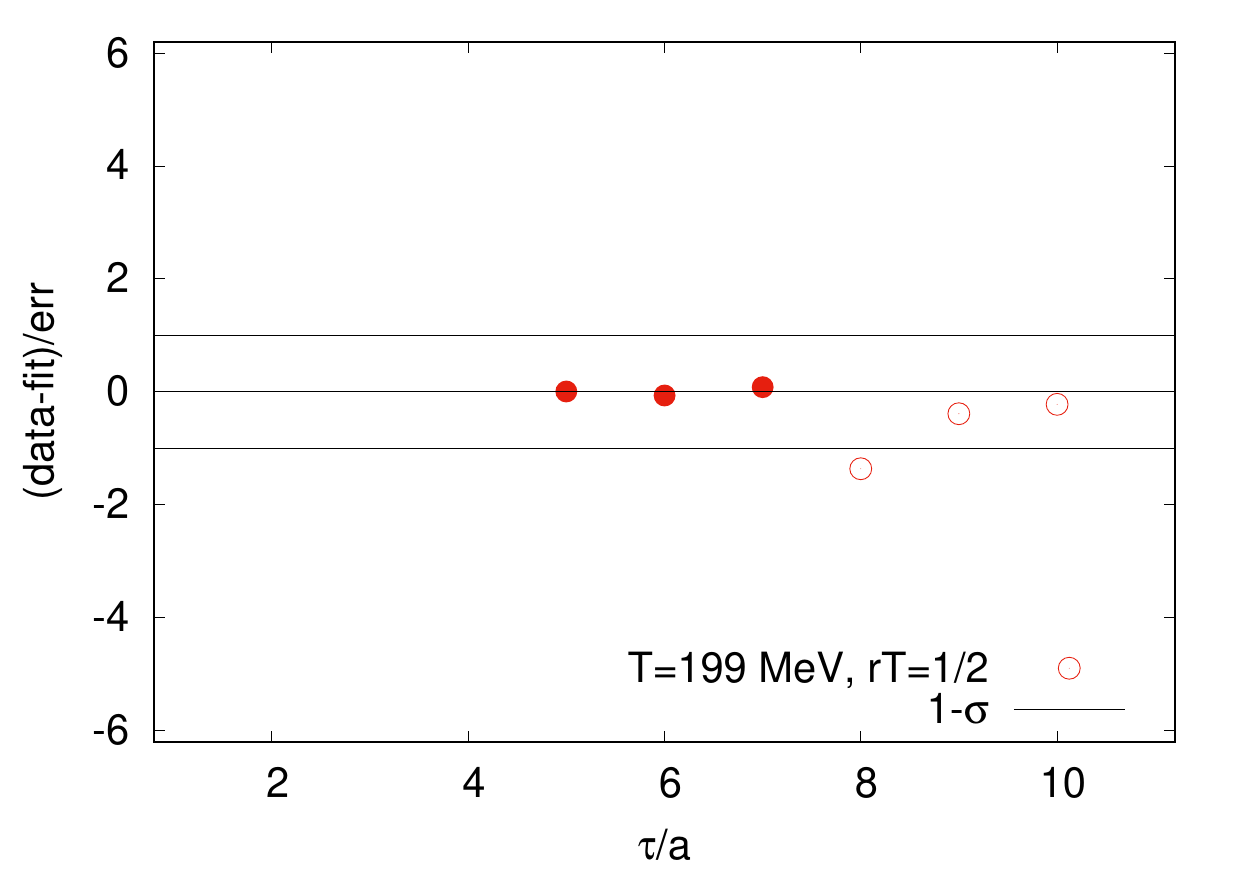}
\includegraphics[width=5cm]{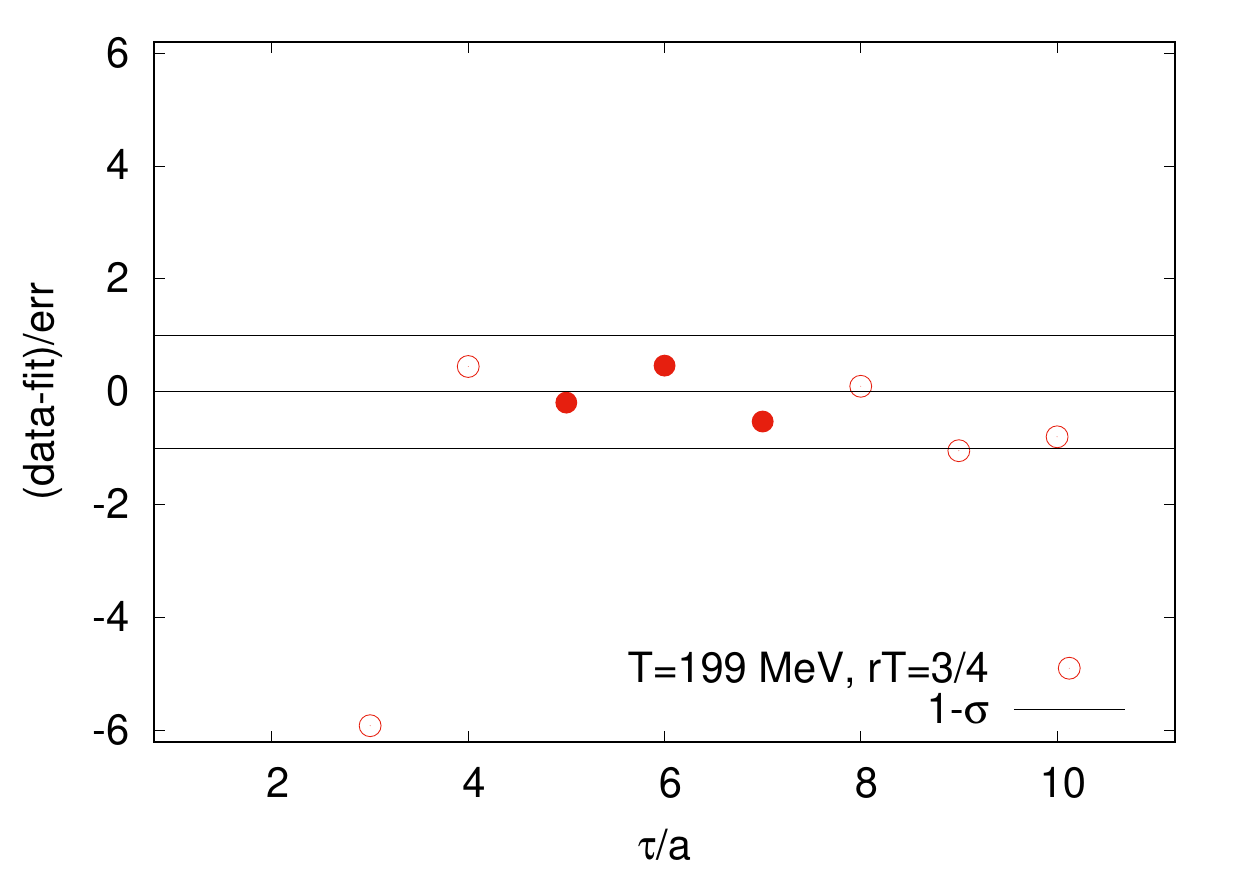}
\includegraphics[width=5cm]{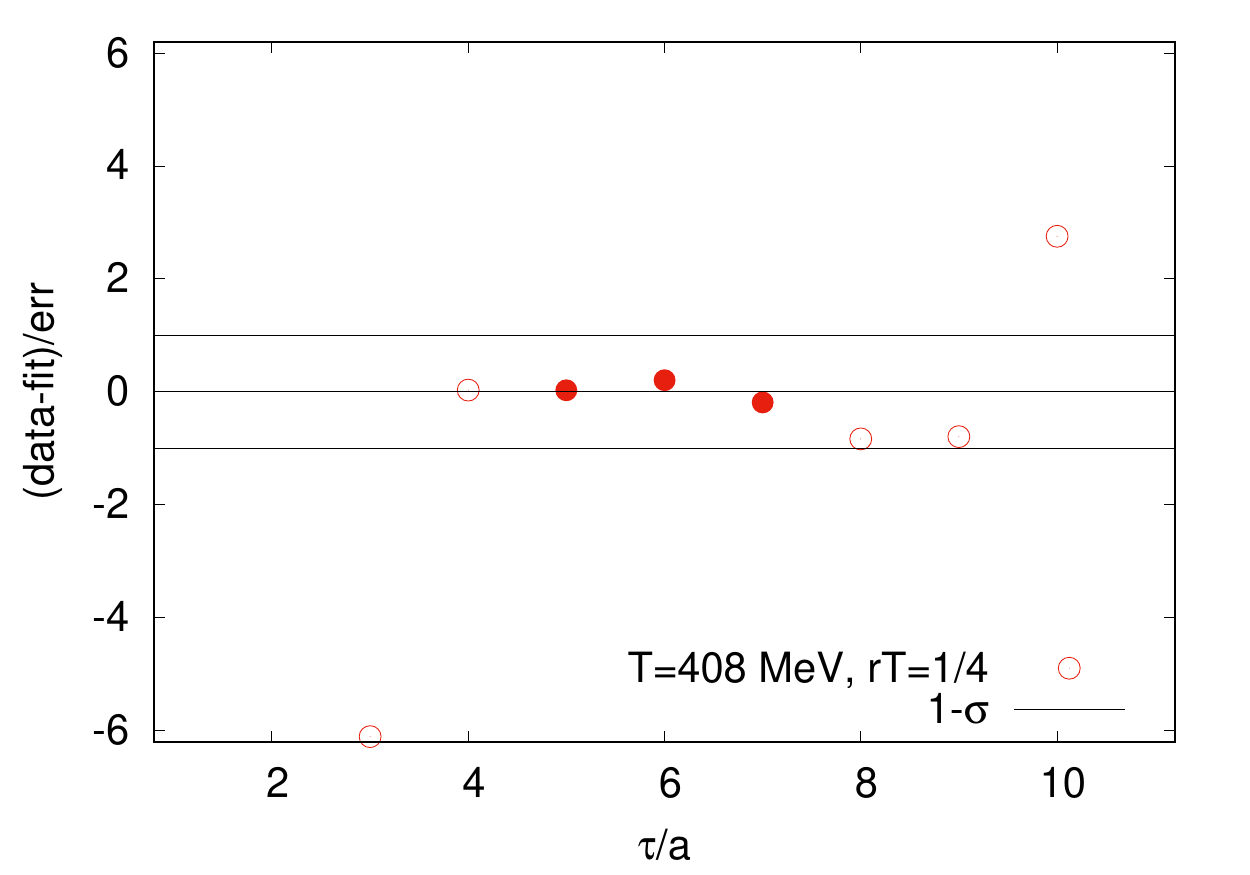}
\includegraphics[width=5cm]{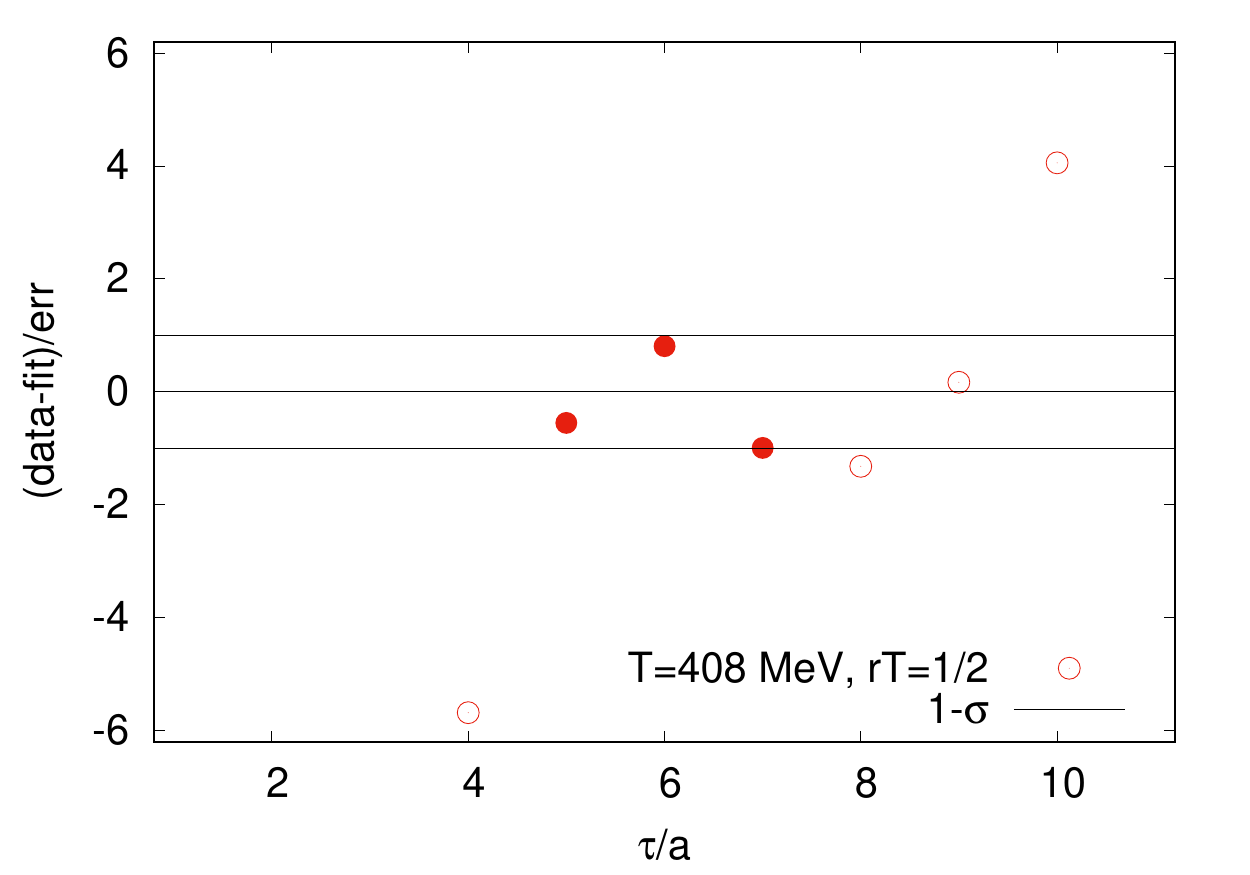}
\includegraphics[width=5cm]{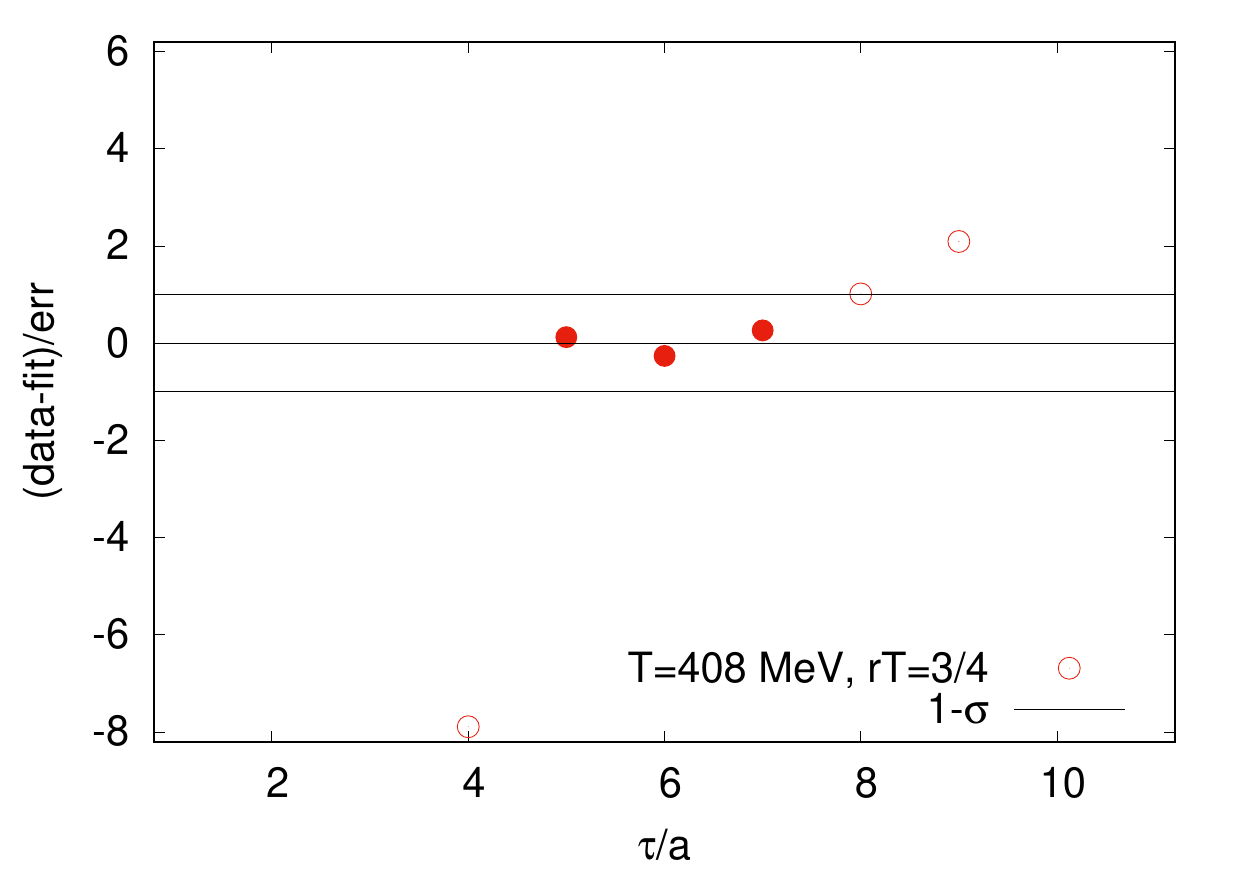}
\includegraphics[width=5cm]{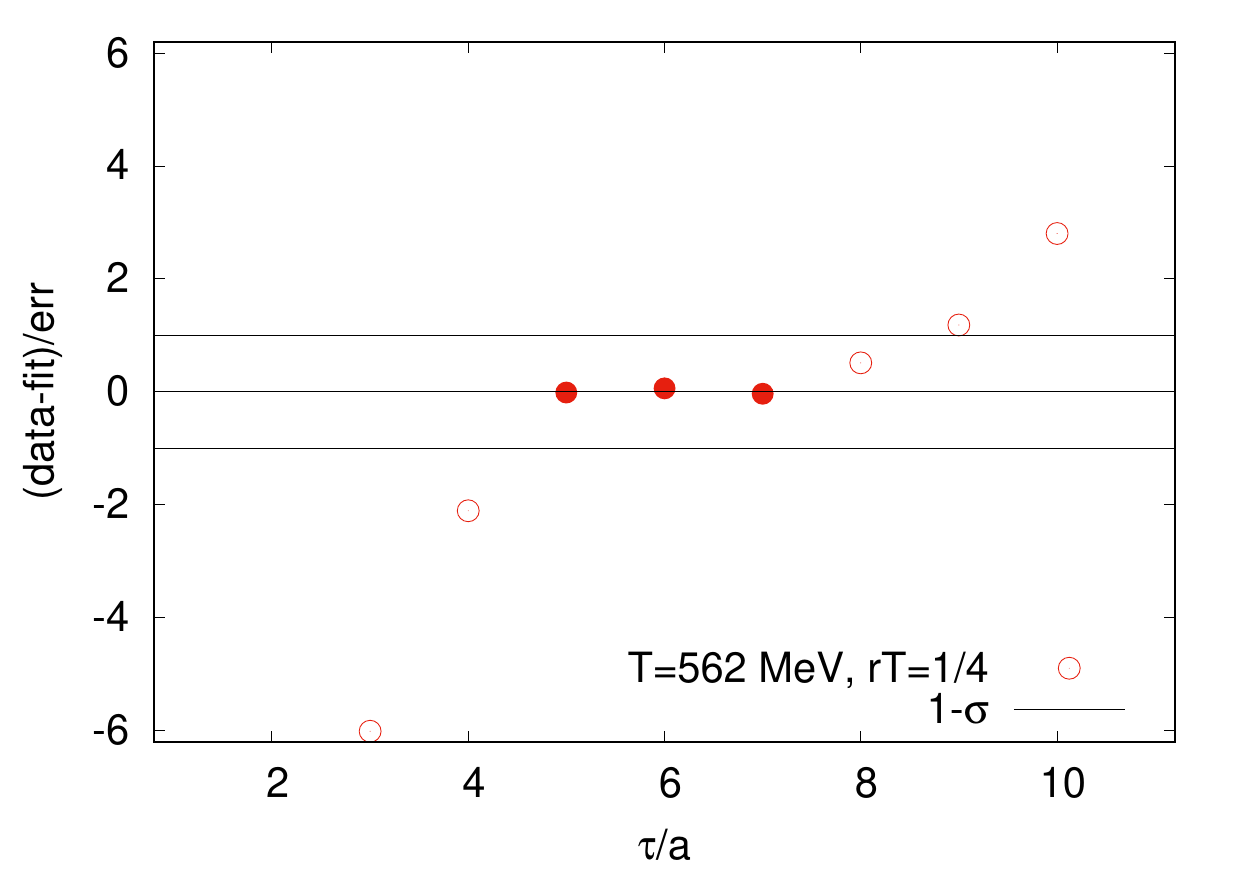}
\includegraphics[width=5cm]{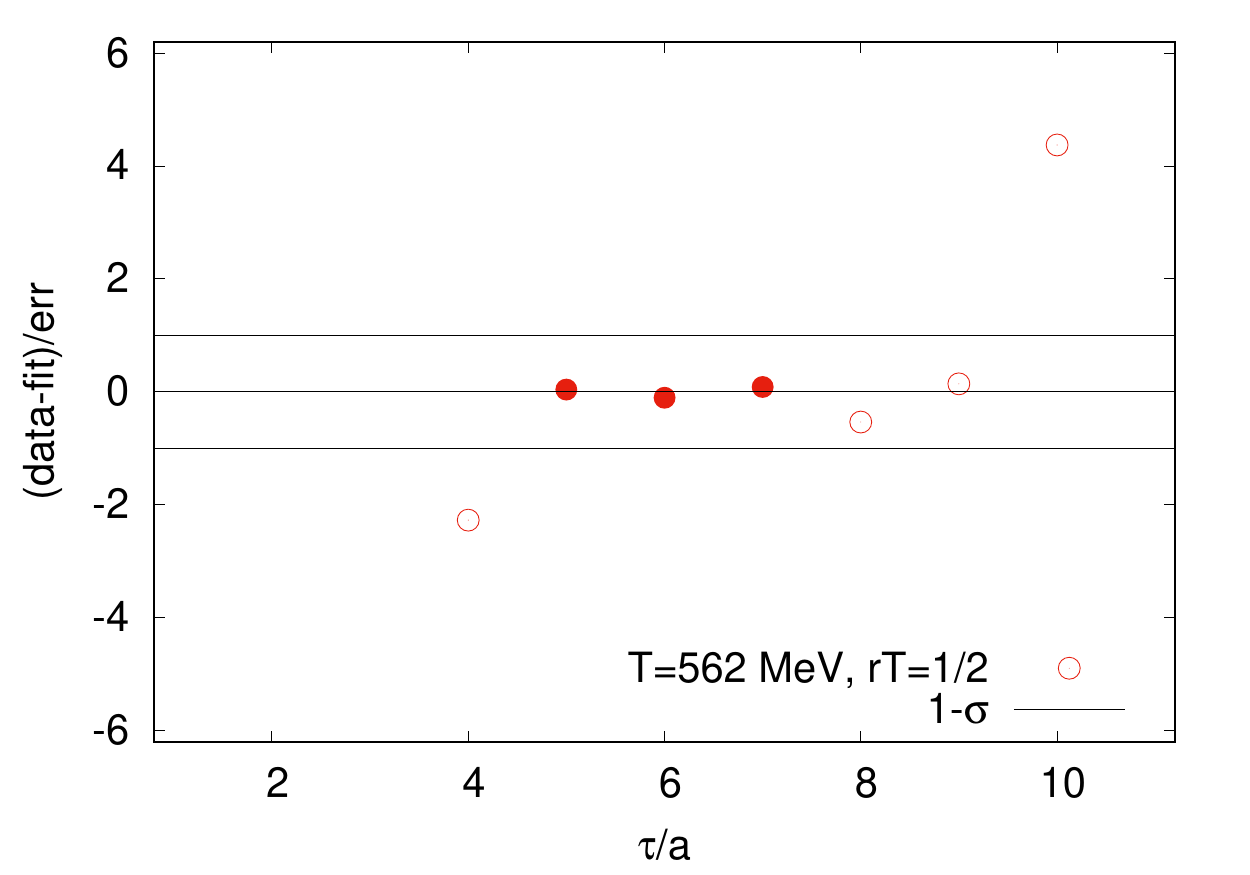}
\includegraphics[width=5cm]{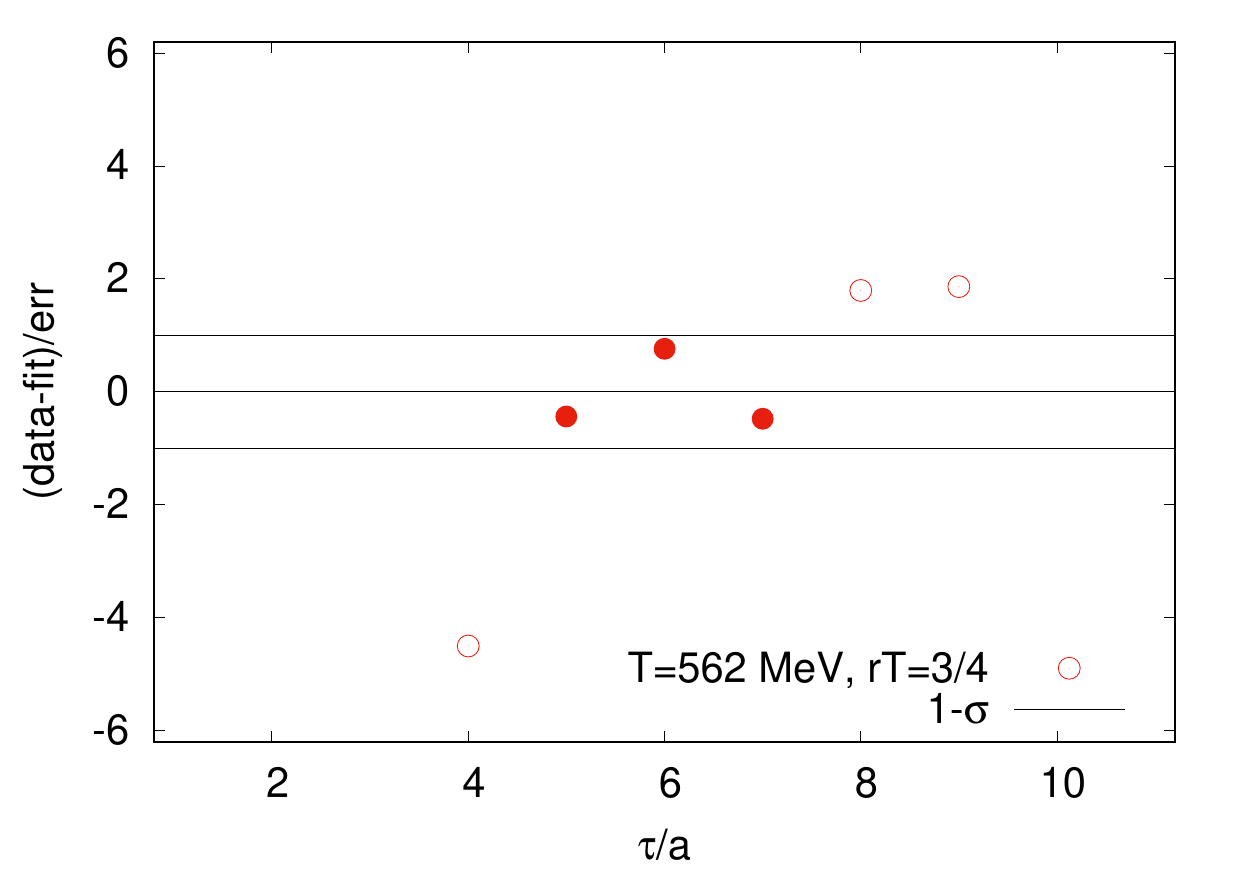}
\caption{The difference between the Bala-Datta fit and the lattice data on the first cumulant for $N_{\tau}=12$ as function of $\tau$ normalized
by the statistical errors for $rT=1/4$ (left), $rT=1/2$ (middle) and $rT=3/4$ (right). The top
row shows the result for $T=199$ MeV, the middle row shows the result for $T=408$ MeV, while the bottom
row corresponds to $T=562$ MeV. The open symbols correspond to the data points not included in
the fit.}
\label{fig:BDfit_res}
\end{figure*}

In Fig. \ref{fig:BDfit_res} we show the residual of the fits for the Bala-Datta form for the first cumulant.
For the shortest distance, $rT=1/4$ the fit captures the data well, except at the shortest and the largest
values of $\tau$. For larger distances this fit only works for $\tau T$ around $1/2$.

\newpage
\bibliography{ref}

%apsrev4-2.bst 2019-01-14 (MD) hand-edited version of apsrev4-1.bst
%Control: key (0)
%Control: author (8) initials jnrlst
%Control: editor formatted (1) identically to author
%Control: production of article title (0) allowed
%Control: page (0) single
%Control: year (1) truncated
%Control: production of eprint (0) enabled
\begin{thebibliography}{69}%
\makeatletter
\providecommand \@ifxundefined [1]{%
 \@ifx{#1\undefined}
}%
\providecommand \@ifnum [1]{%
 \ifnum #1\expandafter \@firstoftwo
 \else \expandafter \@secondoftwo
 \fi
}%
\providecommand \@ifx [1]{%
 \ifx #1\expandafter \@firstoftwo
 \else \expandafter \@secondoftwo
 \fi
}%
\providecommand \natexlab [1]{#1}%
\providecommand \enquote  [1]{``#1''}%
\providecommand \bibnamefont  [1]{#1}%
\providecommand \bibfnamefont [1]{#1}%
\providecommand \citenamefont [1]{#1}%
\providecommand \href@noop [0]{\@secondoftwo}%
\providecommand \href [0]{\begingroup \@sanitize@url \@href}%
\providecommand \@href[1]{\@@startlink{#1}\@@href}%
\providecommand \@@href[1]{\endgroup#1\@@endlink}%
\providecommand \@sanitize@url [0]{\catcode `\\12\catcode `\$12\catcode
  `\&12\catcode `\#12\catcode `\^12\catcode `\_12\catcode `\%12\relax}%
\providecommand \@@startlink[1]{}%
\providecommand \@@endlink[0]{}%
\providecommand \url  [0]{\begingroup\@sanitize@url \@url }%
\providecommand \@url [1]{\endgroup\@href {#1}{\urlprefix }}%
\providecommand \urlprefix  [0]{URL }%
\providecommand \Eprint [0]{\href }%
\providecommand \doibase [0]{https://doi.org/}%
\providecommand \selectlanguage [0]{\@gobble}%
\providecommand \bibinfo  [0]{\@secondoftwo}%
\providecommand \bibfield  [0]{\@secondoftwo}%
\providecommand \translation [1]{[#1]}%
\providecommand \BibitemOpen [0]{}%
\providecommand \bibitemStop [0]{}%
\providecommand \bibitemNoStop [0]{.\EOS\space}%
\providecommand \EOS [0]{\spacefactor3000\relax}%
\providecommand \BibitemShut  [1]{\csname bibitem#1\endcsname}%
\let\auto@bib@innerbib\@empty
%</preamble>
\bibitem [{\citenamefont {Rothkopf}(2020)}]{Rothkopf:2019ipj}%
  \BibitemOpen
  \bibfield  {author} {\bibinfo {author} {\bibfnamefont {A.}~\bibnamefont
  {Rothkopf}},\ }\bibfield  {title} {\bibinfo {title} {{Heavy Quarkonium in
  Extreme Conditions}},\ }\href {https://doi.org/10.1016/j.physrep.2020.02.006}
  {\bibfield  {journal} {\bibinfo  {journal} {Phys. Rept.}\ }\textbf {\bibinfo
  {volume} {858}},\ \bibinfo {pages} {1} (\bibinfo {year} {2020})},\ \Eprint
  {https://arxiv.org/abs/1912.02253} {arXiv:1912.02253 [hep-ph]} \BibitemShut
  {NoStop}%
\bibitem [{\citenamefont {Aarts}\ \emph {et~al.}(2017)\citenamefont {Aarts}
  \emph {et~al.}}]{Aarts:2016hap}%
  \BibitemOpen
  \bibfield  {author} {\bibinfo {author} {\bibfnamefont {G.}~\bibnamefont
  {Aarts}} \emph {et~al.},\ }\bibfield  {title} {\bibinfo {title}
  {{Heavy-flavor production and medium properties in high-energy nuclear
  collisions - What next?}},\ }\bibfield  {booktitle} {\emph {\bibinfo
  {booktitle} {{Lorentz workshop: Tomography of the Quark-Gluon Plasma with
  Heavy Quarks Leiden, Netherlands, October 10-14, 2016}}},\ }\href
  {https://doi.org/10.1140/epja/i2017-12282-9} {\bibfield  {journal} {\bibinfo
  {journal} {Eur. Phys. J.}\ }\textbf {\bibinfo {volume} {A53}},\ \bibinfo
  {pages} {93} (\bibinfo {year} {2017})},\ \Eprint
  {https://arxiv.org/abs/1612.08032} {arXiv:1612.08032 [nucl-th]} \BibitemShut
  {NoStop}%
%%CITATION = ARXIV:1612.08032;%%
\bibitem [{\citenamefont {Mocsy}\ \emph {et~al.}(2013)\citenamefont {Mocsy},
  \citenamefont {Petreczky},\ and\ \citenamefont {Strickland}}]{Mocsy:2013syh}%
  \BibitemOpen
  \bibfield  {author} {\bibinfo {author} {\bibfnamefont {A.}~\bibnamefont
  {Mocsy}}, \bibinfo {author} {\bibfnamefont {P.}~\bibnamefont {Petreczky}},\
  and\ \bibinfo {author} {\bibfnamefont {M.}~\bibnamefont {Strickland}},\
  }\bibfield  {title} {\bibinfo {title} {{Quarkonia in the Quark Gluon
  Plasma}},\ }\href {https://doi.org/10.1142/S0217751X13400125} {\bibfield
  {journal} {\bibinfo  {journal} {Int. J. Mod. Phys.}\ }\textbf {\bibinfo
  {volume} {A28}},\ \bibinfo {pages} {1340012} (\bibinfo {year} {2013})},\
  \Eprint {https://arxiv.org/abs/1302.2180} {arXiv:1302.2180 [hep-ph]}
  \BibitemShut {NoStop}%
%%CITATION = ARXIV:1302.2180;%%
\bibitem [{\citenamefont {Bazavov}\ \emph
  {et~al.}(2010{\natexlab{a}})\citenamefont {Bazavov}, \citenamefont
  {Petreczky},\ and\ \citenamefont {Velytsky}}]{Bazavov:2009us}%
  \BibitemOpen
  \bibfield  {author} {\bibinfo {author} {\bibfnamefont {A.}~\bibnamefont
  {Bazavov}}, \bibinfo {author} {\bibfnamefont {P.}~\bibnamefont {Petreczky}},\
  and\ \bibinfo {author} {\bibfnamefont {A.}~\bibnamefont {Velytsky}},\
  }\bibinfo {title} {{Quarkonium at Finite Temperature}},\ in\ \href
  {https://doi.org/10.1142/9789814293297_0002} {\emph {\bibinfo {booktitle}
  {{Quark-gluon plasma 4}}}},\ \bibinfo {editor} {edited by\ \bibinfo {editor}
  {\bibfnamefont {R.~C.}\ \bibnamefont {Hwa}}\ and\ \bibinfo {editor}
  {\bibfnamefont {X.-N.}\ \bibnamefont {Wang}}}\ (\bibinfo {year} {2010})\ pp.\
  \bibinfo {pages} {61--110},\ \Eprint {https://arxiv.org/abs/0904.1748}
  {arXiv:0904.1748 [hep-ph]} \BibitemShut {NoStop}%
\bibitem [{\citenamefont {Matsui}\ and\ \citenamefont
  {Satz}(1986)}]{Matsui:1986dk}%
  \BibitemOpen
  \bibfield  {author} {\bibinfo {author} {\bibfnamefont {T.}~\bibnamefont
  {Matsui}}\ and\ \bibinfo {author} {\bibfnamefont {H.}~\bibnamefont {Satz}},\
  }\bibfield  {title} {\bibinfo {title} {{$J/\psi$ Suppression by Quark-Gluon
  Plasma Formation}},\ }\href {https://doi.org/10.1016/0370-2693(86)91404-8}
  {\bibfield  {journal} {\bibinfo  {journal} {Phys. Lett.}\ }\textbf {\bibinfo
  {volume} {B178}},\ \bibinfo {pages} {416} (\bibinfo {year}
  {1986})}\BibitemShut {NoStop}%
%%CITATION = PHLTA,B178,416;%%
\bibitem [{\citenamefont {Bazavov}\ and\ \citenamefont
  {Weber}(2021)}]{Bazavov:2020teh}%
  \BibitemOpen
  \bibfield  {author} {\bibinfo {author} {\bibfnamefont {A.}~\bibnamefont
  {Bazavov}}\ and\ \bibinfo {author} {\bibfnamefont {J.~H.}\ \bibnamefont
  {Weber}},\ }\bibfield  {title} {\bibinfo {title} {{Color Screening in Quantum
  Chromodynamics}},\ }\href {https://doi.org/10.1016/j.ppnp.2020.103823}
  {\bibfield  {journal} {\bibinfo  {journal} {Prog. Part. Nucl. Phys.}\
  }\textbf {\bibinfo {volume} {116}},\ \bibinfo {pages} {103823} (\bibinfo
  {year} {2021})},\ \Eprint {https://arxiv.org/abs/2010.01873}
  {arXiv:2010.01873 [hep-lat]} \BibitemShut {NoStop}%
\bibitem [{\citenamefont {Bazavov}\ \emph
  {et~al.}(2018{\natexlab{a}})\citenamefont {Bazavov}, \citenamefont
  {Brambilla}, \citenamefont {Petreczky}, \citenamefont {Vairo},\ and\
  \citenamefont {Weber}}]{Bazavov:2018wmo}%
  \BibitemOpen
  \bibfield  {author} {\bibinfo {author} {\bibfnamefont {A.}~\bibnamefont
  {Bazavov}}, \bibinfo {author} {\bibfnamefont {N.}~\bibnamefont {Brambilla}},
  \bibinfo {author} {\bibfnamefont {P.}~\bibnamefont {Petreczky}}, \bibinfo
  {author} {\bibfnamefont {A.}~\bibnamefont {Vairo}},\ and\ \bibinfo {author}
  {\bibfnamefont {J.~H.}\ \bibnamefont {Weber}} (\bibinfo {collaboration}
  {TUMQCD}),\ }\bibfield  {title} {\bibinfo {title} {{Color screening in
  (2+1)-flavor QCD}},\ }\href {https://doi.org/10.1103/PhysRevD.98.054511}
  {\bibfield  {journal} {\bibinfo  {journal} {Phys. Rev. D}\ }\textbf {\bibinfo
  {volume} {98}},\ \bibinfo {pages} {054511} (\bibinfo {year}
  {2018}{\natexlab{a}})},\ \Eprint {https://arxiv.org/abs/1804.10600}
  {arXiv:1804.10600 [hep-lat]} \BibitemShut {NoStop}%
\bibitem [{\citenamefont {Brambilla}\ \emph {et~al.}(2004)\citenamefont
  {Brambilla} \emph {et~al.}}]{Brambilla:2004wf}%
  \BibitemOpen
  \bibfield  {author} {\bibinfo {author} {\bibfnamefont {N.}~\bibnamefont
  {Brambilla}} \emph {et~al.} (\bibinfo {collaboration} {Quarkonium Working
  Group}),\ }\bibfield  {title} {\bibinfo {title} {{Heavy quarkonium physics}}\
  }\href {https://doi.org/10.5170/CERN-2005-005} {10.5170/CERN-2005-005}
  (\bibinfo {year} {2004}),\ \Eprint {https://arxiv.org/abs/hep-ph/0412158}
  {arXiv:hep-ph/0412158} \BibitemShut {NoStop}%
\bibitem [{\citenamefont {Kawanai}\ and\ \citenamefont
  {Sasaki}(2011)}]{Kawanai:2011xb}%
  \BibitemOpen
  \bibfield  {author} {\bibinfo {author} {\bibfnamefont {T.}~\bibnamefont
  {Kawanai}}\ and\ \bibinfo {author} {\bibfnamefont {S.}~\bibnamefont
  {Sasaki}},\ }\bibfield  {title} {\bibinfo {title} {{Interquark potential with
  finite quark mass from lattice QCD}},\ }\href
  {https://doi.org/10.1103/PhysRevLett.107.091601} {\bibfield  {journal}
  {\bibinfo  {journal} {Phys. Rev. Lett.}\ }\textbf {\bibinfo {volume} {107}},\
  \bibinfo {pages} {091601} (\bibinfo {year} {2011})},\ \Eprint
  {https://arxiv.org/abs/1102.3246} {arXiv:1102.3246 [hep-lat]} \BibitemShut
  {NoStop}%
\bibitem [{\citenamefont {Kawanai}\ and\ \citenamefont
  {Sasaki}(2012)}]{Kawanai:2011jt}%
  \BibitemOpen
  \bibfield  {author} {\bibinfo {author} {\bibfnamefont {T.}~\bibnamefont
  {Kawanai}}\ and\ \bibinfo {author} {\bibfnamefont {S.}~\bibnamefont
  {Sasaki}},\ }\bibfield  {title} {\bibinfo {title} {{Charmonium potential from
  full lattice QCD}},\ }\href {https://doi.org/10.1103/PhysRevD.85.091503}
  {\bibfield  {journal} {\bibinfo  {journal} {Phys. Rev. D}\ }\textbf {\bibinfo
  {volume} {85}},\ \bibinfo {pages} {091503} (\bibinfo {year} {2012})},\
  \Eprint {https://arxiv.org/abs/1110.0888} {arXiv:1110.0888 [hep-lat]}
  \BibitemShut {NoStop}%
\bibitem [{\citenamefont {Kawanai}\ and\ \citenamefont
  {Sasaki}(2014)}]{Kawanai:2013aca}%
  \BibitemOpen
  \bibfield  {author} {\bibinfo {author} {\bibfnamefont {T.}~\bibnamefont
  {Kawanai}}\ and\ \bibinfo {author} {\bibfnamefont {S.}~\bibnamefont
  {Sasaki}},\ }\bibfield  {title} {\bibinfo {title} {{Heavy quarkonium
  potential from Bethe-Salpeter wave function on the lattice}},\ }\href
  {https://doi.org/10.1103/PhysRevD.89.054507} {\bibfield  {journal} {\bibinfo
  {journal} {Phys. Rev. D}\ }\textbf {\bibinfo {volume} {89}},\ \bibinfo
  {pages} {054507} (\bibinfo {year} {2014})},\ \Eprint
  {https://arxiv.org/abs/1311.1253} {arXiv:1311.1253 [hep-lat]} \BibitemShut
  {NoStop}%
\bibitem [{\citenamefont {Nochi}\ \emph {et~al.}(2016)\citenamefont {Nochi},
  \citenamefont {Kawanai},\ and\ \citenamefont {Sasaki}}]{Nochi:2016wqg}%
  \BibitemOpen
  \bibfield  {author} {\bibinfo {author} {\bibfnamefont {K.}~\bibnamefont
  {Nochi}}, \bibinfo {author} {\bibfnamefont {T.}~\bibnamefont {Kawanai}},\
  and\ \bibinfo {author} {\bibfnamefont {S.}~\bibnamefont {Sasaki}},\
  }\bibfield  {title} {\bibinfo {title} {{Bethe-Salpeter wave functions of
  $\eta_c(2S)$ and $\psi(2S)$ states from full lattice QCD}},\ }\href
  {https://doi.org/10.1103/PhysRevD.94.114514} {\bibfield  {journal} {\bibinfo
  {journal} {Phys. Rev. D}\ }\textbf {\bibinfo {volume} {94}},\ \bibinfo
  {pages} {114514} (\bibinfo {year} {2016})},\ \Eprint
  {https://arxiv.org/abs/1608.02340} {arXiv:1608.02340 [hep-lat]} \BibitemShut
  {NoStop}%
\bibitem [{\citenamefont {Larsen}\ \emph
  {et~al.}(2020{\natexlab{a}})\citenamefont {Larsen}, \citenamefont {Meinel},
  \citenamefont {Mukherjee},\ and\ \citenamefont {Petreczky}}]{Larsen:2020rjk}%
  \BibitemOpen
  \bibfield  {author} {\bibinfo {author} {\bibfnamefont {R.}~\bibnamefont
  {Larsen}}, \bibinfo {author} {\bibfnamefont {S.}~\bibnamefont {Meinel}},
  \bibinfo {author} {\bibfnamefont {S.}~\bibnamefont {Mukherjee}},\ and\
  \bibinfo {author} {\bibfnamefont {P.}~\bibnamefont {Petreczky}},\ }\bibfield
  {title} {\bibinfo {title} {{Bethe-Salpeter amplitudes of Upsilons}},\ }\href
  {https://doi.org/10.1103/PhysRevD.102.114508} {\bibfield  {journal} {\bibinfo
   {journal} {Phys. Rev. D}\ }\textbf {\bibinfo {volume} {102}},\ \bibinfo
  {pages} {114508} (\bibinfo {year} {2020}{\natexlab{a}})},\ \Eprint
  {https://arxiv.org/abs/2008.00100} {arXiv:2008.00100 [hep-lat]} \BibitemShut
  {NoStop}%
\bibitem [{\citenamefont {Caswell}\ and\ \citenamefont
  {Lepage}(1986)}]{Caswell:1985ui}%
  \BibitemOpen
  \bibfield  {author} {\bibinfo {author} {\bibfnamefont {W.}~\bibnamefont
  {Caswell}}\ and\ \bibinfo {author} {\bibfnamefont {G.}~\bibnamefont
  {Lepage}},\ }\bibfield  {title} {\bibinfo {title} {{Effective Lagrangians for
  Bound State Problems in QED, QCD, and Other Field Theories}},\ }\href
  {https://doi.org/10.1016/0370-2693(86)91297-9} {\bibfield  {journal}
  {\bibinfo  {journal} {Phys. Lett. B}\ }\textbf {\bibinfo {volume} {167}},\
  \bibinfo {pages} {437} (\bibinfo {year} {1986})}\BibitemShut {NoStop}%
\bibitem [{\citenamefont {Brambilla}\ \emph {et~al.}(2000)\citenamefont
  {Brambilla}, \citenamefont {Pineda}, \citenamefont {Soto},\ and\
  \citenamefont {Vairo}}]{Brambilla:1999xf}%
  \BibitemOpen
  \bibfield  {author} {\bibinfo {author} {\bibfnamefont {N.}~\bibnamefont
  {Brambilla}}, \bibinfo {author} {\bibfnamefont {A.}~\bibnamefont {Pineda}},
  \bibinfo {author} {\bibfnamefont {J.}~\bibnamefont {Soto}},\ and\ \bibinfo
  {author} {\bibfnamefont {A.}~\bibnamefont {Vairo}},\ }\bibfield  {title}
  {\bibinfo {title} {{Potential NRQCD: An Effective theory for heavy
  quarkonium}},\ }\href {https://doi.org/10.1016/S0550-3213(99)00693-8}
  {\bibfield  {journal} {\bibinfo  {journal} {Nucl. Phys. B}\ }\textbf
  {\bibinfo {volume} {566}},\ \bibinfo {pages} {275} (\bibinfo {year}
  {2000})},\ \Eprint {https://arxiv.org/abs/hep-ph/9907240}
  {arXiv:hep-ph/9907240} \BibitemShut {NoStop}%
\bibitem [{\citenamefont {Laine}\ \emph {et~al.}(2007)\citenamefont {Laine},
  \citenamefont {Philipsen}, \citenamefont {Romatschke},\ and\ \citenamefont
  {Tassler}}]{Laine:2006ns}%
  \BibitemOpen
  \bibfield  {author} {\bibinfo {author} {\bibfnamefont {M.}~\bibnamefont
  {Laine}}, \bibinfo {author} {\bibfnamefont {O.}~\bibnamefont {Philipsen}},
  \bibinfo {author} {\bibfnamefont {P.}~\bibnamefont {Romatschke}},\ and\
  \bibinfo {author} {\bibfnamefont {M.}~\bibnamefont {Tassler}},\ }\bibfield
  {title} {\bibinfo {title} {{Real-time static potential in hot QCD}},\ }\href
  {https://doi.org/10.1088/1126-6708/2007/03/054} {\bibfield  {journal}
  {\bibinfo  {journal} {JHEP}\ }\textbf {\bibinfo {volume} {03}},\ \bibinfo
  {pages} {054}},\ \Eprint {https://arxiv.org/abs/hep-ph/0611300}
  {arXiv:hep-ph/0611300 [hep-ph]} \BibitemShut {NoStop}%
%%CITATION = HEP-PH/0611300;%%
\bibitem [{\citenamefont {Brambilla}\ \emph {et~al.}(2008)\citenamefont
  {Brambilla}, \citenamefont {Ghiglieri}, \citenamefont {Vairo},\ and\
  \citenamefont {Petreczky}}]{Brambilla:2008cx}%
  \BibitemOpen
  \bibfield  {author} {\bibinfo {author} {\bibfnamefont {N.}~\bibnamefont
  {Brambilla}}, \bibinfo {author} {\bibfnamefont {J.}~\bibnamefont
  {Ghiglieri}}, \bibinfo {author} {\bibfnamefont {A.}~\bibnamefont {Vairo}},\
  and\ \bibinfo {author} {\bibfnamefont {P.}~\bibnamefont {Petreczky}},\
  }\bibfield  {title} {\bibinfo {title} {{Static quark-antiquark pairs at
  finite temperature}},\ }\href {https://doi.org/10.1103/PhysRevD.78.014017}
  {\bibfield  {journal} {\bibinfo  {journal} {Phys. Rev.}\ }\textbf {\bibinfo
  {volume} {D78}},\ \bibinfo {pages} {014017} (\bibinfo {year} {2008})},\
  \Eprint {https://arxiv.org/abs/0804.0993} {arXiv:0804.0993 [hep-ph]}
  \BibitemShut {NoStop}%
%%CITATION = ARXIV:0804.0993;%%
\bibitem [{\citenamefont {Beraudo}\ \emph {et~al.}(2008)\citenamefont
  {Beraudo}, \citenamefont {Blaizot},\ and\ \citenamefont
  {Ratti}}]{Beraudo:2007ky}%
  \BibitemOpen
  \bibfield  {author} {\bibinfo {author} {\bibfnamefont {A.}~\bibnamefont
  {Beraudo}}, \bibinfo {author} {\bibfnamefont {J.~P.}\ \bibnamefont
  {Blaizot}},\ and\ \bibinfo {author} {\bibfnamefont {C.}~\bibnamefont
  {Ratti}},\ }\bibfield  {title} {\bibinfo {title} {{Real and imaginary-time Q
  anti-Q correlators in a thermal medium}},\ }\href
  {https://doi.org/10.1016/j.nuclphysa.2008.03.001} {\bibfield  {journal}
  {\bibinfo  {journal} {Nucl. Phys. A}\ }\textbf {\bibinfo {volume} {806}},\
  \bibinfo {pages} {312} (\bibinfo {year} {2008})},\ \Eprint
  {https://arxiv.org/abs/0712.4394} {arXiv:0712.4394 [nucl-th]} \BibitemShut
  {NoStop}%
\bibitem [{\citenamefont {Burnier}\ and\ \citenamefont
  {Rothkopf}(2013{\natexlab{a}})}]{Burnier:2013nla}%
  \BibitemOpen
  \bibfield  {author} {\bibinfo {author} {\bibfnamefont {Y.}~\bibnamefont
  {Burnier}}\ and\ \bibinfo {author} {\bibfnamefont {A.}~\bibnamefont
  {Rothkopf}},\ }\bibfield  {title} {\bibinfo {title} {{Bayesian Approach to
  Spectral Function Reconstruction for Euclidean Quantum Field Theories}},\
  }\href {https://doi.org/10.1103/PhysRevLett.111.182003} {\bibfield  {journal}
  {\bibinfo  {journal} {Phys. Rev. Lett.}\ }\textbf {\bibinfo {volume} {111}},\
  \bibinfo {pages} {182003} (\bibinfo {year} {2013}{\natexlab{a}})},\ \Eprint
  {https://arxiv.org/abs/1307.6106} {arXiv:1307.6106 [hep-lat]} \BibitemShut
  {NoStop}%
\bibitem [{\citenamefont {Rothkopf}\ \emph {et~al.}(2012)\citenamefont
  {Rothkopf}, \citenamefont {Hatsuda},\ and\ \citenamefont
  {Sasaki}}]{Rothkopf:2011db}%
  \BibitemOpen
  \bibfield  {author} {\bibinfo {author} {\bibfnamefont {A.}~\bibnamefont
  {Rothkopf}}, \bibinfo {author} {\bibfnamefont {T.}~\bibnamefont {Hatsuda}},\
  and\ \bibinfo {author} {\bibfnamefont {S.}~\bibnamefont {Sasaki}},\
  }\bibfield  {title} {\bibinfo {title} {{Complex Heavy-Quark Potential at
  Finite Temperature from Lattice QCD}},\ }\href
  {https://doi.org/10.1103/PhysRevLett.108.162001} {\bibfield  {journal}
  {\bibinfo  {journal} {Phys. Rev. Lett.}\ }\textbf {\bibinfo {volume} {108}},\
  \bibinfo {pages} {162001} (\bibinfo {year} {2012})},\ \Eprint
  {https://arxiv.org/abs/1108.1579} {arXiv:1108.1579 [hep-lat]} \BibitemShut
  {NoStop}%
%%CITATION = ARXIV:1108.1579;%%
\bibitem [{\citenamefont {Burnier}\ and\ \citenamefont
  {Rothkopf}(2017)}]{Burnier:2016mxc}%
  \BibitemOpen
  \bibfield  {author} {\bibinfo {author} {\bibfnamefont {Y.}~\bibnamefont
  {Burnier}}\ and\ \bibinfo {author} {\bibfnamefont {A.}~\bibnamefont
  {Rothkopf}},\ }\bibfield  {title} {\bibinfo {title} {{Complex heavy-quark
  potential and Debye mass in a gluonic medium from lattice QCD}},\ }\href
  {https://doi.org/10.1103/PhysRevD.95.054511} {\bibfield  {journal} {\bibinfo
  {journal} {Phys. Rev. D}\ }\textbf {\bibinfo {volume} {95}},\ \bibinfo
  {pages} {054511} (\bibinfo {year} {2017})},\ \Eprint
  {https://arxiv.org/abs/1607.04049} {arXiv:1607.04049 [hep-lat]} \BibitemShut
  {NoStop}%
\bibitem [{\citenamefont {Burnier}\ \emph
  {et~al.}(2015{\natexlab{a}})\citenamefont {Burnier}, \citenamefont
  {Kaczmarek},\ and\ \citenamefont {Rothkopf}}]{Burnier:2014ssa}%
  \BibitemOpen
  \bibfield  {author} {\bibinfo {author} {\bibfnamefont {Y.}~\bibnamefont
  {Burnier}}, \bibinfo {author} {\bibfnamefont {O.}~\bibnamefont {Kaczmarek}},\
  and\ \bibinfo {author} {\bibfnamefont {A.}~\bibnamefont {Rothkopf}},\
  }\bibfield  {title} {\bibinfo {title} {{Static quark-antiquark potential in
  the quark-gluon plasma from lattice QCD}},\ }\href
  {https://doi.org/10.1103/PhysRevLett.114.082001} {\bibfield  {journal}
  {\bibinfo  {journal} {Phys. Rev. Lett.}\ }\textbf {\bibinfo {volume} {114}},\
  \bibinfo {pages} {082001} (\bibinfo {year} {2015}{\natexlab{a}})},\ \Eprint
  {https://arxiv.org/abs/1410.2546} {arXiv:1410.2546 [hep-lat]} \BibitemShut
  {NoStop}%
\bibitem [{\citenamefont {Burnier}\ \emph
  {et~al.}(2015{\natexlab{b}})\citenamefont {Burnier}, \citenamefont
  {Kaczmarek},\ and\ \citenamefont {Rothkopf}}]{Burnier:2015tda}%
  \BibitemOpen
  \bibfield  {author} {\bibinfo {author} {\bibfnamefont {Y.}~\bibnamefont
  {Burnier}}, \bibinfo {author} {\bibfnamefont {O.}~\bibnamefont {Kaczmarek}},\
  and\ \bibinfo {author} {\bibfnamefont {A.}~\bibnamefont {Rothkopf}},\
  }\bibfield  {title} {\bibinfo {title} {{Quarkonium at finite temperature:
  Towards realistic phenomenology from first principles}},\ }\href
  {https://doi.org/10.1007/JHEP12(2015)101} {\bibfield  {journal} {\bibinfo
  {journal} {JHEP}\ }\textbf {\bibinfo {volume} {12}},\ \bibinfo {pages}
  {101}},\ \Eprint {https://arxiv.org/abs/1509.07366} {arXiv:1509.07366
  [hep-ph]} \BibitemShut {NoStop}%
%%CITATION = ARXIV:1509.07366;%%
\bibitem [{\citenamefont {Bala}\ and\ \citenamefont
  {Datta}(2020)}]{Bala:2019cqu}%
  \BibitemOpen
  \bibfield  {author} {\bibinfo {author} {\bibfnamefont {D.}~\bibnamefont
  {Bala}}\ and\ \bibinfo {author} {\bibfnamefont {S.}~\bibnamefont {Datta}},\
  }\bibfield  {title} {\bibinfo {title} {{Nonperturbative potential for the
  study of quarkonia in QGP}},\ }\href
  {https://doi.org/10.1103/PhysRevD.101.034507} {\bibfield  {journal} {\bibinfo
   {journal} {Phys. Rev. D}\ }\textbf {\bibinfo {volume} {101}},\ \bibinfo
  {pages} {034507} (\bibinfo {year} {2020})},\ \Eprint
  {https://arxiv.org/abs/1909.10548} {arXiv:1909.10548 [hep-lat]} \BibitemShut
  {NoStop}%
\bibitem [{\citenamefont {Bazavov}\ \emph
  {et~al.}(2014{\natexlab{a}})\citenamefont {Bazavov}, \citenamefont
  {Burnier},\ and\ \citenamefont {Petreczky}}]{Bazavov:2014kva}%
  \BibitemOpen
  \bibfield  {author} {\bibinfo {author} {\bibfnamefont {A.}~\bibnamefont
  {Bazavov}}, \bibinfo {author} {\bibfnamefont {Y.}~\bibnamefont {Burnier}},\
  and\ \bibinfo {author} {\bibfnamefont {P.}~\bibnamefont {Petreczky}},\
  }\bibfield  {title} {\bibinfo {title} {{Lattice calculation of the heavy
  quark potential at non-zero temperature}},\ }\href
  {https://doi.org/10.1016/j.nuclphysa.2014.09.078} {\bibfield  {journal}
  {\bibinfo  {journal} {Nucl. Phys.}\ }\textbf {\bibinfo {volume} {A932}},\
  \bibinfo {pages} {117} (\bibinfo {year} {2014}{\natexlab{a}})},\ \Eprint
  {https://arxiv.org/abs/1404.4267} {arXiv:1404.4267 [hep-lat]} \BibitemShut
  {NoStop}%
%%CITATION = ARXIV:1404.4267;%%
\bibitem [{\citenamefont {Petreczky}\ and\ \citenamefont
  {Weber}(2017)}]{Petreczky:2017aiz}%
  \BibitemOpen
  \bibfield  {author} {\bibinfo {author} {\bibfnamefont {P.}~\bibnamefont
  {Petreczky}}\ and\ \bibinfo {author} {\bibfnamefont {J.}~\bibnamefont
  {Weber}} (\bibinfo {collaboration} {TUMQCD}),\ }\bibfield  {title} {\bibinfo
  {title} {{Lattice Calculations of Heavy Quark Potential at Finite
  Temperature}},\ }\href {https://doi.org/10.1016/j.nuclphysa.2017.04.011}
  {\bibfield  {journal} {\bibinfo  {journal} {Nucl. Phys. A}\ }\textbf
  {\bibinfo {volume} {967}},\ \bibinfo {pages} {592} (\bibinfo {year}
  {2017})},\ \Eprint {https://arxiv.org/abs/1704.08573} {arXiv:1704.08573
  [hep-lat]} \BibitemShut {NoStop}%
\bibitem [{\citenamefont {Brambilla}\ \emph {et~al.}(2017)\citenamefont
  {Brambilla}, \citenamefont {Escobedo}, \citenamefont {Soto},\ and\
  \citenamefont {Vairo}}]{Brambilla:2016wgg}%
  \BibitemOpen
  \bibfield  {author} {\bibinfo {author} {\bibfnamefont {N.}~\bibnamefont
  {Brambilla}}, \bibinfo {author} {\bibfnamefont {M.~A.}\ \bibnamefont
  {Escobedo}}, \bibinfo {author} {\bibfnamefont {J.}~\bibnamefont {Soto}},\
  and\ \bibinfo {author} {\bibfnamefont {A.}~\bibnamefont {Vairo}},\ }\bibfield
   {title} {\bibinfo {title} {{Quarkonium suppression in heavy-ion collisions:
  an open quantum system approach}},\ }\href
  {https://doi.org/10.1103/PhysRevD.96.034021} {\bibfield  {journal} {\bibinfo
  {journal} {Phys. Rev. D}\ }\textbf {\bibinfo {volume} {96}},\ \bibinfo
  {pages} {034021} (\bibinfo {year} {2017})},\ \Eprint
  {https://arxiv.org/abs/1612.07248} {arXiv:1612.07248 [hep-ph]} \BibitemShut
  {NoStop}%
\bibitem [{\citenamefont {Brambilla}\ \emph {et~al.}(2018)\citenamefont
  {Brambilla}, \citenamefont {Escobedo}, \citenamefont {Soto},\ and\
  \citenamefont {Vairo}}]{Brambilla:2017zei}%
  \BibitemOpen
  \bibfield  {author} {\bibinfo {author} {\bibfnamefont {N.}~\bibnamefont
  {Brambilla}}, \bibinfo {author} {\bibfnamefont {M.~A.}\ \bibnamefont
  {Escobedo}}, \bibinfo {author} {\bibfnamefont {J.}~\bibnamefont {Soto}},\
  and\ \bibinfo {author} {\bibfnamefont {A.}~\bibnamefont {Vairo}},\ }\bibfield
   {title} {\bibinfo {title} {{Heavy quarkonium suppression in a fireball}},\
  }\href {https://doi.org/10.1103/PhysRevD.97.074009} {\bibfield  {journal}
  {\bibinfo  {journal} {Phys. Rev. D}\ }\textbf {\bibinfo {volume} {97}},\
  \bibinfo {pages} {074009} (\bibinfo {year} {2018})},\ \Eprint
  {https://arxiv.org/abs/1711.04515} {arXiv:1711.04515 [hep-ph]} \BibitemShut
  {NoStop}%
\bibitem [{\citenamefont {Brambilla}\ \emph {et~al.}(2019)\citenamefont
  {Brambilla}, \citenamefont {Escobedo}, \citenamefont {Vairo},\ and\
  \citenamefont {Vander~Griend}}]{Brambilla:2019tpt}%
  \BibitemOpen
  \bibfield  {author} {\bibinfo {author} {\bibfnamefont {N.}~\bibnamefont
  {Brambilla}}, \bibinfo {author} {\bibfnamefont {M.~A.}\ \bibnamefont
  {Escobedo}}, \bibinfo {author} {\bibfnamefont {A.}~\bibnamefont {Vairo}},\
  and\ \bibinfo {author} {\bibfnamefont {P.}~\bibnamefont {Vander~Griend}},\
  }\bibfield  {title} {\bibinfo {title} {{Transport coefficients from in medium
  quarkonium dynamics}},\ }\href {https://doi.org/10.1103/PhysRevD.100.054025}
  {\bibfield  {journal} {\bibinfo  {journal} {Phys. Rev. D}\ }\textbf {\bibinfo
  {volume} {100}},\ \bibinfo {pages} {054025} (\bibinfo {year} {2019})},\
  \Eprint {https://arxiv.org/abs/1903.08063} {arXiv:1903.08063 [hep-ph]}
  \BibitemShut {NoStop}%
\bibitem [{\citenamefont {Brambilla}\ \emph {et~al.}(2021)\citenamefont
  {Brambilla}, \citenamefont {Escobedo}, \citenamefont {Strickland},
  \citenamefont {Vairo}, \citenamefont {Vander~Griend},\ and\ \citenamefont
  {Weber}}]{Brambilla:2020qwo}%
  \BibitemOpen
  \bibfield  {author} {\bibinfo {author} {\bibfnamefont {N.}~\bibnamefont
  {Brambilla}}, \bibinfo {author} {\bibfnamefont {M.~A.}\ \bibnamefont
  {Escobedo}}, \bibinfo {author} {\bibfnamefont {M.}~\bibnamefont
  {Strickland}}, \bibinfo {author} {\bibfnamefont {A.}~\bibnamefont {Vairo}},
  \bibinfo {author} {\bibfnamefont {P.}~\bibnamefont {Vander~Griend}},\ and\
  \bibinfo {author} {\bibfnamefont {J.~H.}\ \bibnamefont {Weber}},\ }\bibfield
  {title} {\bibinfo {title} {{Bottomonium suppression in an open quantum system
  using the quantum trajectories method}},\ }\href
  {https://doi.org/10.1007/JHEP05(2021)136} {\bibfield  {journal} {\bibinfo
  {journal} {JHEP}\ }\textbf {\bibinfo {volume} {05}},\ \bibinfo {pages}
  {136}},\ \Eprint {https://arxiv.org/abs/2012.01240} {arXiv:2012.01240
  [hep-ph]} \BibitemShut {NoStop}%
\bibitem [{\citenamefont {Akamatsu}(2020)}]{Akamatsu:2020ypb}%
  \BibitemOpen
  \bibfield  {author} {\bibinfo {author} {\bibfnamefont {Y.}~\bibnamefont
  {Akamatsu}},\ }\bibfield  {title} {\bibinfo {title} {{Quarkonium in
  Quark-Gluon Plasma: Open Quantum System Approaches Re-examined}},\
  }\href@noop {} {\  (\bibinfo {year} {2020})},\ \Eprint
  {https://arxiv.org/abs/2009.10559} {arXiv:2009.10559 [nucl-th]} \BibitemShut
  {NoStop}%
\bibitem [{\citenamefont {Kajimoto}\ \emph {et~al.}(2018)\citenamefont
  {Kajimoto}, \citenamefont {Akamatsu}, \citenamefont {Asakawa},\ and\
  \citenamefont {Rothkopf}}]{Kajimoto:2017rel}%
  \BibitemOpen
  \bibfield  {author} {\bibinfo {author} {\bibfnamefont {S.}~\bibnamefont
  {Kajimoto}}, \bibinfo {author} {\bibfnamefont {Y.}~\bibnamefont {Akamatsu}},
  \bibinfo {author} {\bibfnamefont {M.}~\bibnamefont {Asakawa}},\ and\ \bibinfo
  {author} {\bibfnamefont {A.}~\bibnamefont {Rothkopf}},\ }\bibfield  {title}
  {\bibinfo {title} {{Dynamical dissociation of quarkonia by wave function
  decoherence}},\ }\href {https://doi.org/10.1103/PhysRevD.97.014003}
  {\bibfield  {journal} {\bibinfo  {journal} {Phys. Rev. D}\ }\textbf {\bibinfo
  {volume} {97}},\ \bibinfo {pages} {014003} (\bibinfo {year} {2018})},\
  \Eprint {https://arxiv.org/abs/1705.03365} {arXiv:1705.03365 [nucl-th]}
  \BibitemShut {NoStop}%
\bibitem [{\citenamefont {Miura}\ \emph {et~al.}(2020)\citenamefont {Miura},
  \citenamefont {Akamatsu}, \citenamefont {Asakawa},\ and\ \citenamefont
  {Rothkopf}}]{Miura:2019ssi}%
  \BibitemOpen
  \bibfield  {author} {\bibinfo {author} {\bibfnamefont {T.}~\bibnamefont
  {Miura}}, \bibinfo {author} {\bibfnamefont {Y.}~\bibnamefont {Akamatsu}},
  \bibinfo {author} {\bibfnamefont {M.}~\bibnamefont {Asakawa}},\ and\ \bibinfo
  {author} {\bibfnamefont {A.}~\bibnamefont {Rothkopf}},\ }\bibfield  {title}
  {\bibinfo {title} {{Quantum Brownian motion of a heavy quark pair in the
  quark-gluon plasma}},\ }\href {https://doi.org/10.1103/PhysRevD.101.034011}
  {\bibfield  {journal} {\bibinfo  {journal} {Phys. Rev. D}\ }\textbf {\bibinfo
  {volume} {101}},\ \bibinfo {pages} {034011} (\bibinfo {year} {2020})},\
  \Eprint {https://arxiv.org/abs/1908.06293} {arXiv:1908.06293 [nucl-th]}
  \BibitemShut {NoStop}%
\bibitem [{\citenamefont {Rothkopf}\ \emph {et~al.}(2009)\citenamefont
  {Rothkopf}, \citenamefont {Hatsuda},\ and\ \citenamefont
  {Sasaki}}]{Rothkopf:2009pk}%
  \BibitemOpen
  \bibfield  {author} {\bibinfo {author} {\bibfnamefont {A.}~\bibnamefont
  {Rothkopf}}, \bibinfo {author} {\bibfnamefont {T.}~\bibnamefont {Hatsuda}},\
  and\ \bibinfo {author} {\bibfnamefont {S.}~\bibnamefont {Sasaki}},\
  }\bibfield  {title} {\bibinfo {title} {{Proper heavy-quark potential from a
  spectral decomposition of the thermal Wilson loop}},\ }\href
  {https://doi.org/10.22323/1.091.0162} {\bibfield  {journal} {\bibinfo
  {journal} {PoS}\ }\textbf {\bibinfo {volume} {LAT2009}},\ \bibinfo {pages}
  {162} (\bibinfo {year} {2009})},\ \Eprint {https://arxiv.org/abs/0910.2321}
  {arXiv:0910.2321 [hep-lat]} \BibitemShut {NoStop}%
\bibitem [{\citenamefont {Jahn}\ and\ \citenamefont
  {Philipsen}(2004)}]{Jahn:2004qr}%
  \BibitemOpen
  \bibfield  {author} {\bibinfo {author} {\bibfnamefont {O.}~\bibnamefont
  {Jahn}}\ and\ \bibinfo {author} {\bibfnamefont {O.}~\bibnamefont
  {Philipsen}},\ }\bibfield  {title} {\bibinfo {title} {{The Polyakov loop and
  its relation to static quark potentials and free energies}},\ }\href
  {https://doi.org/10.1103/PhysRevD.70.074504} {\bibfield  {journal} {\bibinfo
  {journal} {Phys. Rev. D}\ }\textbf {\bibinfo {volume} {70}},\ \bibinfo
  {pages} {074504} (\bibinfo {year} {2004})},\ \Eprint
  {https://arxiv.org/abs/hep-lat/0407042} {arXiv:hep-lat/0407042} \BibitemShut
  {NoStop}%
\bibitem [{\citenamefont {Bazavov}\ \emph {et~al.}(2008)\citenamefont
  {Bazavov}, \citenamefont {Petreczky},\ and\ \citenamefont
  {Velytsky}}]{Bazavov:2008rw}%
  \BibitemOpen
  \bibfield  {author} {\bibinfo {author} {\bibfnamefont {A.}~\bibnamefont
  {Bazavov}}, \bibinfo {author} {\bibfnamefont {P.}~\bibnamefont {Petreczky}},\
  and\ \bibinfo {author} {\bibfnamefont {A.}~\bibnamefont {Velytsky}},\
  }\bibfield  {title} {\bibinfo {title} {{Static quark anti-quark pair in SU(2)
  gauge theory}},\ }\href {https://doi.org/10.1103/PhysRevD.78.114026}
  {\bibfield  {journal} {\bibinfo  {journal} {Phys. Rev. D}\ }\textbf {\bibinfo
  {volume} {78}},\ \bibinfo {pages} {114026} (\bibinfo {year} {2008})},\
  \Eprint {https://arxiv.org/abs/0809.2062} {arXiv:0809.2062 [hep-lat]}
  \BibitemShut {NoStop}%
\bibitem [{\citenamefont {Burnier}\ and\ \citenamefont
  {Rothkopf}(2013{\natexlab{b}})}]{Burnier:2013fca}%
  \BibitemOpen
  \bibfield  {author} {\bibinfo {author} {\bibfnamefont {Y.}~\bibnamefont
  {Burnier}}\ and\ \bibinfo {author} {\bibfnamefont {A.}~\bibnamefont
  {Rothkopf}},\ }\bibfield  {title} {\bibinfo {title} {{A hard thermal loop
  benchmark for the extraction of the nonperturbative $Q\bar{Q}$ potential}},\
  }\href {https://doi.org/10.1103/PhysRevD.87.114019} {\bibfield  {journal}
  {\bibinfo  {journal} {Phys. Rev.}\ }\textbf {\bibinfo {volume} {D87}},\
  \bibinfo {pages} {114019} (\bibinfo {year} {2013}{\natexlab{b}})},\ \Eprint
  {https://arxiv.org/abs/1304.4154} {arXiv:1304.4154 [hep-ph]} \BibitemShut
  {NoStop}%
%%CITATION = ARXIV:1304.4154;%%
\bibitem [{\citenamefont {Bazavov}\ \emph {et~al.}(2012)\citenamefont {Bazavov}
  \emph {et~al.}}]{Bazavov:2011nk}%
  \BibitemOpen
  \bibfield  {author} {\bibinfo {author} {\bibfnamefont {A.}~\bibnamefont
  {Bazavov}} \emph {et~al.},\ }\bibfield  {title} {\bibinfo {title} {{The
  chiral and deconfinement aspects of the QCD transition}},\ }\href
  {https://doi.org/10.1103/PhysRevD.85.054503} {\bibfield  {journal} {\bibinfo
  {journal} {Phys. Rev.}\ }\textbf {\bibinfo {volume} {D85}},\ \bibinfo {pages}
  {054503} (\bibinfo {year} {2012})},\ \Eprint
  {https://arxiv.org/abs/1111.1710} {arXiv:1111.1710 [hep-lat]} \BibitemShut
  {NoStop}%
%%CITATION = ARXIV:1111.1710;%%
\bibitem [{\citenamefont {Bazavov}\ \emph {et~al.}(2013)\citenamefont
  {Bazavov}, \citenamefont {Ding}, \citenamefont {Hegde}, \citenamefont
  {Karsch}, \citenamefont {Miao}, \citenamefont {Mukherjee}, \citenamefont
  {Petreczky}, \citenamefont {Schmidt},\ and\ \citenamefont
  {Velytsky}}]{Bazavov:2013uja}%
  \BibitemOpen
  \bibfield  {author} {\bibinfo {author} {\bibfnamefont {A.}~\bibnamefont
  {Bazavov}}, \bibinfo {author} {\bibfnamefont {H.~T.}\ \bibnamefont {Ding}},
  \bibinfo {author} {\bibfnamefont {P.}~\bibnamefont {Hegde}}, \bibinfo
  {author} {\bibfnamefont {F.}~\bibnamefont {Karsch}}, \bibinfo {author}
  {\bibfnamefont {C.}~\bibnamefont {Miao}}, \bibinfo {author} {\bibfnamefont
  {S.}~\bibnamefont {Mukherjee}}, \bibinfo {author} {\bibfnamefont
  {P.}~\bibnamefont {Petreczky}}, \bibinfo {author} {\bibfnamefont
  {C.}~\bibnamefont {Schmidt}},\ and\ \bibinfo {author} {\bibfnamefont
  {A.}~\bibnamefont {Velytsky}},\ }\bibfield  {title} {\bibinfo {title} {{Quark
  number susceptibilities at high temperatures}},\ }\href
  {https://doi.org/10.1103/PhysRevD.88.094021} {\bibfield  {journal} {\bibinfo
  {journal} {Phys. Rev. D}\ }\textbf {\bibinfo {volume} {88}},\ \bibinfo
  {pages} {094021} (\bibinfo {year} {2013})},\ \Eprint
  {https://arxiv.org/abs/1309.2317} {arXiv:1309.2317 [hep-lat]} \BibitemShut
  {NoStop}%
\bibitem [{\citenamefont {Bazavov}\ \emph
  {et~al.}(2014{\natexlab{b}})\citenamefont {Bazavov} \emph
  {et~al.}}]{Bazavov:2014pvz}%
  \BibitemOpen
  \bibfield  {author} {\bibinfo {author} {\bibfnamefont {A.}~\bibnamefont
  {Bazavov}} \emph {et~al.} (\bibinfo {collaboration} {HotQCD}),\ }\bibfield
  {title} {\bibinfo {title} {{Equation of state in ( 2+1 )-flavor QCD}},\
  }\href {https://doi.org/10.1103/PhysRevD.90.094503} {\bibfield  {journal}
  {\bibinfo  {journal} {Phys. Rev.}\ }\textbf {\bibinfo {volume} {D90}},\
  \bibinfo {pages} {094503} (\bibinfo {year} {2014}{\natexlab{b}})},\ \Eprint
  {https://arxiv.org/abs/1407.6387} {arXiv:1407.6387 [hep-lat]} \BibitemShut
  {NoStop}%
%%CITATION = ARXIV:1407.6387;%%
\bibitem [{\citenamefont {Ding}\ \emph {et~al.}(2015)\citenamefont {Ding},
  \citenamefont {Mukherjee}, \citenamefont {Ohno}, \citenamefont {Petreczky},\
  and\ \citenamefont {Schadler}}]{Ding:2015fca}%
  \BibitemOpen
  \bibfield  {author} {\bibinfo {author} {\bibfnamefont {H.~T.}\ \bibnamefont
  {Ding}}, \bibinfo {author} {\bibfnamefont {S.}~\bibnamefont {Mukherjee}},
  \bibinfo {author} {\bibfnamefont {H.}~\bibnamefont {Ohno}}, \bibinfo {author}
  {\bibfnamefont {P.}~\bibnamefont {Petreczky}},\ and\ \bibinfo {author}
  {\bibfnamefont {H.~P.}\ \bibnamefont {Schadler}},\ }\bibfield  {title}
  {\bibinfo {title} {{Diagonal and off-diagonal quark number susceptibilities
  at high temperatures}},\ }\href {https://doi.org/10.1103/PhysRevD.92.074043}
  {\bibfield  {journal} {\bibinfo  {journal} {Phys. Rev. D}\ }\textbf {\bibinfo
  {volume} {92}},\ \bibinfo {pages} {074043} (\bibinfo {year} {2015})},\
  \Eprint {https://arxiv.org/abs/1507.06637} {arXiv:1507.06637 [hep-lat]}
  \BibitemShut {NoStop}%
\bibitem [{\citenamefont {Bazavov}\ \emph
  {et~al.}(2018{\natexlab{b}})\citenamefont {Bazavov}, \citenamefont
  {Petreczky},\ and\ \citenamefont {Weber}}]{Bazavov:2017dsy}%
  \BibitemOpen
  \bibfield  {author} {\bibinfo {author} {\bibfnamefont {A.}~\bibnamefont
  {Bazavov}}, \bibinfo {author} {\bibfnamefont {P.}~\bibnamefont {Petreczky}},\
  and\ \bibinfo {author} {\bibfnamefont {J.}~\bibnamefont {Weber}},\ }\bibfield
   {title} {\bibinfo {title} {{Equation of State in 2+1 Flavor QCD at High
  Temperatures}},\ }\href {https://doi.org/10.1103/PhysRevD.97.014510}
  {\bibfield  {journal} {\bibinfo  {journal} {Phys. Rev. D}\ }\textbf {\bibinfo
  {volume} {97}},\ \bibinfo {pages} {014510} (\bibinfo {year}
  {2018}{\natexlab{b}})},\ \Eprint {https://arxiv.org/abs/1710.05024}
  {arXiv:1710.05024 [hep-lat]} \BibitemShut {NoStop}%
\bibitem [{\citenamefont {Bazavov}\ \emph {et~al.}(2019)\citenamefont
  {Bazavov}, \citenamefont {Brambilla}, \citenamefont {Garcia~i Tormo},
  \citenamefont {Petreczky}, \citenamefont {Soto}, \citenamefont {Vairo},\ and\
  \citenamefont {Weber}}]{Bazavov:2019qoo}%
  \BibitemOpen
  \bibfield  {author} {\bibinfo {author} {\bibfnamefont {A.}~\bibnamefont
  {Bazavov}}, \bibinfo {author} {\bibfnamefont {N.}~\bibnamefont {Brambilla}},
  \bibinfo {author} {\bibfnamefont {X.}~\bibnamefont {Garcia~i Tormo}},
  \bibinfo {author} {\bibfnamefont {P.}~\bibnamefont {Petreczky}}, \bibinfo
  {author} {\bibfnamefont {J.}~\bibnamefont {Soto}}, \bibinfo {author}
  {\bibfnamefont {A.}~\bibnamefont {Vairo}},\ and\ \bibinfo {author}
  {\bibfnamefont {J.~H.}\ \bibnamefont {Weber}} (\bibinfo {collaboration}
  {TUMQCD}),\ }\bibfield  {title} {\bibinfo {title} {{Determination of the QCD
  coupling from the static energy and the free energy}},\ }\href
  {https://doi.org/10.1103/PhysRevD.100.114511} {\bibfield  {journal} {\bibinfo
   {journal} {Phys. Rev. D}\ }\textbf {\bibinfo {volume} {100}},\ \bibinfo
  {pages} {114511} (\bibinfo {year} {2019})},\ \Eprint
  {https://arxiv.org/abs/1907.11747} {arXiv:1907.11747 [hep-lat]} \BibitemShut
  {NoStop}%
\bibitem [{Note1()}]{Note1}%
  \BibitemOpen
  \bibinfo {note} {While a few ensembles actually have $N_z=2N_\sigma $, this
  is irrelevant for the considerations in the following.}\BibitemShut {Stop}%
\bibitem [{Note2()}]{Note2}%
  \BibitemOpen
  \bibinfo {note} {It is usually referred to as the potential in the lattice
  literature.}\BibitemShut {Stop}%
\bibitem [{\citenamefont {Bazavov}\ \emph
  {et~al.}(2010{\natexlab{b}})\citenamefont {Bazavov} \emph
  {et~al.}}]{Bazavov:2010hj}%
  \BibitemOpen
  \bibfield  {author} {\bibinfo {author} {\bibfnamefont {A.}~\bibnamefont
  {Bazavov}} \emph {et~al.} (\bibinfo {collaboration} {MILC}),\ }\bibfield
  {title} {\bibinfo {title} {{Results for light pseudoscalar mesons}},\ }\href
  {https://doi.org/10.22323/1.105.0074} {\bibfield  {journal} {\bibinfo
  {journal} {PoS}\ }\textbf {\bibinfo {volume} {LATTICE2010}},\ \bibinfo
  {pages} {074} (\bibinfo {year} {2010}{\natexlab{b}})},\ \Eprint
  {https://arxiv.org/abs/1012.0868} {arXiv:1012.0868 [hep-lat]} \BibitemShut
  {NoStop}%
\bibitem [{\citenamefont {Sommer}(1994)}]{Sommer:1993ce}%
  \BibitemOpen
  \bibfield  {author} {\bibinfo {author} {\bibfnamefont {R.}~\bibnamefont
  {Sommer}},\ }\bibfield  {title} {\bibinfo {title} {{A New way to set the
  energy scale in lattice gauge theories and its applications to the static
  force and alpha-s in SU(2) Yang-Mills theory}},\ }\href
  {https://doi.org/10.1016/0550-3213(94)90473-1} {\bibfield  {journal}
  {\bibinfo  {journal} {Nucl. Phys. B}\ }\textbf {\bibinfo {volume} {411}},\
  \bibinfo {pages} {839} (\bibinfo {year} {1994})},\ \Eprint
  {https://arxiv.org/abs/hep-lat/9310022} {arXiv:hep-lat/9310022} \BibitemShut
  {NoStop}%
\bibitem [{\citenamefont {Bazavov}\ \emph {et~al.}(2016)\citenamefont
  {Bazavov}, \citenamefont {Brambilla}, \citenamefont {Ding}, \citenamefont
  {Petreczky}, \citenamefont {Schadler}, \citenamefont {Vairo},\ and\
  \citenamefont {Weber}}]{Bazavov:2016uvm}%
  \BibitemOpen
  \bibfield  {author} {\bibinfo {author} {\bibfnamefont {A.}~\bibnamefont
  {Bazavov}}, \bibinfo {author} {\bibfnamefont {N.}~\bibnamefont {Brambilla}},
  \bibinfo {author} {\bibfnamefont {H.~T.}\ \bibnamefont {Ding}}, \bibinfo
  {author} {\bibfnamefont {P.}~\bibnamefont {Petreczky}}, \bibinfo {author}
  {\bibfnamefont {H.~P.}\ \bibnamefont {Schadler}}, \bibinfo {author}
  {\bibfnamefont {A.}~\bibnamefont {Vairo}},\ and\ \bibinfo {author}
  {\bibfnamefont {J.~H.}\ \bibnamefont {Weber}},\ }\bibfield  {title} {\bibinfo
  {title} {{Polyakov loop in 2+1 flavor QCD from low to high temperatures}},\
  }\href {https://doi.org/10.1103/PhysRevD.93.114502} {\bibfield  {journal}
  {\bibinfo  {journal} {Phys. Rev. D}\ }\textbf {\bibinfo {volume} {93}},\
  \bibinfo {pages} {114502} (\bibinfo {year} {2016})},\ \Eprint
  {https://arxiv.org/abs/1603.06637} {arXiv:1603.06637 [hep-lat]} \BibitemShut
  {NoStop}%
\bibitem [{\citenamefont {Larsen}\ \emph {et~al.}(2019)\citenamefont {Larsen},
  \citenamefont {Meinel}, \citenamefont {Mukherjee},\ and\ \citenamefont
  {Petreczky}}]{Larsen:2019bwy}%
  \BibitemOpen
  \bibfield  {author} {\bibinfo {author} {\bibfnamefont {R.}~\bibnamefont
  {Larsen}}, \bibinfo {author} {\bibfnamefont {S.}~\bibnamefont {Meinel}},
  \bibinfo {author} {\bibfnamefont {S.}~\bibnamefont {Mukherjee}},\ and\
  \bibinfo {author} {\bibfnamefont {P.}~\bibnamefont {Petreczky}},\ }\bibfield
  {title} {\bibinfo {title} {{Thermal broadening of bottomonia: Lattice
  nonrelativistic QCD with extended operators}},\ }\href
  {https://doi.org/10.1103/PhysRevD.100.074506} {\bibfield  {journal} {\bibinfo
   {journal} {Phys. Rev. D}\ }\textbf {\bibinfo {volume} {100}},\ \bibinfo
  {pages} {074506} (\bibinfo {year} {2019})},\ \Eprint
  {https://arxiv.org/abs/1908.08437} {arXiv:1908.08437 [hep-lat]} \BibitemShut
  {NoStop}%
\bibitem [{\citenamefont {Larsen}\ \emph
  {et~al.}(2020{\natexlab{b}})\citenamefont {Larsen}, \citenamefont {Meinel},
  \citenamefont {Mukherjee},\ and\ \citenamefont {Petreczky}}]{Larsen:2019zqv}%
  \BibitemOpen
  \bibfield  {author} {\bibinfo {author} {\bibfnamefont {R.}~\bibnamefont
  {Larsen}}, \bibinfo {author} {\bibfnamefont {S.}~\bibnamefont {Meinel}},
  \bibinfo {author} {\bibfnamefont {S.}~\bibnamefont {Mukherjee}},\ and\
  \bibinfo {author} {\bibfnamefont {P.}~\bibnamefont {Petreczky}},\ }\bibfield
  {title} {\bibinfo {title} {{Excited bottomonia in quark-gluon plasma from
  lattice QCD}},\ }\href {https://doi.org/10.1016/j.physletb.2019.135119}
  {\bibfield  {journal} {\bibinfo  {journal} {Phys. Lett. B}\ }\textbf
  {\bibinfo {volume} {800}},\ \bibinfo {pages} {135119} (\bibinfo {year}
  {2020}{\natexlab{b}})},\ \Eprint {https://arxiv.org/abs/1910.07374}
  {arXiv:1910.07374 [hep-lat]} \BibitemShut {NoStop}%
\bibitem [{\citenamefont {Bernard}\ \emph {et~al.}(2000)\citenamefont
  {Bernard}, \citenamefont {Burch}, \citenamefont {Orginos}, \citenamefont
  {Toussaint}, \citenamefont {DeGrand}, \citenamefont {DeTar}, \citenamefont
  {Gottlieb}, \citenamefont {Heller}, \citenamefont {Hetrick},\ and\
  \citenamefont {Sugar}}]{Bernard:2000gd}%
  \BibitemOpen
  \bibfield  {author} {\bibinfo {author} {\bibfnamefont {C.~W.}\ \bibnamefont
  {Bernard}}, \bibinfo {author} {\bibfnamefont {T.}~\bibnamefont {Burch}},
  \bibinfo {author} {\bibfnamefont {K.}~\bibnamefont {Orginos}}, \bibinfo
  {author} {\bibfnamefont {D.}~\bibnamefont {Toussaint}}, \bibinfo {author}
  {\bibfnamefont {T.~A.}\ \bibnamefont {DeGrand}}, \bibinfo {author}
  {\bibfnamefont {C.~E.}\ \bibnamefont {DeTar}}, \bibinfo {author}
  {\bibfnamefont {S.~A.}\ \bibnamefont {Gottlieb}}, \bibinfo {author}
  {\bibfnamefont {U.~M.}\ \bibnamefont {Heller}}, \bibinfo {author}
  {\bibfnamefont {J.~E.}\ \bibnamefont {Hetrick}},\ and\ \bibinfo {author}
  {\bibfnamefont {B.}~\bibnamefont {Sugar}},\ }\bibfield  {title} {\bibinfo
  {title} {{The Static quark potential in three flavor QCD}},\ }\href
  {https://doi.org/10.1103/PhysRevD.62.034503} {\bibfield  {journal} {\bibinfo
  {journal} {Phys. Rev. D}\ }\textbf {\bibinfo {volume} {62}},\ \bibinfo
  {pages} {034503} (\bibinfo {year} {2000})},\ \Eprint
  {https://arxiv.org/abs/hep-lat/0002028} {arXiv:hep-lat/0002028} \BibitemShut
  {NoStop}%
\bibitem [{\citenamefont {Cheng}\ \emph {et~al.}(2008)\citenamefont {Cheng}
  \emph {et~al.}}]{Cheng:2007jq}%
  \BibitemOpen
  \bibfield  {author} {\bibinfo {author} {\bibfnamefont {M.}~\bibnamefont
  {Cheng}} \emph {et~al.},\ }\bibfield  {title} {\bibinfo {title} {{The QCD
  equation of state with almost physical quark masses}},\ }\href
  {https://doi.org/10.1103/PhysRevD.77.014511} {\bibfield  {journal} {\bibinfo
  {journal} {Phys. Rev. D}\ }\textbf {\bibinfo {volume} {77}},\ \bibinfo
  {pages} {014511} (\bibinfo {year} {2008})},\ \Eprint
  {https://arxiv.org/abs/0710.0354} {arXiv:0710.0354 [hep-lat]} \BibitemShut
  {NoStop}%
\bibitem [{\citenamefont {Petreczky}\ and\ \citenamefont
  {Weber}(2020)}]{Petreczky:2020tky}%
  \BibitemOpen
  \bibfield  {author} {\bibinfo {author} {\bibfnamefont {P.}~\bibnamefont
  {Petreczky}}\ and\ \bibinfo {author} {\bibfnamefont {J.~H.}\ \bibnamefont
  {Weber}},\ }\bibfield  {title} {\bibinfo {title} {{Strong coupling constant
  from moments of quarkonium correlators revisited}},\ }\href@noop {} {\
  (\bibinfo {year} {2020})},\ \Eprint {https://arxiv.org/abs/2012.06193}
  {arXiv:2012.06193 [hep-lat]} \BibitemShut {NoStop}%
\bibitem [{\citenamefont {Shi}\ \emph {et~al.}(2021)\citenamefont {Shi},
  \citenamefont {Zhou}, \citenamefont {Zhao}, \citenamefont {Mukherjee},\ and\
  \citenamefont {Zhuang}}]{Shi:2021qri}%
  \BibitemOpen
  \bibfield  {author} {\bibinfo {author} {\bibfnamefont {S.}~\bibnamefont
  {Shi}}, \bibinfo {author} {\bibfnamefont {K.}~\bibnamefont {Zhou}}, \bibinfo
  {author} {\bibfnamefont {J.}~\bibnamefont {Zhao}}, \bibinfo {author}
  {\bibfnamefont {S.}~\bibnamefont {Mukherjee}},\ and\ \bibinfo {author}
  {\bibfnamefont {P.}~\bibnamefont {Zhuang}},\ }\bibfield  {title} {\bibinfo
  {title} {{Heavy Quark Potential in QGP: DNN meets LQCD}},\ }\href@noop {} {\
  (\bibinfo {year} {2021})},\ \Eprint {https://arxiv.org/abs/2105.07862}
  {arXiv:2105.07862 [hep-ph]} \BibitemShut {NoStop}%
\bibitem [{\citenamefont {Burnier}\ \emph {et~al.}(2008)\citenamefont
  {Burnier}, \citenamefont {Laine},\ and\ \citenamefont
  {Vepsalainen}}]{Burnier:2007qm}%
  \BibitemOpen
  \bibfield  {author} {\bibinfo {author} {\bibfnamefont {Y.}~\bibnamefont
  {Burnier}}, \bibinfo {author} {\bibfnamefont {M.}~\bibnamefont {Laine}},\
  and\ \bibinfo {author} {\bibfnamefont {M.}~\bibnamefont {Vepsalainen}},\
  }\bibfield  {title} {\bibinfo {title} {{Heavy quarkonium in any channel in
  resummed hot QCD}},\ }\href {https://doi.org/10.1088/1126-6708/2008/01/043}
  {\bibfield  {journal} {\bibinfo  {journal} {JHEP}\ }\textbf {\bibinfo
  {volume} {01}},\ \bibinfo {pages} {043}},\ \Eprint
  {https://arxiv.org/abs/0711.1743} {arXiv:0711.1743 [hep-ph]} \BibitemShut
  {NoStop}%
\bibitem [{\citenamefont {Bala}\ and\ \citenamefont
  {Datta}(2021)}]{Bala:2020tdt}%
  \BibitemOpen
  \bibfield  {author} {\bibinfo {author} {\bibfnamefont {D.}~\bibnamefont
  {Bala}}\ and\ \bibinfo {author} {\bibfnamefont {S.}~\bibnamefont {Datta}},\
  }\bibfield  {title} {\bibinfo {title} {{Interaction potential between heavy
  $Q\overline{Q}$ in a color octet configuration in the QGP from a study of
  hybrid Wilson loops}},\ }\href {https://doi.org/10.1103/PhysRevD.103.014512}
  {\bibfield  {journal} {\bibinfo  {journal} {Phys. Rev. D}\ }\textbf {\bibinfo
  {volume} {103}},\ \bibinfo {pages} {014512} (\bibinfo {year} {2021})},\
  \Eprint {https://arxiv.org/abs/2009.00773} {arXiv:2009.00773 [hep-lat]}
  \BibitemShut {NoStop}%
\bibitem [{\citenamefont {Tripolt}\ \emph {et~al.}(2019)\citenamefont
  {Tripolt}, \citenamefont {Gubler}, \citenamefont {Ulybyshev},\ and\
  \citenamefont {Von~Smekal}}]{Tripolt:2018xeo}%
  \BibitemOpen
  \bibfield  {author} {\bibinfo {author} {\bibfnamefont {R.-A.}\ \bibnamefont
  {Tripolt}}, \bibinfo {author} {\bibfnamefont {P.}~\bibnamefont {Gubler}},
  \bibinfo {author} {\bibfnamefont {M.}~\bibnamefont {Ulybyshev}},\ and\
  \bibinfo {author} {\bibfnamefont {L.}~\bibnamefont {Von~Smekal}},\ }\bibfield
   {title} {\bibinfo {title} {{Numerical analytic continuation of Euclidean
  data}},\ }\href {https://doi.org/10.1016/j.cpc.2018.11.012} {\bibfield
  {journal} {\bibinfo  {journal} {Comput. Phys. Commun.}\ }\textbf {\bibinfo
  {volume} {237}},\ \bibinfo {pages} {129} (\bibinfo {year} {2019})},\ \Eprint
  {https://arxiv.org/abs/1801.10348} {arXiv:1801.10348 [hep-ph]} \BibitemShut
  {NoStop}%
\bibitem [{\citenamefont {Cyrol}\ \emph {et~al.}(2018)\citenamefont {Cyrol},
  \citenamefont {Pawlowski}, \citenamefont {Rothkopf},\ and\ \citenamefont
  {Wink}}]{Cyrol:2018xeq}%
  \BibitemOpen
  \bibfield  {author} {\bibinfo {author} {\bibfnamefont {A.~K.}\ \bibnamefont
  {Cyrol}}, \bibinfo {author} {\bibfnamefont {J.~M.}\ \bibnamefont
  {Pawlowski}}, \bibinfo {author} {\bibfnamefont {A.}~\bibnamefont
  {Rothkopf}},\ and\ \bibinfo {author} {\bibfnamefont {N.}~\bibnamefont
  {Wink}},\ }\bibfield  {title} {\bibinfo {title} {{Reconstructing the
  gluon}},\ }\href {https://doi.org/10.21468/SciPostPhys.5.6.065} {\bibfield
  {journal} {\bibinfo  {journal} {SciPost Phys.}\ }\textbf {\bibinfo {volume}
  {5}},\ \bibinfo {pages} {065} (\bibinfo {year} {2018})},\ \Eprint
  {https://arxiv.org/abs/1804.00945} {arXiv:1804.00945 [hep-ph]} \BibitemShut
  {NoStop}%
\bibitem [{\citenamefont {Burnier}\ \emph {et~al.}(2011)\citenamefont
  {Burnier}, \citenamefont {Laine},\ and\ \citenamefont
  {Mether}}]{Burnier:2011jq}%
  \BibitemOpen
  \bibfield  {author} {\bibinfo {author} {\bibfnamefont {Y.}~\bibnamefont
  {Burnier}}, \bibinfo {author} {\bibfnamefont {M.}~\bibnamefont {Laine}},\
  and\ \bibinfo {author} {\bibfnamefont {L.}~\bibnamefont {Mether}},\
  }\bibfield  {title} {\bibinfo {title} {{A Test on analytic continuation of
  thermal imaginary-time data}},\ }\href
  {https://doi.org/10.1140/epjc/s10052-011-1619-0} {\bibfield  {journal}
  {\bibinfo  {journal} {Eur. Phys. J. C}\ }\textbf {\bibinfo {volume} {71}},\
  \bibinfo {pages} {1619} (\bibinfo {year} {2011})},\ \Eprint
  {https://arxiv.org/abs/1101.5534} {arXiv:1101.5534 [hep-lat]} \BibitemShut
  {NoStop}%
\bibitem [{\citenamefont {Schlessinger}(1968)}]{Schlessinger:1968}%
  \BibitemOpen
  \bibfield  {author} {\bibinfo {author} {\bibfnamefont {L.}~\bibnamefont
  {Schlessinger}},\ }\bibfield  {title} {\bibinfo {title} {Use of analyticity
  in the calculation of nonrelativistic scattering amplitudes},\ }\href
  {https://doi.org/10.1103/PhysRev.167.1411} {\bibfield  {journal} {\bibinfo
  {journal} {Phys. Rev.}\ }\textbf {\bibinfo {volume} {167}},\ \bibinfo {pages}
  {1411} (\bibinfo {year} {1968})}\BibitemShut {NoStop}%
\bibitem [{\citenamefont {Necco}\ and\ \citenamefont
  {Sommer}(2002)}]{Necco:2001xg}%
  \BibitemOpen
  \bibfield  {author} {\bibinfo {author} {\bibfnamefont {S.}~\bibnamefont
  {Necco}}\ and\ \bibinfo {author} {\bibfnamefont {R.}~\bibnamefont {Sommer}},\
  }\bibfield  {title} {\bibinfo {title} {{The N(f) = 0 heavy quark potential
  from short to intermediate distances}},\ }\href
  {https://doi.org/10.1016/S0550-3213(01)00582-X} {\bibfield  {journal}
  {\bibinfo  {journal} {Nucl. Phys. B}\ }\textbf {\bibinfo {volume} {622}},\
  \bibinfo {pages} {328} (\bibinfo {year} {2002})},\ \Eprint
  {https://arxiv.org/abs/hep-lat/0108008} {arXiv:hep-lat/0108008} \BibitemShut
  {NoStop}%
\bibitem [{\citenamefont {Tripolt}\ \emph {et~al.}(2017)\citenamefont
  {Tripolt}, \citenamefont {Haritan}, \citenamefont {Wambach},\ and\
  \citenamefont {Moiseyev}}]{Tripolt:2016cya}%
  \BibitemOpen
  \bibfield  {author} {\bibinfo {author} {\bibfnamefont {R.-A.}\ \bibnamefont
  {Tripolt}}, \bibinfo {author} {\bibfnamefont {I.}~\bibnamefont {Haritan}},
  \bibinfo {author} {\bibfnamefont {J.}~\bibnamefont {Wambach}},\ and\ \bibinfo
  {author} {\bibfnamefont {N.}~\bibnamefont {Moiseyev}},\ }\bibfield  {title}
  {\bibinfo {title} {{Threshold energies and poles for hadron physical problems
  by a model-independent universal algorithm}},\ }\href
  {https://doi.org/10.1016/j.physletb.2017.10.001} {\bibfield  {journal}
  {\bibinfo  {journal} {Phys. Lett. B}\ }\textbf {\bibinfo {volume} {774}},\
  \bibinfo {pages} {411} (\bibinfo {year} {2017})},\ \Eprint
  {https://arxiv.org/abs/1610.03252} {arXiv:1610.03252 [hep-ph]} \BibitemShut
  {NoStop}%
\bibitem [{\citenamefont {Burnier}\ and\ \citenamefont
  {Rothkopf}(2012)}]{Burnier:2012az}%
  \BibitemOpen
  \bibfield  {author} {\bibinfo {author} {\bibfnamefont {Y.}~\bibnamefont
  {Burnier}}\ and\ \bibinfo {author} {\bibfnamefont {A.}~\bibnamefont
  {Rothkopf}},\ }\bibfield  {title} {\bibinfo {title} {{Disentangling the
  timescales behind the non-perturbative heavy quark potential}},\ }\href
  {https://doi.org/10.1103/PhysRevD.86.051503} {\bibfield  {journal} {\bibinfo
  {journal} {Phys. Rev. D}\ }\textbf {\bibinfo {volume} {86}},\ \bibinfo
  {pages} {051503} (\bibinfo {year} {2012})},\ \Eprint
  {https://arxiv.org/abs/1208.1899} {arXiv:1208.1899 [hep-ph]} \BibitemShut
  {NoStop}%
\bibitem [{\citenamefont {Petreczky}\ \emph {et~al.}(2019)\citenamefont
  {Petreczky}, \citenamefont {Rothkopf},\ and\ \citenamefont
  {Weber}}]{Petreczky:2018xuh}%
  \BibitemOpen
  \bibfield  {author} {\bibinfo {author} {\bibfnamefont {P.}~\bibnamefont
  {Petreczky}}, \bibinfo {author} {\bibfnamefont {A.}~\bibnamefont
  {Rothkopf}},\ and\ \bibinfo {author} {\bibfnamefont {J.}~\bibnamefont
  {Weber}},\ }\bibfield  {title} {\bibinfo {title} {{Realistic in-medium
  heavy-quark potential from high statistics lattice QCD simulations}},\ }\href
  {https://doi.org/10.1016/j.nuclphysa.2018.10.012} {\bibfield  {journal}
  {\bibinfo  {journal} {Nucl. Phys. A}\ }\textbf {\bibinfo {volume} {982}},\
  \bibinfo {pages} {735} (\bibinfo {year} {2019})},\ \Eprint
  {https://arxiv.org/abs/1810.02230} {arXiv:1810.02230 [hep-lat]} \BibitemShut
  {NoStop}%
\bibitem [{\citenamefont {Asakawa}\ \emph {et~al.}(2001)\citenamefont
  {Asakawa}, \citenamefont {Hatsuda},\ and\ \citenamefont
  {Nakahara}}]{Asakawa:2000tr}%
  \BibitemOpen
  \bibfield  {author} {\bibinfo {author} {\bibfnamefont {M.}~\bibnamefont
  {Asakawa}}, \bibinfo {author} {\bibfnamefont {T.}~\bibnamefont {Hatsuda}},\
  and\ \bibinfo {author} {\bibfnamefont {Y.}~\bibnamefont {Nakahara}},\
  }\bibfield  {title} {\bibinfo {title} {{Maximum entropy analysis of the
  spectral functions in lattice QCD}},\ }\href
  {https://doi.org/10.1016/S0146-6410(01)00150-8} {\bibfield  {journal}
  {\bibinfo  {journal} {Prog. Part. Nucl. Phys.}\ }\textbf {\bibinfo {volume}
  {46}},\ \bibinfo {pages} {459} (\bibinfo {year} {2001})},\ \Eprint
  {https://arxiv.org/abs/hep-lat/0011040} {arXiv:hep-lat/0011040 [hep-lat]}
  \BibitemShut {NoStop}%
%%CITATION = HEP-LAT/0011040;%%
\bibitem [{\citenamefont {Fischer}\ \emph {et~al.}(2018)\citenamefont
  {Fischer}, \citenamefont {Pawlowski}, \citenamefont {Rothkopf},\ and\
  \citenamefont {Welzbacher}}]{Fischer:2017kbq}%
  \BibitemOpen
  \bibfield  {author} {\bibinfo {author} {\bibfnamefont {C.~S.}\ \bibnamefont
  {Fischer}}, \bibinfo {author} {\bibfnamefont {J.~M.}\ \bibnamefont
  {Pawlowski}}, \bibinfo {author} {\bibfnamefont {A.}~\bibnamefont
  {Rothkopf}},\ and\ \bibinfo {author} {\bibfnamefont {C.~A.}\ \bibnamefont
  {Welzbacher}},\ }\bibfield  {title} {\bibinfo {title} {{Bayesian analysis of
  quark spectral properties from the Dyson-Schwinger equation}},\ }\href
  {https://doi.org/10.1103/PhysRevD.98.014009} {\bibfield  {journal} {\bibinfo
  {journal} {Phys. Rev. D}\ }\textbf {\bibinfo {volume} {98}},\ \bibinfo
  {pages} {014009} (\bibinfo {year} {2018})},\ \Eprint
  {https://arxiv.org/abs/1705.03207} {arXiv:1705.03207 [hep-ph]} \BibitemShut
  {NoStop}%
\bibitem [{\citenamefont {Bala~D}(2022)}]{data}%
  \BibitemOpen
  \bibfield  {author} {\bibinfo {author} {\bibfnamefont {L.~R. M. S. P. G. P.
  P. R. A. W.~J.}\ \bibnamefont {Bala~D}, \bibfnamefont {Kaczmarek~O}},\
  }\bibfield  {title} {\bibinfo {title} {{Data publication: Static quark
  anti-quark interactions at non-zero temperature from lattice QCD}},\
  }\bibfield  {journal} {\bibinfo  {journal} {Bielefeld University}\ }\href
  {https://doi.org/https://doi.org/10.4119/unibi/2961713}
  {https://doi.org/10.4119/unibi/2961713} (\bibinfo {year} {2022})\BibitemShut
  {NoStop}%
\bibitem [{\citenamefont {Mazur}(2021)}]{Mazur:2021zgi}%
  \BibitemOpen
  \bibfield  {author} {\bibinfo {author} {\bibfnamefont {L.}~\bibnamefont
  {Mazur}},\ }\emph {\bibinfo {title} {{Topological Aspects in Lattice QCD}}},\
  \href {https://doi.org/10.4119/unibi/2956493} {Ph.D. thesis},\ \bibinfo
  {school} {Bielefeld U.} (\bibinfo {year} {2021})\BibitemShut {NoStop}%
\bibitem [{\citenamefont {Altenkort}\ \emph {et~al.}(2021)\citenamefont
  {Altenkort}, \citenamefont {Bollweg}, \citenamefont {Clarke}, \citenamefont
  {Kaczmarek}, \citenamefont {Mazur}, \citenamefont {Schmidt}, \citenamefont
  {Scior},\ and\ \citenamefont {Shu}}]{Altenkort:2021fqk}%
  \BibitemOpen
  \bibfield  {author} {\bibinfo {author} {\bibfnamefont {L.}~\bibnamefont
  {Altenkort}}, \bibinfo {author} {\bibfnamefont {D.}~\bibnamefont {Bollweg}},
  \bibinfo {author} {\bibfnamefont {D.~A.}\ \bibnamefont {Clarke}}, \bibinfo
  {author} {\bibfnamefont {O.}~\bibnamefont {Kaczmarek}}, \bibinfo {author}
  {\bibfnamefont {L.}~\bibnamefont {Mazur}}, \bibinfo {author} {\bibfnamefont
  {C.}~\bibnamefont {Schmidt}}, \bibinfo {author} {\bibfnamefont
  {P.}~\bibnamefont {Scior}},\ and\ \bibinfo {author} {\bibfnamefont {H.-T.}\
  \bibnamefont {Shu}},\ }\bibfield  {title} {\bibinfo {title} {{HotQCD on
  Multi-GPU Systems}},\ }in\ \href@noop {} {\emph {\bibinfo {booktitle} {{38th
  International Symposium on Lattice Field Theory}}}}\ (\bibinfo {year}
  {2021})\ \Eprint {https://arxiv.org/abs/2111.10354} {arXiv:2111.10354
  [hep-lat]} \BibitemShut {NoStop}%
\end{thebibliography}%

\end{document}